\documentclass[12pt,a4paper,oneside]{book}
\usepackage{amsmath,amsfonts,amssymb,latexsym,graphicx,lscape,tridefs,hyperref}
\usepackage[nottoc]{tocbibind}

\def \EPSPATH{}   

\newcommand{\N}{\mathcal{N}}
\newcommand{\M}{\mathcal{M}}
\newcommand{\F}{\mathcal{F}}
\newcommand{\K}{K_\varphi{}^\varphi}
\renewcommand{\S}{\mathcal{S}}
\newcommand{\Kf}{\mathcal{K}}
\newcommand{\V}{\mathcal{V}}
\newcommand{\A}{\mathcal{A}}
\newcommand{\C}{\mathcal{C}}
\newcommand{\four}[1]{{^{(4)}#1}} 
\newcommand{\three}[1]{{^{(3)}#1}}
\newcommand{\two}[1]{{^{(2)}#1}}
\newcommand{\diff}{\textrm{d}}
\newcommand{\Lie}[1]{\mathcal{L}_{#1}}
\newcommand{\half}{\textstyle{\frac{1}{2}}}
\newcommand{\third}{\textstyle{\frac{1}{3}}}
\newcommand{\quarter}{\textstyle{\frac{1}{4}}}
\newcommand{\e}{\mathrm{e}}
\newcommand{\p}{\parallel}
\newcommand{\n}{\perp}
\renewcommand{\vec}{\mathbf}
\newcommand{\emptypage}{
  \clearpage
  \newpage 
  \thispagestyle{empty}
  $\,$ 
  \newpage
}

\begin{document}

\pagestyle{empty}

\begin{centering} 
  $\,$
  \vspace{1.5cm}

  \LARGE 
  {\bf Axisymmetric Numerical Relativity}
  
  \vspace{3cm}

  \Large 
  Oliver Rinne \\ Trinity College, Cambridge

  \vspace{3cm}

  \large 
  A dissertation submitted to the\\
  University of Cambridge \\
  for the degree of \\ 
  Doctor of Philosophy
  \vspace{3cm}
  
  13 September 2005

\end{centering}

\emptypage

\pagestyle{plain}
\pagenumbering{roman}

\section*{Preface}

This thesis is based on research carried out under the supervision 
of Dr.~John M.~Stewart at the Department of Applied Mathematics and Theoretical
Physics from November 2002.

Chapters \ref{sec:impl}, \ref{sec:211} and \ref{sec:Z211} contain 
work done in collaboration with my supervisor and published in a joint paper
\cite{RinneStewart05}.
The dynamical shift conditions in chapter \ref{sec:Z211} are a later addition
by myself.
The remaining chapters are my own work.

All computer programmes were written by myself unless otherwise stated.

\bigskip

\copyright Oliver Rinne, 2005


\emptypage

\section*{Abstract}
This thesis is concerned with formulations of the Einstein equations in
axi\-symmetric spacetimes which are suitable for numerical evolutions.
The common basis for our formulations is provided by the (2+1)+1 formalism.
General matter sources and rotational degrees of freedom are included.

A first evolution system adopts elliptic gauge conditions arising from 
maximal slicing and conformal flatness.
The numerical implementation is based on the finite-difference approach,
using a Multigrid algorithm for the elliptic equations and the
method of lines for the hyperbolic evolution equations.
Problems with both constrained and free evolution are explained
from an analytical as well as a numerical viewpoint.

The second half of the thesis is concerned with a strongly hyperbolic
first-order formulation of the axisymmetric Einstein equations.
Hyperbolicity is achieved by combining the (2+1)+1 formalism with the 
Z4 formalism. The system is supplemented with generalized harmonic 
gauge conditions. 
A careful study of the behaviour of regular axisymmetric tensor fields
enables us to cast the equations in a form that is well-behaved on the axis.

A class of exact solutions of linearized theory are used 
as a test problem in order to demonstrate the accuracy of our implementation.
We derive various outer boundary conditions of dissipative and of differential 
type based on the Newman-Penrose scalars and the constraint and gauge 
propagation systems. The stability of these boundary conditions is examined
both analytically and numerically.

The code is applied to the evolution of strong Brill waves close to the 
threshold of black hole formation.  
As a novel ingredient, a nonzero twist is included.
Adaptive mesh refinement is found to be crucial in order to resolve
the highly distorted waveforms that occur if harmonic slicing is used.


\emptypage

\section*{Acknowledgements}

I would like to express my gratitude to Dr.~John Stewart, my research 
supervisor, for his advice and encouragement throughout this project.

I would also like to thank Dr.~Nikolaos Nikiforakis of the Laboratory of
Computational Dynamics and fellow students Dr.~Anita Barnes and Joshua Horwood
for helpful discussions, and Dr.~Stuart Rankin, the Relativity Group's 
computer officer, for helping me with my computing problems.

I appreciated the hospitality of the numerical relativity groups at the 
Max Planck Institute for Gravitational Physics (Albert Einstein Institute), 
Golm, Germany, and at the University of Southampton.

Financial support from the Gates Cambridge Trust, the Engineering and
Physical Sciences Research Council and Trinity College Cambridge 
is gratefully acknowledged.

My final thanks go to my friends at Cambridge and beyond and, above all, 
to my parents, for all their sympathy and support.


\emptypage

\section*{Errata}

The following typos in the submitted version of the thesis have been corrected 
in the present version. 
I am grateful to Sergio Dain, Carsten Gundlach and John Stewart for pointing
out some of these.

\begin{itemize}
  \item equation \eqref{eq:Geroch-Einstein3}:
    the exponent of $\lambda$ has been changed from $3$ to $-3$.
  \item equation \eqref{eq:tauabconservation}:
    the right-hand side has been multiplied by $\lambda$.
  \item equation \eqref{eq:axid0X}:
    replaced $\psi_{-1}$ with $\psi^{-1}$ in the fourth line 
    and $r s_{,z}$ with $(r s_{,r} + s)$ in the first term in the
    fifth line.
  \item equations \eqref{eq:divs}--\eqref{eq:apphoreqn}:
    for consistency with the notation of chapter \ref{sec:211}, $f$ has been 
    replaced with $\lambda$.
  \item equation \eqref{eq:K3tochi2}: the exponent of $\lambda$
    in the second term on the right-hand side has been changed from $-2$
    to $-3$.
\end{itemize}

Oliver Rinne

December 2013


\emptypage

\tableofcontents
\emptypage

\listoffigures

\listoftables
\emptypage

\chapter{Introduction}
\label{sec:Intro}

\pagestyle{headings}
\pagenumbering{arabic}

\section{Numerical relativity}

Albert Einstein's 1915 theory of general relativity has radically changed
our understanding of space and time. Whereas in previous field theories the
spacetime geometry was regarded as being fixed, with the other fields evolving
on top of it, the geometry is now part of the field equations themselves and
has thus entered the dynamical arena.
Spacetime is described as a four-dimensional manifold endowed with a
Lorentzian metric, which sets the lengths and angles measured 
between spacetime events. According to general relativity, the metric is 
determined by the matter content of spacetime via the field equations, 
and in turn the motion of the matter is determined by the metric. 
Thus gravitation becomes a purely geometric concept.

Despite their elegant tensorial form, Einstein's equations turn out to be a
complicated set of coupled nonlinear second-order partial differential
equations when written out explicitly in terms of the metric. Only under
rather restrictive assumptions has one been able to find analytical solutions
to these equations, e.g., by imposing symmetries or considering weak
perturbations about a fixed background solution.

One of the most promising routes towards a deeper understanding of the 
full implications of general relativity appears to be the use of numerical 
methods to solve the field equations.
Since its first steps in the 1960s \cite{HahnLindquist64}, 
numerical relativity has sparked many new insights, 
including the discovery of critical phenomena by Choptuik \cite{Choptuik93}. 
With gravitational wave observatories such as LIGO, VIRGO,
GEO, and TAMA soon expected to be fully operating, there is today 
a strong demand for waveform templates from numerical simulations of
astrophysical scenarios such as the collision of black holes or neutron stars.

Current research focuses on two main branches, which are
increasingly moving towards each other. The general relativistic side of the
field is mainly concerned with the geometry of spacetime, studying, 
for example, vacuum black hole spacetimes, most notably the binary black 
hole problem. The astrophysical side concentrates on the general relativistic
motion of matter, e.g., stellar collapse, and tries to incorporate physically
realistic forms of matter, equations of state and interactions.
This thesis is almost entirely concerned with the first approach. 

For a comprehensive review of numerical relativity, we refer the reader to
the review article by Lehner \cite{Lehner01}.

\section{Axisymmetry}

Most of the early calculations in numerical relativity were concerned 
with spherically symmetric spacetimes. 
Because this is effectively a one-dimensional problem, it could be tackled 
with the modest computational resources available at that time.
Powered by the rapid increase in the capacity and speed of hardware, 
attention has now almost entirely turned to the case without any symmetries.

The intermediate case, axisymmetry, has not been studied to the same extent.
However, axisymmetric situations occur frequently in astrophysics and there
are many interesting problems one can study: e.g., the head-on collision 
of two black holes, rotating stars and accretion disks.
In contrast to spherical symmetry (as a consequence of Birkhoff's theorem
\cite{Birkhoff}), axisymmetric spacetimes admit gravitational waves. 
Evolving axisymmetric spacetimes is less computationally expensive than 
the case without symmetries because there are only two (rather than three) 
effective spatial dimensions. Thus many questions in numerical relativity 
can be investigated much more directly.

The main difficulty with axisymmetric spacetimes is the coordinate singularity
on the axis in the coordinate system that is adapted to the symmetry,
cylindrical polar coordinates. Many attempts to deal with this
proved unsuccessful and numerical evolutions became unstable.
There are many ways to address this problem, of which we only outline the
two most often used.

The \emph{cartoon method} of Alcubierre et al. \cite{Alcubierre01} uses
Cartesian coordinates and thereby avoids the coordinate singularity.
In one of the three spatial dimensions, the numerical grid consists
only of a few (typically three) grid points, and the axisymmetry is imposed by
appropriate boundary conditions in that direction.
This method has been used successfully in practice, 
although its stability properties are somewhat dubious 
\cite{Frauendiener02}. An interesting variant of the method that avoids 
the use of a third dimension altogether can be found in \cite{Pretorius05}.

The second approach, which we shall adopt in this thesis, uses cylindrical
polar coordinates and imposes appropriate \emph{regularity conditions} on the
variables at the axis $r = 0$ so that the equations are well-behaved there.
This is the method used by Nakamura et al.~\cite{Nakamura87} and later by 
Garfinkle and Duncan \cite{Garfinkle00} and Choptuik et al.~\cite{Choptuik03a} 
for a relatively simple formulation of the axisymmetric Einstein equations.
One of the main objectives of this thesis is to develop a systematic
regularization procedure for a rather general (and complicated) system such as
the one studied in chapter \ref{sec:Z211}. 
Our regularity conditions are based on the small-$r$ behaviour of various 
axisymmetric tensor fields derived in chapter \ref{sec:impl}.

We apply a differential geometric trick in order to reduce the number of
variables we need to evolve: the axisymmetry is essentially ``divided out''
and the Einstein equations are expressed entirely within the three-dimensional 
manifold formed by the trajectories of the Killing vector generating the
symmetry. This was invented in this context by Geroch \cite{Geroch71}
although the idea resembles the famous Kaluza-Klein reduction 
\cite{Kaluza21,Klein26}.
In contrast to previous numerical studies \cite{Choptuik03a,Garfinkle00},
we do not restrict ourselves to the case in which the Killing vector is
hypersurface-orthogonal. Thus we are able to evolve rotating spacetimes.

\section{Evolution formalisms}

To make the Einstein equations suitable for numerical treatment, one typically
introduces a foliation of spacetime into three-dimensional hypersurfaces.
The most frequently used approach is to choose the hypersurfaces to be
spacelike, which leads to the \emph{ADM} or \emph{3+1} or \emph{Cauchy
  formulation} of general relativity \cite{ADM62}. Another possibility is to
take the hypersurfaces to be null, which leads to the \emph{characteristic
  formulation} \cite{Bondi62,Sachs61a}. A third approach is known as the
\emph{conformal Einstein equations} \cite{Friedrich81}, in which one applies
the ADM approach to a larger unphysical spacetime which contains the 
physical one in a finite region. In this thesis, we will adopt the ADM 
formulation and combine it with the aforementioned dimensional reduction, 
resulting in what is known as the \emph{(2+1)+1 formalism} \cite{Maeda80}.
In this case, the slices of the foliation are two-dimensional.

In the ADM approach, the Einstein equations split into elliptic
\emph{constraint equations} within the spacelike hypersurfaces and 
\emph{hyperbolic evolution equations} governing the time evolution normal to
the hypersurfaces. The constraints are conserved by the evolution equations.
In addition, certain \emph{gauge variables} appear that can be freely
specified and that reflect the general covariance of general relativity --
the field equations are invariant under transformations of the spacetime
coordinates. These basic properties immediately raise two questions:
firstly, how to choose the gauge (i.e., the coordinates) 
and secondly, how to deal with the constraints during the evolution.

Since one does not normally know in advance what spacetime the initial data
one specifies on the initial hypersurface will evolve to, one would not like
to specify the gauge {\it a priori} as a fixed function of spacetime.
Rather, one would like to tie it to the dynamics so that it can
adapt itself to the solution. A first class of gauge conditions we consider
in chapter \ref{sec:hypell} are based on geometrical considerations: 
we choose the foliation such that the slices have maximal proper volume 
and their induced metric is conformally flat. This leads to 
\emph{elliptic} equations for the gauge variables. The resulting system 
is similar to the ones considered in \cite{Choptuik03a,Garfinkle00} 
but our version is more general in that it includes rotation.

There are two different ways of dealing with the constraints. One can either
solve them during the evolution to update some of the variables
(\emph{constrained evolution}) or one can evolve all the variables via the
evolution equations, leaving the constraints unsolved (\emph{free evolution}).
As we shall see, both approaches have certain advantages and disadvantages.

If one decides to adopt the free evolution approach, one can look for
\emph{strongly} (or even \emph{symmetric}) \emph{hyperbolic} formulations 
of the Einstein equations, in a sense made mathematically precise in 
section \ref{sec:hyperbolicity}.
This requires certain modifications to the ADM system, for that system is
itself only weakly hyperbolic \cite{Kidder01}.
Hyperbolic formulations of the Einstein equations have been a very active area
of research over the past few years and there exist today a plethora of
examples (see \cite{ReulaLivRev} for a recent review).

Of the many techniques for obtaining hyperbolic systems, a particularly simple
and beautiful one is the so-called \emph{Z4 extension} of the field equations
developed by Bona et al. \cite{Bona03a}. This involves adding a covariant extra
term to the Einstein equations such that the enlarged ADM system is 
automatically hyperbolic (subject to certain provisos). 
We apply this technique to the (2+1)+1 formalism to obtain a new
strongly hyperbolic formulation of the Einstein equations tailored to 
axisymmetric spacetimes. Of course, we now have to replace our elliptic gauge
conditions with hyperbolic ones, and the ones we use are a generalization of
\emph{harmonic gauge}, in which the spacetime coordinates obey the wave
equation.

In contrast to mixed hyperbolic-elliptic formulations, hyperbolic ones have
the advantage that there is a well-developed mathematical machinery for
analyzing the well-posedness of the initial boundary value problem.
This depends crucially on the boundary conditions that one imposes at the
outer boundaries of the (finite) computational domain. Obtaining stable
boundary conditions that avoid spurious reflections is another objective
of this thesis.

\section{Numerical methods and implementation}

Once one has decided on a particular evolution formalism, the next question is
how to solve the system numerically.

The first step is to discretize the spatial domain. The approach that is used
most often in numerical relativity and that we shall follow here is the 
\emph{finite difference technique} \cite{GKO}. 
Thereby the domain is covered by a discrete grid and 
the numerical approximation is represented by its values at the grid points.
Differential operators are translated into finite differences by means of
Taylor expansions. 
A different approach is based on an expansion of the numerical solution
with respect to a given set of basis functions. This leads to methods such as
\emph{finite element, spectral} and \emph{pseudo-spectral methods}.
Considerable progress for the vacuum Einstein equations has been obtained 
with the latter \cite{Scheel02,Gourgoulhon02}. 
Finite element methods have been used to
construct initial data for black hole and Brill wave spacetimes
\cite{Arnold97, Holst05}.

As mentioned in the previous section, the ADM formalism generically leads to
two different types of PDEs: hyperbolic and elliptic ones. Accordingly, their
solution requires two rather different classes of numerical methods.

The framework we use for the hyperbolic equations is the \emph{method of
  lines}, whereby the PDE is first only discretized in space, leaving the time
dependence continuous. A suitable ODE integrator is then used for the time
integration.
The method of lines combined with straightforward finite differencing 
works well for smooth solutions such as those of vacuum general relativity
considered in this thesis.
Once matter is included, e.g., a perfect fluid, one has to
deal with discontinuities such as shocks and more sophisticated methods
from computational fluid dynamics are needed (see \cite{LeVeque} for a
comprehensive introduction).

Elliptic equations are generally thought to be expensive to solve numerically
because they typically require $O(N^2)$ operations ($N$ being the number
of unknowns, i.e., grid points) as opposed to $O(N)$ operations for hyperbolic
equations. A class of elliptic methods that achieve a complexity of $O(N)$ as
well are \emph{Multigrid methods} \cite{Briggs}. This makes them the ideal 
method for numerical relativity if elliptic equations need to be solved 
at each time step of the evolution. However, as we shall see in this thesis, 
Multigrid is not suitable for the solution of certain indefinite elliptic 
equations such as one of the constraint equations (the Hamiltonian constraint).
\emph{Conjugate gradient methods} \cite{Shewchuk94} provide an alternative but
are much more computationally expensive.

Solutions of partial differential equations often exhibit a variety of
relevant length scales. For example, certain hyperbolic gauge
conditions tend to produce highly distorted slices. As a consequence, 
steep gradients and peaks appear in the metric variables that propagate 
through the numerical grid, whereas the solution is completely
smooth elsewhere.
\emph{Adaptive mesh refinement (AMR)} is a numerical technique
that addresses this problem in a computationally efficient way.
More resolution is added in regions where and when it is needed, and
discarded when it becomes obsolete. We use Berger and Oliger's
\cite{BergerOliger84} classic version of the algorithm for hyperbolic
partial differential equations.

All algorithms employed in this thesis have been implemented in C++. 
In order to manipulate the equations to be solved, we make 
extensive use of the computer algebra language REDUCE \cite{REDUCE}, 
from which we also generate C code automatically.
Gnuplot and the Data-Vault \cite{PretoriusPhD} are used for visualization.

\section{Gravitational waves and critical collapse}

The main application we consider in this thesis is the numerical evolution of 
gravitational waves. 
Axisymmetric gravitational waves (with time-symmetric initial data) are known
as \emph{Brill waves} \cite{Brill59}.

For weak perturbations of flat space, one can construct analytical solutions
of the linearized field equations. We use some of these as test problems for
our code (chapter \ref{sec:lin}).

No analytical solutions are known in the nonlinear case and one has to resort
to numerical methods. The earliest numerical study of Brill waves we know of
is the one by Eppley \cite{Eppley77, Eppley79}, which uses a similar gauge 
as the one described in section \ref{sec:ellgauge} of this thesis.
Those early experiments indicated that while weak Brill waves disperse to leave
flat space behind, sufficiently strong ones collapse to form a black hole.
Abrahams and Evans \cite{Abrahams93, Abrahams94} looked closer at the
threshold of black hole formation and found what is known as \emph{critical
  behaviour} in gravitational collapse.

Critical behaviour was first discovered by Choptuik \cite{Choptuik93}, albeit
for a very different system: the massless scalar field in spherical symmetry.
Choptuik considered a one-parameter family of asymptotically flat smooth
initial data. Let $p_\ast$ be the critical value of the parameter $p$
separating dispersal of the field and black hole formation. For slightly
supercritical evolutions, Choptuik found a \emph{scaling relation} for the
mass of the black holes formed,
\begin{equation}
  \label{eq:scaling}
  M_\mathrm{BH} \propto (p - p_\ast)^\gamma \,,
\end{equation}
where the \emph{critical exponent} $\gamma$ appeared to be independent of the
particular family of initial data chosen. Note the similarity of
\eqref{eq:scaling} with the scaling relations found in thermodynamic phase
transitions. Moreover, the critical solution $Z^\ast$ appeared
to be \emph{universal}, i.e., independent of the initial data, and
\emph{discretely self-similar} with \emph{echoing exponent} $\Delta$:
\begin{equation}
  Z^\ast(x, \tau) = Z^\ast(x, \tau + \Delta) \,, \qquad
  x \equiv r/(-t) \,, \qquad
  \tau \equiv - \ln(-t) \,.
\end{equation}
(Here $r$ is an areal radial coordinate and $t$ is proper time 
of the central observer such that $t = 0$ coincides with the
\emph{accumulation point} \cite{Choptuik93}.)
For a review of critical phenomena in gravitational collapse see 
Gundlach \cite{Gundlach99}.

Abrahams and Evans found the same type of critical behaviour as described 
above (commonly referred to as \emph{Type II}) in Brill wave collapse 
and estimated the constants $\gamma$ and $\Delta$. 
This is important because it suggests that critical behaviour is a
property of general relativity alone rather than of the specific
matter model used. It would be important to confirm their calculation,
possibly with greater precision and longer run times (only $\approx 4$ 
echos of the critical solution were tracked in \cite{Abrahams93}), 
which has not been done yet as far as I can determine. 
In addition, Abrahams and Evans have only
considered the case in which the Killing vector is hypersurface-orthogonal.
The formalism presented in this thesis does not rely on that restriction,
and it would be interesting to see how a nonzero twist might influence 
the critical behaviour.

Even if our code is not yet capable of addressing these questions
quantitatively, we have come across a variety of problems along the way
which appear to be ubiquitous in current research in numerical relativity.
This thesis documents our efforts towards a solution of those problems.

\section{Outline of the thesis}

We begin by deriving the regularity conditions that various axisymmetric
tensor fields must obey on the axis (chapter \ref{sec:impl}).
This will be used throughout the thesis in order to cast the equations to be
solved in a regular form. 
The evolution formalism that forms the basis of all later developments,
the (2+1)+1 formalism, is derived in chapter \ref{sec:211}.
The numerical methods we use are described in chapter \ref{sec:num}.
We then construct in chapter \ref{sec:hypell} a first evolution system 
that uses elliptic gauge conditions and (partially) constrained evolution. 
Some preliminary results on the simulation of Brill waves are presented.
To my knowledge, this is the first time twisting Brill waves have been evolved.

The remainder of this thesis is concerned entirely with a strongly
hyperbolic formulation of the Einstein equations for axisymmetric spacetimes.
This so-called Z(2+1)+1 system is derived in chapter \ref{sec:Z211}.
By a careful choice of variables we write the equations in a form that is
well-behaved on the axis, which is one of our main results.
Exact solutions of linearized theory are constructed in chapter \ref{sec:lin}
and are used in order to demonstrate the accuracy of our numerical 
implementation.
Chapter \ref{sec:outerbcs} is concerned with various ways of imposing boundary
conditions at the outer boundaries of the computational domain.
Their stability is analyzed both analytically and numerically.
AMR evolutions of strong Brill waves close to the critical point are
presented in chapter \ref{sec:brill}, including twist.
We conclude and give an outlook on future work in chapter \ref{sec:concl}.

\section{Notation and conventions}

The Einstein summation
convention is used for tensor indices, i.e., repeated indices are summed over.
Round (square) brackets enclosing tensor indices 
denote (anti-)symmetrization, i.e., 
$T_{(\alpha \beta)} = \half (T_{\alpha\beta} + T_{\beta\alpha})$ and
$T_{[\alpha \beta]} = \half (T_{\alpha\beta} - T_{\beta\alpha})$.
Ordinary (partial) differentiation is denoted by a comma ($,$). 
Sometimes the comma is
left out if no ambiguity arises, e.g., $f_\alpha \equiv f_{,\alpha}$ for a 
scalar $f$. 

We use a Lorentzian metric of signature $(-+++)$. 
Our curvature convention is
\begin{equation}
  R_{\alpha\beta\gamma\delta} v^\delta = 2 \nabla_{[\alpha} \nabla_{\beta]} 
  v^\gamma \,.
\end{equation}
Geometric units are chosen,
in which Newton's constant $G$ and the speed of light $c$ are equal to $1$
so that $\kappa = 8 \pi$ in Einstein's equations 
\begin{equation}
  G_{\alpha\beta} = \kappa T_{\alpha \beta} \,.
\end{equation}


\chapter{Implications of axisymmetry}
\label{sec:impl}

Many problems with axisymmetry arise from the fact that the coordinate
system adapted to the symmetry, cylindrical polar coordinates, is
singular on the axis of symmetry. As a consequence, axisymmetric
tensor fields that are regular on the axis (in a sense made precise
below) may take strange forms when expressed in those coordinates.
In this chapter, we derive the regularity conditions that various
axisymmetric tensor fields must obey on the axis.

We want to use a $(t, z, r, \varphi)$ chart adapted to the Killing
vector $\xi = \partial/\partial\varphi$.
We shall assume \emph{elementary flatness}: in a neighbourhood of the
axis we can introduce local Cartesian coordinates $x^A = (x, y)$ such
that
\begin{equation}
  x = r\cos\varphi, \quad y = r\sin\varphi\quad \Longleftrightarrow
  \quad r = \sqrt{x^2 + y^2}, \quad \varphi = \arctan\frac{y}{x} \,.
\end{equation}
Note that the Cartesian chart is regular on the axis $r = 0$, while the polar
chart is not. With respect to Cartesian coordinates, the Killing vector is
\begin{equation}
  \xi = -y \frac{\partial}{\partial x} + x \frac{\partial}{\partial y} \,.
\end{equation}
This representation is valid everywhere, while
$\xi = \partial/\partial\varphi$ is valid only for $r>0$.

We say that a tensor field $T$ is \emph{regular on axis} if its Cartesian
components have a Taylor expansion with respect to $x$ and $y$ about $x^A=0$
convergent in some neighbourhood of $r=0$.
(Throughout this chapter we are ignoring $t$ and $z$ dependencies, which
are implicit in all calculations.)
It is \emph{axisymmetric} if its Lie derivative with respect to the Killing vector 
vanishes, 
\begin{equation}
  \Lie{\xi} T = 0 \,.
\end{equation}

\section{Functions}

Let us start with an axisymmetric function $f$ that is regular on axis.
Axi\-symmetry implies that
\begin{equation}
  \label{eq:kill1}
  k \equiv \Lie{\xi} f = -yf_{,x} + xf_{,y} = 0,
\end{equation}
which is valid everywhere.
In particular, we may differentiate \eqref{eq:kill1} an arbitrary
number of times with respect to $x$ and $y$ and require all the derivatives to
vanish on axis:
\begin{eqnarray}
  0 = k_{,x} = x f_{,xy} - y f_{,xx} + f_{,y} 
  &\Rightarrow& f_{,y} \dot = 0 \, , \nonumber\\
  0 = k_{,y} = -y f_{,xy} - f_{,x} + x f_{,yy} 
  &\Rightarrow& f_{,x} \dot = 0 \, , \nonumber\\
  0 = k_{,xx} = 2 f_{,xy} - y f_{,xxx} + x f_{,xxy}
  &\Rightarrow& f_{,xy} \dot = 0 \, , \\
  0 = k_{,xy} = x f_{,xyy} - y f_{,xxy} - f_{,xx} + f_{,yy}
  &\Rightarrow& f_{,xx} \dot = f_{,yy} \equiv f_2 \, , \nonumber\\
  0 = k_{,yy} = -y f_{,xyy} - 2 f_{,xy} + x f_{,yyy} && \nonumber\\
  &\vdots& \nonumber
\end{eqnarray}
where $\dot =$ denotes equality on axis. We find that the Taylor
expansion of $f$ in a neighbourhood of the axis has the form
\begin{equation} 
  f = \sum_{n=0}^{\infty} \frac{f_{2n}}{(2n)!} (x^2+y^2)^n\,, 
\end{equation}
i.e., $f$ is an even function of $r$:
\begin{equation}
  \label{eq:regcond1}
  f = f(t, z, r^2) \,.
\end{equation}

\section{Vectors and covectors}

Next consider a vector field $u^\alpha$.
For $a = (t,z)$,  $\Lie{\xi}u^a = 0$ implies 
$\partial u^a/\partial\varphi = 0$.
This reduces to the scalar field case and we may deduce $u^a = u^a(r^2)$.
For $u^x$ and $u^y$ we have
\begin{equation}
  \label{eq:kill2}
  \frac{\partial u^x}{\partial\varphi} + u^y = 0, \quad
  \frac{\partial u^y}{\partial\varphi} - u^x = 0.
\end{equation}
The general solution for $r>0$ is
\begin{equation}
  \label{eq:kill3}
  u^x = \widehat{a}(r) \cos\varphi - \widehat{b}(r)\sin\varphi,\quad
  u^y = \widehat{a}(r) \sin\varphi + \widehat{b}(r)\cos\varphi.
\end{equation}
However in the Cartesian chart, \eqref{eq:kill2} takes the form
\begin{equation}
  \label{eq:kill5}
  -yu^x{}_{,x} + x u^x{}_{,y} + u^y = 0, \quad
  -yu^y{}_{,x} + x u^y{}_{,y} - u^x = 0.
\end{equation}
Setting $x^A = 0$ we see that $u^A = 0$ on axis. We may thus write  
$\widehat{a} = ra, \, \widehat{b} = rb$ so that \eqref{eq:kill3} becomes
\begin{equation}
  \label{eq:kill6}
  u^x = xa - yb, \quad u^y = ya + xb.
\end{equation}
We now regard $a$ and $b$ as unknown functions of $x$ and $y$ to be
determined by substituting \eqref{eq:kill6} into \eqref{eq:kill5},
differentiating the latter an arbitrary number of times, and then
solving the recurrence relations for the Taylor coefficients of $a$
and $b$.
Again we find that $a$ and $b$ are even functions of $r$.
Finally, the polar components of $u$ are obtained from \eqref{eq:kill6} as 
\begin{equation}
  u^r = \frac{\partial r}{\partial x^A} u^A = r a(r^2) \, , \quad
  u^\varphi = \frac{\partial \varphi}{\partial x^A} u^A = b(r^2) \,.
\end{equation}
Thus in the $(t, z, r, \varphi)$ chart an axisymmetric vector field
which is regular on axis must take the form
\begin{equation}
  \label{eq:regcond2}
  u^\alpha = (A, B, rC, D),
\end{equation}
where $A$, $B$, $C$ and $D$ are functions of $t$, $z$ and $r^2$.

Next consider an axisymmetric covector field $\omega_\alpha$
which is regular on axis.
For $a = (t,z)$,  $\Lie{\xi}\omega_a = 0$ implies 
$\partial \omega_a/\partial\varphi = 0$.
This reduces to the scalar field case and we may deduce 
$\omega_a = \omega_a(r^2)$.
For the other indices we find
\begin{equation}
  \label{eq:kill10}
  -y\omega_{x,x} + x\omega_{x,y} + \omega_y = 0, \quad
  -y\omega_{y,x} + x\omega_{y,y} - \omega_x = 0,
\end{equation}
which is equivalent to \eqref{eq:kill5}, interchanging $u^A$ and
$\omega_A$.
We therefore deduce the analogue of \eqref{eq:kill6} and hence, 
in polar coordinates,
\begin{equation}
  \label{eq:regcond3}
  \omega_\alpha = (A, B, rC, r^2 D),
\end{equation}
where $A$, $B$, $C$ and $D$ are functions of $t$, $z$ and $r^2$.

\section{Symmetric 2-tensors}

Finally we consider a symmetric valence 2 tensor field $M_{\alpha\beta}$ 
which is both axisymmetric and regular on axis.
For $(a, b) = (t, z)$ we have $\Lie{\xi}M_{ab}=0$ and so 
$M_{ab} = M_{ab}(r^2)$. The mixed $(aA)$ components obey the Killing 
equations
\begin{equation}
  \label{eq:kill11}
  -yM_{ax,x} + xM_{ax,y} + M_{ay} = 0, \quad
  -yM_{ay,x} + xM_{ay,y} - M_{ax} = 0.
\end{equation}
This is essentially the same as \eqref{eq:kill10} and we may deduce
$M_{ar} = r A_a(r^2)$ and $M_{a\varphi} = r^2 B_a(r^2)$.
The remaining Killing equations are
\begin{equation*}
  -yM_{xx,x} + xM_{xx,y} + 2M_{xy} = 0,\quad
  -yM_{yy,x} + xM_{yy,y} - 2M_{xy} = 0,
\end{equation*}
\begin{equation}
  -yM_{xy,x} + xM_{xy,y} + M_{yy} - M_{xx} = 0\,. 
\end{equation}
If we introduce new variables
$u = \half(M_{xx} + M_{yy})$, $v = \half(M_{xx} - M_{yy})$ and 
$w=M_{xy}$ then
\begin{equation}
  -yu_{,x} + xu_{,y} = 0
\end{equation}
which implies $u = u(r^2)$.
The remaining equations are
\begin{equation}
  \label{eq:kill12}
  -yv_{,x} + xv_{,y} + 2w=0, \quad
  -yw_{,x} + xw_{,y} - 2v=0.
\end{equation}
For $r>0$ these can be written as
\begin{equation}
  v_{,\varphi} + 2w=0, \quad w_{,\varphi} - 2w=0,
\end{equation}
so that
\begin{equation}
  v = \widehat a(r)\cos2\varphi - \widehat b(r)\sin2\varphi,\quad
  w = \widehat a(r)\sin2\varphi + \widehat b(r)\cos2\varphi,
\end{equation}
where $\widehat a$ and $\widehat b$ are arbitrary functions of $r$.
But \eqref{eq:kill12} and its first derivatives imply that
$v$, $w$ and their first derivatives vanish on axis so that we may set
$\widehat a = r^2 a$ and $\widehat b = r^2 b$ to obtain
\begin{equation}
  \label{eq:kill13}
  v = (x^2-y^2)a - 2xyb, \quad w = 2xya + (x^2-y^2)b.
\end{equation}
Substituting \eqref{eq:kill13} into \eqref{eq:kill12} gives
\begin{eqnarray}
  x^3a_{,y} - x^2y(a_{,x} + 2b_{,y}) - 
  xy^2(a_{,y} - 2b_{,x}) + y^3a_{,x} =& 0,\nonumber\\
  x^3b_{,y} + x^2y(-b_{,x} + 2a_{,y}) + 
  xy^2(-2a_{,x} - b_{,y}) + y^3b_{,x} =& 0.
\end{eqnarray}
Differentiating these many times and proceeding as in the scalar and
vector cases, we conclude that $a$ and $b$ are functions of $r^2$.
Thus
\begin{equation*}
  M_{xx} = u + (x^2-y^2)a -2xyb, \quad
  M_{yy} = u - (x^2-y^2)a + 2xyb,
\end{equation*}
\begin{equation}
  M_{xy} = 2xya + (x^2-y^2)b\,.
\end{equation}
Re-expressing these as polar components we obtain
\begin{equation}
  M_{rr} = u + r^2a, \quad M_{r\varphi} = r^3b, \quad
  M_{\varphi\varphi} = r^2(u - r^2a).
\end{equation}
Finally combining all of the results we have
\begin{equation}
  \label{eq:regcond4}
  M_{\alpha\beta} = \left( \begin{array}{cccc} 
      A & B & r D & r^2 F \\
      B & C & r E & r^2 G \\ r D & r E & H + r^2 J & r^3 K \\ 
      r^2 F & r^2 G & r^3 K & r^2 \left( H - r^2 J \right) 
  \end{array} \right) \,,
\end{equation}
where $A,B,\ldots,K$ are functions of $t$, $z$ and $r^2$.

One should remark that one could relax our definition of regularity
on axis. If we only required functions to be $C^0$, vectors and
covectors to be $C^1$ and 2-tensors to be $C^2$ in a neighbourhood of
the axis, the above analysis would still go through and we would
arrive at the same regularity conditions. 
However, the coefficients $A, B, \ldots$ in \eqref{eq:regcond2}, 
\eqref{eq:regcond3} and \eqref{eq:regcond4} would then only be continuous 
and not necessarily even functions of $r$. 
For numerical purposes of course, it is not unduly restrictive to
assume analyticity if the solutions are smooth.

Summarizing, we have derived the regularity conditions that
axisymmetric functions
\eqref{eq:regcond1}, vectors \eqref{eq:regcond2}, covectors
\eqref{eq:regcond3} and symmetric 2-tensors \eqref{eq:regcond4}
must obey in order to be regular on the axis of symmetry.
Note in particular the subtle relation between the $(rr)$ and
$(\varphi\varphi)$ components in \eqref{eq:regcond4}. 
If a numerical evolution scheme fails to preserve precisely the
indicated $r$-dependencies then the fields become irregular on axis
and instability is inevitable. 
In chapters \ref{sec:hypell} and \ref{sec:Z211}, we will use the
regularity conditions in order to cast reductions of Einstein's
equations into forms that are free of any divergencies on the axis.


\chapter{The (2+1)+1 formalism}
\label{sec:211}

Quite generally, the existence of a symmetry can be used to reduce the
dimensionality of the problem under consideration. This should be
exploited whenever possible in order not only to simplify the problem
mathematically but also to save computational resources when
attempting a numerical solution.

We shall see how in the case of axisymmetry, the Einstein equations can be 
reduced from four-dimensional spacetime $\M$ to a three-dimensional
Lorentzian manifold $\N$ (section \ref{sec:Geroch}).
This was first performed for vacuum spacetimes by Geroch \cite{Geroch71},
although the original idea goes back to the famous papers by
Kaluza \cite{Kaluza21} and Klein \cite{Klein26}. We extend the reduction to 
include general matter sources. 

The three-dimensional Lorentzian manifold $\N$ then undergoes an
ADM-like decomposition (cf. \cite{ADM62} for the standard $3+1$
version), i.e., it is foliated into level surfaces of a time function
(section \ref{sec:ADM}).
The Einstein equations split into elliptic constraint
equations to be solved within the hypersurfaces and hyperbolic 
evolution equations governing the evolution normal to the hypersurfaces,
making the problem suitable for numerical simulations.
This procedure was first applied in this context by Maeda et al. 
\cite{Maeda80} and is commonly referred to as the \emph{(2+1)+1 formalism}.

We use energy-momentum and number conservation to derive evolution 
equations for the matter variables in this formalism (section 
\ref{sec:generalmatter}).
No particular matter model is chosen at this stage.

We mainly follow the notation of Maeda et al. \cite{Maeda80} with some 
clearly stated changes. In the following, Greek indices range over
$t, r, z, \varphi$, lower-case Latin indices over $t, r, z$, 
and upper-case Latin indices over $r, z$.

\section{The Geroch decomposition}
\label{sec:Geroch}

\begin{figure}[t]
 \setlength{\unitlength}{1cm}  
 \begin{center}
  \begin{picture}(6,8)
    \normalcolor
    \multiput(2,0)(0,6){2}{\line(3,1){3}}
    \multiput(2,0)(3,1){2}{\line(0,1){6}}
    \multiput(2,1.5)(0,1.5){3}{\oval(2,1)[b]\oval(2,1)[tl]}
    \multiput(1,1.5)(0,1.5){3}{\vector(0,1){0}}
    \put(0.5,2.8){$\xi$}
    \put(4,5.5){$\N$}
  \end{picture}   
 \end{center}
 \caption[The manifold $\N$]{\label{fig:Geroch}  
   \footnotesize The manifold $\N$}
\end{figure}

Spacetime is assumed to be a four-dimensional manifold 
$(\M, g_{\alpha\beta})$ with signature $(-+++)$ and a preferred polar
coordinate chart $(t, r, z, \varphi)$.
Axi\-symmetry means that there is an everywhere spacelike Killing
vector field $\xi = \partial / \partial \varphi$ with closed orbits.
Let $\N$ be the collection of the orbits of $\xi$. We assume that $\N$
is a differentiable 3-manifold and that there is a smooth mapping from
$\M$ into $\N$ mapping a point in $\M$ to the orbit passing through it.

Geroch \cite{Geroch71} has shown that there is a one-to-one
correspondence between tensor fields $M'\,^{\alpha ...}{}_{\beta ...}$ in
$\N$ and tensor fields $M^{\alpha ...}{}_{\beta ...}$ in $\M$ that are
both orthogonal to the Killing vector,
\begin{equation} 
  M^{\alpha...}{}_{\beta...} \xi^{\beta} =  M^{\alpha...}{}_{\beta...}
  \xi_{\alpha} = \ldots = 0\,,
\end{equation}
and axisymmetric,
\begin{equation}
  \Lie{\xi} M^{\alpha...}{}_{\beta...} = 0\,.
\end{equation}
As a shorthand, tensors satisfying these conditions are said to be
\emph{in $\N$}.

Some basic tensor fields in $\N$ are
the norm of the Killing vector 
\begin{equation}
  \label{eq:norm}
  \lambda^2 = g_{\alpha\beta} \xi^\alpha \xi^\beta > 0 \, , 
\end{equation}
the (Lorentzian) metric in $\N$,
\begin{equation}
  h_{\alpha\beta} = g_{\alpha\beta} - \lambda^{-2}\xi_\alpha \xi_\beta \, , 
\end{equation}
the Levi-Civita tensor
\begin{equation}
  \epsilon_{\alpha\beta\gamma} =
  \lambda^{-1}\epsilon_{\alpha\beta\gamma\delta} \xi^\delta \, , 
\end{equation}
and the \emph{twist vector}
\begin{equation}
  \label{eq:twist}
  \omega_\alpha = \epsilon_{\alpha\beta\gamma\delta} \xi^\beta \nabla^\gamma
  \xi^\delta \, , 
\end{equation}
which encodes the rotational degrees of freedom.
Here $\nabla$ is the covariant derivative of $g_{\alpha\beta}$.
The covariant derivative $D$ associated with
$h_{\alpha\beta}$ is obtained by projecting $\nabla$ into $\N$,
\begin{equation}
  \label{eq:D3}
  D_\alpha v^\beta = h_\alpha{}^\mu h_\nu{}^\beta \nabla_\mu \left(
  h_\rho{}^\nu v^\rho \right)  \, .
\end{equation}
The Riemann tensor of $h_{\alpha\beta}$ is denoted by 
$\three{R}_{\alpha\beta\gamma\delta}$.

We wish to reduce Einstein's theory in $\M$ to a set of equations that
only involve tensor fields in $\N$.
Starting with the Ricci identity in $\N$,
\begin{equation}
  \label{eq:RicciId}
  D_{[\alpha} D_{\beta]} v_\gamma = 
  \half \three{R}_{\alpha\beta\gamma\delta}v^\delta \,,
\end{equation}
we evaluate the left-hand-side using the definition of $D$
(\ref{eq:D3}) to find
\begin{equation}
  \label{eq:Riemann4}
  ^{(3)}R_{\alpha\beta\gamma\delta} = h_{[\alpha}{}^\mu h_{\beta]}{}^\nu
  h_{[\gamma}{}^\rho h_{\delta]}{}^\sigma \left[
    ^{(4)}R_{\mu\nu\rho\sigma} + 2 \lambda^{-2} \left( Q_{\mu\nu}
  Q_{\rho\sigma} + Q_{\mu\rho} Q_{\nu\sigma} \right) \right] \, , 
\end{equation}
where the four-dimensional Ricci identity has been used to produce
the Riemann tensor $\four{R}_{\alpha\beta\gamma\delta}$ of $\M$. 
The quantity $Q_{\alpha\beta} \equiv \nabla_\alpha \xi_\beta =
\nabla_{[\alpha} \xi_{\beta]} $ can be expressed as
\begin{equation}
  \label{eq:Q}
  Q_{\alpha\beta} = \half \lambda^{-2}
  \epsilon_{\alpha\beta\gamma\delta} \xi^\gamma \omega^\delta -
  2 \lambda^{-1} \xi_{[\alpha} \lambda_{\beta]} \, .
\end{equation}
Contracting \eqref{eq:Riemann4} with $h^{\beta\delta}$ and using
\eqref{eq:Q} we obtain
\begin{equation}
  \label{eq:Ricci3}
  \three{R}_{\alpha\gamma} = \perp \four{R}_{\alpha\gamma} 
    + \lambda^{-1} D_\alpha D_\gamma \lambda + \half \lambda^{-4}
  \left( \omega_\alpha \omega_\gamma - h_{\alpha\gamma} \omega^\tau
  \omega_\tau \right) \, .
\end{equation}
Here and in the following, the symbol $\perp$ means projection of the
free indices with $h$, and an index $\xi$ denotes contraction
with $\xi$.

Some more equations are needed in order to reflect all the Einstein
equations of the original manifold $\M$. Taking the curl and divergence
of \eqref{eq:twist}, we obtain, respectively,
\begin{equation}
  \label{eq:curl}
  D_{[\alpha} \omega_{\beta]} = 
     \lambda \epsilon_{\alpha\beta\gamma} \perp \four{R}^\gamma{}_\xi
\end{equation}
and
\begin{equation}
  \label{eq:divergence}
  D_\alpha \omega^\alpha = 3 \lambda^{-1} \lambda_{,\alpha}
  \omega^{\alpha} \, , 
\end{equation}
where we have used \eqref{eq:Q} and a standard identity for Killing
vectors,
\begin{equation}
  \label{eq:KillingId}
  \nabla_\alpha \nabla_\beta \xi_\gamma =
  \four{R}_{\delta\alpha\beta\gamma} \xi^\delta \, .
\end{equation}
Finally, applying $D^2 = D^\alpha D_\alpha$ to \eqref{eq:norm}, we
obtain
\begin{equation}
  \label{eq:D2lam}
  D^2 \lambda = -\frac{1}{2} \lambda^{-3} \omega_\alpha \omega^\alpha 
  - \lambda^{-1} \four{R}_{\xi\xi} \, .
\end{equation}

Next, we include an energy-momentum tensor $T_{\alpha\beta}$, which is
decomposed into
\begin{eqnarray}
  \label{eq:Tprojections}
  \tau &\equiv& \lambda^{-2} \xi^\mu \xi^\nu T_{\mu\nu}\, , \nonumber\\
  \tau_\alpha &\equiv& \lambda^{-2} h_\alpha{}^\mu \xi^\nu
  T_{\mu\nu} \, ,  \\
  \tau_{\alpha\beta} &\equiv& h_\alpha{}^\mu h_\beta{}^\nu 
  T_{\mu\nu} \, . \nonumber
\end{eqnarray}
(The powers of $\lambda$ included in the above definitions
differ from the ones in \cite{Maeda80}. Our choice guarantees that
$\tau$ and $\tau_\alpha$ have the correct small-$r$ behaviour of
scalars \eqref{eq:regcond1} and covectors \eqref{eq:regcond3}. 
Note that $\lambda = O(r)$ near the axis.)
It follows from axisymmetry
\begin{equation}
  \Lie{\xi} T_{\alpha\beta} = 0
\end{equation}
that the fields $\tau, \tau_\alpha$ and $\tau_{\alpha\beta}$ are in $\N$.
The Einstein equations
\begin{equation}
  \label{eq:Einstein}
  \four{R}_{\alpha\beta} = \kappa \left( T_{\alpha\beta} - \half T
  g_{\alpha\beta} \right) 
\end{equation}
can be used to express the projections of the Ricci tensor in terms of
those of the energy-momentum tensor,
\begin{eqnarray}
  \label{eq:RTparts}
  \perp \four{R}_{\alpha\beta} &=& 
     \kappa \left[ \tau_{\alpha\beta} - \half h_{\alpha\beta} 
     \left( \tau + \tau_\gamma{}^\gamma \right)
     \right] \, , \nonumber\\
  \perp \four{R}_{\alpha \xi} &=& 
     \kappa \lambda^2 \tau_\alpha \,,\\
  \four{R}_{\xi\xi} &=& \half \kappa \lambda^2 
     \left( \tau - \tau_\gamma{}^\gamma \right) \, . \nonumber
\end{eqnarray}

Inserting \eqref{eq:RTparts} into (\ref{eq:Ricci3}--\ref{eq:divergence}) 
and \eqref{eq:D2lam}, we arrive at the Geroch-Einstein equations
\begin{eqnarray}
  \label{eq:Geroch-Einstein1}
  \three{R}_{ab} &=& \kappa \left[ \tau_{ab} - \half h_{ab} \left( \tau +
  \tau_c{}^c \right) \right] + \half \lambda^{-4} \left(
  \omega_a \omega_b - h_{ab} \omega^c \omega_c \right)
  \nonumber\\ && + \lambda^{-1} \lambda_{|ab} \, , \\ 
  \label{eq:Geroch-Einstein2}
  \omega_{[a|b]} &=& -\kappa \lambda^3 \epsilon_{abc} \tau^c \, , \\
  \label{eq:Geroch-Einstein3}
  (\lambda^{-3} \omega^c){}_{|c} &=& 0\, , \\
  \label{eq:Geroch-Einstein4}
  \lambda_{|a}{}^a &=& -\half \lambda^{-3} \omega_c \omega^c 
    - \half \kappa \lambda \left( \tau - \tau_c{}^c \right) \, .
\end{eqnarray}
Here $|$ stands for the covariant derivative $D$. All terms in
(\ref{eq:Geroch-Einstein1}--\ref{eq:Geroch-Einstein4}) are in $\N$, which
justifies the use of lower-case Latin indices ranging over $t, r$ and
$z$ only.

Geroch \cite{Geroch71} has also shown how the
original four-dimensional manifold $(\mathcal{M},g_{\alpha\beta})$ can be
recovered from the three-dimensional manifold $(\mathcal{N},h_{ab})$
and the fields $\omega_\alpha$ and $\lambda$.
To begin with, choose an arbitrary four-dimensional manifold $\M$
along with a nowhere-vanishing vector field 
$\xi = \partial/\partial\varphi$ on $\M$. 
Consider the following skew 2-form in $\N$:
\begin{equation}
  F_{ab} \equiv - \half \lambda^{-3} \epsilon_{abc}\omega^c \, .
\end{equation}
By equation \eqref{eq:Geroch-Einstein3}, it is curl-free:
\begin{equation}
  \epsilon^{abc} D_a F_{bc} = 0 \, .
\end{equation}
If we pull it back to $\M$, we obtain a curl-free
2-form $F_{\alpha\beta}$ because
the pull-back commutes with differentiation. By Frobenius' theorem
there exists a covector field $\eta_\beta$ such that 
\begin{equation}
  \partial_{[\alpha} \eta_{\beta]} = F_{\alpha\beta} \,.
\end{equation}
There is a gauge freedom
\begin{equation}
  \eta_\alpha \rightarrow \eta_\alpha + \partial_\alpha \sigma 
\end{equation}
for an arbitrary function $\sigma$.
We exploit the $\varphi$-dependence of $\sigma$ to set 
\begin{equation}
  \eta_\alpha \xi^\alpha = 1 \,.
\end{equation}
Next we define
\begin{equation}
  g_{\alpha\beta} = h_{\alpha\beta} + \lambda^2 \eta_\alpha \eta_\beta 
\end{equation}
so that $\xi_\alpha = g_{\alpha\beta} \xi^\beta = \lambda^2 \eta_\alpha$ and
hence
\begin{equation}
  g_{\alpha\beta} = h_{\alpha\beta} + \lambda^{-2} \xi_\alpha \xi_\beta \, ,
\end{equation}
as desired. 
It can be verified from \eqref{eq:Q} that 
\begin{equation}
  \nabla_{(\alpha} \xi_{\beta )} = 0 \,, 
\end{equation} 
i.e., that $\xi$ is a Killing vector of $g_{\alpha\beta}$.
It is not clear how to implement this procedure numerically, and
indeed there is no need to do so. All physically interesting quantities
in $\M$ have their counterparts in $\N$ and so we choose to work
entirely within $\N$.
For instance, it suffices to know the variables in $\N$ in order to
form the Newman-Penrose scalar $\Psi_4$ containing the
gravitational wave information (see section \ref{sec:radbcs} for an
explicit derivation of $\Psi_0$; the $\Psi_4$ calculation is similar).


\section{The ADM decomposition}
\label{sec:ADM}

\begin{figure}[h]
  \setlength{\unitlength}{1cm}
  \begin{center}
  \begin{picture}(9,7)
    \qbezier(0,0)(4,1)(8,0)
    \qbezier(1,2)(4,3)(9,2)
    \qbezier(0,0)(1,1)(1,2)
    \qbezier(8,0)(9,1)(9,2)
    \qbezier(0,4)(4,5)(8,4)
    \qbezier(1,6)(4,7)(9,6)
    \qbezier(0,4)(1,5)(1,6)
    \qbezier(8,4)(9,5)(9,6)
    \put(4,1.5){\vector(0,1){4}}
    \put(4,1.5){\vector(1,2){2}}
    \put(4,1.5){\vector(1,0){2}}
    \put(4,1.5){\dashbox{0.2}(2,4)}
    \put(3.1,3.5){$\alpha \, \diff t$}
    \put(4.8,1){$\vec{\beta} \, \diff t$}
    \put(3.9,5.7){$\vec{n}$}
    \put(6,5.7){$\vec{t}$}
    \put(1.1,1){$\Sigma(t)$}
    \put(1.1,5){$\Sigma(t + \diff t)$}
  \end{picture} 
  \end{center}
  \caption[The ADM setup]{\label{fig:ADM}  \footnotesize The ADM setup}
\end{figure}
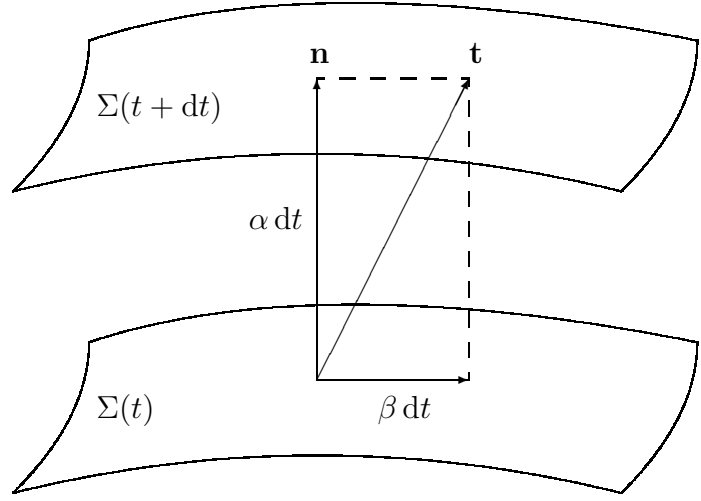

Following the standard ADM \cite{ADM62} procedure, a time function $t$
is introduced and the three-dimensional manifold $\N$ is foliated into 
two-dimensional spacelike hypersurfaces $\Sigma(t)$ of constant $t$.
The future-directed unit timelike normal to the hypersurfaces is 
\begin{equation}
  \label{eq:normal}
  n_a = - \alpha \, \partial_a t \,
\end{equation}
normalized such that $n_a n^a = -1$. Here $\alpha$ is the 
\emph{lapse function},
which describes the amount of proper time elapsing when
passing from one hypersurface $\Sigma(t)$ to a nearby one 
$\Sigma(t + \diff{t})$. The direction of time is not unique, however:
the spatial coordinate origin in $\Sigma(t + \diff{t})$ can be shifted
with respect to the origin in $\Sigma(t)$ by an arbitrary \emph{shift vector} 
$\beta^a$ (figure \ref{fig:ADM}),
\begin{equation}
\label{eq:timevector}
  t^a = \alpha n^a + \beta^a .
\end{equation}
The induced 2-metric on the hypersurfaces $\Sigma$ is
\begin{equation}
  \label{eq:H2}
  H_{ab} = h_{ab} + n_a n_b \, ,
\end{equation}
satisfying $H_{ab} n^b = 0$, i.e., it is indeed a tensor in $\Sigma$
and can thus be written as $H_{AB}$, where capital Latin indices $A,B$
run over $r$ and $z$.
With those definitions, the line element of $\N$ takes the form
\begin{equation}
  \label{eq:lineelement}
  {\diff s}^2 = -\alpha^2 {\diff t}^2 + H_{AB} \left( \diff x^A +
  \beta^A \diff t \right) \left( \diff x^B + \beta^B \diff t \right) \, .
\end{equation}
How the hypersurfaces $\Sigma(t)$ are imbedded in $\N$ is
described by the \emph{second fundamental form} or \emph{extrinsic curvature}
\begin{equation}
  \label{eq:chi2}
  \chi_{ab} = - H_a{}^c H_b{}^d n_{(c|d)} = -\half \Lie{n} H_{ab} \, ,
\end{equation}
which because of \eqref{eq:timevector} is essentially the
time-derivative of the 2-metric.

The Riemann tensor $\two{R}_{ABCD}$ in $\Sigma(t)$ is related to the one 
in $\N$ by the well-known \emph{Gauss-Codazzi equations}
\begin{eqnarray}
  \label{eq:Gauss-Codazzi1}
  \perp \three{R}_{ABCD} &=& \two{R}_{ABCD} + \chi_{AC} \chi_{BD} - \chi_{AD}
  \chi_{BC} \, ,\\ 
  \label{eq:Gauss-Codazzi2}
  \perp \three{R}_{ABCn} &=& \chi_{AC \parallel B} - \chi_{BC \parallel A}
  \, ,
\end{eqnarray}
where $\perp$ means projection of the free indices with $H$, and an
index $n$ stands for contraction with $n$.
The symbol $\parallel$ denotes the covariant derivative $\diff$ of $H_{ab}$,
\begin{equation}
  \label{eq:d}
  \diff_a v^b = H_a{}^c H_d{}^b D_c \left( H_e{}^d v^e \right) \, . 
\end{equation}
The derivation of (\ref{eq:Gauss-Codazzi1}) and (\ref{eq:Gauss-Codazzi2}) 
is completely analogous to the reduction of $\four{R}$ to $\three{R}$
presented in the previous section. 
The projections of the three-dimensional Einstein tensor
\begin{equation}
  \three{G}_{ab} = \three{R}_{ab} - \half \three{R} h_{ab}
\end{equation}
are found to be
\begin{eqnarray}
  \label{eq:hamcons1}
  \three{G}_{nn} &=& 2 \three{R}_{nn} + \three{R} = \half \left( \two{R}
    + (\chi_A{}^A)^2 - \chi_{AB} \chi^{AB} \right) \, , \\ 
  \label{eq:momcons1}
  \perp \three{G}^{An} &=& \perp \three{R}^{An} = \diff_B \left( -
  \chi^{AB} + H^{AB} \chi_C{}^C \right) \, .
\end{eqnarray}
The above two equations form so-called \emph{constraint equations} on
the hypersurfaces $\Sigma(t)$ because they do not involve any time 
derivatives. \eqref{eq:hamcons1} is called the \emph{Hamiltonian} or 
\emph{energy constraint}, and \eqref{eq:momcons1} are the 
\emph{momentum constraints}. 
Let us also calculate the time derivative of the extrinsic
curvature. Using the Ricci identity, 
\begin{equation}
  \Lie{N} \chi_{AB} = \alpha \perp \three{R}_{AnBn} - \alpha
  \chi_{AC} \chi_B{}^C - \alpha_{\parallel AB} \, ,
\end{equation}
and decomposing the Riemann tensor using the Gauss-Codazzi equations
({\ref{eq:Gauss-Codazzi1}--\ref{eq:Gauss-Codazzi2}), 
\begin{eqnarray}
  \label{eq:d0chi1}
  \Lie{n} \chi_{AB} &=& -\perp \three{R}_{AB} + \left(
  \two{R}_{AB} + \chi_C{}^C \chi_{AB} \right) - 2 \chi_{AC}
  \chi_B{}^C \nonumber\\&&- \alpha^{-1} \alpha_{\parallel AB} \,.
\end{eqnarray}
Equations (\ref{eq:chi2}) and (\ref{eq:d0chi1}) form a set of
\emph{evolution equations} for the 2-metric and extrinsic curvature.

We are now ready to insert the results of our $2+1$ decomposition
into the Geroch-Einstein equations
(\ref{eq:Geroch-Einstein1}-\ref{eq:Geroch-Einstein4}).
The trace of the extrinsic curvature is abbreviated as $\chi = \chi_A{}^A$.
Further variables defined in each $\Sigma(t)$ are
the alternating symbol
 \begin{equation}
  \epsilon_{AB} = n^c \epsilon_{cAB} \, ,
\end{equation}
the ($\varphi \varphi$)-component of the extrinsic curvature,
\begin{equation}
  \label{eq:Kpp}
  \K = - \lambda^{-1} n^a \lambda_{a} \, ,
\end{equation}
and the projections of the twist vector,
\begin{eqnarray}
  \label{eq:E}
  E^A &=& \lambda^{-3} \epsilon^{Ab} \omega_b \, ,  \\
  \label{eq:B}
  B^\varphi &=& \lambda^{-3} n_a \omega^a \, . 
\end{eqnarray}
(Again, the last two definitions differ from those in \cite{Maeda80}
by factors of $\lambda$. This ensures that $E^A$ has the correct
small-$r$ behaviour of a vector \eqref{eq:regcond2}. Our
$B^\varphi$ is $O(r)$ on the axis, which is easier to enforce
numerically than the $O(r^2)$ behaviour of $B_\varphi$ in \cite{Maeda80}). 
The various projections of the energy-momentum tensor are
\begin{eqnarray}
  \label{eq:tauprojections}
  J^\varphi &=& -n_a \tau^a \, , \nonumber\\
  S^A &=& H_a{}^A \tau^a \, ,  \nonumber\\
  \rho_H &=& n^a n^b \tau_{ab} \, , \\
  J_A &=& - H_A{}^a n^b \tau_{ab} \, ,  \nonumber\\
  S_{AB} &=& H_A{}^a H_B{}^b \tau_{ab} \, ,\nonumber
\end{eqnarray}
and, of course, $\tau$ defined in \eqref{eq:Tprojections}.

\paragraph{The constraint equations.} 
Inserting the Geroch result (\ref{eq:Geroch-Einstein1}) for $\three{R}_{ab}$ 
into the constraint equations 
(\ref{eq:hamcons1}, \ref{eq:momcons1}), we obtain 
\begin{eqnarray}
  \label{eq:hamcons} 
  \mathcal{C} &\equiv& \half (\chi^2 - \chi_{AB} \chi^{AB} + \two{R}) -
  \lambda^{-1} \lambda_{\parallel A}{}^A + \chi \K \nonumber\\ && 
  - \quarter \lambda^2 \left(E_A E^A + {B^\varphi}^2 \right) 
  - \kappa \rho_H  = 0 \, ,\\
  \label{eq:momcons}
    \mathcal{C}_A &\equiv& {\chi_A{}^B}_{|| B} - (\chi + \K)_{,A} 
  + \lambda^{-1} \lambda^B \chi_{AB} - \lambda^{-1} \lambda_A \K  
  \nonumber\\&&
  - \half \lambda^2 B^\varphi \epsilon_{AB} E^B - \kappa J_A = 0 \, .
\end{eqnarray}
The Geroch equation (\ref{eq:Geroch-Einstein2}) forms an additional
constraint equation, which we call the \emph{Geroch constraint}:
\begin{equation}
  \label{eq:gercons}  
  \mathcal{C}_\varphi \equiv \half E^A{}_{|| A} 
  + \frac{3}{2} \lambda^{-1} \lambda_A E^A - \kappa J^\varphi = 0 \, .
\end{equation}
Hence there are four constraint equations, as in the standard $3+1$ 
ADM decomposition.

\paragraph{The evolution equations.}
These are expressed in terms of the Lie derivative along the normal
lines, which by \eqref{eq:timevector} is
\begin{equation}
  \Lie{n} = \alpha^{-1} \partial_t - \Lie{\beta} \, .
\end{equation}
Definitions \eqref{eq:chi2} and \eqref{eq:Kpp} form evolution
equations for the 2-metric and the norm of the Killing vector,
\begin{eqnarray}
  \label{eq:d0H}
  \Lie{n} H_{AB} &=& - 2 \, \chi_{AB} \, ,\\
  \label{eq:d0lambda}
  \Lie{n} \lambda &=& - \lambda \K \, .
\end{eqnarray}
In the evolution equation for the extrinsic curvature
(\ref{eq:d0chi1}), we substitute (\ref{eq:Geroch-Einstein1}) for
$\three{R}_{ab}$ to obtain
\begin{eqnarray}  
  \label{eq:d0chi}
  \Lie{n} \chi_{AB} &=& \two{R}_{AB} - \lambda^{-1} \lambda_{A||B} 
   - \alpha^{-1} \alpha_{A||B} \nonumber\\&&
   + (\chi + \K) \chi_{AB} - 2 \chi_A{}^C \chi_{CB}  \\ && 
  - \half \lambda^2 \left[ \epsilon_{AC} \epsilon_{BD} E^C E^D 
  - H_{AB} \left( E_C E^C - {B^\varphi}^2 \right) \right] \nonumber\\ && 
   - \kappa \left[ S_{AB} + \half
  H_{AB} \left( \rho_H - S_C{}^C - \tau \right) \right] \, .\nonumber
\end{eqnarray}
The Geroch equation (\ref{eq:Geroch-Einstein4}) can be rewritten as
\begin{eqnarray}
  \label{eq:d0Kpp}
  \Lie{n} \K &=& 
  - \lambda^{-1} \lambda^A{}_{||A} - (\alpha \lambda)^{-1} \lambda_A \alpha^A 
  + \K \left( \chi + \K \right) \nonumber\\ && 
  - \half \lambda^2 \left(E_A E^A - {B^\varphi}^2 \right) 
  - \half \kappa \left(\rho_H - S_A{}^A + \tau \right) \, .
\end{eqnarray}
Finally using \eqref{eq:Geroch-Einstein3} together with
\eqref{eq:Geroch-Einstein4}, we obtain the following evoluton
equations for the twist variables,
\begin{eqnarray}
 \label{eq:d0E} 
  \Lie{n} E^A &=&  \epsilon^{AB} B^\varphi{}_{,B} + (\chi + 3 \K) E^A 
    \nonumber\\&& + \epsilon^{AB} (\alpha^{-1} \alpha_B
   + 3 \lambda^{-1} \lambda_B) B^\varphi  - 2 \kappa S^A \, ,\\
  \label{eq:d0B}
  \Lie{n} B^\varphi &=& \epsilon^{AB} E_{A || B} + \chi B^\varphi 
  + \alpha^{-1} \epsilon^{AB} E_A \alpha_B  \, .
\end{eqnarray}

Equations \eqref{eq:gercons} and (\ref{eq:d0E}--\ref{eq:d0B})
are remarkably similar to the axisymmetric Maxwell equations 
for an $\vec E$ field in the 
$(r, z)$ plane and a $\vec B$ field in the $\varphi$ direction,
which justifies the notation:
\begin{eqnarray}
  \nabla \cdot \vec E &=& \rho \,,\\
  \partial_t \vec E &=& \nabla \times \vec B + \vec j \,,\\
  \partial_t \vec B &=& -\nabla \times \vec E \,.
\end{eqnarray}
This is not surprising, of course, as the original Kaluza-Klein theory
\cite{Kaluza21,Klein26} was designed to incorporate electrodynamics into
four-dimensional general relativity by assuming an additional
compactified spacelike dimension.


\section{Matter evolution equations}
\label{sec:generalmatter}

Evolution equations for the matter variables can be obtained from
energy-momentum conservation
\begin{equation}
  \label{eq:Tconservation}
  \nabla_\alpha T^{\alpha\beta} = 0 \,.
\end{equation}
In our case, the energy-momentum tensor $T^{\alpha\beta}$ 
is also axisymmetric,
\begin{equation}
  \label{eq:Taxisymmetry}
  \Lie{\xi} T^{\alpha\beta} = 0 \,.
\end{equation}

First we decompose $T^{\alpha\beta}$ with respect to the Killing
vector $\xi^\alpha$ (Geroch decomposition),
\begin{equation}
  \label{eq:TGerochdecomposition}
  T^{\alpha\beta} = \lambda^{-2} \xi^\alpha \xi^\beta \tau
     + 2 \tau^{(\alpha} \xi^{\beta)} + \tau^{\alpha\beta}\,,
\end{equation}
where $\tau, \tau^\alpha$ and $\tau^{\alpha\beta}$ were defined in
\eqref{eq:Tprojections}.
Contracting \eqref{eq:Tconservation} with $\xi_\beta$, we obtain
the following conservation law in $\N$,
\begin{equation}
  \label{eq:tauaconservation}
  D_a (\lambda^3 \tau^a) = 0 \,,
\end{equation}
and projecting \eqref{eq:Tconservation} with $h_\beta{}^b$ yields 
\begin{equation}
  \label{eq:tauabconservation}
  D_a (\lambda \tau^{ab}) = \lambda^b \tau 
    - \epsilon^{bcd} \tau_c \omega_d \,.
\end{equation}

Next we decompose $\tau^{ab}$ further with respect to the unit timelike 
normal $n_a$ (ADM decomposition),
\begin{equation}
  \label{tauADMdecomposition}
  \tau^{ab} = \rho_H n^a n^b + 2 J^{(a} n^{b)} + S^{ab} \,,
\end{equation}
where $\rho_H, J^a$ and $S^{ab}$ were defined in \eqref{eq:tauprojections}.
Contracting \eqref{eq:tauabconservation} with $n_b$, we obtain an
evolution equation for $\rho_H$,
\begin{eqnarray}
  \label{eq:d0rhoH}
  \Lie{n} \rho_H &=& - J^A{}_{||A} 
    - J^A ( 2 \alpha^{-1} \alpha_A + \lambda^{-1} \lambda_A ) 
    + \chi \rho_H + \chi_{AB} S^{AB}  \nonumber\\&&
    + \K (\tau + \rho_H) + \lambda^2 E^A S_A \, ,
\end{eqnarray}
and projecting \eqref{eq:tauabconservation} with $H_b{}^B$ yields an
evolution equation for $J^A$,
\begin{eqnarray}
  \label{eq:d0JA}
  \Lie{n} J_A &=& - S_{AB}{}^{||B} + J_A (\chi + \K) 
  - S_{AB} ( \alpha^{-1} \alpha^B + \lambda^{-1} \lambda^B)  \nonumber\\&&
  - \alpha^{-1} \alpha_A \rho_H + \lambda^{-1} \lambda_A \tau 
  + \lambda^2 (E_A J^\varphi + \epsilon_{AB} S^B B^\varphi) \, .
\end{eqnarray}
Equation \eqref{eq:tauaconservation} can be rewritten as an evolution
equation for $J^\varphi$ (also defined in \eqref{eq:Tprojections}),
\begin{equation}
  \label{eq:d0Jp}
  \Lie{n} J^\varphi = - S^A{}_{||A} - S^A (\alpha^{-1} \alpha_A 
  + 3 \lambda^{-1} \lambda_A) + J^\varphi (\chi + 3 \K) \, .
\end{equation}
We recognize in the above the general relativistic version of the
Euler equations: \eqref{eq:d0rhoH} expresses energy conservation, 
\eqref{eq:d0JA} linear momentum conservation 
and \eqref{eq:d0Jp} angular momentum conservation.

In some situations (e.g., for a perfect fluid) there may be a
conserved particle number density $N^\alpha$ satisfying
\begin{equation}
  \label{eq:Nconservation}
  \nabla_\alpha N^\alpha = 0 
\end{equation}
and which is also axisymmetric,
\begin{equation}
  \label{eq:Naxisymmetry}
  \Lie{\xi} N^\alpha = 0 \,.
\end{equation}

Again, we can obtain an evolution equation from \eqref{eq:Nconservation} by
performing the same dimensional reductions as above.
First we decompose $N^\alpha$ with respect to the Killing vector $\xi^\alpha$
(Geroch decomposition)
\begin{equation}
  \label{NGerochdecomposition}
  N^\alpha = \lambda^{-1} \xi^\alpha \nu + \nu^\alpha \,,
\end{equation}
where we have defined
\begin{equation}
  \nu \equiv \xi_\alpha N^\alpha 
\end{equation}
and
\begin{equation}
  \nu^\alpha \equiv h^\alpha{}_\beta N^\beta \,.
\end{equation}
It follows from axisymmetry \eqref{eq:Naxisymmetry} that $\nu$ and
$\nu^\alpha$ are in $\N$.
Particle number conservation \eqref{eq:Nconservation} can be expressed
in $\N$ as
\begin{equation}
  \label{eq:nuaconservation}
  D_a (\lambda \nu^a) = 0 \,.
\end{equation}

Next we decompose $\nu^\alpha$ with respect to the unit timelike normal $n_a$
(ADM decomposition),
\begin{equation}
  \label{eq:NADMdecomposition}
  \nu^a = n^a \sigma + \Sigma^a \,,
\end{equation}
where we have defined
\begin{equation}
  \sigma \equiv -n_a \nu^a
\end{equation}
and
\begin{equation}
  \Sigma^a \equiv H^a{}_b \nu^b \,.
\end{equation}
From \eqref{eq:nuaconservation} we obtain an evolution equation for $\sigma$,
\begin{equation}
    \label{eq:d0sigma}
  \Lie{n} \sigma = - \Sigma^A{}_{||A} - \Sigma^A (\alpha^{-1} \alpha_A 
  + \lambda^{-1} \lambda_A) + \sigma (\chi + \K) \,.
\end{equation}

So far we have not chosen any specific matter model. In appendix
\ref{sec:fluid}, a perfect fluid is discussed. For the main part of
this thesis, however, we will focus on vacuum spacetimes.


\emptypage

 \chapter{Numerical methods}
\label{sec:num}

We have seen in the previous chapter that the Einstein equations split into
hyperbolic evolution equations and elliptic constraint equations when an ADM
(or ``space + time'') decomposition is applied.
In this chapter we describe the numerical methods we use to solve these two
classes of PDEs.

The basis for all the methods is the finite
difference technique, which is briefly described in section \ref{sec:FD},
along with the ghost cell technique for implementing boundary conditions.

The basic framework we adopt for solving the hyperbolic evolution equations is
the method of lines (section \ref{sec:MOL}). We explain the properties of
explicit Runge-Kutta and iterative Crank-Nicholson schemes and briefly comment
on the use of numerical dissipation. 

We then turn to elliptic equations and describe the Multigrid method (section
\ref{sec:MG}), starting from linear scalar equations in one dimension and
extending the algorithm to nonlinear problems, systems of equations and
multidimensions.

Some alternative methods for hyperbolic and elliptic PDEs and their 
applicability to our problem are discussed in section \ref{sec:altnum}.

Finally, we describe the adaptive mesh refinement technique for hyperbolic
PDEs in section \ref{sec:AMR}, pointing out some of the changes we have made
to the algorithm.

\section{The finite difference technique}
\label{sec:FD}

\subsection{The numerical grid}
\label{sec:FDgrid}

 The spatial domain we evolve is a finite rectangular region in the
 $(r, z)$ plane,
 \begin{equation}
   \Omega = [0, r_\mathrm{max}] \times [0, z_\mathrm{max}] \,.
 \end{equation}
 We restrict ourselves to the upper half of the $(r, z)$ plane 
 ($z \geqslant 0$) because we will impose reflection symmetry about 
 $z = 0$ for all the numerical evolutions carried out in this thesis.
 (This is not an essential restriction and one could equally well work 
 with a general domain that extends into the lower half of the $(r, z)$
 plane.)

 The numerical domain is covered by an equidistant \emph{grid} with
 grid points $(r_i, z_j)$, where
 \begin{eqnarray}
   r_i = (i - \half) h \,, \qquad 1 \leqslant i \leqslant N_r \,,\\
   z_j = (j - \half) h \,, \qquad 1 \leqslant j \leqslant N_z \,.
 \end{eqnarray}
 Here $N_r$ and $N_z$ are the number of grid points in the $r$ and $z$
 direction and the grid spacing $h$ is the same in both dimensions,
 \begin{equation}
   h = \frac{r_\mathrm{max}}{N_r} = \frac{z_\mathrm{max}}{N_z} \,.
 \end{equation}
 In all applications, we choose $r_\mathrm{max} = z_\mathrm{max}$ and hence
 $N_r = N_z \equiv N$. The Multigrid method (section \ref{sec:MG})
 further requires that $N$ be of the form $N = k \times 2^{l-1}$, where 
 $l$ is the number of Multigrid levels and $k$ is the size of the coarsest
 grid.

 Note that the first grid point in the $r$ direction is at $r = h/2$,
 not at $r = 0$. Thereby we avoid dividing by zero in terms formally
containing factors of $r^{-1}$.
 The grid points $(r_i,z_j)$ can also be viewed as the centres of the 
\emph{cells}
 \begin{equation}
   C_{ij} = [(i-1)h, ih] \times [(j-1)h, jh] \,, 
 \end{equation}
 which cover the entire spatial domain $\Omega$. For this reason, we
 call the type of grid we use a \emph{cell-centred} one.
 Using such a grid has certain additional advantages if matter is included,
 for example in the form of a perfect fluid. This is usually evolved
 using the \emph{finite volume} technique (e.g., \cite{LeVeque}), 
 where the solution is represented by the cell averages in the cells $C_{ij}$.

\subsection{Centred finite difference operators}

 The vacuum equations, however, are discretized using the \emph{finite
 difference} technique. Thereby continuum functions $u(r,z)$ are
 represented by their values $u_{ij}$ at the grid points $(r_i, z_j)$.
 Differential operators are translated into finite difference operators
 acting on the discrete grid function. These can be derived by means of
 Taylor expansions: as an example, consider
 \begin{eqnarray}
   u_{i+1,j} &=& u_{ij} + h (u_{,r})_{ij} + O(h^2) \, , \nonumber\\
   u_{i-1,j} &=& u_{ij} - h (u_{,r})_{ij} + O(h^2) \, , \nonumber\\
   \Rightarrow (u_{,r})_{ij} &=& \tfrac{1}{2h} (u_{i+1,j} - u_{i-1,j}) 
   + O(h^2) \, . 
 \end{eqnarray}
 All the finite-difference operators we use are centred and
 second-order accurate in the grid spacing, as in the above example. 
 The first-order derivatives are represented by
 \begin{eqnarray}
   u_{,r} &\rightarrow& \tfrac{1}{2h} (u_{i+1,j} - u_{i-1,j}) \,,\\
   u_{,z} &\rightarrow& \tfrac{1}{2h} (u_{i,j+1} - u_{i,j-1}) \,,\\
   (r^{-1} u)_{,r} &\rightarrow& \tfrac{1}{2h} (
   r_{i+1}^{-1} u_{i+1,j} - r_{i-1}^{-1} u_{i-1,j} ) \, .
 \end{eqnarray}
 The second-order finite differences needed for the hyperbolic-elliptic system
 (chapter \ref{sec:hypell}) are 
 \begin{eqnarray}
   u_{,rr} &\rightarrow& \tfrac{1}{h^2} (u_{i+1,j} - 2 u_{ij} +
     u_{i-1,j}) \, , \\
   u_{,zz} &\rightarrow& \tfrac{1}{h^2} (u_{i,j+1} - 2 u_{ij} +
     u_{i,j-1}) \, , \\
   u_{,rz} &\rightarrow& \tfrac{1}{4 h^2} (u_{i+1,j+1} - u_{i+1,j-1} -
     u_{i-1,j+1} + u_{i-1,j-1}) \, ,  \\
   \label{eq:fdrm1dr}
   (r^{-1} u_{,r})_{,r} &\rightarrow& \tfrac{1}{4 h^2} \left[
     r_{i+1}^{-1} (u_{i+2,j} - u_{ij}) - r_{i-1}^{-1} (u_{ij} - u_{i-2,j}) 
     \right] \, .
 \end{eqnarray}

\subsection{The ghost cell technique}
\label{sec:ghosts}

 Boundary conditions are implemented using the \emph{ghost cell} technique.
 Ghost cells are unphysical cells just outside the numerical domain.
 We add two layers of ghost cells at each boundary, at
 \begin{eqnarray}
   r_0 = -\tfrac{h}{2} \,, &\qquad&
   r_{N_r+1} = r_\mathrm{max} + \tfrac{h}{2} \,,\nonumber\\
   r_{-1} = -\tfrac{3h}{2} \,, &\qquad&
   r_{N_r+2} = r_\mathrm{max} + \tfrac{3h}{2} \,, 
 \end{eqnarray}
 and similarly in the $z$ direction.   
 The ghost cells are filled with values according to the boundary
 conditions that one would like to impose. The same finite difference
 operators can then be applied at all interior points, without having
 to modify them close to the boundaries. Two layers of ghost cells are
 required because the stencils we use
 have a width of up to 5 grid points (the second-order derivative 
 \eqref{eq:fdrm1dr} and the fourth-order dissipation operator
 \eqref{eq:diss4} have width 5).

 In order to implement a \emph{Neumann} condition 
 \begin{equation}
   u_{,r} \vert_{r = 0} = 0 \,,
 \end{equation}
 we note that
 \begin{eqnarray}
     u_{,r}(0,z_j) &=& \tfrac{1}{h}(u_{1j} - u_{0j}) + O(h^2) \nonumber\\
     &=&  \tfrac{1}{3h}(u_{2j} - u_{-1,j}) + O(h^2) \,, 
 \end{eqnarray}
 so that we choose the values of the ghost cells to be
 \begin{equation}
   u_{0j} = u_{1j} \,,\qquad u_{-1,j} = u_{2j} \,.
 \end{equation} 
 For a \emph{Dirichlet} condition
 \begin{equation}
   u\vert_{r = 0} = 0
 \end{equation}
 we note that
 \begin{eqnarray}
     u(0,z_j) &=& \half(u_{1j} + u_{0j}) + O(h^2) \nonumber\\
     &=&  \half (u_{2j} + u_{-1,j}) + O(h^2) \,, 
 \end{eqnarray}
 so that we need
 \begin{equation}
   u_{0j} = -u_{1j} \,,\qquad u_{-1,j} = -u_{2j} \,.
 \end{equation} 
 For \emph{linear extrapolation} at $r = r_\mathrm{max}$, we set
 \begin{equation}
   u_{N_r+1,j} = 2 u_{N_r j} - u_{N_r-1,j} \,, \qquad
   u_{N_r+2,j} = 2 u_{N_r+1, j} - u_{N_r j} \,.
 \end{equation}
 The \emph{differential} boundary conditions used in this thesis (the
 ``$1/R$ fall-off'' condition \eqref{eq:invRfalloff} and the
 differential boundary conditions for the incoming modes in chapter
 \ref{sec:outerbcs}) can be written in the general form
 \begin{equation}
   \label{eq:modeldiffbc}
   u_{,r} + f_{,z} + s = 0 
 \end{equation}
 where $f$ and $s$ may depend on $u$. (This is sometimes called a 
 \emph{Robin} condition.)
 To implement \eqref{eq:modeldiffbc} at the outer $r$ boundary, 
 we use the discretization
 \begin{eqnarray}
   \label{eq:modeldiffbcdiscr}
   \tfrac{1}{h} (u_{N_r+1,j} - u_{N_r j}) + \tfrac{1}{2h} (f_{N_r, j+1}
   - f_{N_r, j-1}) + s_{N_r j} &=& 0 \,, \nonumber\\
   \tfrac{1}{2h} (u_{N_r+2,j} - u_{N_r j}) + \tfrac{1}{2h} (f_{N_r, j+1}
   - f_{N_r, j-1}) + s_{N_r j} &=& 0 
 \end{eqnarray}
 (except at $j = 1$ and $j = N_z$, where the $z$-derivatives are
 replaced with
 \begin{equation}
   \tfrac{1}{h} (f_{N_r 2} - f_{N_r 1}) \,, \qquad
   \tfrac{1}{h} (f_{N_r N_z} - f_{N_r, N_z-1}) \,,
 \end{equation}
 respectively).
 Hence the ghosts are filled with
 \begin{eqnarray}
   u_{N_r+1,j} &=& u_{N_r j} - \half (f_{N_r,j+1} - f_{N_r,j-1}) - h
   s_{N_r j} \,,\nonumber\\
   u_{N_r+2,j} &=& u_{N_r j} - (f_{N_r,j+1} - f_{N_r,j-1}) - 2 h
   s_{N_r j} \,.
 \end{eqnarray}
 This discretization is only first-order accurate because a one-sided
 derivative is used in the $r$ direction.
 We have also tried a second-order accurate discretization
 \begin{eqnarray}
   \label{eq:modeldiffbcdiscr2}
   \tfrac{1}{2h} (u_{N_r+1,j} - u_{N_r-1, j}) + \tfrac{1}{2h} (f_{N_r, j+1}
   - f_{N_r, j-1}) + s_{N_r j} &=& 0 \,, \nonumber\\
   \tfrac{1}{4h} (u_{N_r+2,j} - u_{N_r-2, j}) + \tfrac{1}{2h} (f_{N_r, j+1}
   - f_{N_r, j-1}) + s_{N_r j} &=& 0 \,,
 \end{eqnarray}
 but this turned out to be less stable in some cases (section
 \ref{sec:bcnum}).

 The implementation of the boundary conditions at $z = 0$ and $z =
 z_\mathrm{max}$ follows by symmetry.

 We fill the ghost cells at each step of the time integration scheme
 discussed in the following section. This is crucial for the on-axis
 boundary conditions -- if these are not enforced at all stages of the
 algorithm, instabilities quickly develop.


 \section{The method of lines}
\label{sec:MOL}

 The basic framework we adopt in order to integrate the hyperbolic
 evolution equations forward in time is the \emph{method of lines (MOL)} 
 \cite{SchiesserMOL}. The basic idea is to finite-difference the 
 spatial derivatives of the PDE as described in the previous section,
 leaving the time derivatives continuous.
 This leads to a coupled set of ODEs for the time dependence of the
 variables $\vec u = (u_{ij})$ at the spatial grid points,
 \begin{equation}
   \label{eq:MOLODEs}
   \partial_t \vec u = \vec f (t, \vec u) 
 \end{equation}
 (in our case, there is no explicit time-dependence on the
 right-hand-side but we include it here for generality).
 A suitable ODE integrator is then used to integrate these ODEs forward
 in time.

\subsection{Properties of Runge-Kutta and ICN schemes}

 The ODE integrators we consider here belong to the class of
 \emph{explicit Runge-Kutta schemes}. Given the unknowns $\vec u^n$ at
 time $t^n$, these compute an approximation $\vec u^{n+1}$ at time
 $t^{n+1} = t^n + \Delta t$ in $s$ stages as follows:
 \begin{eqnarray}
   \label{eq:RKfirst}
   \vec k_1 &=& \vec f (t^n, \vec u^n) \,,\\
   \vec k_2 &=& \vec f(t^n + c_2 \Delta t, 
                  \vec u^n +  a_{21} \Delta t \vec k_1) \,,\\
   \vec k_3 &=& \vec f(t^n + c_3 \Delta t,
                  \vec u^n + a_{31} \Delta t \vec k_1 
                  + a_{32} \Delta t  \vec k_2)\,,\\
   &\vdots&\nonumber\\               
   \vec k_s &=& \vec f(t^n + c_s h,
                  \vec u^n + a_{s1} \Delta t \vec k_1
                  + a_{s2} \Delta t  \vec k_2 + \ldots
                  \nonumber\\&&\qquad\qquad\qquad\,
                  + a_{s,s-1} \Delta t \vec k_{s-1}) \,,\\
   \label{eq:RKlast}               
   \vec u^{n+1} &=& \vec u^n + \Delta t (b_1 \vec k_1 + \ldots 
                + b_s \vec k_s) \,.
 \end{eqnarray}
 Obviously a particular scheme is uniquely defined by the coefficients
 $a_{ij}$, $b_i$ and $c_i$, which are conveniently written as a tableau
 \begin{equation}
   \begin{array}{c|ccccc}
     0 & & & & & \\
     c_2 & a_{21} & & & & \\
     c_3 & a_{31} & a_{32} & & & \\
     \vdots & \vdots & \vdots & \ddots & & \\
     c_s & a_{s1} & a_{s2} & \cdots & a_{s,s-1} & \\
     \hline
     & b_1 & b_2 & \cdots & b_{s-1} & b_s
   \end{array}
 \end{equation}
 The method is said to be $p$th order if
 \begin{equation}
   \lVert u^{n+1} - u^n \lVert = O(\Delta t^{p+1})
 \end{equation}
 for sufficiently smooth $\vec f$.

 The simplest Runge-Kutta method is the \emph{Euler method}
 \begin{equation}
   \vec u^{n+1} = \vec u^n + \Delta t \, \vec f(t^n, \vec u^n) \,,
 \end{equation}
 which is first-order and is represented by the tableau
 \begin{equation}
   \label{eq:RK1}
   \begin{array}{c|c} 
     0 & \\ \hline & 1 
   \end{array} 
 \end{equation}  
 Two second-order Runge-Kutta methods are given by the tableaux
 \begin{equation}
   \label{eq:RK2}
   \begin{array}{c|cc}
     0 & & \\ 1 & 1 & \\ \hline & \half & \half 
   \end{array} 
   \qquad
   \begin{array}{c|cc}
     0 & & \\ \half & \half & \\ \hline & 0 & 1 
   \end{array} 
 \end{equation}
 The first is known as the \emph{trapezoidal rule}, the second as the
 \emph{midpoint rule}.
 Two examples of third-order methods are
 \begin{equation}
   \label{eq:RK3}
   \begin{array}{c|ccc}
     0 & & & \\ 
     \tfrac{1}{3} & \tfrac{1}{3} & & \\
     \tfrac{2}{3} & 0 & \tfrac{2}{3} & \\\hline
     & \quarter & 0 & \quarter
   \end{array} 
   \qquad
   \begin{array}{c|ccc}
     0 & & & \\
     1 & 1 & & \\
     \half & \quarter & \quarter & \\\hline
     & \tfrac{1}{6} & \tfrac{1}{6} & \tfrac{2}{3}
   \end{array}
 \end{equation}
 The first is \emph{Heun's (third order) method}, the second
 can be found in Shu and Osher \cite{Shu88}.
 Of the many known fourth-order Runge-Kutta methods we only state
 the most popular one:
 \begin{equation}
   \label{eq:RK4}
   \begin{array}{c|cccc}
     0 & & & & \\
     \half & \half & & & \\
     \half & 0 & \half & & \\
     1 & 0 & 0 & 1 & \\\hline
     & \tfrac{1}{6} & \tfrac{1}{3} & \tfrac{1}{3} & \tfrac{1}{6}
   \end{array} 
 \end{equation}
 For orders $p > 4$, it is no longer possible to construct a $p$th 
 order method with $s = p$ stages.
 For a survey of higher-order Runge-Kutta methods, see for
 example Butcher \cite{ButcherODEs}.

 A method that has become very popular in numerical relativity is the
 \emph{iterative Crank-Nicholson (ICN) method} \cite{Teukolsky00}
 given by
 \begin{eqnarray}
   \label{eq:ICNfirst}
   \vec k_1 &=& \vec f(t^n, \vec u^n) \,,\\
   \vec k_2 &=& \vec f(t^n + \half \Delta t, 
                       \vec u^n + \half \Delta t \vec k_1) \,,\\
   \vec k_3 &=& \vec f(t^n + \half \Delta t, 
                       \vec u^n + \half \Delta t \vec k_2) \,,\\
   &\vdots& \nonumber\\
   \vec k_s &=& \vec f(t^n + \half \Delta t, 
                       \vec u^n + \half \Delta t \vec k_{s-1}) \,,\\
   \label{eq:ICNlast}
   \vec u^{n+1} &=& \vec u^n + \Delta t \vec k_s \,.
 \end{eqnarray}
 In the limit $s \rightarrow \infty$ this yields the well-known implicit
 \emph{Crank-Nicholson method} 
\begin{equation}
  \frac{\vec u^{n+1} - \vec u^n}{\Delta t} = \vec f \left( \frac{\vec u^n +
  \vec u^{n+1}}{2} \right) \,.
\end{equation}
The iterative version (\ref{eq:ICNfirst}--\ref{eq:ICNlast}) 
can also be viewed \cite{Frauendiener02}
as an explicit Runge-Kutta scheme with tableau
 \begin{equation}
   \begin{array}{c|ccccc}
     0 & & & & &\\
     \half & \half & & & & \\
     \half & 0 & \half & & & \\
     \vdots & \vdots & \vdots & \ddots & & \\
     \half & 0 & 0 & \cdots & \half & \\\hline
     & 0 & 0 & \cdots & 0 & 1
   \end{array}
 \end{equation}
 The ICN method is always second-order accurate regardless of the
 number of steps $s$.

 To analyze the stability of the above ODE integrators,
 one applies them to the model equation
 \begin{equation}
   \partial_t \vec u = \lambda \vec u \,, \qquad \lambda \in \mathbb{C} \,.
 \end{equation}
 From the form of the general Runge-Kutta scheme
 (\ref{eq:RKfirst}--\ref{eq:RKlast}) it is clear that 
 \begin{equation}
   \vec u^{n+1} = P(\lambda \Delta t) \vec u^n
 \end{equation}
 for a certain complex polynomial $P$ of degree $s$, the \emph{stability
   function} of the method. Since $\lVert \vec u^{n+1} \rVert = \lvert
   P(\lambda \Delta t) \rvert \, \lVert \vec u^n \rVert$, the numerical
   approximation will remain bounded if and only if 
 \begin{equation}
   \label{eq:ODEstabcond}
    \lvert P(\lambda \Delta t) \rvert \leqslant 1\,.
\end{equation}
The set
 \begin{equation}
   S \equiv \{ z \in \mathbb{C} \, : \, \lvert P(z) \rvert \leqslant 1 \}
 \end{equation}
 is called the \emph{stability region} of the method.
 The Runge-Kutta schemes presented above have the stability function
\begin{equation}
  P_{\mathrm{RK}[s]} (z) = \sum_{n = 0}^{s} \frac{z^n}{n!} 
\end{equation}
(the result is the same for all schemes of a given order $s \leqslant 4$
\cite{ButcherODEs}).
The stability function of the ICN method is found to be
\begin{equation}
  P_{\mathrm{ICN}[s]} (z) = 1 + z \sum_{n = 0}^{s-1} \left(
    \frac{z}{2} \right)^n \,.
\end{equation}
Graphs of the stability regions are shown in figures \ref{fig:rkstab}
and \ref{fig:icnstab}.

\begin{figure}[t]
  \centering
  \includegraphics[scale = 0.7]{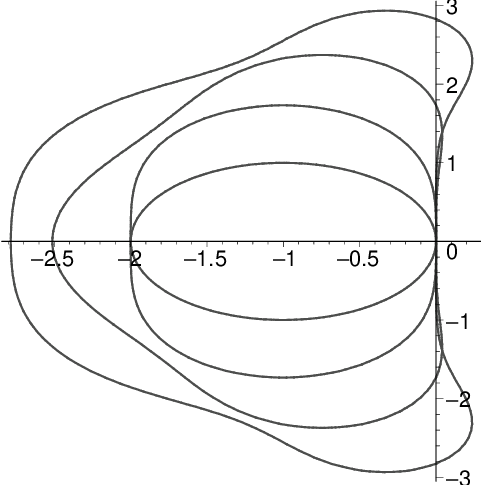}
  \bigskip
  \caption[Stability regions of the Runge-Kutta methods]
  {\label{fig:rkstab}  \footnotesize
    Stability regions of the Runge-Kutta methods with 
    $s = 1,2,3,4$ stages (from inward to outward)}
\end{figure}

\begin{figure}[p]
  \centering
  \begin{minipage}[t]{0.49\textwidth}
    \includegraphics[scale = 0.7]{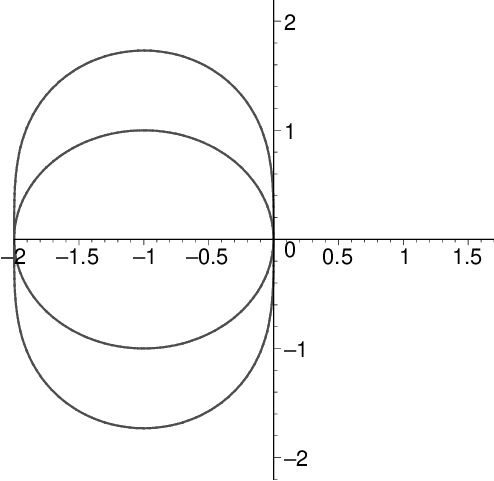}
    \centerline{$s = 1,2$}
  \end{minipage}
  \hfill  
  \begin{minipage}[t]{0.49\textwidth}
    \includegraphics[scale = 0.7]{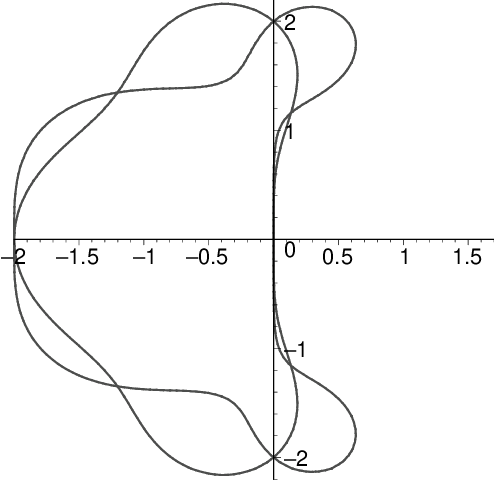}
    \centerline{$s = 3,4$}
  \end{minipage}
  \hfill  
  \bigskip

  \bigskip
  \begin{minipage}[t]{0.49\textwidth}
    \includegraphics[scale = 0.7]{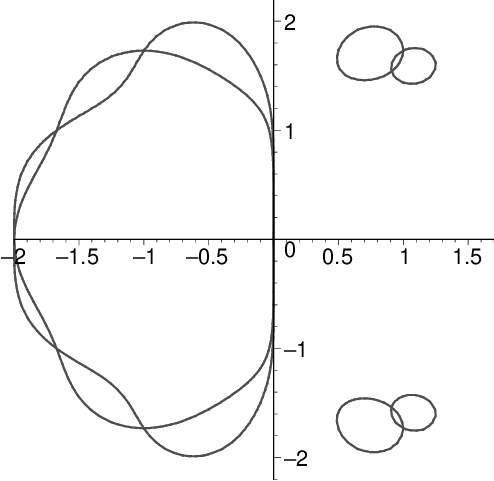}
    \centerline{$s = 5,6$}
  \end{minipage}
  \hfill  
  \begin{minipage}[t]{0.49\textwidth}
    \includegraphics[scale = 0.7]{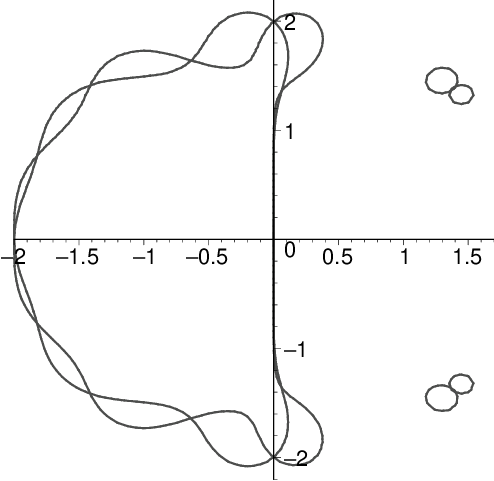}
    \centerline{$s = 7,8$}
  \end{minipage}
  \hfill  
  \bigskip
  \caption[Stability regions of the ICN method]
  {\label{fig:icnstab}  \footnotesize
    Stability regions of the ICN method for iteration numbers $1
    \leqslant s \leqslant 8$}
\end{figure}

As an example of the method of lines, let us consider the scalar
advection equation in one spatial dimension,
\begin{equation}
  \partial_t u + c \partial_x u = 0 \,, 
\end{equation}
where the speed $c \in \mathbb{R}$ is a constant.
Discretizing this in space using second-order centred finite differences,
we obtain the system of ODEs
\begin{equation}
  \label{eq:advdiscr}
  \partial_t u_j = -\tfrac{c}{2h} (u_{j+1} - u_{j-1}) \,.
\end{equation}
Suppose we represent the numerical approximation as a superposition of
Fourier modes (assuming periodic boundary conditions)
\begin{equation}
  \label{eq:MOLFT}
  u_j = \hat u (i \xi) \e^{i \omega x_j} \,,
\end{equation}
where we set $\xi \equiv \omega h$.
Inserting this into \eqref{eq:advdiscr} yields
\begin{equation}
  \partial_t \hat u(i \xi) = -i \tfrac{c}{h} \sin \xi \, \hat u(i \xi)  \,.
\end{equation}
Hence the eigenvalues of the system \eqref{eq:advdiscr} lie in the
interval $[-i h^{-1} c, i h^{-1} c]$ on the imaginary axis.

A method for integrating \eqref{eq:advdiscr} will be stable iff this
interval lies within the stability region of that method.
Setting $z = iq$ with $q \in \mathbb{R}$ we find for the Runge-Kutta methods
\begin{eqnarray}
  \lvert P_{\mathrm{RK[1]}}(iq)\rvert^2 - 1 &=& q^2 \,,\\
  \lvert P_{\mathrm{RK[2]}}(iq)\rvert^2 - 1 &=& \quarter q^4 \,,\\
  \lvert P_{\mathrm{RK[3]}}(iq)\rvert^2 - 1 &=& \tfrac{1}{36} q^4 
    (q^2 - 3) \,,\\
  \lvert P_{\mathrm{RK[4]}}(iq)\rvert^2 - 1 &=& \tfrac{1}{576} q^6
    (q^2 - 8) \,.
\end{eqnarray}  
Hence the stability regions of the first- and second-order schemes
only touch the imaginary axis in the origin. For the third-order
schemes, the boundary of the stability region intersects the imaginary
axis at $z = \pm \sqrt{3} \, i$, and for the fourth-order schemes at 
$z = \pm 2 \sqrt{2} \, i$.
We conclude that the first- and second-order Runge-Kutta methods are
unstable for the advection equation discretized as in \eqref{eq:advdiscr}.
This result is well-known in the first-order case, which is identical
with the \emph{FTCS (forward-time central-space) method}.
Setting $\lambda = \pm i h^{-1} c$ for the extremal eigenvalues in 
\eqref{eq:ODEstabcond} and thus 
$z = \lambda \Delta t = \pm i h^{-1} c \Delta t$,
we conclude that the third-order methods are stable iff
\begin{equation}
  \label{eq:CFL1}
  \Big\lvert\frac{\Delta t}{h} \, c\Big\rvert \leqslant \sqrt{3}\,,
\end{equation}
and the fourth-order methods are stable iff
\begin{equation}
  \label{eq:CFL2}
  \Big\lvert\frac{\Delta t}{h} \, c\Big\rvert \leqslant 2 \sqrt{2} \,.
\end{equation}
The restriction on the time step $\Delta t$ as expressed in equations 
(\ref{eq:CFL1}--\ref{eq:CFL2}) is known as the  
\emph{Courant-Friedrichs-Lewy (CFL) condition},
and the quotient $\Delta t / h$ is called the \emph{CFL} or 
\emph{Courant number}.
A simple interpretation of the CFL condition is that the time step
must be chosen small enough such that the numerical domain of
dependence (as given by the stencil  $[x_{j-1}, x_j, x_{j+1}]$) 
contains the physical domain of dependence (figure \ref{fig:CFL}). 

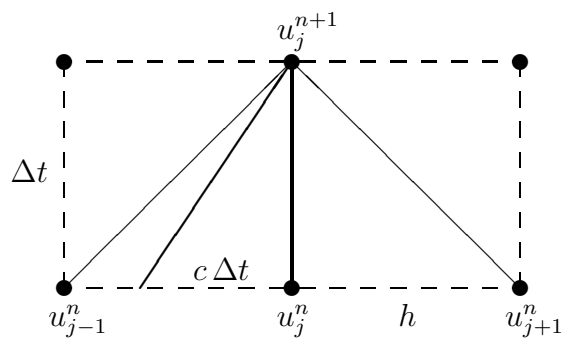
\begin{figure}[t]
  \setlength{\unitlength}{1cm}
  \centering
  \begin{picture}(8,5)
    \put(1,1){\circle*{0.2}}
    \put(4,1){\circle*{0.2}}
    \put(7,1){\circle*{0.2}}
    \put(1,4){\circle*{0.2}}
    \put(4,4){\circle*{0.2}}
    \put(7,4){\circle*{0.2}}
    \put(1,1){\dashbox{0.2}(6,3)}
    \put(1,1){\line(1,1){3}}
    \put(7,1){\line(-1,1){3}}
    \thicklines
    \put(2,1){\line(2,3){2}}
    \put(4,1){\line(0,1){3}}
    \thinlines
    \put(0.8,0.5){$u^n_{j-1}$}
    \put(3.8,0.5){$u^n_j$}
    \put(6.8,0.5){$u^n_{j+1}$}
    \put(3.8,4.3){$u^{n+1}_j$}
    \put(5.4,0.5){$h$}
    \put(2.7,1.1){$c \, \Delta t$}
    \put(0.3,2.4){$\Delta t$}
  \end{picture}  
  \caption[The CFL condition]{\label{fig:CFL} 
  \footnotesize The CFL condition. The thick solid
  lines mark the boundary of the physical past domain of dependence of 
  $u_j^{n+1}$ for the case $c > 0$, the thin solid lines the numerical 
  domain of dependence.}
\end{figure}

A similar analysis for the ICN method shows that the method is always
unstable for $s = 1,2,5,6,9,10,\ldots$ and stable for $s =
3,4,7,8,11,12,\ldots$ provided that 
  \begin{equation}
  \label{eq:CFL3}
  \Big\lvert\frac{\Delta t}{h} \, c\Big\rvert \leqslant 2 \,.
\end{equation}
Since neither the order of accuracy nor the size of the interval on
the imaginary axis contained in the stability region increase as 
$s$ is increased, the optimal choice for $s$ is $s = 3$. This was
pointed out by Teukolsky \cite{Teukolsky00} and confirmed with
the method used here by Frauendiener \cite{Frauendiener02}.
Since the Runge-Kutta method with the same number of steps ($s = 3$)
is third-order while the ICN method is only second-order, Runge-Kutta 
is always superior to ICN.

For the numerical experiments presented in this thesis, we have chosen
the third-order Runge-Kutta scheme of Shu (\ref{eq:RK3}b).
(Unlike the other third-order method (\ref{eq:RK3}a) this is
\emph{total variation diminishing (TVD)} \cite{Shu88}, a property that 
plays an important role in avoiding spurious oscillations around shocks 
if a perfect fluid is evolved with the same time integrator.)
We found that the fourth-order method \eqref{eq:RK4} (although not TVD)
has similar stability properties to (\ref{eq:RK3}a) for our system of
equations. We decided not to use the fourth-order method because it is
slightly more expensive\footnote{However, the fourth-order method has a larger
stability region and thus permits a larger time step.}, 
and it does not improve the overall accuracy because the spatial accuracy 
of the finite differencing is only 2.

\subsection{Numerical dissipation}

For the scalar advection equation with periodic boundary conditions,
we found that the system of ODEs \eqref{eq:MOLODEs} solved in the MOL
framework had purely imaginary eigenvalues. For more complicated
(systems of) PDEs and in particular for non-periodic boundary
conditions, this may no longer be the case. 
There may exist solutions that grow like $\exp (a t/h)$ with $a > 0$.
These are not present in the continuum problem but are a mere consequence 
of the spatial finite differencing. 

In many cases, these spurious modes can be eliminated by adding
\emph{Kreiss-Oliger dissipation} \cite{GKO, KO} to the right-hand-side 
of \eqref{eq:MOLODEs}. An example of a dissipation operator is
\begin{equation}
  \label{eq:diss4}
  (D_4 u)_j = -\tfrac{1}{16} h^{-1} (u_{j-2} - 4 u_{j-1} + 6 u_j 
  - 4 u_{j+1} + u_{j+2} ) \, .
\end{equation}
This operator has a Taylor expansion
\begin{equation}
  (D_4 u)_j = -\tfrac{1}{16} h^3 (u'''')_j + O(h^5) \,.
\end{equation}
Because our finite-differencing is second-order accurate,
the order of accuracy is not changed when adding $D_4 \vec u$ to
the right-hand-side,
\begin{equation}
     \partial_t \vec u = \vec f (t, \vec u) + \epsilon_D D_4 \vec u \,.
\end{equation}
Inserting a Fourier mode \eqref{eq:MOLFT} into \eqref{eq:diss4} gives us
\begin{equation}
  \label{eq:diss4FT}
  D_4 \hat u(i \xi) = - h^{-1}  \sin^4 \tfrac{\xi}{2} \, \hat u(i \xi) \,.
\end{equation}
We see that adding dissipation will decrease the amplification factor of
high-frequency modes ($\xi$ close to $\pi$).
The same argument as above for the advection equation shows that 
\begin{equation}
  \Big\lvert\frac{\Delta t}{h} \, \epsilon_D \Big\rvert \lesssim  1
\end{equation}
is needed for stability (the precise bound depending on the ODE integrator).
In practice we mostly use $0.1 \lesssim \epsilon_D \lesssim 0.5$.

Our numerical experiments indicate that a small amount of artificial 
dissipation is essential in order to avoid high-frequency instabilities 
that occur at very late times during the evolutions, particularly close 
to the boundaries.
For a theoretical justification for a simple model problem see Oliger
\cite[p 255]{Oliger}.

We apply the fourth-order operator \eqref{eq:diss4} both in the $r$ and
the $z$ direction and add it to the right-hand-side of the discretized
evolution equations at all interior grid points.
In order to eliminate (or at least postpone) instabilities sometimes 
caused by outer boundary conditions of the differential type 
\eqref{eq:modeldiffbc}, we have tried replacing \eqref{eq:diss4} 
with the second-order operator
\begin{equation}
  \label{eq:diss2}
  (D_2 u)_j = \quarter h^{-1} (u_{j+1} - 2 u_j + u_{j-1}) 
  = \quarter h (u'')_j + O(h^3) \,.
\end{equation}
This is only applied at the outermost layer of grid points.
Since our discretization \eqref{eq:modeldiffbcdiscr} of the 
differential boundary conditions is only first-order accurate, 
the leading-order accuracy is again unaffected.


 \section{The Multigrid method}
\label{sec:MG}

Solving elliptic equations is generally thought to be expensive because
typically $O(N^2)$ operations are required, where $N$ is the number of 
grid points. In contrast, hyperbolic equations only require $O(N)$ operations
per time step. However, the Multigrid method developed by Brandt 
\cite{Brandt77} gets away with $O(N)$ operations and is thus competitive.
This section serves as an introduction to this method.
We begin by describing basic relaxation methods for elliptic differential
equations and how the Multigrid idea
can accelerate those methods significantly. We then
generalize from linear to non-linear problems, systems of equations
and two-dimensional problems.
The first half of this section is mainly based on the book by Briggs
et al. \cite{Briggs}, which gives an excellent introduction to Multigrid
methods. Further details can be found in Wesseling \cite{Wesseling} 
and Hackbusch \cite{Hackbusch}.

\subsection{Relaxation Methods}
\label{sec:MGrelax}

Consider the simple one-dimensional model problem
\begin{eqnarray}
  \label{eq:modelDE}
  -u''(x) &=& f(x) \, , \qquad 0 < x < 1 \, , \nonumber\\
  u(0) = u(1) &=& 0 \, .
\end{eqnarray}
The domain $\Omega = [0,1]$ is discretized\footnote{This discretization is
  vertex-centred rather then cell-centred. We return to cell-centred
  discretizations in section \ref{sec:MGextensions}.}
 by introducing the grid points 
\nolinebreak $x_j = j h, \, 0 \leqslant j \leqslant N$, where $h = 1/N$ is the
constant width of the cells. The original differential equation 
(\ref{eq:modelDE}) is replaced by a system of difference equations 
\begin{eqnarray}
  \label{eq:modelFD}
  h^{-2} (-u_{j-1} + 2 u_j - u_{j+1}) &=& f(x_j) \, , \qquad
  1 \leqslant j \leqslant N-1 \, , \nonumber\\
  u_0 = u_N &=& 0 \, ,
\end{eqnarray}
where $u_j \equiv u(x_j)$. 
In matrix form, this system of linear equations is written as 
\begin{equation}
  \label{eq:mateqn}
  A \vec u = \vec f
\end{equation}
where
\begin{equation}
  A = \frac{1}{h^2} \left[ 
    \begin{array}{ccccc}
       2 & -1 & & &  \\
       -1 & 2 & -1 & &  \\
       & \ddots & \ddots & \ddots & \\
       & &  -1 & 2 & -1 \\
       & &  & -1 & 2 
     \end{array} 
   \right]
\end{equation}

One of the simplest schemes to solve (\ref{eq:modelFD}) iteratively is
the \emph{Jacobi} method. It is obtained by solving the $j$th equation
of (\ref{eq:modelFD}) for $u_j$, using the current approximation for
$u_{j-1}$ and $u_{j+1}$:
\begin{equation}
  \label{eq:Jacobi}
  v^{m+1}_j = \half ( v^m_{j-1} + v^m_{j+1} + h^2 f_j) \, , \qquad
  1 \leqslant j \leqslant N-1 \, ,
\end{equation}
where $\vec v^m$ denotes the $m$th step approximation to the unknown
$\vec u$.
The \emph{Gauss-Seidel} method differs from the Jacobi method in that
the components of the new approximation are used as soon as they are
available, i.e., we successively replace
\begin{equation}
  \label{eq:Gauss-Seidel}
  v_j \leftarrow \half ( v_{j-1} + v_{j+1} + h^2 f_j ) \, , \qquad
  1 \leqslant j \leqslant N-1 \, ,
\end{equation}
in ascending order. Alternatively, one can sweep through the grid
points in descending order, or one can first update the even
components of $\vec v$ and then the odd components. The latter
variant is known as \emph{red-black} Gauss-Seidel. 

Now consider a general matrix $A = D - L - U$ where $D$ denotes the
diagonal and $-L$ and $-U$ the strictly lower and upper triangular
parts of $A$. 
The Jacobi method can be written as 
\begin{equation}
  \vec v^{m+1} = R_J \vec v^m + D^{-1} \vec f
\end{equation}
where the \emph{iteration matrix} $R_J$ is given by
\begin{equation}
  R_J = D^{-1} ( L + U ) \, .
\end{equation}
Similarly, the Gauss-Seidel method takes the form
\begin{equation}
  \vec v^{m+1} = R_G \vec v^{m} + (D-L)^{-1} \vec f
\end{equation}
where
\begin{equation}
  R_G = (D-L)^{-1} U \, .
\end{equation}
The convergence properties of the above relaxations depend crucially
on the size of the largest eigenvalue of the iteration matrix $R$,
which is known as the \emph{spectral radius}
\begin{equation}
  \rho(R) = \max |\lambda(R)| \, .
\end{equation}
The iteration converges if and only if $\rho(R) < 1$. 
This is satisfied if the original matrix $A = (a_{ij})$ is 
\emph{diagonal dominant}, i.e. 
\begin{equation}
  \label{eq:diagdom}
  \sum_{1 \leqslant j \leqslant N-1, \, j \neq i} 
  |a_{ij}| \leqslant |a_{ii}| \,, \qquad 
  1 \leqslant i \leqslant N-1 \,.
\end{equation}
This condition imposes quite a severe restriction on realistic
problems, as we shall see in section \ref{sec:ellsolvnum}.

To analyze the convergence properties in more detail, it is useful to
look at the characteristic structure of the iteration matrix $R$. We
illustrate this for the Jacobi method applied to the model problem 
(\ref{eq:modelFD}). 
The eigenvalues of $R_J$ are found to be 
\begin{equation}
  \label{eq:spectrum}
  \lambda_k(R_J) = 1 - 2 \sin^2 \frac{k \pi}{2 N} \, ,
  \qquad 1 \leqslant k \leqslant N-1 \, ,
\end{equation}
and the eigenvectors $\vec w_k$ are given by
\begin{equation}
  w_{k,j} = \sin \frac{j k \pi}{2 N} \, , \qquad  
  1 \leqslant k \leqslant N-1 \, , \quad 0 \leqslant j \leqslant N \,.
\end{equation}
We see that the eigenvectors are simply Fourier modes.
Let $\vec e = \vec u - \vec v$ denote the \emph{error} of the
approximation $\vec v$. Suppose we choose the initial error to be one
of the Fourier modes, $\vec e^0 = \vec w_k$.
Then the error after $m$ steps of the iteration is
$\vec e^m = R_J^m \vec e^0 = \lambda_k^m (R_J) \vec w_k$.
After $m$ iterations, the $k$th mode of the initial error has been
reduced by a factor of $\lambda_k^m (R_J)$. From (\ref{eq:spectrum}), 
we see that high-frequency or oscillatory modes of the
error are damped much more effectively than low-frequency or
smooth modes. This so-called \emph{smoothing property} is generic for
all classical relaxation schemes. These schemes are very effective in
reducing the oscillatory components of the error but the smooth components
persist.

\subsection{The Multigrid idea}

The idea of Multigrid is based on a simple observation: a smooth mode on
a fine grid appears more oscillatory on a coarser grid. This suggests
that when the relaxation on a fine grid begins to stall, we transfer
the error to a coarser grid, e.g.~with twice the step size of the fine
grid, where it can be damped much more effectively. Then the error is
transferred back to the fine grid and we relax again. This idea will
be made precise below.
 
The equation we use to reduce the error on the coarser grid is the 
\emph{residual equation} 
\begin{equation}
  \label{eq:reseqn}
  A \vec e = \vec r \, ,
\end{equation}
which follows from \eqref{eq:mateqn} if we define the error to be
$\vec e = \vec u - \vec v$ and the \emph{residual} to be 
$\vec r = \vec f - A \vec v$. 

Let $\vec v^h$ denote the approximation on the fine grid $\Omega^h$ with
step size $h$ and $\vec v^{2h}$ the approximation on the coarse grid
$\Omega^{2h}$ with step size $2h$. 
Suppose we have a \emph{restriction operator} $I^{2h}_h$ which
transfers vectors from $\Omega^h$ to $\Omega^{2h}$ and a
\emph{prolongation operator} $I^h_{2h}$ 
which transfers vectors from $\Omega^{2h}$
to $\Omega^h$. Then we can define the following algorithm
\cite[chapter 3]{Briggs}:

\paragraph{Two-Grid Correction Scheme} 
\begin{itemize}
  \item Relax $\nu_1$ times on $A^h \vec u^h = \vec f^h$ on $\Omega^h$
    with initial guess $\vec v^h$.
  \item Compute the fine-grid residual $\vec r^h = \vec f^h - A^h \vec
    v^h$ and restrict it to the coarser grid by $\vec r^{2h} =
    I_h^{2h} \vec r^h$.
  \item Relax on $A^{2h} \vec e^{2h} = \vec r^{2h}$ on $\Omega^{2h}$.
  \item Prolong the coarse-grid error to the fine grid by $\vec e^h =
    I^h_{2h} \vec e^{2h}$ and correct the fine-grid approximation by
    $\vec v^h \leftarrow \vec v^h + \vec e^h$.
  \item Relax $\nu_2$ times on $A^h \vec u^h = \vec f^h$ on $\Omega^h$
    with initial guess $\vec v^h$.  
\end{itemize}

\paragraph{Intergrid transfer.}
There are many ways to choose the intergrid transfer operators
$I^h_{2h}$ and $I^{2h}_h$. The simplest (and very effective) choice
for the prolongation operator is linear interpolation,
\begin{eqnarray}
  \label{eq:lininterp}
  v^h_{2j} &=& v^{2h}_j \, , \nonumber\\
  v^h_{2j+1} &=& \half (v^{2h}_j + v^{2h}_{j+1}) \, , \qquad 0 \leqslant j
  \leqslant \frac{N}{2} - 1 \, .
\end{eqnarray}
A systematic way of defining the restriction operator is given by
\begin{equation}
  I^h_{2h} = c \, (I^{2h}_h)^T \, , \qquad c \in \mathbb{R} \, ,
\end{equation}
with $c = 1$ for one-dimensional problems (for arbitrary dimension, 
$c$ is determined by the requirement that restriction should preserve 
constant vectors).
For the prolongation operator (\ref{eq:lininterp}), this leads to
\begin{equation}
  v_j^{2h} = \quarter ( v^h_{2j-1} + 2 v^h_{2j} + v^h_{2j+1} ) \, ,
  \qquad 1 \leqslant j \leqslant \frac{N}{2} - 1 \, ,
\end{equation}
which is known as \emph{full weighting}. The coarse-grid operator may
then be defined by the \emph{Galerkin condition}
\begin{equation}
  A^{2h} = I^{2h}_h A^h I^h_{2h} \, ,
\end{equation}
which arises from a minimization principle \cite[chapter 10]{Briggs}.
For the model problem at hand, it turns out that the coarse-grid
matrix $A^{2h}$ defined in this way is just the fine-grid matrix $A^h$
with $h$ replaced by $2h$, which makes it easy to implement.
For more complicated problems, this will not be the case and the 
Galerkin condition is no longer useful in practice. Instead, we will
\emph{choose} the coarse grid operator $A^{2h}$ to have the same form as the
fine-grid operator $A^h$. The choice of intergrid transfer operators is
restricted by a simple criterion, given by Hackbusch \cite{Hackbusch}:
to obtain a mesh-size independent rate of convergence of Multigrid, it
is sufficient that
\begin{equation}
  \label{eq:RPcriterion}
  m_r + m_p \geqslant 2m \, ,
\end{equation}
where $m_r$ is the interpolation order for restriction, $m_p$ is
the interpolation order for prolongation, and $2m$ is the order of the
differential equation to be solved. Hence for second-order
differential equations, it is sufficient to choose $m_p = 1$
(piecewise-constant interpolation) and $m_r = 2$ (linear
interpolation), which is indeed what we have found perfectly satisfactory. 

\paragraph{Multigrid cycles.}
Looking at the two-grid correction scheme again, we see that the
coarse-grid problem is not much different from the fine-grid
problem. Therefore, we can apply the two-grid correction scheme to the
residual equation on $\Omega^{2h}$, i.e. we relax there and move to
$\Omega^{4h}$ for the correction step. We can repeat this process
recursively:
\pagebreak

\paragraph{$\mu$-Cycle Scheme} 
                      $\vec v^h \leftarrow {M \mu}^h (\vec v^h, \vec f^h)$
\begin{enumerate}
  \item Relax $\nu_1$ times on $A^h \vec u^h = \vec f^h$ with a given
    initial guess $v^h$.
  \item If $\Omega^h$ is the coarsest grid, then go to step 4.
    Else
    \begin{eqnarray}
      \vec f^{2h} &\leftarrow& I^{2h}_h (\vec f^h - A^h \vec v^h) \,
      , \nonumber\\
      \vec v^{2h} &\leftarrow& \vec 0 \quad \textrm{(initial guess)}, 
      \nonumber\\
      \vec v^{2h} &\leftarrow& {M \mu}^{2h} (\vec v^{2h}, \vec f^{2h})
      \quad \mu \, \textrm{times.} \nonumber
    \end{eqnarray}
   \item Correct $\vec v^h \leftarrow \vec v^h + I^h_{2h} \vec v^{2h}$.
   \item Relax $\nu_2$ times on $A^h \vec u^h = \vec f^h$ with initial
     guess $\vec v^h$.   
\end{enumerate}

For $\mu = 1$ we obtain the \emph{V-cycle} scheme, for $\mu = 2$ the
\emph{W-cycle} scheme. The choice of the names becomes obvious from
figure \ref{fig:cycles}, which shows the order in which the grids are visited.
For the elliptic equations considered in this thesis, we found
that W-cycles were more efficient than V-cycles in driving the residual
below a given threshold.

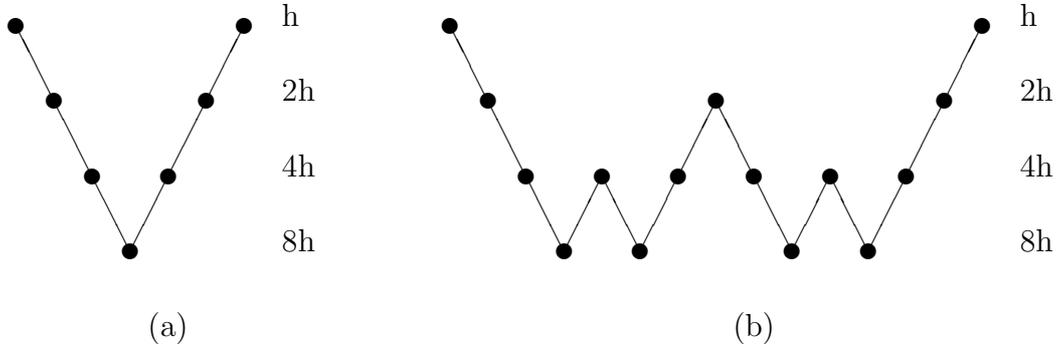
\begin{figure}[t]
  \setlength{\unitlength}{1cm}
  \begin{center}
    \begin{minipage}[t]{4.0cm}
      \begin{picture}(5,4)
        \put(0,3){\circle*{0.2}}
        \put(0.5,2){\circle*{0.2}}
        \put(1,1){\circle*{0.2}}
        \put(1.5,0){\circle*{0.2}}
        \put(2,1){\circle*{0.2}}
        \put(2.5,2){\circle*{0.2}}
        \put(3,3){\circle*{0.2}}
        \put(1.5,0){\line(-1,2){1.5}}
        \put(1.5,0){\line(1,2){1.5}}
        \put(3.5,0){8h}
        \put(3.5,1){4h}
        \put(3.5,2){2h}
        \put(3.5,3){h}
      \end{picture} 
      \begin{center} (a) \end{center}
    \end{minipage}  
    \hfill
    \begin{minipage}[t]{8.0cm}
      \begin{picture}(8,4)
        \put(0,3){\circle*{0.2}}
        \put(0.5,2){\circle*{0.2}}
        \put(1,1){\circle*{0.2}}
        \put(1.5,0){\circle*{0.2}}
        \put(2,1){\circle*{0.2}}
        \put(2.5,0){\circle*{0.2}}
        \put(3,1){\circle*{0.2}}
        \put(3.5,2){\circle*{0.2}}
        \put(4,1){\circle*{0.2}}
        \put(4.5,0){\circle*{0.2}}
        \put(5,1){\circle*{0.2}}
        \put(5.5,0){\circle*{0.2}}
        \put(6,1){\circle*{0.2}}
        \put(6.5,2){\circle*{0.2}}
        \put(7,3){\circle*{0.2}}
        \put(1.5,0){\line(-1,2){1.5}} 
        \put(1.5,0){\line(1,2){0.5}}
        \put(2.5,0){\line(-1,2){0.5}}
        \put(2.5,0){\line(1,2){1}}
        \put(4.5,0){\line(-1,2){1}}
        \put(4.5,0){\line(1,2){0.5}}
        \put(5.5,0){\line(-1,2){0.5}}
        \put(5.5,0){\line(1,2){1.5}}
        \put(7.5,0){8h}
        \put(7.5,1){4h}
        \put(7.5,2){2h}
        \put(7.5,3){h}
      \end{picture}
      \begin{center} (b) \end{center}
    \end{minipage}  
  \end{center}
  \caption[Multigrid cycles]{\label{fig:cycles}\footnotesize 
    Schedule of grids for a (a) V-cycle and (b) W-cycle on
    four levels}
\end{figure}

\subsection{Nonlinear Multigrid}
\label{sec:NonlinMG}

Nonlinear PDEs lead to nonlinear finite difference discretizations of
the form
\begin{equation}
  \label{eq:nonlin}
  A(\vec u) = \vec f \, , \qquad \vec u, \vec f \in \mathbb{R}^N \, .
\end{equation}
As before, we define the error of an approximation $\vec v$ to be 
$\vec e = \vec u - \vec v$ and the residual to be $\vec r = \vec f -
A(\vec v)$. 
By subtracting (\ref{eq:nonlin}) from the definition of the residual,
we obtain
\begin{equation}
  \label{eq:nonlinreseqn}
  A(\vec u) - A(\vec v) = \vec r \, .
\end{equation}
The crucial difference between linear and nonlinear problems is that 
if the operator $A$ is nonlinear, we \emph{cannot} conclude that 
$A(\vec u) - A(\vec v) = A(\vec u - \vec v) = A(\vec e)$. 
We no longer have the simple linear residual equation (\ref{eq:reseqn}) 
but have to use (\ref{eq:nonlinreseqn}) instead.

\emph{One} way to solve (\ref{eq:nonlinreseqn}) is to replace 
$\vec u = \vec v + \vec e$ and to Taylor-expand about $\vec v$. 
Keeping only the linear term, we arrive at the \emph{linear} system
\begin{equation}
  \label{eq:Newtonlin}
  J(\vec v) \, \vec e = \vec r \, ,
\end{equation}
where $J_{ij} = \partial A(\vec v)_i / \partial v_j$ is the $N \times N$
Jabobian matrix. Iteration of this step leads to the well-known 
\emph{Newton-Raphson method}. To solve the linear equation
(\ref{eq:Newtonlin}), we could use the linear Multigrid methods 
presented above, and this combination is usually
called \emph{Newton-Multigrid}.
  
However, Multigrid can treat the nonlinearity directly! Let us write
down the residual equation (\ref{eq:nonlinreseqn}) on the coarse grid
$\Omega^{2h}$,
\begin{equation}
  \label{eq:coarsereseqn}
  A^{2h} (\vec u^{2h}) = A^{2h} (\vec v^{2h}) + \vec r^{2h} \, .
\end{equation}
Suppose we have some approximation $\vec v^h$ on $\Omega^h$. We can
restrict that approximation to $\Omega^{2h}$, 
\begin{equation}
  \vec v^{2h} = I^{2h}_h \vec v^h \, ,
\end{equation}
and similarly restrict the residual, 
\begin{equation}
  \vec r^{2h} = I^{2h}_h \vec r^h =  I^{2h}_h (\vec f^h - A^h(\vec
  v^h)) \, . 
\end{equation}
Then the right-hand-side of (\ref{eq:coarsereseqn}) is known, and we
can relax that equation with respect to $\vec u^{2h}$ to obtain a new 
approximation on $\Omega^{2h}$. This leads to the following algorithm 
\cite[chapter 6]{Briggs}:

\paragraph{Full Approximation Scheme (FAS)}
\begin{itemize}
  \item Restrict the current approximation and its fine-grid residual
    to the coarse grid: $\vec r^{2h} = I^{2h}_h (\vec f^h - A^h(\vec
    v^h))$ and $\vec v^{2h} = I^{2h}_h \vec v^h$.
  \item Relax on the coarse-grid problem $A^{2h}(\vec u^{2h}) =
    A^{2h}(\vec v^{2h}) + \vec r^{2h}$.
  \item Compute the coarse-grid approximation to the error: $\vec
    e^{2h} = \vec u^{2h} - \vec v^{2h}$.
  \item Prolong the error approximation to the fine grid and correct
    the current fine-grid approximation: $\vec v^h \leftarrow \vec v^h
    + I^h_{2h} \vec e^{2h}$.  
\end{itemize}

This is called \emph{full approximation scheme} because apart from the vector
that is iterated in the relaxation steps, it requires the additional
storage of the current approximation coming from the finer grid.
The extension of the above two-grid scheme to the Multigrid
$\mu$-cycles is obvious.

The same relaxation schemes as in the linear case can be applied:
e.g. for the Gauss-Seidel relaxation, one solves the equation $A(\vec u) =
\vec f$ for $u_j$ at grid point $j$. However, since the underlying PDE
is nonlinear, the Gauss-Seidel step may require solving a nonlinear
equation instead of a linear one as in (\ref{eq:Gauss-Seidel}).
This can be done with Newton's method, as discussed above in the case
of Newton-Multigrid. Note, however, that we only need to solve a
single nonlinear equation now instead of an $N \times N$ system as in
(\ref{eq:Newtonlin})!

\subsection{Extension to systems and multidimensions}
\label{sec:MGextensions}

\paragraph{Systems of PDEs.} It is straightforward to generalize the
above algorithms to systems of equations. The vector of unknowns now
has the form $\vec u = (\vec u^1, \, \vec u^2, \, \ldots , \, \vec
u^k)^T$ where $k$ is the number of equations and each subvector $\vec u^i
\in \mathbb{R}^N$. 
For prolongation and restriction, one simply treats the subvectors
separately. 

For non-static relaxations such as Gauss-Seidel (as opposed to Jacobi), 
the order in which the unknowns are updated matters. 
One possibility is to update all the unknowns 
$u_j^1, \, u_j^2, \, \ldots , \, u_j^k$ simultaneously at
each grid point $j$. Another possibility is to first update $u_j^1$
over the entire grid, then $u_j^2$, and so forth. 
We found that the first method is more efficient for the elliptic
equations occurring in our problem.

\paragraph{Multidimensional problems.} The extension to two (or more)
dimensions is straightforward, too. The vector of unknowns is now
$\vec u = (u_{ij})$, 1 $\leqslant i \leqslant N_r$, 
$\quad 1 \leqslant j \leqslant N_z$ for a grid with $N_r$ 
points in the $r$ direction and $N_z$ points in the $z$ direction.
We consider a cell-centred grid here as described in section \ref{sec:FD}. 
The restriction and prolongation operators are now based on 
\emph{two-dimensional} interpolation. 
We use piecewise-constant interpolation for prolongation,
\begin{eqnarray}
  u^h_{2i-1,2j} = u^h_{2i, 2j-1} = u^h_{2i-1,2j-1} = u^h_{2i,2j} =
  u^{2h}_{ij} \, , \nonumber\\
  1 \leqslant i \leqslant \tfrac{N_r}{2} \, , \quad 
  1 \leqslant j \leqslant \tfrac{N_z}{2} \, ,
\end{eqnarray}
and (triangular) linear interpolation for restriction,
\begin{eqnarray}
  u^{2h}_{ij} &=& \tfrac{1}{16} \left( u^h_{2i,2j-2} + u^h_{2i+1,2j-2} +
  2 u^h_{2i-1,2j-1} + 3 u^h_{2i,2j-1}  \right. \nonumber\\
  && \qquad  \left. + u^h_{2i+1,2j-1} + u^h_{2i-2,2j} + 3 u^h_{2i-1,2j} 
  + 2 u^h_{2i,2j} \right. \nonumber\\
  && \qquad  \left. + u^h_{2i-2, 2j+1} + u^h_{2i-1,2j+1} \right) \, , 
  \nonumber\\
  && 1 \leqslant i \leqslant \tfrac{N_r}{2} \, , \quad 
  1 \leqslant j \leqslant \tfrac{N_z}{2} \, .
\end{eqnarray}
 
Again, there is an ambiguity in the order of update in
the relaxation step for the Gauss-Seidel method. The
\emph{lexicographical} Gauss-Seidel method first sweeps through the 
index $i$ (in ascending order) in an outer iteration and then through 
$j$ in an inner iteration. The \emph{red-black} Gauss-Seidel method
colours the grid points in chessboard manner and first sweeps through
all red points (in lexicographical order), then through all black points.  
We have found the red-black version to be more efficient for most problems.


 \section{Alternative methods}
\label{sec:altnum}

In this section we discuss some alternative methods -- finite volume methods
for hyperbolic equations and conjugate gradient methods for elliptic
equations.

\subsection{Finite volume methods}
\label{sec:altnumFV}

Centred finite-differencing in conjunction with the method of lines
works well if the solution one tries to approximate is smooth.
This is the case for the vacuum gravitational waves considered in this
thesis. Once matter is included, however, discontinuous solutions have
to be taken into account. For example, the formation of \emph{shocks}
is a common phenomenon in fluid dynamics. In \cite{Barnes04} it was
shown for the coupled Euler-Einstein equations in plane-symmetric
Gowdy spacetimes that discontinuities can also show up in the first- and
second-order derivatives of the metric.  

The method presented in the previous sections is unsuitable for
discontinuous solutions because it produces large oscillations around 
the discontinuities, known as the \emph{Gibbs phenomenon}. 
A huge variety of methods that are capable of dealing with discontinuities
have been developed in computational fluid dynamics 
\cite{Laney, LeVeque, Toro}.
These methods are typically based on a formulation of the underlying
hyperbolic PDEs as \emph{conservation laws} (possibly with sources)
\begin{equation}
  \label{eq:FVconsform}
  \partial_t \vec u + \partial_A \vec f^A (\vec u) = \vec S( \vec u) \,,
\end{equation}
and they adopt the finite volume approach.
Most methods consist of three stages: starting from the cell averages,
the numerical solution is first \emph{reconstructed} at the cell
interfaces. A \emph{numerical flux} is then computed from the
reconstructed values. Finally, the numerical solution is integrated
forward in time.
 
Since the Z(2+1)+1 system discussed in chapter \ref{sec:Z211} is
written precisely in the form \eqref{eq:FVconsform}, one might try to
apply finite volume methods to that system as well. Our experiments
indicated that this is not feasible for the following reasons.

Consider first the reconstruction procedure. We found that 
reconstructions which adapt themselves to discontinuities, such as
the \emph{weighted essentially non-oscillatory (WENO)} reconstruction
\cite{Shu97} and the \emph{slope-limited} reconstruction \cite{Toro}, 
led to instabilities unless an extremely high resolution was used. 

Consider next the numerical flux. Many numerical fluxes used in
high-resolution shock capturing methods are based on the solution of
the Riemann problem and require that one can compute the
characteristic variables from the conserved variables and {\it vice versa}. 
It turns out that this cannot be done in a regular way close to the
axis for our system (section \ref{sec:charregularity}).
Therefore we have to resort to numerical fluxes that do not require
any knowledge of the characteristic structure. Possible candidates include 
the basic Lax-Friedrichs flux and the FORCE flux used in the
SLIC and FLIC methods of Toro \cite{Toro}. Unfortunately, these fluxes are
very viscous and led to a severe damping of the waves in our experiments.
The recent higher-order MUSTA flux of Toro and Titarev \cite{ToroMUSTA1, 
ToroMUSTA2} might be a promising alternative.

We have decided to stick with centred finite-differencing combined
with the method of lines as described in the previous sections 
because it can be applied to second-order as well as first-order
(in space) PDEs, it does not include any artificial viscosity, 
and it appears to be very stable.

However, our long-term goal is to include a perfect fluid, and then we
will have to use finite-volume methods at least for the matter equations.
Such methods have been successfully applied to general relativistic
hydrodynamics (see \cite{Font00} for a comprehensive review),
including a study of perfect fluid critical collapse in
Friedmann-Robertson-Walker spacetimes \cite{Hawke02, HawkePhD}.

\subsection{Conjugate gradient methods}
\label{sec:altnumCG}

The convergence of the Multigrid method depends crucially on the underlying
relaxation. If the diagonal dominance condition (\ref{eq:diagdom}) is strongly
violated so that the relaxation diverges then Multigrid will also diverge.

A different class of linear relaxation methods that do not suffer from this 
problem
are based on the \emph{conjugate gradient (CG) method} 
(see \cite{Shewchuk94} for an elementary introduction). 
This can be interpreted as a function minimization algorithm applied to the 
norm of the residual,
\begin{equation}
  r(\vec v) \equiv \lVert \vec f - A \vec v \rVert_2
\end{equation}
in the notation of section \ref{sec:MG}.
The advantage of these methods as compared with direct matrix solvers is that
the matrix $A$ is only ever referred to in the form $A \vec v$.
Since $A$ is sparse for finite difference discretizations, this matrix-vector
product can be implemented very efficiently.

While the original conjugate gradient method only converges for symmetric
positive definite matrices $A$, the \emph{bi-conjugate gradient method}
in principle converges for any non-singular $A$.
An improved version of that algorithm is
the bi-conjugate gradient stabilized (BiCGStab) method
of van der Vorst \cite{Vorst92}.
The convergence of CG methods depends crucially on the use
of the so-called \emph{preconditioner}, a matrix $B$ that approximates $A$ and
is easy to invert. One then applies the conjugate gradient method to 
$B^{-1} A$, which is closer to the unit matrix than $A$ and exhibits 
a much faster convergence.
A preconditioner that is straightforward to implement and that worked 
well in our experiments is the 
SSOR preconditioner (e.g., \cite{StoerBulirsch}).

Still, the complexity of CG methods is greater than that of Multigrid:
at least $O(N \ln N)$ operations 
are required, as compared with $O(N)$ operations for Multigrid.
In principle, one can also use CG as a relaxation \emph{within} 
Multigrid but this is not very efficient because CG methods do not share the
smoothing property of the classical relaxation schemes (section
\ref{sec:MGrelax}).
In addition, CG methods are not capable of dealing with nonlinear problems
directly. Hence an outer Newton-Raphson iteration has to be applied, which
multiplies the workload.

If Multigrid converges, it is always more advisable to use it rather than CG
methods. We have successfully implemented the BiCGStab algorithm for the 
Hamiltonian constraint \eqref{eq:axihamcons} in situations when Multigrid 
fails for that equation (e.g., in strong Brill wave evolutions).
However, as explained later in section \ref{sec:ellsolvana}, the Hamiltonian
constraint might have multiple solutions in those situations. Preliminary 
results indicate that the CG method sometimes appears to converge to a 
``wrong solution'', i.e., one that is not compatible with the remaining 
equations (see also the remarks in section \ref{sec:hypellconcl}).


 \section{Adaptive mesh refinement}
\label{sec:AMR}

So far we have only considered grids that are uniform across the whole
computational domain. The solution of hyperbolic partial differential
equations on such a grid can be very inefficient if the solution contains a
wide range of relevant length scales. To resolve the small-scale features, a
high resolution is needed, whereas a much lower resolution would be sufficient
in smooth regions. Since the computational workload scales (roughly) linearly
with the number of grid points, having a single grid with a uniformly high
resolution is impractical. One way out would be to use a non-uniform grid.
The main disadvantage of this is that because of the CFL condition (section
\ref{sec:MOL}), the whole grid needs to be evolved with a time step restricted
by the smallest grid spacing. This is avoided by the \emph{adaptive mesh 
refinement (AMR)} technique, whereby the computational domain is 
covered by a dynamical hierarchy of uniform grids of increasing resolution,
each advanced with its own time step.
During a numerical simulation, refined grids are added in regions
where and when they are needed, and discarded when they become obsolete.

AMR was invented in 1984 by Berger and Oliger \cite{BergerOliger84}.
It was first applied to numerical relativity by Choptuik \cite{Choptuik93}
to study the critical collapse of a massless scalar field in spherical
symmetry. Further one-dimensional applications include the same problem in
double-null coordinates by Hamad\'e and Stewart \cite{Hamade96} and 
perfect fluid collapse in Friedmann-Robertson-Walker universes by 
Hawke and Stewart \cite{Hawke02}. Two-dimensional examples include studies of
inhomogeneous cosmologies by Hern \cite{HernPhD} and scalar field 
critical collapse by Choptuik et al.~\cite{Choptuik03b}.
Three-dimensional AMR has been used in single and binary black hole
simulations, see e.g. Pretorius \cite{Pretorius05a} for a promising 
recent attempt.

In this work, we mainly use Berger and Oliger's classic algorithm (applied to 
two spatial dimensions), with certain simplifications and modifications 
described below. The implementation is based on a code originally 
written by Stewart and Hern (see \cite{HernPhD} for a detailed description). 
Changes have been made mainly to the boundary treatment (section
\ref{sec:AMRtimestep}), regridding procedure and refinement criteria
(section \ref{sec:AMRregrid}). One of the outstanding features of the AMR part
of the code is that it is to a large extent independent of the other parts
such as the time integrator, the initial data solver, and the actual
implementation of the equations being solved.

We have only implemented AMR for hyperbolic systems of partial differential
equations. The extension to mixed hyperbolic-elliptic systems and the
combination of AMR with the Multigrid method for elliptic equations (section
\ref{sec:MG}) is rather complicated, although some progress has been made
(see \cite{Pretorius05b} for an implementation with numerical 
relativity in mind).

\subsection{The grid hierarchy}
\label{sec:AMRhierarchy}

The building blocks of the AMR algorithm are the uniform \emph{grids} 
described in section \ref{sec:FDgrid}. We use cell-centred grids rather 
than vertex-centred ones (this is not an essential restriction -- it only 
affects the details of the interpolation between grids). The grids are 
arranged in a hierarchy in the following way: the hierarchy consists of
$l_\mathrm{max}$ \emph{levels}. Each level contains grids of the same 
resolution. Level 1 only contains a single grid, the \emph{base grid}, 
the next coarsest level is 2 and so on until level $l_\mathrm{max}$, 
which contains the finest grids in the hierarchy.

The grid spacings of two consecutive levels are related by
$h_l / h_{l+1} = \rho$, where $\rho \geqslant 2$ is an integer.
We choose $\rho = 2$ in our applications.

Each \emph{child grid} on level $l+1$ is entirely contained within its 
\emph{parent grid} on level $l$, a property called \emph{proper nesting}.
Unlike in Berger and Oliger's original work \cite{BergerOliger84}, 
we require that grids on a given level do not overlap.
Furthermore, we require that all grids be aligned with the boundaries of the
computational domain (Berger and Oliger allowed for rotated grids).
Figure \ref{fig:AMRhierarchy} shows an example of a typical AMR hierarchy.

\begin{figure}[t]
 \setlength{\unitlength}{1cm}  
 \begin{center}
  \begin{picture}(10,5)
    \put(0,0){\framebox(10,5)}
    \multiput(0,0)(1,0){10}{\dashbox{0.05}(1,1)}
    \multiput(0,1)(1,0){10}{\dashbox{0.05}(1,1)}
    \multiput(0,2)(1,0){10}{\dashbox{0.05}(1,1)}
    \multiput(0,3)(1,0){10}{\dashbox{0.05}(1,1)}
    \multiput(0,4)(1,0){10}{\dashbox{0.05}(1,1)}
    \put(0,0){\framebox(3,3)}
    \multiput(0,0)(0.5,0){6}{\dashbox{0.05}(0.5,0.5)}
    \multiput(0,0.5)(0.5,0){6}{\dashbox{0.05}(0.5,0.5)}
    \multiput(0,1)(0.5,0){6}{\dashbox{0.05}(0.5,0.5)}
    \multiput(0,1.5)(0.5,0){6}{\dashbox{0.05}(0.5,0.5)}
    \multiput(0,2)(0.5,0){6}{\dashbox{0.05}(0.5,0.5)}
    \multiput(0,2.5)(0.5,0){6}{\dashbox{0.05}(0.5,0.5)}
    \put(7,1){\framebox(2,3)}
    \multiput(7,1)(0.5,0){4}{\dashbox{0.05}(0.5,0.5)}
    \multiput(7,1.5)(0.5,0){4}{\dashbox{0.05}(0.5,0.5)}
    \multiput(7,2)(0.5,0){4}{\dashbox{0.05}(0.5,0.5)}
    \multiput(7,2.5)(0.5,0){4}{\dashbox{0.05}(0.5,0.5)}
    \multiput(7,3)(0.5,0){4}{\dashbox{0.05}(0.5,0.5)}
    \multiput(7,3.5)(0.5,0){4}{\dashbox{0.05}(0.5,0.5)}
    \put(5,3){\framebox(2,2)}
    \multiput(5,3)(0.5,0){4}{\dashbox{0.05}(0.5,0.5)}
    \multiput(5,3.5)(0.5,0){4}{\dashbox{0.05}(0.5,0.5)}
    \multiput(5,4)(0.5,0){4}{\dashbox{0.05}(0.5,0.5)}
    \multiput(5,4.5)(0.5,0){4}{\dashbox{0.05}(0.5,0.5)}
    \put(0,0){\framebox(1.5,1.5)}
    \multiput(0,0)(0.25,0){6}{\dashbox{0.05}(0.25,0.25)}
    \multiput(0,0.25)(0.25,0){6}{\dashbox{0.05}(0.25,0.25)}
    \multiput(0,.5)(0.25,0){6}{\dashbox{0.05}(0.25,0.25)}
    \multiput(0,0.75)(0.25,0){6}{\dashbox{0.05}(0.25,0.25)}
    \multiput(0,1)(0.25,0){6}{\dashbox{0.05}(0.25,0.25)}
    \multiput(0,1.25)(0.25,0){6}{\dashbox{0.05}(0.25,0.25)}
    \put(7.5,1.5){\framebox(1,1.5)}
    \multiput(7.5,1.5)(0.25,0){4}{\dashbox{0.05}(0.25,0.25)}
    \multiput(7.5,1.75)(0.25,0){4}{\dashbox{0.05}(0.25,0.25)}
    \multiput(7.5,2)(0.25,0){4}{\dashbox{0.05}(0.25,0.25)}
    \multiput(7.5,2.25)(0.25,0){4}{\dashbox{0.05}(0.25,0.25)}
    \multiput(7.5,2.5)(0.25,0){4}{\dashbox{0.05}(0.25,0.25)}
    \multiput(7.5,2.75)(0.25,0){4}{\dashbox{0.05}(0.25,0.25)}
  \end{picture}   
 \end{center}
  \caption[AMR grid hierarchy]
  {\label{fig:AMRhierarchy}  \footnotesize
  Example of an AMR grid hierarchy with 3 levels. The grid functions are
  defined at the cell centres. Realistic grids contain much more cells.}
\end{figure}
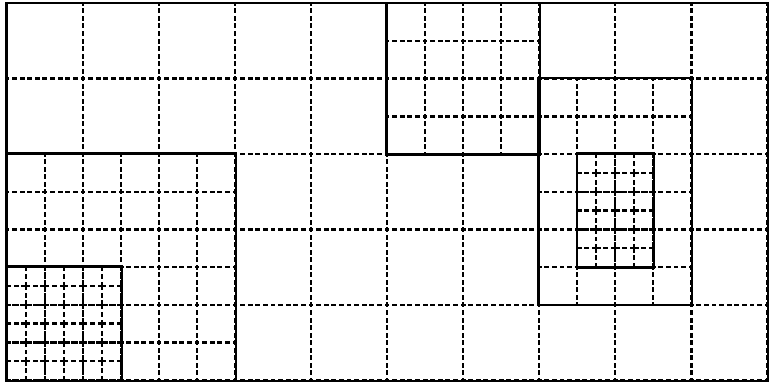

\subsection{Time-stepping the grid hierarchy}
\label{sec:AMRtimestep}

The grids are advanced in time by the following recursive procedure:
first all grids on level $l$ are advanced one time step to time $t + \Delta
t_l$, then all grids on level $l+1$ are advanced to the same time.
To be consistent with the CFL condition (\ref{eq:CFL1}--\ref{eq:CFL3}), 
the time step on level $l+1$ has to be
refined by the same ratio as the grid spacing: $\Delta t_l / \Delta t_{l+1} =
\rho$. This means that $\rho$ time steps have to be taken on level $l+1$ 
until it has caught up with level $l$.

Boundary data has to provided before a time step can be carried out.
We distinguish between \emph{external} boundaries, which coincide with the 
boundaries of the computational domain, and \emph{internal} ones, which lie in
the interior of the computational domain. External boundaries of a given grid
are dealt with outside of the AMR algorithm -- the boundary conditions to be 
enforced there depend on the problem being solved.
To provide data at internal boundaries, we exploit the fact that when
a grid is to be advanced in time, its parent grid has already been
advanced. Interpolation in time and space is used to
interpolate the data from the parent grid to the ghost cells of the child grid
(see section \ref{sec:ghosts} for the ghost cell technique). 
This requires storage of data from both 
the current and the previous time step on the parent grid. 
The interpolation scheme we use is trilinear interpolation.
We have not experienced any problems with high-frequency noise at the grid
boundaries as reported by some authors. The same fourth-order Kreiss-Oliger
dissipation operator \eqref{eq:diss4} is applied at all interior grid points,
with no modifications necessary close to the boundaries.

Once a child grid on level $l+1$ has reached the same time as its parent grid
on level $l$, the data on the child grid is \emph{injected} into the parent
grid. For cell-centred grids, this involves some sort of interpolation because
the cell centres of the parent grid do not coincide with cell centres of the
child grid (we use bilinear interpolation). By this update step, the most
accurate solution available at a given point is propagated to all grids in the
hierarchy.

When implementing systems of hyperbolic conservation laws using the
finite volume method, one has to ensure that the scheme is
conservative across the grid boundaries. This implies that certain
modifications have to be made to the fluxes at the 
boundaries \cite{BergerColella88}, a technique known as \emph{refluxing}.
We have not implemented this because we are using the finite difference 
rather than the finite volume method, and the ``conserved'' quantities
have no physical significance in our problem (cf.~section 
\ref{sec:firstorder}).

\subsection{Adapting the grid hierarchy}
\label{sec:AMRregrid}

The most powerful feature of the AMR algorithm is its ability to automatically
adapt the grid hierarchy in order to maintain an appropriate resolution of the
data at all times. The process of adding, removing or extending grids is 
called \emph{regridding}. On each level, regridding takes place every
$T_r$ time steps, where $T_r$ is a user-defined integer (we choose $T_r = 4$).
We need to address two questions: firstly, how to decide where on a given grid
refinement is needed and secondly, how to rearrange the grid hierarchy.

At the start of the regridding phase, grid cells that fail to 
meet a certain accuracy criterion are \emph{flagged}. 
A common criterion is based
on an estimate of the local truncation error via \emph{Richardson
  extrapolation}. In the original Berger and Oliger algorithm, this is
implemented in the following way. Two copies of the grid under investigation
are made, the first one, $G^h$, being an identical copy, the second one,
$G^{2h}$, a version coarsened by a factor of 2. Two time steps of size 
$\Delta t$ are taken on $G^h$ and a single step of size $2 \Delta t$ on 
$G^{2h}$. Let us write the resulting approximation on $G^h$ as
\begin{equation}
  \label{eq:rich1}
  \vec u^h = \vec u_0 + \vec e^h
\end{equation}
and the one on $G^{2h}$ as
\begin{equation}
  \label{eq:rich2}
  \vec u^{2h} = \vec u_0 + \vec e^{2h} \,,
\end{equation}
where $\vec u_0$ denotes the (generally unknown) exact solution. For 
second-order 
accurate finite differencing, the error should behave like $\sim h^2$ so 
that the errors on the two grids are related by
\begin{equation}
  \label{eq:rich3}
  \vec e^{2h} \approx 4 \vec e^{h} \,. 
\end{equation}
Subtracting \eqref{eq:rich1} from \eqref{eq:rich2} and using \eqref{eq:rich3},
we obtain an estimate for the error on $G^h$:
\begin{equation}
  \label{eq:errorestimate}
  \vec e^h \approx \third (\vec u^{2h} - \vec u^h) \,.
\end{equation}
A grid cell is flagged if the absolute value (or some norm, in the
case of systems) of the estimated error there is bigger than a 
user-defined threshold.

We can simplify Berger and Oliger's implementation by recalling that at the
time when a child level and its parent level are in synchrony, just before
the child$\rightarrow$parent update step, the information required for
computing the error estimate is readily available. All we need to do is
interpolate the data from the parent grid to a copy of the child grid and form
the difference \eqref{eq:errorestimate}.\footnote{Strictly speaking, the fine
  grid solution might have been modified by injections from even finer grids
  in the meantime. The correction to the estimate \eqref{eq:errorestimate}
is of order $(h/\rho)^2$, i.e., it is suppressed by a factor of $\rho^2$.}
Because the error is $O(h^2)$, it is
important that one uses more than second-order accurate interpolation here (we
choose biquadratic interpolation). This simplified scheme has been called a
\emph{self-shadow hierarchy} \cite{Pretorius05b} because it is no longer
necessary to create a separate ``shadow'', i.e., a copy of a grid that
is merely advanced for truncation error estimation. 
The only level where this procedure does not
work is the coarsest one, for this does not have a parent. We therefore
require that the coarsest level always be fully refined. The resolution on
level 2 should then be chosen to match the desired coarsest resolution;
level 1 is merely used for error estimation (the overhead that this causes is
small because the number of grid points on level 1 is only a fourth of that
on level 2).

Error estimation via Richardson extrapolation is by no means the only 
possible refinement criterion. For instance, an indication of how well
a function is resolved can be obtained by evaluating its (suitably
normalized) second spatial derivative \cite{Loehner87, Fryxell00}.
We have experimented with similar criteria but did not find them
appropriate for the wavelike solutions considered in this thesis.
Alternatively, one could use certain physically motivated quantities as
refinement indicators, e.g., the distortion of a spacelike slice 
or the residuals of the constraints. More details on this will follow in
section \ref{sec:brillcollapse} when we discuss our particular application.

Suppose now we have obtained an array of error flags. Next, the flagged 
regions are enlarged by so-called \emph{buffer zones}. These ensure that 
high-error features of the
solution do not propagate out of the refined regions before the next
regridding step is performed. The width of the buffer regions required 
depends on the choice of $T_r$ and the Courant number $\lambda$: signals 
that propagate at speed $c$ can travel at most $c \lambda T_r$ 
grid spacings in $T_r$ time steps. Because $\lambda \lesssim 1$ is needed 
for stability and $c \leqslant 1$ (the speed of light), a safe choice is 
a buffer width of $T_r$ cells.\footnote{Depending on the width of the stencil
  used for intergrid interpolation, a slightly larger buffer zone may be 
  required.} 
Also, regions of the grid on level $l$ that 
are covered by \emph{grandchildren} (i.e., level $l+2$ grids) are flagged. 
This ensures that (the union of) the newly created child grids contain 
the grandchild grids. 
The modified array of flags is then passed to a \emph{clustering} algorithm, 
which creates a set of rectangular child grids containing the flagged points.
The algorithm we use is based on techniques common in computer vision and
pattern recognition and is described in detail in Berger and 
Rigoutsos \cite{BergerRigoutsos91}. The user can specify a target
value for the approximate \emph{filling factor} $F$, the ratio of flagged 
cells to the total number of cells in a child grid. A high value for 
this will produce many small child grids, a low value few large ones.
Both extremes are computationally unefficient; a good compromise
appears to be $F \approx 0.7$. 
The clustering algorithm had to be modified slightly in order to ensure that
each existing grandchild grid is contained within a single newly
created child grid, so that proper nesting is maintained.

After clustering, the child grids are filled with data interpolated from 
their parent (we use bilinear interpolation for this). 
If there are existing grids on level
$l+1$ that overlap with the newly created grids, their data is used instead
before they are destroyed.

When it is time to regrid level $l$, we carry out the above operations 
first on the finest level $l_\mathrm{max}$, then on level 
$l_\mathrm{max} - 1$ and so on until level $l$ itself.
This is Berger and Oliger's original ``top-down'' approach; 
a different ``down-top'' approach is used by Hern \cite{HernPhD}.


  \chapter{A mixed hyperbolic-elliptic system}
\label{sec:hypell}

In this chapter we present a first evolution scheme based on
the (2+1)+1 formalism. Apart from hyperbolic evolution equations, it
contains elliptic equations that need to be solved at each time
step, which distinguishes it from the completely hyperbolic Z(2+1)+1 
system considered in the following chapters.
This chapter is mainly based on an earlier essay \cite{RinneEssay}.

We begin by explaining our gauge conditions (section \ref{sec:ellgauge}), 
which are motivated by geometric considerations and by simplifying the
system as much as possible. They lead to elliptic equations for the
lapse function and the shift vector. 
We then turn to the issue of regularity on axis and explain how the
regularity conditions of chapter \ref{sec:impl} can be enforced by
an appropriate choice of variables (section \ref{sec:hypellreg}).
The final equations to be solved are written out in terms of these 
variables in section \ref{sec:axieqns}. 
Three alternate evolution schemes are then discussed
that differ in the way the constraint equations are treated, ranging
from a fully constrained scheme to a free evolution scheme
(section \ref{sec:hypellschemes}).
The elliptic equations occurring in the different schemes 
are investigated further with regard to
uniqueness of solutions and numerical solvability with Multigrid methods
(section \ref{sec:ellsolv}).
The constraint evolution system is derived in section \ref{sec:hypellsubs}
and its well-posedness is analyzed for the three evolution schemes.
Numerical evolutions of generalized Brill waves, including twist,
are presented in section \ref{sec:hypellnum}.
Evidence for the existence of a critical point
separating dispersal and black hole formation is given, and the
present limitations of the code are indicated.

\section{Elliptic gauge conditions}
\label{sec:ellgauge}

To complete our evolution formalism, we have to come up with a
prescription for the gauge variables, which specify the coordinate system.
We would like to tie them to the evolution so that they can adapt 
themselves, for instance when a spacetime singularity is approached.

Consider first the prescription for the lapse function, also
called the \emph{slicing condition}. The one we choose here is
\emph{maximal slicing}. Its name arises from the fact that it
maximizes the proper volume of the individual spacelike slices \cite{York79}.
This suggests that when a high-curvature region of spacetime is
approached, the slices try to avoid that region and ``pile up'' in
front of it. In this way, a large part of spacetime can be explored
without hitting a potential singularity.
Maximal slicings have been constructed analytically for Schwarzschild
spacetime \cite{Estabrook72} but the singularity avoiding
property has also been shown to hold in more general 
situations \cite{Eardley79}.

For maximal slices, the trace of the second fundamental form of the
\emph{three-}dimensional (i.e., including the dimension of the Killing
vector) $t = \mathrm{const.}$ surfaces vanishes.
In $(2+1)+1$ language, this translates into
\begin{equation}
  \label{eq:maximalslicing}
  \chi + \K = 0 \, .
\end{equation}
As we want this condition to hold at all times, we also require its
Lie derivative along the normal lines to vanish.
$\Lie{n} \chi = \Lie{n} ( H^{AB} \chi_{AB} )$ can be computed using the
evolutions equations \eqref{eq:d0H} and \eqref{eq:d0chi}, and
$\Lie{n} \K$ is given by \eqref{eq:d0Kpp}. Together we obtain
\begin{eqnarray}
  \Lie{n} \left( \chi + \K \right) &=& - \alpha^{-1}
  \alpha_{\p A}{}^A + \two{R} + \left( \chi + \K \right)^2  
  - \lambda^{-1} \alpha^{-1} \lambda_{,A} \alpha^{,A} \nonumber\\&&
  - 2 \lambda^{-1} \lambda_{\p A}{}^{A} - \half \lambda^2 {B^\varphi}^2 
  + \half \kappa \left( \tau + S_A{}^A - 3 \rho_H \right) \,.
\end{eqnarray}
If we now impose (\ref{eq:maximalslicing}) and subtract twice the
Hamiltonian constraint \eqref{eq:hamcons} from the right-hand-side, 
we end up with the following elliptic equation for $\alpha$:
\begin{eqnarray}
  \label{eq:slicing}
  \S &\equiv& \alpha_{\p A}{}^A 
  - \alpha \big[ \chi^{AB} \chi_{AB} + {\K}^2 
  - \lambda^{-1} \alpha^{-1} \lambda_{,A} \alpha^{,A}
  + \half \lambda^2 E_A E^A \nonumber\\&& \qquad \qquad
  + \half \kappa \left( \rho_H + \tau + S_A{}^A \right) \big] = 0 \, .
\end{eqnarray}

Next we deal with the shift vector. We exploit the fact that every
two-dimensional Riemannian manifold is conformally flat. 
Hence we can choose coordinates such that at all times
\begin{equation}
  \label{eq:confflat}
  H_{rr} = H_{zz} \equiv \sqrt{H} \,, \qquad H_{rz} = 0 \,.
\end{equation}
This conformal flatness condition thus simplifies the 
system by reducing the number of variables to be evolved: $H_{AB}$ now
has only 1 instead of 3 degrees of freedom. This gauge is also known
as \emph{Wilson gauge} \cite{Wilson79} in the literature.

Imposing \eqref{eq:confflat} on the time derivative of $H_{AB}$
\eqref{eq:d0H} now implies that
\begin{eqnarray}
  \label{eq:betam}
  \beta_- \equiv 
  \beta^r{}_{,r} - \beta^z{}_{,z} &=& \alpha (\chi_r{}^r - \chi_z{}^z)  \,,\\
  \label{eq:betap}
  \beta_+ \equiv 
  \beta^r{}_{,z} + \beta^z{}_{,r} &=& 2 \alpha \chi_r{}^z \,.
\end{eqnarray}
By forming the combinations $\partial_r$ \eqref{eq:betam} $+ \, \partial_z$
\eqref{eq:betap} and $-\partial_z$ \eqref{eq:betam} $+ \, \partial_r$
\eqref{eq:betap}, we arrive at the following Poisson equations for the
components of the shift vector:
\begin{eqnarray}
  \label{eq:shiftr}
  \S_r &\equiv& \beta^r{}_{,rr} + \beta^r{}_{,zz} 
  - \left[ 2 \alpha \chi_r{}^z \right]_{,z}
  - \left[ \alpha (\chi_r{}^r - \chi_z{}^z) \right]_{,r} = 0\, ,\\
  \label{eq:shiftz}
  \S_z &\equiv& \beta^z{}_{,rr} + \beta^z{}_{,zz} 
  - \left[ 2 \alpha \chi_r{}^z \right]_{,r}
  + \left[ \alpha (\chi_r{}^r - \chi_z{}^z) \right]_{,z} = 0\, .
\end{eqnarray} 
An alternate way of solving for the shift vector using the momentum
constraints will be explained in section \ref{sec:hypellschemes}.


  \section{Regularity on axis}
\label{sec:hypellreg}

As emphasized in chapter \ref{sec:impl}, great care must be taken to
enforce the regularity conditions for axisymmetric tensor fields
during a numerical evolution because otherwise certain terms in the
evolution equations become singular on the axis.

Let us first deal with the regularity condition \eqref{eq:regcond4}
applied to the spacetime metric $g_{\alpha\beta}$. It implies that
\begin{equation}
  \frac{g_{\varphi\varphi}}{r^2 g_{rr}} = \frac{\lambda^2}{r^2 \sqrt{H}}
  = 1 + O(r^2) \,.
\end{equation}
If we evolved $\sqrt{H}$ and $\lambda$ independently, this subtle
relation would soon be violated during the evolution 
due to numerical errors. Instead, we replace $\sqrt{H}$ and $\lambda$
with new variables $\psi$ and $s$ defined by
\begin{equation}
  \label{eq:psisdef}
  \sqrt{H} = \psi^4 \e^{2rs} \,, \qquad \lambda = r \psi^2 \,,
\end{equation}
where $s = O(r)$ and $\psi = O(1)$ on the axis. 
Indeed, this guarantees that
\begin{equation}
   \label{eq:regok}
   \frac{\lambda^2}{r^2 \sqrt{H}} = \e^{-2rs} \approx 1 - 2rs = 1 + O(r^2)
\end{equation}
for small $r$. 
Equation \eqref{eq:psisdef} is the choice of variables made in
Garfinkle and Duncan \cite{Garfinkle00}. Another possibility
(satisfying \eqref{eq:regok} with $s$ replaced by $-s$) is
\begin{equation}
  \label{eq:psisdef2}
  \sqrt{H} = \psi^4  \,, \qquad \lambda = r \psi^2 \e^{rs} \,,
\end{equation}
which are the variables of Choptuik et al. \cite{Choptuik03a}.
We have also considered replacing $\psi^4$ in the above with $\exp 2\psi$, 
which has the advantage that no logarithmic derivatives
$\psi^{-1} \psi_{,A}$ appear in the final equations and thus many
nonlinearities drop out.
All the various choices have been implemented and it was found that
strong Brill wave evolutions were slightly more stable for
\eqref{eq:psisdef} than for the other possibilities. We therefore adopt
that choice in the following. 

The corresponding regularity condition for the extrinsic curvature
implies that
\begin{equation}
  \K = r^{-2} K_{\varphi\varphi} + O(r^2) = \chi_{rr}
  + O(r^2) = \chi_r{}^r + O(r^2) \,.
\end{equation}
Hence $\K - \chi_r{}^r$ is $O(r^2)$ on axis, while each single term is
$O(1)$. This will not hold numerically if we evolve $\K$ and
$\chi_r{}^r$ separately and so we introduce a new variable $Y$ defined
by\footnote{This definition is the same as in \cite{Choptuik03a}. The
  variable $W$ in \cite{Garfinkle00} is $W = -Y$.}
\begin{equation}
  \label{eq:axiYdef}
  Y \equiv r^{-1} (\K - \chi_r{}^r) \,,
\end{equation}
where $Y = O(r)$ near the axis.
Because of the maximal slicing condition $\chi + \K = 0$, the
extrinsic curvature has only two more independent components, which
in agreement with \cite{Garfinkle00} are taken to be
\begin{equation}
  U \equiv \chi_r{}^r - \chi_z{}^z \,,\qquad X \equiv \chi_r{}^z =
  \chi_z{}^r \,.
\end{equation}

Similarly, the regularity condition for the energy-momentum tensor implies
that $\tau - S_r{}^r$ is $O(r^2)$ on axis and so we redefine $\tau$ by
\begin{equation}
  \tilde \tau \equiv r^{-1} (\tau - S_r{}^r) \,.
\end{equation}

The set of variables we evolve is summarized in table \ref{tab:axirpar},
which also states their small-$r$ behaviour. 
This determines the boundary conditions we impose on the axis: 
for a variable $\underline{u}$ that is $O(r)$ on the axis, 
a Dirichlet condition
\begin{equation}
 \underline{u} \vert_{r = 0} = 0
\end{equation}
is needed, and for a variable $u$ that is $O(1)$ on axis,
we enforce a Neumann condition 
\begin{equation}
  \partial_r u \vert_{r = 0} = 0 \,.
\end{equation}
Now we also see why our definitions of the twist variables
(\ref{eq:E}--\ref{eq:B}) differ from those in Maeda et al. \cite{Maeda80} 
by factors of $\lambda$ (which is $O(r)$ on the axis): if we had adopted
their definition then the variables $B^\varphi$ and $E^r$ would have
been $O(r^2)$ on the axis, which is difficult to enforce numerically
because we can easily impose a Dirichlet condition or a Neumann
condition, but not both at the same time.

\begin{table}
  \begin{eqnarray*}
    \alpha, \underline{\beta^r}, \beta^z, \psi, \underline{s}, 
    \underline{Y}, U, \underline{X}, \underline{B^\varphi},
    \underline{E^r}, E^z , \\
    \rho_H, \sigma, J_\varphi, \underline{J_r}, J_z,
    \underline{\tilde \tau}, \underline{S^r}, S^z,
    \underline{\Sigma^r}, \Sigma^z, S_r{}^r, 
    \underline{S_r{}^z}, S_z{}^z . 
  \end{eqnarray*}
  \caption[Hyperbolic-elliptic system: $r$-parity of the variables]
  {\label{tab:axirpar} \footnotesize
    Variables of the hyperbolic-elliptic
    system and their small-$r$ behaviour. 
    Underlined variables are $O(r)$, the remaining variables are 
    $O(1)$ on the axis.}
\end{table}


  \section{Final equations}
\label{sec:axieqns}

We are now ready to write out the equations to be solved in our gauge 
and variables.

\paragraph{The gauge conditions\\}
The slicing condition \eqref{eq:slicing} is
\begin{eqnarray}
  \label{eq:axislicing}
  \hat \S &\equiv& \alpha_{,rr} + r^{-1} \alpha_{,r} + \alpha_{,zz} 
  + 2 \psi^{-1} (\psi_{,r} \alpha_{,r} + \psi_{,z} \alpha_{,z}) \nonumber\\&&
  - \alpha \psi^4 \e^{2rs} \big[ \tfrac{2}{3} (U - \half r Y)^2 
  + \half r^2 Y^2 + 2 X^2 \nonumber\\&&\qquad\qquad
  + \half r^2 \psi^8 \e^{2rs} ({E^r}^2 + {E^z}^2) \\&&\qquad\qquad
  + \half \kappa ( \rho_H + r \tilde \tau + 2 S_r{}^r + S_z{}^z )
  \big] = 0 \,.\nonumber
\end{eqnarray}
Equations (\ref{eq:betam}--\ref{eq:betap})
imply that $U$ and $X$ are given in terms of the shift by
\begin{eqnarray}
  \label{eq:axiU}
  U &=& - \alpha^{-1} \beta_- \,,\\
  \label{eq:axiX}
  X &=& \half \alpha^{-1} \beta_+ \,.
\end{eqnarray}
The elliptic shift conditions (\ref{eq:shiftr}--\ref{eq:shiftz}) read
\begin{eqnarray}
  \label{eq:axishiftr}
  \hat \S_r &\equiv& \beta^r{}_{,rr} + \beta^r{}_{,zz} 
    - (\alpha X)_{,z} + (\alpha U)_{,r}  = 0 \,,\\
  \label{eq:axishiftz}  
  \hat \S_z &\equiv& \beta^z{}_{,rr} + \beta^z{}_{,zz} -
    2 (\alpha X)_{,r} - (\alpha U)_{,z}  = 0 \,.  
\end{eqnarray}

\paragraph{The constraint equations\\}
The Hamiltonian constraint \eqref{eq:hamcons} becomes
\begin{eqnarray}
  \label{eq:axihamcons}
  \hat \C &\equiv& \psi_{,rr} + r^{-1} \psi_{,r} + \psi_{,zz} 
  + \quarter \psi (r s_{,rr} + 2 s_{,r} + r s_{,zz}) \nonumber\\&&
  + \psi^5 \e^{2rs} \big[ \tfrac{1}{3} (U - \half r Y)^2 
  + \quarter r^2 Y^2  + X^2 + \quarter \kappa \rho_H \big] \\&&
  + \tfrac{1}{16} r^2 \psi^9 \e^{2rs} \big[ {B^\varphi}^2
  + \psi^4 \e^{2rs} ({E^r}^2 + {E^z}^2) \big]  = 0 \,.\nonumber
\end{eqnarray}
The momentum constraints \eqref{eq:momcons} are
\begin{eqnarray}
  \label{eq:aximomconsr}
  \hat \C_r &\equiv& -\tfrac{1}{3} U_{,r} + X_{,z} 
   - \tfrac{1}{3} r Y_{,r} - 2 \psi^{-1}
   \psi_{,r} (r Y + U) + 6 \psi^{-1} \psi_{,z} X \nonumber\\&&
   - (r s_{,r} + s) U +
   2 r s_{,z} X - \tfrac{4}{3} Y - \half r^2 \psi^8 \e^{2rs} B^\varphi
   E^z \\&&- \kappa J_r  = 0\,,\nonumber\\
  \label{eq:aximomconsz}
  \hat \C_z  &\equiv& \tfrac{2}{3} U_{,z} + X_{,r} 
   - \tfrac{1}{3} r Y_{,z}
   + 6 \psi^{-1} \psi_{,r} X + 2 \psi^{-1} \psi_{,z} (2 U - r Y) \nonumber\\&&
   + (2 r s_{,r} + 2 s + r^{-1}) X + r s_{,z} U
   + \half r^2 \psi^8  \e^{2rs} B^\varphi E^r \\&&
   - \kappa J_z  = 0\,.\nonumber
\end{eqnarray}
The Geroch constraint \eqref{eq:gercons} is
\begin{eqnarray}
  \label{eq:axigercons}
  \hat \C_\varphi &\equiv& E^r{}_{,r} + E^z{}_{,z} 
  + E^r (10 \psi^{-1} \psi_{,r} + 2 r s_{,r} + 2 s + 3 r^{-1}) \nonumber\\&&
  + E^z (10 \psi^{-1} \psi_{,z} + 2 r s_{,z}) 
  - 2 \kappa J^\varphi = 0\,.
\end{eqnarray}

\paragraph{The evolution equations\\}
Equations (\ref{eq:d0H}--\ref{eq:d0lambda}) imply the following
evolution equations for $\psi$ and $s$:
\begin{eqnarray}
  \label{eq:axid0psi}
  \partial_t \psi &=& \beta^r \psi_{,r} + \beta^z \psi_{,z} + \psi
  \big[ \half r^{-1} \beta^r + \tfrac{1}{6} (U - 2 r Y) \big] \,,\\
  \label{eq:axid0s}
  \partial_t s &=& \beta^r s_{,r} + \beta^z s_{,z} + \alpha Y
  + (r^{-1} \beta^r)_{,r} + r^{-1} \beta^r s \,.
\end{eqnarray}
The evolution equations for the extrinsic curvature 
(\ref{eq:d0chi}--\ref{eq:d0Kpp}) become
\begin{eqnarray}
  \label{eq:axid0Y}   
  \partial_t Y &=& \beta^r Y_{,r} + \beta^z Y_{,z} + r^{-1} \beta^r Y
    - r^{-1} X (\beta^z{}_{,r} - \beta^r{}_{,z}) \nonumber\\&&
    + \psi^{-4} \e^{-2rs} \big[ (r^{-1} \alpha_{,r})_{,r} - r^{-1}
    \alpha_{,r} (r s_{,r} + s + 4 \psi^{-1} \psi_{,r}) 
    \nonumber\\&&\qquad\qquad\quad + \alpha_{,z} s_{,z} \big] \nonumber\\&&
    + \alpha \psi^{-4} \e^{-2rs} \big[ 2 \psi^{-1} (r^{-1}
    \psi_{,r})_{,r} + s_{,rr} + s_{,zz} + (r^{-1} s)_{,r} 
    \\&&\qquad\qquad\quad
    - 2 r^{-1} \psi^{-1} \psi_{,r} ( r s_{,r} + s + 3
    \psi^{-1} \psi_{,r}) \nonumber\\&&\qquad\qquad\quad
    + 2 \psi^{-1} \psi_{,z} s_{,z}   \big] \nonumber\\&&
    - r \alpha \psi^4 \big[ {B^\varphi}^2 + \psi^4 \e^{2rs} ({E^r}^2 +
    {E^z}^2) \big] - \kappa \alpha \tilde \tau\,,
\end{eqnarray}
\begin{eqnarray}
  \label{eq:axid0U}
  \partial_t U &=& \beta^r U_{,r} + \beta^z U_{,z} + 2 X
    (\beta^r{}_{,z} - \beta^z{}_{,r}) \nonumber\\&&
    + \psi^{-4} \e^{-2rs} \big[ 2 \alpha_{,z} (2 \psi^{-1} \psi_{,z} +
    r s_{,z}) - 2 \alpha_{,r} (2 \psi^{-1} \psi_{,r} + r s_{,r} + s) 
    \nonumber\\&&\qquad\qquad\quad + \alpha_{,rr} + \alpha_{,zz} 
    \big] \nonumber\\&&
    - 2 \alpha \psi^{-4} \e^{-2rs} \big[ - \psi^{-1} \psi_{,rr} +
    \psi^{-1} \psi_{,zz} + s_{,r} + r^{-1} s \\&& \qquad\qquad\qquad
    + \psi^{-1} \psi_{,r} ( 3 \psi^{-1} \psi_{,r} + 2 r s_{,r} + 2 s)
    \nonumber\\&&\qquad\qquad\qquad
    - \psi^{-1} \psi_{,z} (3 \psi^{-1} \psi_{,z} + 2 r s_{,z}) \big]
    \nonumber\\&&
    - \half r^2 \alpha \psi^8 \e^{2rs} ({E^r}^2 - {E^z}^2)
    + \kappa \alpha (S_r{}^r - S_z{}^z) \,, \nonumber\\
  \label{eq:axid0X}
  \partial_t X &=& \beta^r X_{,r} + \beta^z X_{,z} + \half U
    (\beta^z{}_{,r} - \beta^r{}_{,z}) \nonumber\\&&
    + \psi^{-4} \e^{-2rs} \big[ -\alpha_{,rz} + \alpha_{,r} (r s_{,z} + 2
    \psi^{-1} \psi_{,z}) \nonumber\\&&\qquad\qquad\quad
    + \alpha_{,z} (r s_{,r} + s + 2 \psi^{-1} \psi_{,r}) \big] \\&&
    + \alpha \psi^{-4} \e^{-2rs} \big[ - 2 \psi^{-1} \psi_{,rz} +
    \psi^{-1} \psi_{,r} (3 \psi^{-1} \psi_{,z} + 2 r s_{,z})
    \nonumber\\&& \qquad \qquad \qquad 
    + 2 \psi^{-1} \psi_{,z} (r s_{,r} + s) + s_{,z} \big] \nonumber\\&&
    + \half r^2 \alpha \psi^8 \e^{2rs} E^r E^z - \kappa \alpha S_r{}^z 
    \,.\nonumber
\end{eqnarray}}
The evolution equations for the twist variables
(\ref{eq:d0E}--\ref{eq:d0B}) are
\begin{eqnarray}
  \label{eq:axid0Er}
  \partial_t E^r &=& \beta^r E^r{}_{,r} + \beta^z E^r{}_{,z}
   \nonumber\\&&
    + \psi^{-4} \e^{-2rs} \big[ \alpha_{,z} B^\varphi + \alpha
    B^\varphi{}_{,z} + 6 \alpha \psi^{-1} \psi_{,z} B^\varphi \big] \\&& 
    + E^r \big[ \tfrac{2}{3} \alpha (2 r Y - U) - \beta^r{}_{,r} \big]
    - E^z \beta^r{}_{,z} - 2 \kappa \alpha S^r \,, \nonumber\\
  \label{eq:axid0Ez}
  \partial_t E^z &=& \beta^r E^z{}_{,r} + \beta^z E^z{}_{,z}
    \nonumber\\&&
    - \psi^{-4} \e^{-2rs} \big[ \alpha_{,r} B^\varphi + \alpha
    B^\varphi{}_{,r} + 3 \alpha B^\varphi (2 \psi^{-1} \psi_{,1} +
    r^{-1}) \big] \\&&
     - E^r \beta^z{}_{,r} + E^z \big[ \tfrac{1}{3} \alpha (4 r Y - 5 U) 
    - \beta^r{}_{,r} \big]  - 2 \kappa \alpha S^z \,,\nonumber\\
  \label{eq:axid0Bp}
  \partial_t B^\varphi &=& \beta^r B^\varphi{}_{,r} + \beta^z
    B^\varphi{}_{,z} + \tfrac{1}{3}\alpha B^\varphi (U - 2 r Y)
    \nonumber\\&&
    + 2 \alpha E^r (2 \psi^{-1} \psi_{,z} + r s_{,z})
    - 2 \alpha E^z (2 \psi^{-1} \psi_{,r} + r s_{,r} + s) \\&&
    + \alpha_{,z} E^r - \alpha_{,r} E^z + \alpha (E^r{}_{,z} -
    E^z{}_{,r}) \,. \nonumber
\end{eqnarray}

We focus on vacuum spacetimes in this chapter and so we do not include
the matter evolution equations here (see appendix \ref{sec:fluid} for
a discussion of a perfect fluid).

The above equations have been derived with the help of a programme
written in the computer algebra language REDUCE \cite{REDUCE}. 
Note that they are all regular on axis provided that the variables
have the correct small-$r$ behaviour (table \ref{tab:axirpar}).


  \section{Alternate evolution schemes}
\label{sec:hypellschemes}

In this section we explain in more detail how the various variables
are evolved.
The variables $s$, $Y$, $E^r$, $E^z$ and $B^\varphi$ 
are always evolved using the evolution equations 
(\ref{eq:axid0s}, \ref{eq:axid0Y}, \ref{eq:axid0Er}, \ref{eq:axid0Ez}, 
\ref{eq:axid0Bp}).
For the remaining variables, there are several possibilities,
and we discuss three alternate evolution schemes here.

\subsection{A free evolution scheme}
\label{sec:freescheme}

In the first scheme, one solves the elliptic gauge conditions
(\ref{eq:axislicing}, \ref{eq:axishiftr}--\ref{eq:axishiftz}) for the gauge 
variables $\alpha, \beta^r$ and $\beta^z$ and evolves the variables 
$\psi, U$ and $X$ using their evolution equations (\ref{eq:axid0psi}, 
\ref{eq:axid0U}, \ref{eq:axid0X}).

This scheme uses the maximum number of evolution equations to
update the variables. None of the constraints are solved during the
evolution, which is why this scheme is called a \emph{free evolution
  scheme}.

This is essentially the scheme of Garfinkle and Duncan \cite{Garfinkle00}
(although their scheme does not include the twist variables)
with the exception that they use the Hamiltonian constraint
\eqref{eq:axihamcons} to solve for $\psi$.

\subsection{A constrained evolution scheme}
\label{sec:consscheme}

In the second scheme, one eliminates the variables $U$ and $X$
completely using the relations (\ref{eq:axiU}--\ref{eq:axiX}).
The slicing condition \eqref{eq:axislicing} then takes the form
\begin{eqnarray}
  \label{eq:axislicingc}
  \hat \S^\mathrm{(C)} &\equiv& \alpha_{,rr} + r^{-1} \alpha_{,r} 
  + \alpha_{,zz} + 2 \psi^{-1}
  (\psi_{,r} \alpha_{,r} + \psi_{,z} \alpha_{,z}) \nonumber\\&&
  - \psi^4 \e^{2rs} \big[ \tfrac{2}{3} (\alpha^{-1} \beta_-^2 
  + \alpha r^2 Y^2 + r Y \beta_-) +
  \half \alpha^{-1} \beta_+^2 \nonumber\\&&\qquad\qquad
  + \half \alpha r^2 \psi^8 \e^{2rs} ({E^r}^2 + {E^z}^2) \\&&\qquad\qquad
  + \half \kappa \alpha ( \rho_H + r \tilde \tau + 2 S_r{}^r + S_z{}^z ) \big] 
  = 0 \,.\nonumber
\end{eqnarray}
To solve for the shift vector, we use the momentum constraints
(\ref{eq:aximomconsr}--\ref{eq:aximomconsz}), which can be written as
\begin{eqnarray}
  \label{eq:aximomconsrc}
  \hat \C_r^\mathrm{(C)} &=& \tfrac{2}{3} \beta^r{}_{,rr} + \beta^r{}_{,zz} 
  + \tfrac{1}{3}
  \beta^z{}_{,rz} + \beta_+ (6 \psi^{-1} \psi_{,z} + 2 r s_{,z} -
  \alpha^{-1} \alpha_{,z}) \nonumber\\&&
  + \tfrac{2}{3} \beta_- (6 \psi^{-1} \psi_{,r} + 3 r s_{,r} + 3 s -
  \alpha^{-1} \alpha_{,r}) - \tfrac{8}{3} \alpha Y \\&&
  - \tfrac{2}{3} \alpha r ( 6 \psi^{-1} \psi_{,r} Y + Y_{,r}) 
  - \alpha r^2 \psi^8 \e^{2rs} B^\varphi E^z - 2 \kappa \alpha J_r = 0
  \,, \nonumber\\
  \label{eq:aximomconszc}
  \hat \C_z^\mathrm{(C)} &=& \beta^z{}_{,rr} + \tfrac{4}{3} \beta^z{}_{,zz} 
  - \tfrac{1}{3}
  \beta^r{}_{,rz} - 2 \beta_- (4 \psi^{-1} \psi_{,z} + r s_{,z} - \tfrac{2}{3}
  \alpha^{-1} \alpha_{,z}) \nonumber\\&&
  + \beta_+ (6 \psi^{-1} \psi_{,r} + 2 r s_{,r} + 2 s + r^{-1} 
  - \alpha^{-1} \alpha_{,r}) \\&&
  - \tfrac{2}{3} \alpha r (6 \psi^{-1} \psi_{,z} Y + Y_{,z}) 
  + \alpha r^2 \psi^8 \e^{2rs} B^\varphi E^r 
  - 2 \kappa \alpha J_z = 0 \,.\nonumber
\end{eqnarray}
The Hamiltonian constraint \eqref{eq:axihamcons} becomes
\begin{eqnarray}
  \label{eq:axihamconsc}
  \hat \C^\mathrm{(C)} &=& \psi_{,rr} + r^{-1} \psi_{,r} + \psi_{,zz} 
  + \quarter \psi (r s_{,rr} + 2 s_{,r} + r s_{,zz}) \nonumber\\&&
  + \psi^5 \e^{2rs} \big[ \tfrac{1}{3} (\alpha^{-2} \beta_-^2 
  + r^2 Y^2 + r Y \alpha^{-1} \beta_-) + \quarter \alpha^{-2} \beta_+^2
  \nonumber\\&&\qquad \qquad  + \quarter \kappa \rho_H \big] \\&&
  + \tfrac{1}{16} r^2 \psi^9 \e^{2rs} \big[ {B^\varphi}^2
  + \psi^4 \e^{2rs} ({E^r}^2 + {E^z}^2) \big] = 0 \,.\nonumber
\end{eqnarray}
Equations (\ref{eq:axislicingc}--\ref{eq:axihamconsc}) form a system
of coupled elliptic equations that is solved for the variables $\alpha,
\beta^r, \beta^z$ and $\psi$.

This scheme uses the maximum number of elliptic equations. All the
constraints are enforced during the evolution except for the Geroch
constraint \eqref{eq:axigercons}. Hence we call this scheme a
\emph{constrained evolution scheme}.

This is essentially the scheme of Choptuik et al. \cite{Choptuik03a}
(although their scheme does not include the twist variables).

\subsection{A partially constrained evolution scheme}
\label{sec:modscheme}

Finally, we propose a new scheme that can be viewed as a
compromise between the two previous ones.
Here the variables $U$ and $X$ are first evolved to the next time
level using the evolution equations (\ref{eq:axid0U}--\ref{eq:axid0X}). 
The slicing condition \eqref{eq:axislicing} is then solved for
$\alpha$ as in the free evolution scheme. To solve for the shift,
however, we use the momentum constraints
(\ref{eq:aximomconsrc}--\ref{eq:aximomconszc}).
After solving for the lapse and shift, 
the variables $U$ and $X$ are immediately
overwritten by  (\ref{eq:axiU}--\ref{eq:axiX}).
Then the slicing condition and the momentum constraints are
again solved for the lapse and shift, and the procedure is iterated until
convergence. In this way, we enforce both the momentum constraints 
and the gauge conditions.
However, we do not solve the Hamiltonian constraint but evolve $\psi$ 
using its evolution equation \eqref{eq:axid0psi}.

This is the scheme we will use for the numerical evolutions in section
(\ref{sec:hypellnum}). We will explain its advantages over the other
schemes in the following sections.


  \section{Solvability of the elliptic equations}
\label{sec:ellsolv}

All the evolution systems presented in the previous section involve
(to varying extent) the solution of elliptic equations. The question
arises whether these equations are well-posed, i.e., whether a unique
solution exists. We would also like to know which numerical methods
can by used to solve the equations, in particular whether the
Multigrid method (section \ref{sec:MG}) will work.

\subsection{Analytical considerations}
\label{sec:ellsolvana}

The elliptic equations we encounter are of the general type
\begin{equation}
  \label{eq:ellmodel}
  L u \equiv a^{AB} \partial_A \partial_B u + b^A \partial_A u 
  + H(u, x^A) = f \,,
\end{equation}
where the coefficients $a^{AB}, b^A$ and the right-hand-side $f$
depend on the coordinates $x^A$ only and $H(u, x^A)$ may be a
nonlinear function. We assume that all of these are smooth.
The boundary consists of a part where we impose a 
Neumann condition $\partial_\n u = 0$ (on the axis $r = 0$) and a part
where we impose a Dirichlet condition $u = 0$ (the outer
boundaries)\footnote{Strictly speaking, $u = 0$ may only be imposed at
  infinity but the following analysis requires a bounded domain. We
  choose it to be sufficiently large such that the Dirichlet boundary 
  conditions are a good approximation.}.

Proving existence and uniqueness of solutions of \eqref{eq:ellmodel}
can be decidedly nontrivial. However, it is relatively easy to obtain
a necessary condition for the solution (should it exist) to be unique.
Suppose we are given a solution $u_0$ and we consider a small
perturbation $u = u_0 + \delta u$. For $u$ to be a solution as well, 
$\delta u$ must satisfy the linearized equation
\begin{equation}
  \label{eq:linellmodel}
    a^{AB} \partial_A \partial_B \delta u + b^A \partial_A
    \delta u + c \delta u = \tilde f \,,
\end{equation}
where
\begin{equation}
  c \equiv \frac{\partial H}{\partial u} \Big\vert_{u = u_0} \,.
\end{equation}
For the solution of \eqref{eq:ellmodel} to be unique, the solution of 
\eqref{eq:linellmodel} must also be unique (\emph{linearization stability}).
One can prove using the maximum principle \cite{ProtterWeinberger, 
GilbargTrudinger} that equation \eqref{eq:linellmodel} has a unique
solution satisfying the mixed Dirichlet-Neumann boundary conditions 
if the following conditions hold:
\begin{enumerate}
  \item the operator $L$ is \emph{elliptic}, i.e.,
  \begin{equation}
    \label{eq:ellipticity}
    a^{AB} \xi_A \xi_B > 0 \quad \forall \xi \in \mathbb{R}^2
    \setminus 0 \,,
  \end{equation}
  \item 
  \begin{equation}
    \label{eq:cnegative} 
    c \leqslant 0 \,.
  \end{equation}
\end{enumerate}

Let us check whether these conditions are satisfied for the elliptic
equations we would like to solve. 

\paragraph{The slicing condition.}

Consider first the slicing condition in the form \eqref{eq:axislicing}
as used in the free evolution scheme (section \ref{sec:freescheme}) and in
the partially constrained scheme (section \ref{sec:modscheme}). 
Its principal part is the Laplace operator, $a^{AB} = \delta^{AB}$, which
clearly satisfies \eqref{eq:ellipticity}. 
The equation is already linear with the coefficient $c$ in
\eqref{eq:linellmodel} given by 
\begin{eqnarray}
  c &=& - \psi^4 \e^{2rs} \big[ \tfrac{2}{3} (U - \half r Y)^2 + \half
  r^2 Y^2 + 2 X^2 \nonumber\\&&\qquad\qquad
  + \half r^2 \psi^8 \e^{2rs} ({E^r}^2 + {E^z}^2) \\&&\qquad\qquad
  + \half \kappa ( \rho_H + r \tilde \tau + 2 S_r{}^r + S_z{}^z ) \big] \,.
\end{eqnarray}
At least in vacuum, we have $c < 0$ (note that $\psi > 0$).
Hence a solution of the slicing condition (should it exist) is unique.

Suppose now that we use the constrained evolution scheme (section
\ref{sec:consscheme}) so that the slicing condition has the form 
\eqref{eq:axislicingc}. That equation has a nonlinear source term
$H(\alpha, x^A)$ as in \eqref{eq:ellmodel}, and we find
\begin{eqnarray}
  \frac{\partial H}{\partial \alpha} &=& - \psi^4 \e^{2rs} \big[
  - \alpha^{-2} (\tfrac{2}{3} \beta_-^2 + \half \beta_+^2)
  \nonumber\\&&\qquad\qquad
  + r^2 Y^2 + \half r^2 \psi^8 \e^{2rs} ({E^r}^2 + {E^z}^2) \\&&\qquad\qquad
  + \half \kappa ( \rho_H + r \tilde \tau + 2 S_r{}^r + S_z{}^z ) \big] \,.
\end{eqnarray}
Now the first line inside the square bracket has the wrong sign and so
$\partial H / \partial \alpha$ can be non-negative. 
Hence we cannot prove uniqueness for the linearized equation, 
and solutions to the nonlinear slicing condition \eqref{eq:axislicingc}
could \emph{potentially} be non-unique as well.

\paragraph{The shift conditions.}

The shift conditions (\ref{eq:axishiftr}--\ref{eq:axishiftz}) are
simple Poisson equations ($a^{AB} = \delta^{AB}$, $b^A = H = 0$ 
in \eqref{eq:ellmodel}) and therefore have a unique solution.

\paragraph{The momentum constraints.}

The momentum constraints (\ref{eq:aximomconsrc}--\ref{eq:aximomconszc})
are linear, $H = 0$. The only condition that is not
immediately obvious is the ellipticity of the differential operator,
for now we have a coupled system 
\begin{equation}
  L_{AB} (\partial_r, \partial_z) \beta^B = f_A \,.
\end{equation}
The principal symbol is given by
\begin{equation}
  L_{AB} (x, y) = \left( \begin{array}{cc} \tfrac{2}{3} x^2 + y^2 &
  \tfrac{1}{3} x y \\ -\tfrac{1}{3} x y & x^2 + \tfrac{4}{3} y^2
  \end{array} \right).
\end{equation}
Its determinant is
\begin{equation}
  \det L (x, y) = \tfrac{2}{3} (x^4 + 3 x^2 y^2 + 2 y^4) > 0 \quad
  \forall \, (x, y) \in \mathbb{R}^2 \setminus (0,0) \,.
\end{equation}
We have shown that the system is elliptic and hence the momentum 
constraints have a unique solution.

\paragraph{The Hamiltonian constraint.}

Finally we turn to the Hamiltonian constraint \eqref{eq:axihamcons} or
\eqref{eq:axihamconsc}. This has the form \eqref{eq:ellmodel} with
\begin{eqnarray}
  \frac{\partial H}{\partial \psi} &=& \quarter (r s_{,rr} + 2 s_{,r}
  + r s_{,zz}) \nonumber\\&&
  + 5 \psi^4 \e^{2rs} \big[ \tfrac{1}{3} (U - \half r Y)^2 + \quarter
  r^2 Y^2 + \quarter \kappa \rho_H \big] \nonumber\\&&
  + \tfrac{1}{16} r^2 \psi^8 \e^{2rs} \big[ 9 {B^\varphi}^2 + 13
  \psi^4 \e^{2rs} ({E^r}^2 + {E^z}^2) \big] \,.
\end{eqnarray}
All the terms in the square brackets are positive, and the terms in the
first line are oscillatory. Hence condition \eqref{eq:cnegative} 
is \emph{not} satisfied everywhere.
We conclude that quite possibly the Hamiltonian constraint
\eqref{eq:axihamcons} does not have a unique solution in general.

A few more remarks are about the Hamiltonian constraint are in order.
This equation is essentially the \emph{Yamabe equation}
\begin{equation}
  \label{eq:yamabe}
  \Delta \psi + K^2 \psi^p = 0 \,,
\end{equation}
where $p = 5$, $K^2$ is the square of the extrinsic curvature (which
we assume to be a smooth function), and we disregard the twist and 
matter here.
If we set $u \equiv \psi - 1$ and $f(u) \equiv K^2 (1 + u)^p$,
equation \eqref{eq:yamabe} can be written as
\begin{equation}
  \label{eq:Evans1}
  - \Delta u = f(u) \,,
\end{equation}
and we consider the boundary conditions
\begin{equation}
  \label{eq:Evans2}
  u \rvert_{\partial \Omega} = 0 \,.
\end{equation}
By a theorem in Evans \cite[sec.~8.5.2]{EvansPDEs} based on the
Mountain Pass Theorem, the boundary value problem 
(\ref{eq:Evans1}--\ref{eq:Evans2}) has at least one weak solution 
$u \neq 0$ provided that 
\begin{equation}
  1 < p < \frac{n+2}{n-2} \,,
\end{equation}
where $n$ is the spatial dimension.
In our case ($n = 2$), $p$ is not restricted from above and we deduce
that a solution to the Hamiltonian constraint does exist.
(Note that in $n = 3$ dimensions, $p = 5$ is the critical case and
the theorem is not applicable.)

However, nothing is being said about the uniqueness of the solution.
In fact, our argument above indicates that the Hamiltonian constraint 
might not be linearization stable. We can improve on this by applying 
York's \cite{York79} \emph{conformal rescaling} procedure:
let us redefine the extrinsic curvature $K$ by setting
\begin{equation}
  K = \psi^{-q} \tilde K \,.
\end{equation}
Then \eqref{eq:yamabe} reads (with $p = 5$)
\begin{equation}
  \label{eq:yamabe2}  
  \Delta \psi + \tilde K^2 \psi^{5 - 2q} = 0 \,,
\end{equation}
the linearization of which is
\begin{equation}
  \Delta \delta \psi + (5 - 2q) \tilde K^2 \delta \psi^{4 - 2q} = f \,.
\end{equation}
If we choose $q \geqslant 5/2$ then our analysis above shows that 
the modified Hamiltonian constraint \eqref{eq:yamabe2}  
is linearization stable.
York applies this trick only at the initial time in order to set up
a well-posed elliptic problem for the initial data. However, we want to solve
the Hamiltonian constraint at each time step, which means that we have
to evolve $\tilde K$ instead of $K$. We have implemented this 
but unfortunately the numerical evolutions quickly became unstable. 
A somewhat heuristic explanation for this might lie in the fact 
that a rescaling of the extrinsic curvature is also applied in the 
\emph{BSSN system} \cite{Baumgarte98}, which is known to be much
more stable than the standard ADM system.\footnote{On the other hand, 
some formulations such as NOR \cite{NOR04} appear to be stable without 
conformal rescalings.}
It turns out that our choice of extrinsic curvature $K$ corresponds 
precisely to the BSSN variables, whereas the rescaled $\tilde K$ corresponds 
to the ADM variables, with $q = 4$.
 
\paragraph{Summary.} 
We have indicated that the Hamiltonian constraint and the version of 
the slicing condition that is used in the constrained evolution scheme 
might not have a unique solution in general. 
In contrast, all the elliptic equations of the free evolution scheme 
and the partially constrained scheme are well-posed.

\subsection{Numerical considerations}
\label{sec:ellsolvnum}

There is a close connection between the above analytical results and
the numerical solvability of the elliptic equations under discussion 
using the Multigrid method.
The Newton-Gauss-Seidel relaxation employed in that method
effectively linearizes the elliptic equation so that it suffices to 
deal with a linear model problem here. For simplicity, we consider the
\emph{Helmholtz equation}
\begin{equation}
    \label{eq:helmholtz}
    u_{,rr} + u_{,zz} + c u = f \,.
\end{equation}
This equation is discretized as 
\begin{equation}
  \label{eq:discrhelmholtz}
  \tfrac{1}{h^2} (u_{i+1,j} + u_{i-1,j} + u_{i,j+1} + u_{i,j-1} - 4 u_{ij})
  + c_{ij} u_{ij} = f_{ij} \,.
\end{equation}
As explained in section \ref{sec:MG}, the Gauss-Seidel relaxation
converges if the matrix on the left-hand-side of 
\eqref{eq:discrhelmholtz} is diagonal dominant. 
The (absolute values of the) off-diagonal terms in 
\eqref{eq:discrhelmholtz} add up to $4 h^{-2}$ 
and the diagonal term is $-4 h^{-2} + c_{ij}$.
Hence the matrix is diagonal dominant if and only if $c_{ij} \leqslant 0$,
which is again condition \eqref{eq:cnegative}.
(If first-order derivatives are included in \eqref{eq:helmholtz}
or if the principal part is not the Laplace operator, this condition
may not be sufficient to guarantee diagonal dominance.)

In practice, the relaxation still converges if $c > 0$ and $c$ is 
sufficiently small. For larger and larger positive $c$, however, 
the relaxation first stalls and ultimately diverges. 
The failure of Multigrid for such indefinite Helmholtz equations 
has been reported many times in the literature (e.g., \cite{Bramble88}). 
Cures of the problem usually involve some kind of conjugate gradient 
or other Krylov subspace iterations, which are very slow as compared 
with standard Multigrid (e.g., \cite{Elman01}; see also section 
\ref{sec:altnumCG}).

We conclude that the Multigrid method is suitable for solving all the
equations that occur in the free evolution scheme and the partially
constrained scheme but that it might fail for the Hamiltonian
constraint and the slicing condition used in the constrained scheme. 
This is indeed what we have observed when trying to evolve strong Brill
waves (section \ref{sec:hypellnum}) with the constrained scheme.
Similar observations have been reported by Choptuik et al. \cite{Choptuik03a} 
and Barnes \cite{BarnesPhD}. The former try to avoid the problem by
evolving the conformal factor $\psi$ using its evolution equation
\eqref{eq:axid0psi} instead of solving the Hamiltonian constraint for
it. However, they find that their Multigrid solver still fails for 
strong Brill waves. A likely explanation for this is the argument
given above for the slicing condition.


  \section{Evolution of the constraints}
\label{sec:hypellsubs}

We have seen how in the (2+1)+1 formalism (as in all ADM-like
formalisms), the Einstein equations split into elliptic constraint
equations and hyperbolic evolution equations.
Analytically, the constraints are preserved by the evolution
equations. However, if in a numerical evolution the constraints are
only solved initially, they might get violated during the evolution
due to numerical errors. Catastrophic growth of the constraints in
free evolution schemes is a very common plague in numerical
relativity and to-date one of the major limitations to the runtime of
simulations.

In this section, we take a closer look at the evolution of the
constraints and assess the schemes presented in section
\ref{sec:hypellschemes} with regard to their stability against
constraint violations. 

Suppose first we adopt the free evolution scheme (section
\ref{sec:freescheme}). The constraints
(\ref{eq:axihamcons}--\ref{eq:axigercons}) are found to obey the 
following evolution equations:

\begin{eqnarray}
  \label{eq:hamconsev}
  \partial_t \hat \C &\simeq& 
    - \beta^r \partial_r \hat \C - \beta^z \partial_z \hat \C
    - \quarter \alpha \psi (\partial_r \hat \C_r + \partial_z \hat \C_z)
    \,,\\
  \label{eq:momconsrev}
  \partial_t \hat \C_r &\simeq&
    - \beta^r \partial_r \hat \C_r - \beta^z \partial_z \hat \C_r
    + 4 \alpha \psi^{-5} \e^{-2rs}  (\partial_r \hat \C + \quarter
    \alpha^{-1} \psi \, \partial_r \hat \S) \,,\\
  \label{eq:momconszev}
  \partial_t \hat \C_z &\simeq&
    - \beta^r \partial_r \hat \C_z - \beta^z \partial_z \hat \C_z
    - 4 \alpha \psi^{-5} \e^{-2rs}  \partial_z \hat \C \,,\\
  \label{eq:gerconsev}
  \partial_t \hat \C_\varphi &\simeq&
    - \beta^r \partial_r \hat \C_\varphi - \beta^z \partial_z \hat
  \C_\varphi  \,.  
\end{eqnarray}
Here $\hat \S$ is the slicing condition \eqref{eq:axislicing}, and
$\simeq$ denotes equality to principal parts. The terms we have left
out are all linear and homogeneous in the constraints and the gauge 
conditions, so that the constraints are indeed conserved (equations
(\ref{eq:hamconsev}--\ref{eq:gerconsev}) are satisfied if all the
constraints and gauge conditions vanish at all times).

We enforce the slicing condition during the numerical evolution, i.e.,
we may set $\hat S = 0$ in \eqref{eq:momconsrev}.
Then the constraint evolution system can be written in closed form as
\begin{equation}
  \label{eq:consevsys}
  \partial_t \vec c = A^A \partial_A \vec c + B \vec c\,,
\end{equation}
where $\vec c = (\hat \C, \hat \C_r, \hat \C_z, \hat \C_\varphi)^T$ 
and the matrices $A^A$ are given by
\begin{eqnarray}
  A^r &=& \left( \begin{array}{cccc} - \beta^r & - \quarter \alpha \psi
  & 0 & 0 \\ 4 \alpha \psi^{-5} \e^{-2rs} & -\beta^r & 0 & 0 \\
  0 & 0 & -\beta^r & 0 \\ 0 & 0 & 0 & -\beta^r \end{array} \right) \,,\\
  A^z &=& \left( \begin{array}{cccc} - \beta^z & 0 & - \quarter \alpha
  \psi & 0 \\ 0 & - \beta^z & 0 & 0 \\ -4 \alpha \psi^{-5} \e^{-2rs} &
  0 & -\beta^z & 0 \\ 0 & 0 & 0 & -\beta^z \end{array} \right) \,.
\end{eqnarray}
The matrix $A^r$ has complex eigenvalues 
$-\beta^r \pm i \alpha \psi^{-2} \e^{-rs}$ 
whereas $A^z$ has real eigenvalues 
$-\beta^z \pm \alpha \psi^{-2} \e^{-rs}$.
This means that $A^r$ is not real diagonalizable, and so the system is
not hyperbolic (see section \ref{sec:hyperbolicity} for a precise
definition of hyperbolicity and its implications).
Hence the initial value problem (IVP) for the constraint evolution system is
ill-posed, and small violations of the constraints may grow without
bound.

The reason for the lack of hyperbolicity lies in the slicing
condition \eqref{eq:slicing}. Recall that when we derived it, 
we added a multiple of the Hamiltonian constraint.
If we undo this and replace
\begin{equation}
  \label{eq:modslicing}
  \hat \S \rightarrow \hat \S' \equiv \hat \S + 8 \alpha \psi^{-1} \hat \C 
\end{equation}
then the evolution equation for the $r$-momentum constraint
\eqref{eq:momconsrev} becomes
\begin{equation}
    \partial_t \hat \C_r \simeq
    - \beta^r \partial_r \hat \C_r - \beta^z \partial_z \hat \C_r
    - 4 \alpha \psi^{-5} \e^{-2rs} ( \partial_r \hat \C - \quarter
    \alpha^{-1} \psi \, \partial_r \hat \S') \,.
\end{equation}
Hence 
\begin{equation}
  A^r = \left( \begin{array}{cccc} - \beta^r & - \quarter \alpha \psi
  & 0 & 0 \\ -4 \alpha \psi^{-5} \e^{-2rs} & -\beta^r & 0 & 0 \\
  0 & 0 & -\beta^r & 0 \\ 0 & 0 & 0 & -\beta^r \end{array} \right) \,,
\end{equation}
which has real eigenvalues $-\beta^r \pm \alpha \psi^{-2} \e^{-rs}$ 
and so the system is hyperbolic and the IVP is well-posed.
However, the modified slicing condition \eqref{eq:modslicing} can
easily become indefinite, depending on the sign of $\hat \C$. 
Indeed, the Multigrid method turns out to 
fail for the modified slicing condition even for relatively weak
perturbations of flat space.

Fortunately there is a way out: suppose we enforce the momentum
constraints by using either the constrained scheme (section
\ref{sec:consscheme}) or the partially constrained scheme (section
\ref{sec:modscheme}). Then the offending equation
\eqref{eq:momconsrev} is discarded and the remaining system is
clearly hyperbolic.

We have thus given a strong argument for solving the
momentum constraints if a maximal slicing condition is used that is
manipulated by adding a multiple of the Hamiltonian constraint. 
Together with the results of
section \ref{sec:ellsolv}, we conclude that the partially constrained
scheme \ref{sec:modscheme} is the only one of the schemes presented 
in section \ref{sec:hypellschemes} that may be suitable for the numerical
evolution of strong gravitational waves.

One should remark that hyperbolicity of the constraint evolution
system is not sufficient to rule out growth of the constraints.
Depending on the matrix $B$ in \eqref{eq:consevsys}, there could well be
exponentially growing solutions.
One could try to adjust $B$ by adding suitable multiples of the
constraints to the main evolution equations, as done, for example, 
in \cite{YonedaShinkai01}.
We have not investigated this possibility because the constraints appear to 
be bounded in our numerical evolutions (section \ref{sec:hypellnum}).


  \section[Evolutions of Brill waves]
{Numerical evolutions of generalized Brill waves}
\label{sec:hypellnum}

In this section we present some numerical results on the evolution of
time-symmetric axisymmetric gravitational waves in vacuum, 
also known as \emph{Brill waves} \cite{Brill59}. 
As a new ingredient, we include a nonzero twist.

\subsection{Initial data}

The initial time $t = 0$ is chosen to be a moment of 
time symmetry, i.e., under the coordinate transformation 
$t \rightarrow t' \equiv -t$ the metric transforms as
\begin{equation}
  g'_{\alpha\beta}(t) = g_{\alpha\beta}(-t) \,.
\end{equation}
This implies that the spatial metric is an even function of $t$ and so
the extrinsic curvature (the time derivative of the spatial metric) 
vanishes at $t = 0$:
\begin{equation}
  Y = U = X = 0 \,.
\end{equation}
From definition (\ref{eq:twist}) we infer that the spatial components
of the twist vector are odd functions of time and the time component
is even. Now definition \eqref{eq:E} implies that
\begin{equation}
  E^r = E^z = 0
\end{equation}
at $t = 0$. 
The initial data for the variables $s$ and $B^\varphi$ is taken to be
\begin{equation}
  s = A_s \, r \, \e^{-r^2 - z^2} \,, \qquad
  B^\varphi = A_B \, r \, z \, \e^{-r^2 - z^2} 
\end{equation}
with constant amplitudes $A_s$ and $A_B$.
The Hamiltonian constraint \eqref{eq:axihamcons} is then solved for 
the conformal factor $\psi$. For this initial data the momentum constraints
(\ref{eq:aximomconsr}--\ref{eq:aximomconsz}) and the Geroch constraint 
\eqref{eq:axigercons} are automatically satisfied, and the unique
solution of the gauge conditions (\ref{eq:axislicing}, 
\ref{eq:axishiftr}--\ref{eq:axishiftz}) is
\begin{equation}
  \alpha = 1 \,, \qquad \beta^r = \beta^z = 0 \,.
\end{equation}

In Brill's original work \cite{Brill59} and in all subsequent studies
we know of, the twist was assumed to vanish. However, we would
like to stress that a nonzero $B^\varphi$ is consistent with time symmetry
so that the term \emph{generalized Brill waves} is justified for the
problem considered here.

\subsection{Boundary conditions}

On the axis $r = 0$, we enforce the appropriate Dirichlet or Neumann
conditions as stated in table \ref{tab:axirpar}.
Since our initial data is reflection-symmetric about the $z = 0$ plane 
and the evolution equations preserve that symmetry, we can save 
computational time by only evolving the upper half of the $(r, z)$ plane.
Reflection symmetry means that under the coordinate transformation
$z \rightarrow z' \equiv -z$ the metric transforms as
\begin{equation}
  g'_{\alpha\beta}(z) = g_{\alpha\beta}(-z) \,.
\end{equation}
This implies that the variables we evolve are either odd or even
functions of $z$. For an odd variable $\underline{u}$, we impose a Dirichlet
condition
\begin{equation}
  \underline{u} \vert_{z = 0} = 0 \,,
\end{equation}
and for an even variable $u$, a Neumann condition
\begin{equation}
  \partial_z u \vert_{z = 0} = 0 
\end{equation}
is needed. The $z$-parity of all the variables we evolve 
is summarized in table \ref{tab:axizpar}.

\begin{table}[t]
  \begin{eqnarray*}
    \alpha, \beta^r, \underline{\beta^z}, \psi, s, 
    Y, U, \underline{X}, \underline{B^\varphi},
    E^r, \underline{E^z} , \\
    \rho_H, \sigma, J^\varphi, J_r, \underline{J_z},
    \tilde \tau, S^r, \underline{S^z},
    \Sigma^r, \underline{\Sigma^z}, S_r{}^r, 
    \underline{S_r{}^z}, S_z{}^z . 
  \end{eqnarray*}
  \caption[Hyperbolic-elliptic system: $z$-parity of the variables]
  {\label{tab:axizpar} \footnotesize
    $z$-parity of the variables of the hyperbolic-elliptic
    system if reflection symmetry is assumed. 
    Underlined variables are odd functions of $z$, the remaining ones
    are even.}
\end{table}

Throughout this thesis we focus on asymptotically flat spacetimes. 
We therefore assume a fall-off of all the variables like
\begin{equation}
  u = u_\infty + \frac{c}{R}
\end{equation}
for large $R \equiv \sqrt{r^2 + z^2}$, where $u_\infty$ is the 
flat-space value of the variable $u$ and $c$ is independent of $R$.
This implies that
\begin{equation}
  \label{eq:invRfalloff}
  0 = \partial_R [ R(u - u_\infty)] = u - u_\infty + r u_{,r} + z
  u_{,z} \,.
\end{equation}
We use this as a boundary condition at $r = r_\mathrm{max}$ and 
$z = z_\mathrm{max}$ for the elliptic equations, i.e., for the
variables $\alpha, \beta^r, \beta^z$ and initially $\psi$.

The remaining variables are evolved by hyperbolic evolution equations,
and for them we impose an \emph{outgoing wave} or \emph{Sommerfeld condition}
\begin{equation}
  u = u_\infty + \frac{f(t - R)}{R} \,,
\end{equation}
which we rewrite as 
\begin{equation}
  0 = (\partial_R + \partial_t) [R(u - u_\infty)] 
  = u - u_\infty + r u_{,r} + z u_{,z} + R u_{,t}  \,.
\end{equation}

These boundary conditions appear to work well in practice, although 
they are a rather crude choice which is not 
fully justified theoretically. We refer the reader to 
chapter \ref{sec:outerbcs}, where outer boundary conditions are
discussed at length for a completely hyperbolic formulation of
Einstein's equations.

\subsection{Numerical method}

The equations are discretized using second-order accurate centred
finite differencing (section \ref{sec:FD}) on a single uniform cell-centred 
grid. Unlike Garfinkle and Duncan \cite{Garfinkle00}, we do
not compactify the spatial coordinates, for fear that we might fail to
resolve the waves as they travel out to infinity.

For the time integration, we use the method of lines with the 
third-order Runge-Kutta scheme (\ref{eq:RK3}b).
The second-order Runge-Kutta schemes \eqref{eq:RK2} and the three-step 
iterative Crank-Nicholson method (\ref{eq:ICNfirst}--\ref{eq:ICNlast})
were also tried but were found to be
substantially less stable in strong Brill wave evolutions.
In particular, the simulations with those schemes suffered from an
unbounded growth of the constraint residuals.
The fourth-order Runge-Kutta scheme \eqref{eq:RK4} gave results
comparable to the third-order one but is computationally more expensive.
The Courant number is taken to be $\Delta t / h = 0.5$ in all the evolutions
presented here.

Fourth-order Kreiss-Oliger dissipation \eqref{eq:diss4} with amplitude 
$\epsilon_D = 0.5$ is added to the
right-hand-side of the evolution equations. We found that without
dissipation, a high-frequency instability occurred at very late times,
in particular close to the boundaries.

The boundary conditions are implemented via the method of ghost cells
as explained in section \ref{sec:ghosts}.

We adopt the partially constrained evolution scheme (section
\ref{sec:modscheme}), for the reasons discussed in the previous two
sections. The elliptic equations are solved using the Multigrid method
(section \ref{sec:MG}) with red-black Gauss-Seidel relaxation.
The FAS version of the method is used (although the equations are
linear and so linear Multigrid would work just as well).
Typically five W-cycles are needed to drive the residual well below 
the discretization error.

\subsection{Weak Brill waves with twist}

We first consider Brill waves with an amplitude $A_s = 1$, which is
well in the subcritical r\'egime. To study the influence of the twist,
we perform simulations with three different amplitudes $A_B = 0, 2, 4$. 
The resolution is taken to be 128 points in both the $r$ and the $z$ 
direction and the outer boundaries are placed at $r_\mathrm{max} =
z_\mathrm{max} = 10$.

\begin{figure}[t]
  \centering
  \includegraphics[scale = 1]{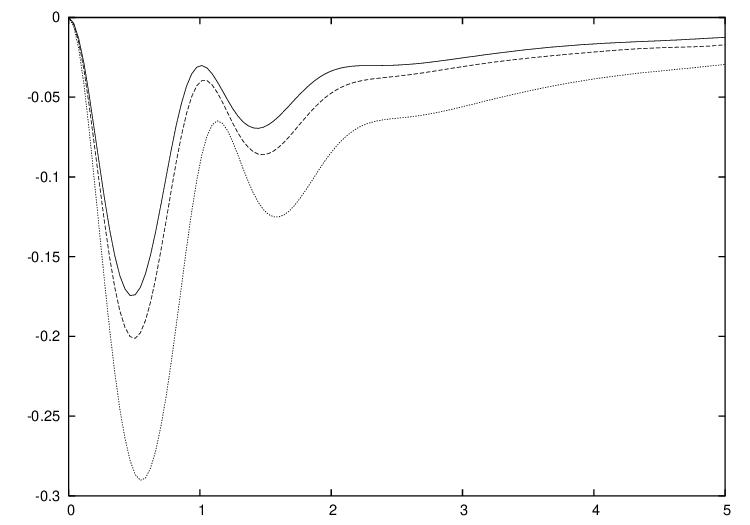}
  \caption[Weak Brill waves with twist: central lapse]
  {\label{fig:axitwistclapse}  \footnotesize
  $\ln \alpha_0$ (logarithm of the lapse function at the origin) as a 
  function of time for a Brill wave with amplitude $A_s = 1$ and three
  different amplitudes of the twist: $A_B = 0$ (solid line), 
  2 (dashed line) and 4 (dotted line)}
\end{figure}

Figure \ref{fig:axitwistclapse} shows the lapse function at the origin $r =
z = 0$ as a function of time. When a high-curvature region of spacetime
is approached, we expect the lapse function to collapse because the
(maximal) slices try to avoid that region and pile up. Because the minimum
of the lapse is always found to lie in the origin, the value of the
lapse there serves as a good ``curvature indicator''.

We see that the lapse performs a few damped oscillations and
eventually returns to its flat-space value. The amplitude of the
extrema is found to increase with increasing $A_B$, while the extrema
occur almost at the same times.

\begin{figure}[t]
  \centering
  \begin{minipage}[t]{0.49\textwidth}
    \includegraphics[scale = 1]{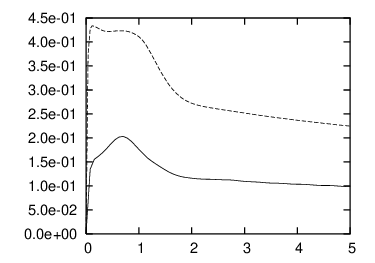}
    \centerline{$\hat \C$}
  \end{minipage}
  \hfill  
  \begin{minipage}[t]{0.49\textwidth}
    \includegraphics[scale = 1]{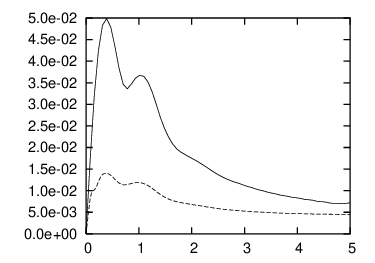}
    \centerline{$\hat \C_r$}
  \end{minipage}
  \hfill  
  \bigskip
  \begin{minipage}[t]{0.49\textwidth}
    \includegraphics[scale = 1]{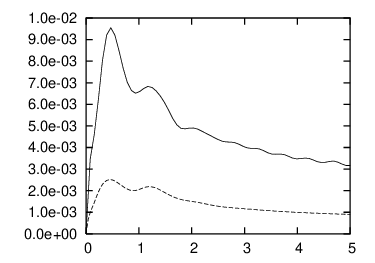}
    \centerline{$\hat \C_z$}
  \end{minipage}
  \hfill  
  \begin{minipage}[t]{0.49\textwidth}
    \includegraphics[scale = 1]{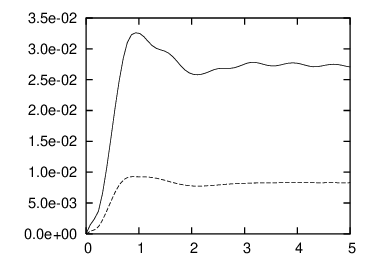}
    \centerline{$\hat \C_\varphi$}
  \end{minipage}
  \hfill  
  \bigskip
  \caption[Weak Brill wave with twist: constraint convergence]
  {\label{fig:axiconsconv}  \footnotesize
  $L^2$ norm of the constraint residuals as a function of time for a
  Brill wave with amplitudes $A_s = 1$ and $A_B = 2$ and two different
  resolutions: 64 points (solid lines) and 128 points (dashed lines)}
\end{figure}

To check the accuracy of our code, we perform a convergence test:
figure \ref{fig:axiconsconv} shows the $L^2$ norm\footnote{see
  equation \eqref{eq:L2norm} for a definition of the discrete $L^2$ norm}
of the constraint residuals 
as a function of time for two different resolutions (here the
amplitude of $B^\varphi$ is taken to be $A_B = 2$).
Because the finite-differencing we use is second-order accurate, the
residual of the constraints should decrease by a factor of four as the
resolution is doubled.
The numerical results indicate that we do not quite achieve
second-order convergence (the decrease lies between a factor of 2 and 3).
This is probably due to reflections caused by the imperfect outer boundary
conditions, which do not appear to converge away with increasing resolution.

\subsection{Strong Brill waves}

Next we turn to strong Brill waves with amplitudes $A_s \gtrsim 4$.
Thanks to the modified evolution scheme we use 
(section \ref{sec:modscheme}), we are able to evolve much stronger 
Brill waves than with the constrained scheme 
(section \ref{sec:consscheme}) used by both Choptuik et
al.~\cite{Choptuik03a} and Barnes \cite{BarnesPhD}.
The constrained scheme failed for amplitudes $A_s \gtrsim 3$ due to a
breakdown of the Multigrid solver, as explained in 
section \ref{sec:ellsolvnum}. 
We also found that the free evolution scheme (section
\ref{sec:freescheme}) suffered from an unbounded growth of the
constraints particularly for strong Brill wave evolutions, as
predicted in section \ref{sec:hypellsubs}.

\begin{figure}[t]
  \centering
  \includegraphics[scale = 1]{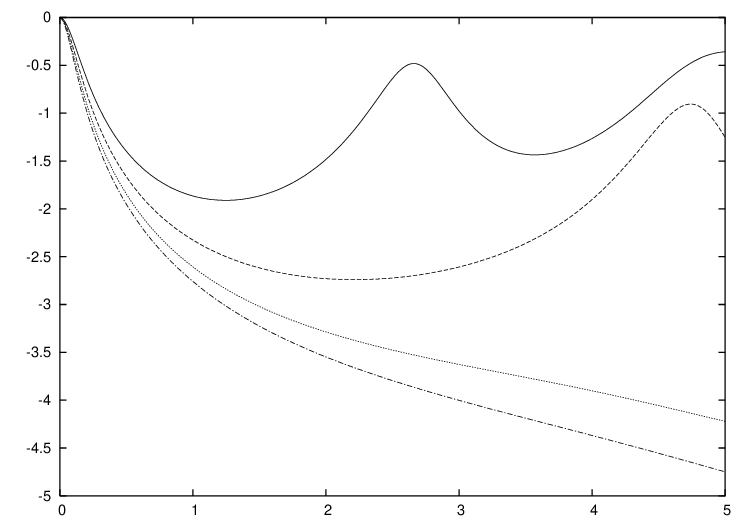}
  \caption[Strong Brill waves: central lapse]
  {\label{fig:axistrongclapse}  \footnotesize
  $\ln \alpha_0$ (logarithm of the lapse function at the origin) as a 
  function of time for non-twisting Brill waves with amplitudes 
  $A_s = 4$ (solid line), 5 (dashed line), 6 (dotted line) and
  7 (dot-dashed line)}
\end{figure}

Figure \ref{fig:axistrongclapse} shows again the lapse function at the origin
as a function of time for four different values of the amplitude $A_s$. 
In order to compare our results with those of Garfinkle and Duncan
\cite{Garfinkle00}, we choose the twist to vanish here.
As $A_s$ is increased, the oscillations of the lapse become larger 
and larger and their frequency decreases. 
For $A_s \geqslant 6$, the lapse function continues to collapse 
and the formation of a black hole is expected.
The interval 
\begin{equation}
  5 < A_s^\ast < 6
\end{equation} 
for the critical amplitude is in agreement with \cite{Garfinkle00}.

\begin{figure}
  \centering
  \begin{minipage}[t]{0.49\textwidth}
    \includegraphics[scale = 0.49]{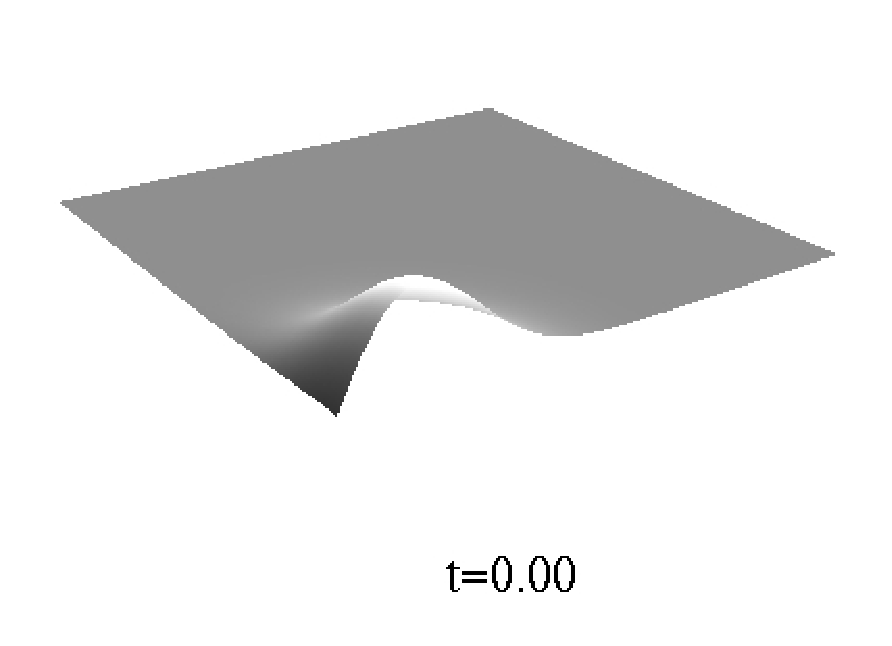}
  \end{minipage}
  \hfill  
  \begin{minipage}[t]{0.49\textwidth}
    \includegraphics[scale = 0.49]{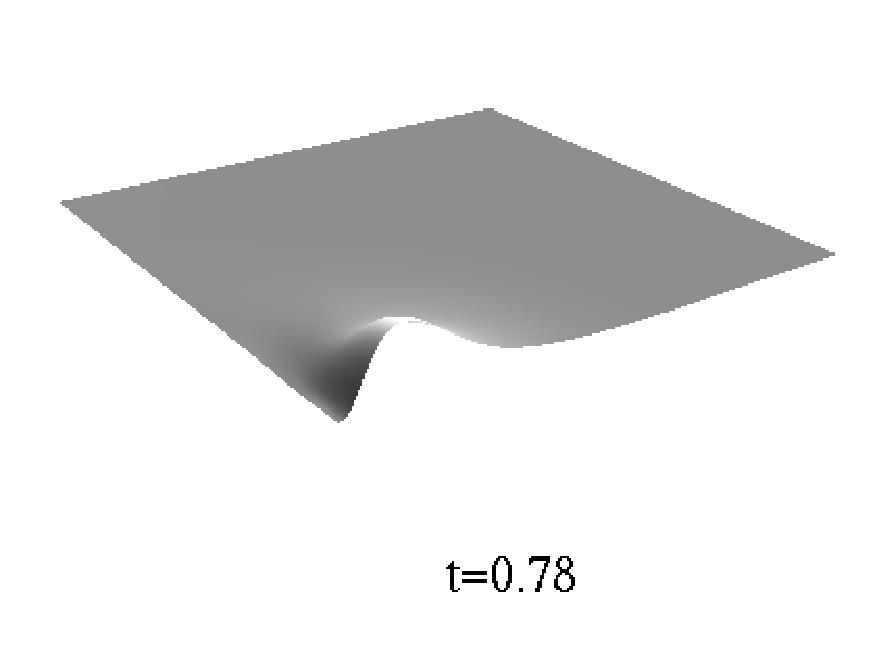}
  \end{minipage}
  \hfill  
  \begin{minipage}[t]{0.49\textwidth}
    \includegraphics[scale = 0.49]{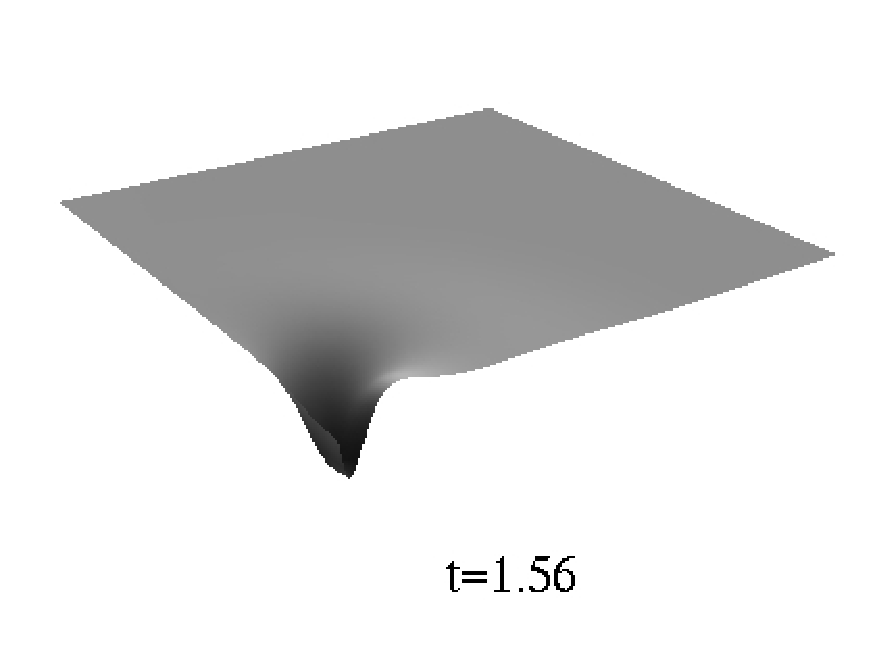}
  \end{minipage}
  \hfill  
  \begin{minipage}[t]{0.49\textwidth}
    \includegraphics[scale = 0.49]{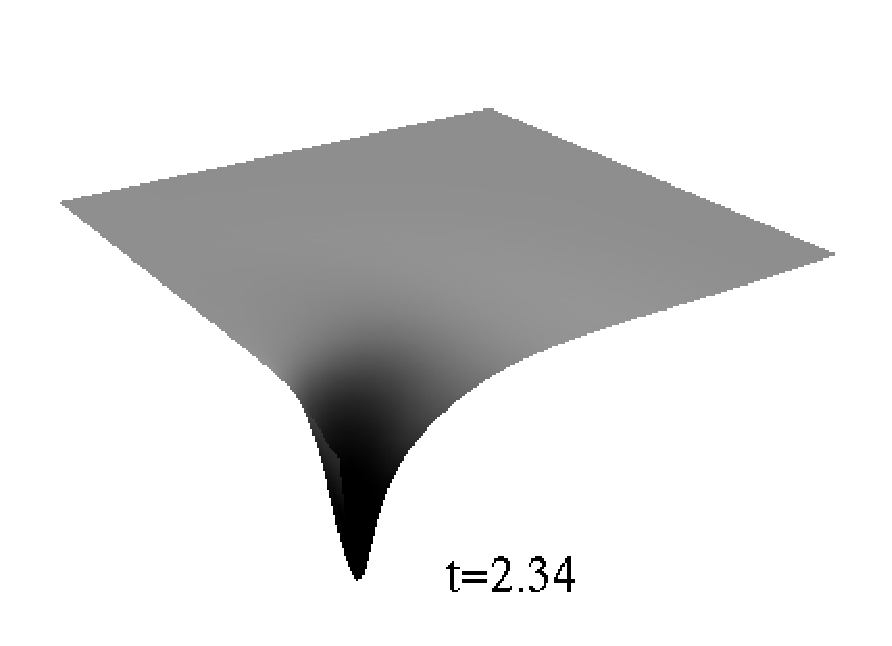}
  \end{minipage}
  \hfill  
  \begin{minipage}[t]{0.49\textwidth}
    \includegraphics[scale = 0.49]{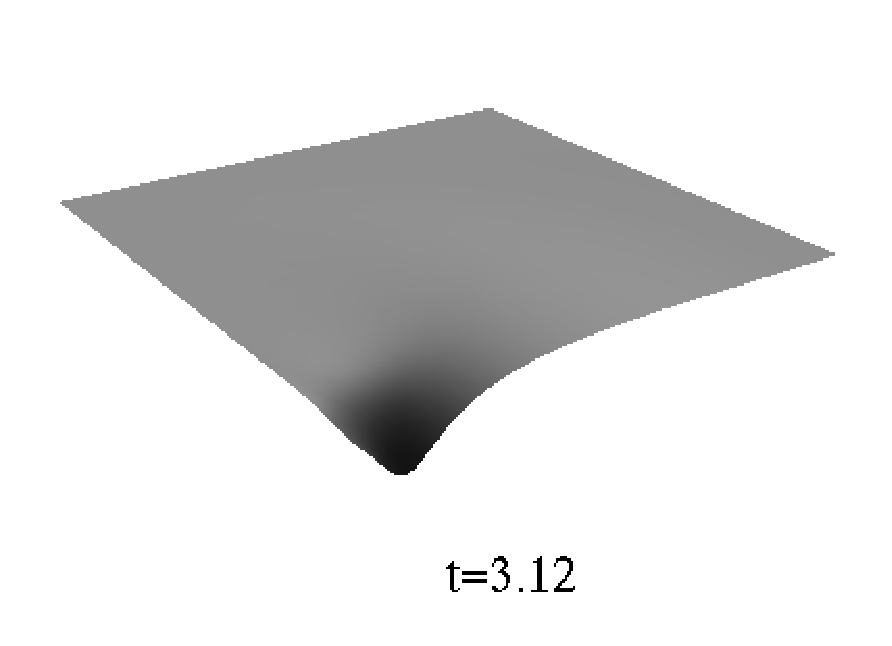}
  \end{minipage}
  \hfill  
  \begin{minipage}[t]{0.49\textwidth}
    \includegraphics[scale = 0.49]{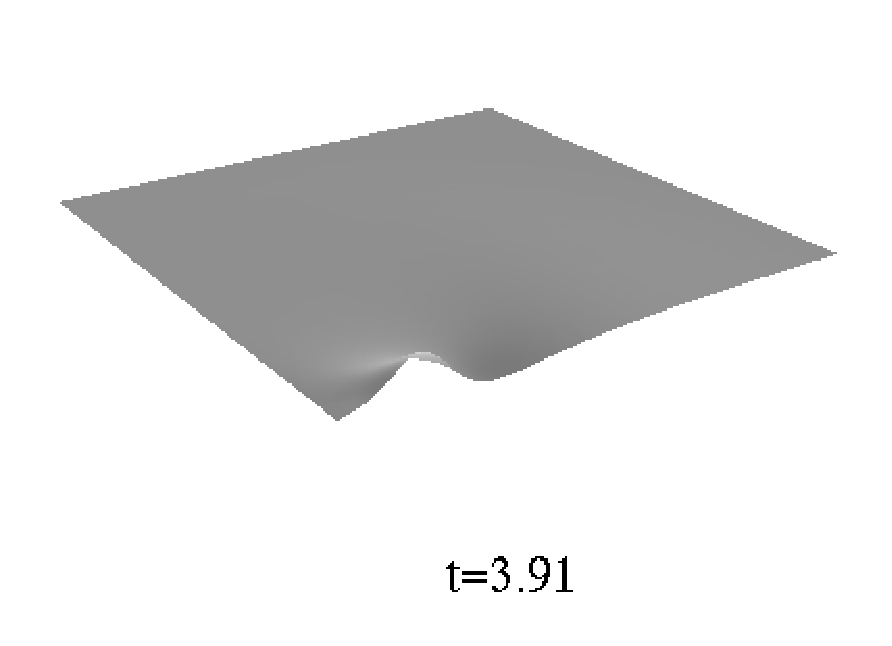}
  \end{minipage}
  \hfill  
  \caption[Strong Brill wave: snapshots of $s$ for $A_s = 4$]
  {\label{fig:axis4}  \footnotesize
  Snapshots of the variable $s$ for a subcritical Brill wave with
  amplitude $A_s = 4$. The resolution is 128 points in each dimension and
  the outer boundaries are placed at $r_\mathrm{max} = z_\mathrm{max} = 5$.
  In all plots of this thesis, the axis $r = 0$ is the bottom left boundary.} 
\end{figure}

\begin{figure}
  \centering
  \begin{minipage}[t]{0.49\textwidth}
    \includegraphics[scale = 0.49]{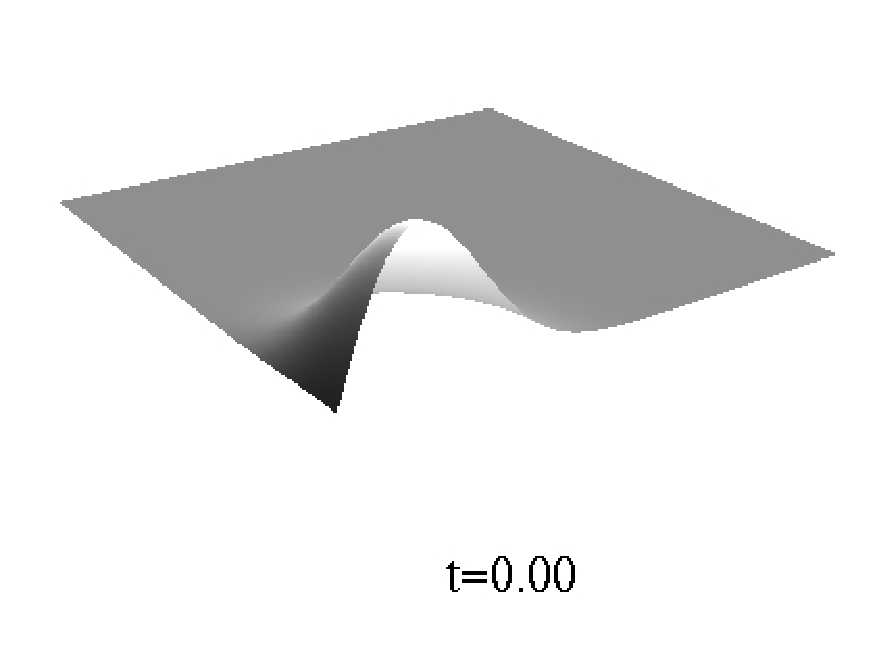}
  \end{minipage}
  \hfill  
  \begin{minipage}[t]{0.49\textwidth}
    \includegraphics[scale = 0.49]{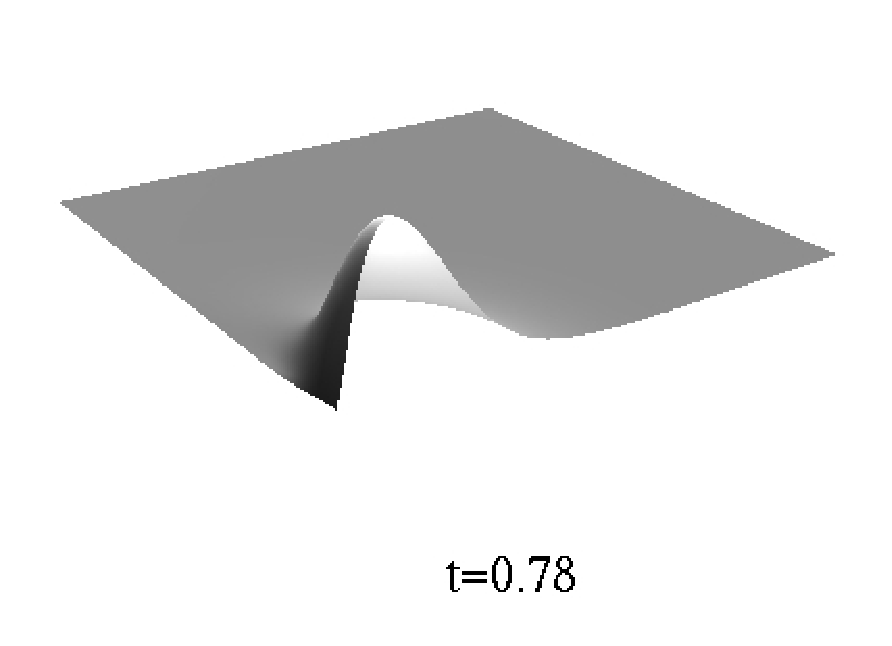}
  \end{minipage}
  \hfill  
  \begin{minipage}[t]{0.49\textwidth}
    \includegraphics[scale = 0.49]{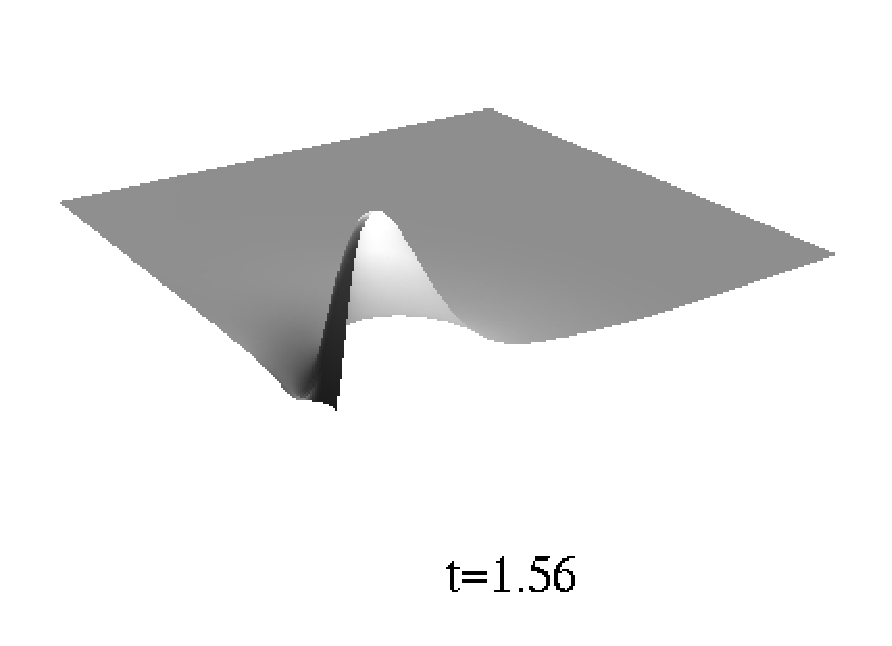}
  \end{minipage}
  \hfill  
  \begin{minipage}[t]{0.49\textwidth}
    \includegraphics[scale = 0.49]{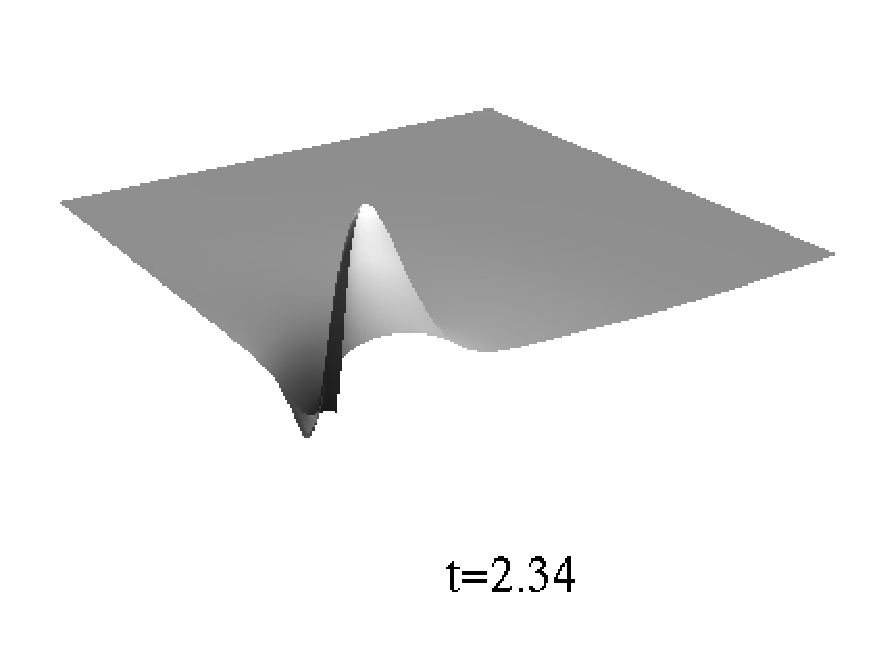}
  \end{minipage}
  \hfill  
  \begin{minipage}[t]{0.49\textwidth}
    \includegraphics[scale = 0.49]{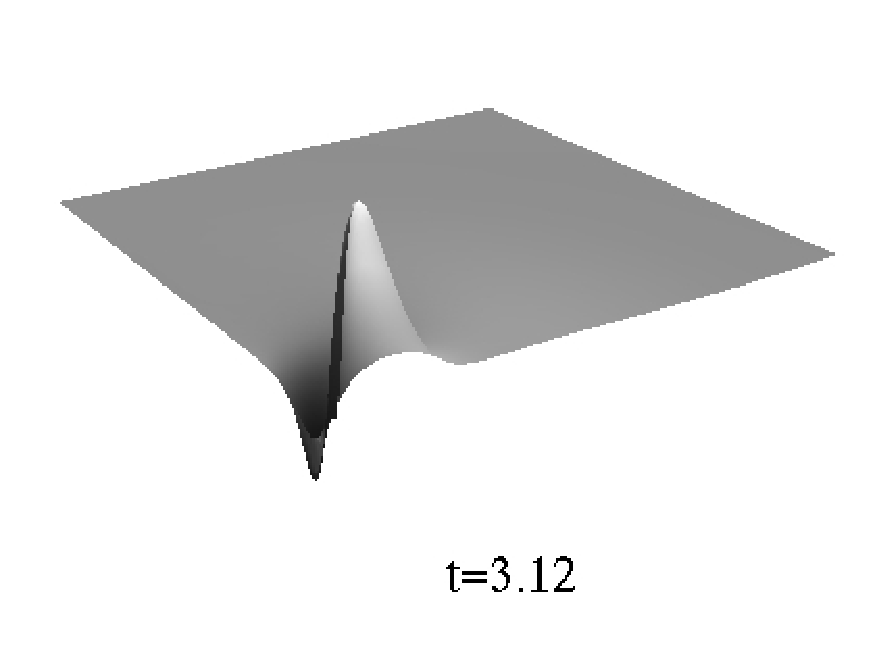}
  \end{minipage}
  \hfill  
  \begin{minipage}[t]{0.49\textwidth}
    \includegraphics[scale = 0.49]{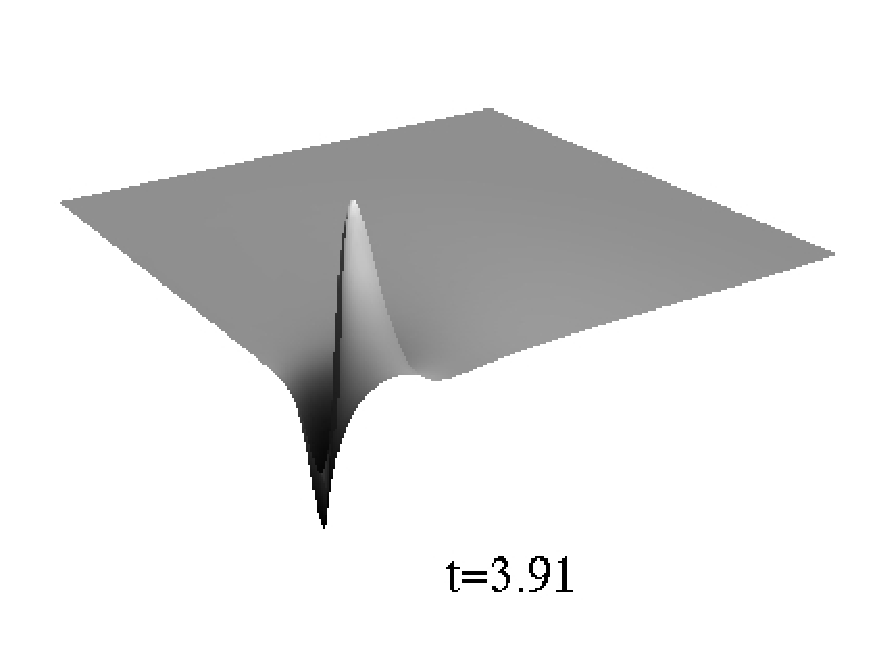}
  \end{minipage}
  \hfill  
  \caption[Strong Brill wave: snapshots of $s$ for $A_s = 6$]
  {\label{fig:axis6}  \footnotesize
  Snapshots of the variable $s$ for a supercritical Brill wave with
  amplitude $A_s = 6$. Same parameters as in figure \ref{fig:axis4}.} 
\end{figure}

To get some idea of what happens at the ``phase transition'', we show
a few snapshots of the variable $s$ for a slightly subcritical evolution
($A_s = 4$, figure \ref{fig:axis4}) and a slightly supercritical one
($A_s = 6$, figure \ref{fig:axis6}).
While we see an outgoing wave form in the subcritical evolution, 
the marginally supercritical solution contracts rather than disperses.
The $A_s = 6$ run crashed at $t \approx 6$ because the resolution was
insufficient to resolve the small and highly dynamical features close
to the origin. 

Critical collapse thus poses a major computational problem: 
more and more resolution is needed close to the origin as one
approaches the critical point.
At the same time, the solution is very smooth further away from the 
origin, so it would be a waste of computational resources to have a high
resolution across the entire grid.
This is a classic case for adaptive mesh refinement (AMR) (section 
\ref{sec:AMR}): we would like to add resolution only in regions where and 
when it is needed.
We have not implemented AMR for mixed hyperbolic-elliptic systems yet
but will use it in chapter \ref{sec:brill} for the completely hyperbolic
system derived in the following chapter.


  \section{Conclusions}
\label{sec:hypellconcl}

This is a good place to draw some preliminary conclusions before we
move on to the second part of the thesis.

In this chapter, we considered a mixed hyperbolic-elliptic system
that involved solving elliptic gauge conditions as well as (some or
all) constraint equations. Two major problems with such systems were
indicated, which we expect to be fairly generic in many formulations
of the Einstein equations used in numerical relativity.

Firstly, it is not always clear whether the elliptic equations one
tries to solve have unique solutions. In particular, the Hamiltonian 
constraint in the form used here is problematic. Suppose that one
attempts to solve this constraint during the evolution. 
Even if the numerical solver finds a solution, 
that solution could be non-unique. It could be a solution
that is not compatible with the evolution equation (\ref{eq:axid0psi}) 
that the conformal factor must also obey. Therefore it is not sufficient 
to enforce the Hamiltonian constraint alone -- one must also check the 
residual of the evolution equation. Because the Multigrid method we use is
unsuitable for the Hamiltonian constraint, we decided to evolve the 
conformal factor freely and monitor the constraint residual instead.

Secondly, we saw that if one uses free evolution (none of the
constraints are solved), the constraint evolution system can
become ill-posed if the maximal slicing condition is simplified by 
adding a multiple of the Hamiltonian constraint, as usually done
in the literature.
To cure this problem, we proposed a modified evolution scheme which
solves the momentum constraints but not the Hamiltonian constraint
and which has a well-posed constraint evolution system.

With our modified evolution system we were able to evolve both weak
and strong Brill waves. We included a nonzero twist, which to our knowledge
is the first time this has been done.
The existence of a critical amplitude that separates dispersal of the
waves from black hole formation was indicated. At present, we cannot
study the critical behaviour more closely because we run out of
resolution to resolve the features that occur near the
origin on smaller and smaller scales. 
Adaptive mesh refinement would be needed to tackle this
problem in a computationally efficient way. 

We decided not to continue to work on this formulation for the time
being for a variety of reasons: 
well-posedness of the initial boundary value problem is difficult to 
prove for mixed hyperbolic-elliptic systems, 
it is not clear what the characteristics of the system are because 
part of the dynamics resides in the variables that are solved for 
using elliptic equations, 
and the outer boundary conditions are ill-understood.
All these questions will be addressed in the second part of this
thesis for a strongly hyperbolic reduction of Einstein's equations 
in axisymmetric spacetimes.


   \chapter{The Z(2+1)+1 system}
\label{sec:Z211}

Whereas in chapter \ref{sec:hypell} we considered a mixed hyperbolic-elliptic
system, we construct in this chapter a completely hyperbolic
formulation of the Einstein equations for axisymmetric spacetimes.
In contrast to elliptic systems, hyperbolic systems of equations have the
property that information propagates with finite speed along the
characteristics. This makes them amenable to mathematical analysis
more easily than mixed hyperbolic-elliptic systems in which because of
the elliptic sector the solution at a given point depends on the
solution in the entire spatial domain. 
In particular, one can use the characteristic structure to set up
boundary conditions at the outer boundary of the
computational domain. For certain types of hyperbolic systems and
boundary conditions, theorems exist that guarantee the well-posedness
of the initial boundary value problem (IBVP). By well-posedness we broadly
mean that a unique solution exists at least for some finite time and that it
depends continuously on the initial and boundary data.

There are many ways of obtaining hyperbolic formulations of the
Einstein equations. Most approaches are based on the ADM decomposition
outlined in section \ref{sec:ADM}. Unfortunately, without further
modifications the ADM system is only weakly hyperbolic and thus does
not have a well-posed IBVP (e.g., \cite{Kidder01}).
Strongly hyperbolic systems can be obtained by adding certain multiples of the
constraints to the evolution equations. Among the variety of such
systems are the ones of Frittelli and Reula \cite{Frittelli96} and
Kidder, Scheel and Teukolsky \cite{Kidder01}. Whereas those authors
assume an arbitrary but fixed gauge, dynamical gauge conditions were
incorporated later (e.g., Lindblom and Scheel \cite{Lindblom03}).
A particularly simple and beautiful way of producing the required
constraint additions ``automatically'' is a covariant extension of the
Einstein equations first introduced by Bona et al. \cite{Bona03a}
called the \emph{Z4 system}.
That formulation has the additional advantage of a simpler constraint 
structure, as we shall see in the following.

This chapter is mainly based on Rinne \& Stewart \cite{RinneStewart05}.
We apply the Z4 extension to the (2+1)+1 formalism
presented in chapter \ref{sec:211} (section \ref{sec:Z4}). 
The evolution system is completed by dynamical gauge conditions 
that generalize harmonic gauge (section \ref{sec:dyngauge}).
We cast the system in first-order form (section \ref{sec:firstorder})
and analyze its hyperbolicity (section \ref{sec:hyperbolicity}).
The characteristic variables and speeds are worked out explicitly.
Particular emphasis is placed on the treatment of the coordinate
singularity on the axis (section \ref{sec:regularity}).
By a judicious choice of new dependent variables
we can write our first-order strongly hyperbolic system in a form where
each and every term is manifestly regular on axis.
Some exact solutions are used to check the equations using a programme
written in the computer algebra language REDUCE (section \ref{sec:algebra}). 
This programme was also used to generate functions written in C for the 
numerical evolution.

\section{The Z4 extension of the (2+1)+1 formalism}
\label{sec:Z4}

Bona et al. \cite{Bona03a} suggested adding a covariant term
$\nabla_{(\alpha} Z_{\beta)}$ to the Einstein equations,
\begin{equation}
  \label{eq:Z4}
  R_{\alpha\beta} + 2 \nabla_{(\alpha} Z_{\beta)} = 
  \kappa \left( T_{\alpha\beta} - \half T g_{\alpha\beta} \right) \,.
\end{equation}
Clearly this reduces to the Einstein equations if and only if
$Z_\alpha = 0$.\footnote{Strictly speaking, it is sufficient if $Z_\alpha$ is
Killing, but from a numerical point of view that is a very special case.}
For the extended equations to be axisymmetric, $Z_\alpha$ has to share
the axisymmetry,
\begin{equation}
  \Lie{\xi} Z_\alpha = 0 \,.
\end{equation}

We would now like to apply the (2+1)+1 formalism directly to \eqref{eq:Z4}
rather than to the original Einstein equations. To do this, it is
convenient to rewrite \eqref{eq:Z4} as Einstein's equations 
\begin{equation}
  G_{\alpha\beta} = \kappa \widetilde T_{\alpha\beta}
\end{equation}
with a modified energy-momentum tensor
\begin{equation}
  \label{eq:Ttilde}
  \widetilde T_{\alpha\beta} = T_{\alpha\beta} - \tfrac{2}{\kappa} \left( 
    \nabla_{(\alpha} Z_{\beta)} - \half g_{\alpha\beta}\nabla_\gamma Z^\gamma 
    \right) \, .
\end{equation}
We then compute the (2+1)+1 matter variables corresponding to
\eqref{eq:Ttilde} and insert them into the (2+1)+1 equations 
(\ref{eq:hamcons}--\ref{eq:d0B}).

First, we decompose $Z_\alpha$ with respect to the Killing vector $\xi^\alpha$
(Geroch decomposition),
\begin{equation}
  Z_\alpha = \hat Z_\alpha + \xi_\alpha Z^\varphi \,,
\end{equation}
where we have defined
\begin{equation}
  \label{eq:ZinN}
  \hat Z_\alpha \equiv h_\alpha{}^\beta Z_\beta
\end{equation}
and
\begin{equation}
  Z^\varphi \equiv \lambda^{-2} \xi^\alpha Z_\alpha \,.
\end{equation}
The projections of
\begin{equation}
  X_{\alpha \beta} \equiv \nabla_{(\alpha} Z_{\beta)}
\end{equation}
are found to be
\begin{eqnarray}
  X_{\xi\xi} &=& \lambda \lambda_a \hat Z^a \,,\\
  X_{\xi a} &=& \half \epsilon_{abc} \hat Z^b \omega^c 
     + \half \lambda^2 Z^\varphi{}_{,a} \,,\\
  X_{ab} &=& D_{(a} \hat Z_{b)} \,.
\end{eqnarray}
Using this, we can easily compute the modified matter variables 
\eqref{eq:Tprojections} corresponding to \eqref{eq:Ttilde},
\begin{eqnarray}
  \tilde \tau &=& \tau - \kappa^{-1} (\lambda^{-1} \lambda_a Z^a
     - D_a Z^a) \,\\
  \tilde \tau_a &=& \tau_a - \kappa^{-1} (\lambda^{-3} \epsilon_{abc}
     Z^b \omega^c + Z^\varphi{}_{,a}) \,\\
  \tilde \tau_{ab} &=& \tau_{ab} - \kappa^{-1} \left[ 2 D_{(a} Z_{b)}
     - (\lambda^{-1} \lambda_c Z^c + D_c Z^c) h_{ab} \right] \,.
\end{eqnarray}
Here and in the following, we leave out the hat in $\hat Z_a$
(there should be no ambiguity because it carries a Latin index). 

Next, we decompose $Z_a$ with respect to the timelike normal $n_a$
(ADM decomposition),
\begin{equation}
   Z_a = \hat {\hat Z}_a + n_a \theta \,,
\end{equation}
where we have defined
\begin{equation}
  \theta \equiv - n^a Z_a
\end{equation}
and
\begin{equation}
  \hat{\hat Z}_a \equiv H_a{}^b Z_b \,.
\end{equation}
The projections of
\begin{equation}
  X_{ab} \equiv D_{(a} Z_{b)}
\end{equation}
are found to be
\begin{eqnarray}
  X_{nn} &=& -\alpha^{-1} \alpha_A \hat{\hat Z}^A - \Lie{n} \theta \,,\\
  X_{nA} &=& \half \Lie{n} \hat {\hat Z}_A + \chi_{AB} \hat{\hat Z}^B + \half
     \alpha^{-1} \alpha_A \theta - \half \theta_{A} \,,\\
  X_{AB} &=& d_{(A} \hat{\hat Z}_{B)} - \chi_{AB} \theta \,.
\end{eqnarray}
Further identities we need are
\begin{eqnarray}
  \lambda_a Z^a &=& \lambda_A \hat{\hat Z}^A - \lambda \K \theta \,,\\
  n^a \epsilon_{abc} Z^b \omega^c &=& \lambda^3 E^A \hat{\hat Z}_A \,,\\
  H^a{}_A \epsilon_{abc} Z^b \omega^c &=& \lambda^3 B^\varphi
    \epsilon_{AB} \hat{\hat Z}^B - \lambda^3 E_A \theta \,,
\end{eqnarray}
where the definitions of $\K$ \eqref{eq:Kpp}, $E^A$ \eqref{eq:E} 
and $B^\varphi$ \eqref{eq:B} have been used.
The modified (2+1)+1 matter variables (\ref{eq:tauprojections})
are then computed as
\begin{eqnarray}
  \label{eq:modmat}
  \widetilde \tau &=& \tau + \kappa^{-1} \left[ \Lie{n} \theta 
  + Z^A{}_{||A} + (A_A - L_A ) Z^A + (\K - \chi) \theta \right] \,, \\
  \widetilde S_A &=& S_A + \kappa^{-1} \left[ - Z^\varphi{}_{,A} 
    + B^\varphi \epsilon_{AB} Z^B + E_A \theta  \right]
  \,, \\
  \widetilde J^\varphi &=& J^\varphi + \kappa^{-1} \left[ \Lie{n} Z^\varphi 
    + E^A Z_A \right] \,, \\
  \widetilde S_{AB} &=& S_{AB} + \kappa^{-1} \big[ -2 Z_{(A||B)} 
    + 2 \chi_{AB} \theta + H_{AB} \left\{ \Lie{n} \theta + Z^C{}_{||C} \right.
    \nonumber\\ && \left. \qquad \qquad \qquad 
    + (A_C + L_C) Z^C - (\chi + \K) \theta \right\} \big] \,, \\
  \widetilde J_A &=& J_A + \kappa^{-1} \left[ \Lie{n} Z_A - \theta_{,A} 
    + 2 \chi_{AB} Z^B + A_A \theta  \right] \, , \\
  \widetilde \rho_H &=& \rho_H + \kappa^{-1} \left[ \Lie{n} \theta 
    - Z^A{}_{||A} + (A_A - L_A) Z^A + (\chi + \K) \theta \right] 
\end{eqnarray}
where we again leave out the double hat in $\hat{\hat Z}_A$.

Inserting the modified matter variables into the (2+1)+1 equations, we
arrive at what we call the \emph{Z(2+1)+1 equations}.
The constraints (\ref{eq:hamcons}--\ref{eq:gercons}) are turned into
evolution equations for the $Z$ vector,
\begin{eqnarray}
  \label{eq:d0theta}
  \Lie{n} \theta &=& \mathcal{C} + (\lambda^{-1} \lambda_A -
  \alpha^{-1} \alpha_A) Z^A 
  + Z^A{}_{||A} - (\chi + \K) \theta \,,\\
  \label{eq:d0ZA}
  \Lie{n} Z_A &=& \mathcal{C}_A
  - 2 \chi_{AB} Z^B - \alpha^{-1} \alpha_A \theta  + \theta_{,A} \,,\\
  \label{eq:d0Zphi}
  \Lie{n} Z^\varphi &=& \mathcal{C}_\varphi - E^A Z_A \,.
\end{eqnarray}
We see from (\ref{eq:d0theta}--\ref{eq:d0Zphi}) that if the $Z$ vector
vanishes at all times, then $\mathcal{C}, \mathcal{C}_A$ and 
$\mathcal{C}_\varphi$ also vanish at all times. In this sense, the
original constraints $\mathcal{C} = \mathcal{C}_A =
\mathcal{C}_\varphi = 0$, which involve derivatives of the metric and
extrinsic curvature, are replaced with the purely \emph{algebraic} 
constraints $\theta = Z_A = Z^\varphi = 0$.

The evolution equations are modified in the following way:
\begin{eqnarray}
  \label{eq:Zd0chi}
  \Lie{n} \chi_{AB} &=& \ldots + 2 Z_{(A||B)} - 2 \chi_{AB} \theta \,,\\
  \label{eq:Zd0K}
  \Lie{n} \K &=& \ldots + 2 L_A Z^A - 2 \K
    \theta \,,\\
  \label{eq:Zd0E}  
  \Lie{n} E^A &=& \ldots  + 2 Z^\varphi{}^{,A} - 2 E^A \theta 
    - 2 B^\varphi \epsilon^{AB} Z_B \,,
\end{eqnarray}
where $\ldots$ denote the right-hand-sides of \eqref{eq:d0chi},
\eqref{eq:d0Kpp} and \eqref{eq:d0E}, respectively.
The remaining evolution equations are unchanged.
Thus terms homogeneous in the constraints are added to the evolution
equations, a feature common to many hyperbolic reductions of the
Einstein equations. Here it occurs in a completely natural way --
there is no need to add the constraints ``by hand''.


   \section{Dynamical gauge conditions}
\label{sec:dyngauge}

To complete our evolution formalism, we need to prescribe the gauge
variables $\alpha$ and $\beta^A$. Since we are aiming for a completely
hyperbolic system, we would like to impose a hyperbolic gauge
condition as well. The prototype of such a condition is \emph{harmonic
  gauge}, which can be derived as follows.
The principal part of the Einstein equations can be written as 
\cite{DeDonder21, Fock59}
\begin{equation}
  - g^{\gamma\delta}g_{\alpha\beta,\gamma\delta} + 2 \Gamma_{(\alpha, \beta)}
  \simeq 0 \,,
\end{equation}
where we have defined
\begin{equation}
  \Gamma_\alpha \equiv \Gamma_{\alpha\gamma}{}^\gamma \equiv
  g_{\alpha\delta} g^{\beta\gamma} \Gamma^\delta{}_{\beta\gamma}
\end{equation}
and $\simeq$ denotes equality to principal parts.
If we now adopt the harmonic gauge condition
\begin{equation}
  \label{eq:harmgauge}
  g^{\gamma\delta}x^\alpha{}_{;\gamma\delta} = -\Gamma^\alpha = 0 \, ,
\end{equation}
where the coordinates $x^\alpha$ are to be treated as scalar fields,
the Einstein equations reduce to a wave equation for the metric,
 \begin{equation}
   \label{eq:Einsteinwaveeqn}
   g^{\gamma\delta}g_{\alpha\beta,\gamma\delta} \simeq 0 \,.
\end{equation}
This system of PDEs is clearly \emph{symmetric hyperbolic} (section
\ref{sec:hyperbolicity}), a property 
used by Bruhat \cite{Bruhat52} in the first well-posedness 
theorem for the initial-value problem of the Einstein equations.

The principal part of the Z4-Einstein equations \eqref{eq:Z4} takes the form
\begin{equation}
  - g^{\gamma\delta}g_{\alpha\beta,\gamma\delta} + 2 \Gamma_{(\alpha, \beta)}
  + 4 Z_{(\alpha, \beta)} \simeq 0 \,.
\end{equation}
In order to retain \eqref{eq:Einsteinwaveeqn}, we have to
replace \eqref{eq:harmgauge} with
\begin{equation}
  \label{eq:Zharmgauge}
  g^{\gamma\delta}x^\alpha{}_{;\gamma\delta} = - \Gamma^\alpha 
     = 2 Z^\alpha \,.
\end{equation}

This condition can be translated into (2+1)+1 language by going through the
Geroch and ADM decompositions as in chapter \ref{sec:211}.
We arrive at the following evolution equations for the lapse and shift:
\begin{eqnarray}
  \label{eq:d0alphaharm}
  \eth_t \alpha &=& -\alpha^2 ( \chi + \K - 2 \theta) \,,\\
  \label{eq:d0betaharm}
  \eth_t \beta^A &=& -\alpha^2 (\partial^A \ln(\alpha\lambda\sqrt{H})
                            + \partial_B H^{AB} - 2 Z^A) \,,
\end{eqnarray}
where here and in the following we set
\begin{equation}
  \eth_t \equiv \partial_t - \beta^B \partial_B \,.
\end{equation}

Bona et al. \cite{Bona03b, Bona04a} have generalized 
(the 3+1 analogue of) these conditions by inserting some free constant 
parameters $f, m, \mu, d$ and $a$,
\begin{eqnarray}
  \label{eq:d0alpha}
  \eth_t \alpha &=& - \alpha^2 \left[ f( \chi + \K - m \theta) \right] \,,\\
  \label{eq:d0beta}                                  
  \eth_t \beta^A &=& -\alpha^2 \left[ 2 \mu \left( 
      \partial^A \ln(\lambda\sqrt{H}) + \half \partial_B H^{AB} - Z^A \right)
    \right. \nonumber\\ &&\qquad\quad \left.
    - d \partial^A \ln(\lambda\sqrt{H}) 
    + a \partial^A \ln \alpha  \right]\,.
\end{eqnarray}
Clearly, we recover the original harmonic gauge conditions
(\ref{eq:d0alphaharm}--\ref{eq:d0betaharm}) if we set $f = \mu = d = a = 1$
and $m = 2$. 

For even more generality, one could add to the right-hand-side of 
\eqref{eq:Zharmgauge} an
arbitrary \emph{gauge source function} $G^\mu$, which may depend on
the coordinates and the metric but not on its derivatives, so that the
principal parts of the Einstein equations are unaffected. 
Such a modification corresponds to adding
\begin{eqnarray}
  \label{eq:dyngaugewithsource}
  \eth_t \alpha &=& \ldots - \alpha^2 G^0 \,,\nonumber\\
  \eth_t \beta^A &=& \ldots -\alpha^2 G^A
\end{eqnarray}
in (\ref{eq:d0alpha}--{\ref{eq:d0beta}), where
\begin{eqnarray}
  \label{eq:gaugesource}
  G^0 &=& G^0 (x^A, H_{AB}, \lambda, \alpha, \beta^A) \,,\nonumber\\
  G^A &=& G^A (x^A, H_{AB}, \lambda, \alpha, \beta^A) \,.
\end{eqnarray}
Such gauge source functions were first introduced by Friedrich
\cite{Friedrich96} and have recently been applied to numerical
relativity \cite{Garfinkle02, Pretorius05}.
Here we argue that they are particularly important in the context of
axisymmetry: notice that the $r$-component of the right-hand-side of 
\eqref{eq:d0beta} is
singular on the axis because $\lambda = O(r)$ there and so 
$\partial^r \ln \lambda = O(r^{-1})$. One might hope that by choosing
the gauge source function $G^r$ appropriately, one might be able to
cancel the offending term. We will see in section \ref{sec:regularity}
that this is indeed possible.

As an alternative to \eqref{eq:d0beta}, one could choose the shift vector 
to vanish,
\begin{equation}
  \beta^A = 0 \,,
\end{equation}
and this is the choice we made in \cite{RinneStewart05}.
More generally, one could set $\beta^A$ to some arbitrary but fixed
functions.

In both cases, we apply the harmonic slicing condition \eqref{eq:d0alpha}.
Harmonic slicing has been shown to have similar singularity avoidance
properties as maximal slicing \cite{BonaMasso88}. It has been
successfully used in stable evolutions of black hole spacetimes
\cite{Anninos95}. Claims have been made \cite{Alcubierre97} 
that for $f \neq 1$ in \eqref{eq:d0alpha}, coordinate pathologies might arise. 
Another reason for choosing $f = 1$ is the symmetric hyperbolicity of
the system in the zero-shift case (section \ref{sec:hyperbolicity}).


   \section{First-order reduction}
\label{sec:firstorder}

The Z(2+1)+1 equations (\ref{eq:d0theta}--\ref{eq:Zd0E},
\ref{eq:d0H}--\ref{eq:d0lambda}, \ref{eq:d0B}),
supplemented with the dynamical gauge conditions 
(\ref{eq:d0alpha}--\ref{eq:d0beta}), form a system
of pure evolution equations. They contain only first-order time derivatives
but up to second-order spatial derivatives.
Whilst methods for analyzing the hyperbolicity of such second-order
systems have recently been developed (e.g., \cite{NOR04, Gundlach04a, 
Gundlach04b}), the most straightforward way is to perform a reduction
to a set of evolution equations that are first-order in space and time.

To eliminate the second-order spatial derivatives, we introduce new
variables for the first-order spatial derivatives of the metric and gauge:
\begin{eqnarray}
  \label{eq:Ddef}
  D_{ABC} &\equiv& \half \partial_A H_{BC} \,,\\
  \label{eq:Ldef}
  L_A &\equiv& \lambda^{-1} \partial_A \lambda \,,\\
  \label{eq:Adef}
  A_A &\equiv& \alpha^{-1} \partial_A \alpha \,,\\
  \label{eq:Bdef}
  B_A{}^B &\equiv& \alpha^{-1} \partial_A \beta^B \,.
\end{eqnarray}
Evolution equations for these can
be obtained from (\ref{eq:d0H}--\ref{eq:d0lambda}) and 
(\ref{eq:d0alpha}--\ref{eq:d0beta}) by commuting space and time
derivatives\footnote{There is an ordering ambiguity for the second-order
  spatial derivatives on the right-hand-side of the evolution
  equations for the first-order variables. 
  We always use the ordering that produces an
  advection term along the shift, equation \eqref{eq:consform}.},
\begin{equation}
  \partial_t D_{ABC} = \half (\partial_t H_{AB})_{,C} \quad \mathrm{etc.}
\end{equation}
and noting that $\Lie{n} = \alpha^{-1} (\partial_t - \Lie{\beta})$.
Indices are raised and lowered with the 2-metric $H_{AB}$
(formally, for $D_{ABC}$ and $B_A{}^B$ are not tensors).
The two independent traces of $D_{ABC}$ are denoted by
\begin{equation*}
  D^\mathrm{I}{}_A \equiv D_{AB}{}^B \,, \qquad 
  D^\mathrm{II}{}_A \equiv D^B{}_{BA} \,.
\end{equation*}

A crucial step for obtaining a hyperbolic system is the reduction of
the Ricci tensor. We use the De Donder--Fock decomposition 
\cite{DeDonder21, Fock59}
\begin{eqnarray}
  \label{eq:Riccidecomp}
  \two{R}_{AB} &=& - D^C{}_{AB,C} + 2 D^\mathrm{II}{}_{(A,B)} 
  - D^\mathrm{I}{}_{(A,B)} \nonumber \\ &&- 2 D_{CAB} D^{\mathrm{II}\,C}  
  - \Gamma_{CAB} ( 2 D^{\mathrm{II}\,C} - D^{\mathrm{I}\,C} )  \\&&
  + 4 D_{CDA} D^{CD}{}_B- \Gamma_{ACD} \Gamma_B{}^{CD} \,,\nonumber
\end{eqnarray}
where of course the Christoffel symbols are given by
\begin{equation}
  \label{eq:Chris}
  \Gamma_{ABC} = D_{CAB} + D_{BCA} - D_{ABC}.
\end{equation}
A different possibility would be the standard Ricci decomposition
\begin{equation}
  \two{R}_{AB} = \Gamma^C{}_{AB,C} - \Gamma^C{}_{CB,A} 
    + \Gamma^D{}_{DC} \Gamma^C{}_{AB} - \Gamma^C{}_{DA}
    \Gamma^D{}_{CB} 
\end{equation}
or linear combinations of the two \cite{Bona03b}, but only the choice
\eqref{eq:Riccidecomp} leads to a symmetric hyperbolic system for
$f=1$ (section \ref{sec:hyperbolicity}).

It is now straightforward to write the Z(2+1)+1 equations in
\emph{conservation form} with sources,
\begin{equation}
  \label{eq:consform}
  \vec u_{,t} 
  + \left[ - \beta^D \vec u + \alpha \vec \F^D (\vec u) \right]_{,D} 
  = \alpha \vec \S (\vec u) \,.
\end{equation}
Here, $\vec u$ is the vector of \emph{conserved variables},
\begin{eqnarray}
  \vec u &=& (H_{AB}, \lambda, \alpha, \beta^A, D_{ABC}, L_A, A_A,
  B_A{}^B, \chi_{AB}, \K, E^A, B^\varphi, \nonumber\\&&\quad
  \theta, Z_A, Z^\varphi)^T \,.
\end{eqnarray}
(These variables do \emph{not} have any physical interpretation 
as conserved quantities such as mass, angular momentum etc.)
$\F^D (\vec u)$ are \emph{flux vectors}, whose components are given by
{\allowdisplaybreaks\begin{eqnarray}
  \F^D{}_{H_{AB}} &=& 0 \, ,\\
  \F^D{}_{\lambda} &=& 0 \, ,\\
  \F^D{}_{\alpha} &=& 0 \, ,\\
  \F^D{}_{\beta^A} &=& 0 \, ,\\
  \label{eq:Dflux}
  \F^D{}_{D_{ABC}} &=& \delta_A{}^D (\chi_{BC} - 2 B_{(BC)}) \, ,\\
  \F^D{}_{L_A} &=& \delta_A{}^D \K \, ,\\
  \F^D{}_{A_A} &=& \delta_A{}^D f (\chi + \K - m \theta) \,,\\
  \label{eq:Bflux}
  \F^D{}_{B_A{}^B} &=& \half \delta_A{}^D \left[ 2 \mu (L^B + D^{\mathrm{I}\,B}
      - D^{\mathrm{II}\,B} - Z^B) - d(L^B + D^{\mathrm{I}\,B}) 
      \right. \nonumber\\&&\qquad\quad \left. + a A^B \right] \,,\\
  \F^D{}_{\chi_{AB}} &=& D^D{}_{AB} - \delta_{(A}{}^D \left( 
    2 D^{\mathrm{II}}{}_{B)} + 2 Z_{B)} -  D^{\mathrm{I}}{}_{B)} 
    - L_{B)} - A_{B)} \right) \,,\\
  \F^D{}_{\K} &=& L^D \,,\\
  \F^D{}_{E^A} &=& - 2 H^{AD} Z^\varphi - \epsilon^{AD} B^\varphi \,,\\
  \F^D{}_{B^\varphi} &=& -\epsilon^{AD} E_A \,,\\
  \F^D{}_{\theta} &=& D^{\mathrm{I}\,D} - D^{\mathrm{II}\,D} + L^D - Z^D \,,\\
  \F^D{}_{Z_A} &=& -\chi_A{}^D + \delta_A^D ( \chi + \K 
    - \theta ) \,,\\
  \F^D{}_{Z^\varphi} &=& - \half E^D \,.
\end{eqnarray}}
We have separated the common advection term along the shift vector in
(\ref{eq:consform}) from the fluxes. Because this is a diagonal term, it does
not affect the eigenvectors presented below in section \ref{sec:hyperbolicity}
(it merely shifts the eigenvalues).

$\S(\vec u)$ is a \emph{source term} containing no derivatives,
apart from those of the gauge source functions $G^0$ and $G^A$, but
because of \eqref{eq:gaugesource} those can be written as first-order
variables without derivatives. The sources are given by
{\allowdisplaybreaks\begin{eqnarray}
  \S_{H_{AB}} &=& -2 \chi_{AB} + 4 B_{(AB)} - 2 B_D{}^D H_{AB}\,,\\
  \S_\lambda &=& -\lambda \K - 2 B_D{}^D \lambda \,,\\
  \S_\alpha &=& - \alpha \left[ f(\chi + \K - m \theta) + G^0 \right]
     - 2 B_D{}^D \alpha \,,\\
  \S_{\beta^A} &=&  -\alpha \left[ 2 \mu (L^A + D^{\mathrm{I}\,A}
     - D^{\mathrm{II}\,A} - Z^A) - d(L^A + D^{\mathrm{I}\,A})
     \right. \nonumber\\ && \left. \qquad + a A^A + G^A \right] 
     - 2 B_D{}^D \beta^A \,,\\
  \S_{D_{ABC}} &=& 2 B_A{}^D D_{DBC} - 2 B_D{}^D D_{ABC} \,,\\
  \S_{L_A} &=& 2 B_A{}^B A_B - 2 B_D{}^D L_A \,,\\
  \S_{A_A} &=& 2 B_A{}^B L_B - G^0{}_{,A} - A_A G^0 - 2 B_D{}^D A_A \,,\\
  \S_{B_A{}^B} &=& 2 B_A{}^C B_C{}^B - 2 B_D{}^D B_A{}^B
     + f(\chi + \K - m \theta) B_A{}^B  \nonumber\\&&
     - \half A_A \left[ 2 \mu (L^B + D^{\mathrm{I}\,B}
      - D^{\mathrm{II}\,B} - Z^B) \right. \nonumber\\&&\qquad\quad \left.
      - d(L^B + D^{\mathrm{I}\,B}) + a A^B + 2 G^B \right]
     - \half G^B{}_{,A}  \,,\\
  \S_{\chi_{AB}} &=& A_{(A} \left( -2 D^{\mathrm{II}}{}_{B)} 
     D^\mathrm{I}{}_{B)} + L_{B)} - 2 Z_{B)} \right) \nonumber\\&& 
      - L_A L_B + D_{CAB} (A^C - 2 D^{\mathrm{II}\,C})
       \nonumber\\&&  
       - \Gamma_{CAB} (2 Z^C + 2 D^{\mathrm{II}\,C} - D^{\mathrm{I}\,C} 
       - L^C - A^C)  \nonumber\\&&
     + 4 D_{CDA} D^{CD}{}_B - \Gamma_{ACD} \Gamma_B{}^{CD} 
     - 2 B_D{}^D \chi_{AB} \\&&
     + 2 (2 B_{(A}{}^C - \chi_{(A}{}^C) \chi_{B)C}   
     + \chi_{AB} (\chi + \K - 2 \theta) \nonumber\\&&
     - \half \lambda^2 \left[ \epsilon_{AC} \epsilon_{BD} E^C E^D 
       - H_{AB} (E_C E^C - B^\varphi{}^2 ) \right] \nonumber\\&& 
     - \kappa \left[ S_{AB} + \half H_{AB} (\rho_H - S_C{}^C - \tau)
     \right] \,,\nonumber\\
  \S_{\K} &=&  L_A (2 Z^A - L^A - D^{\mathrm{I}\,A})
     + \K (\chi + \K - 2 \theta )  \nonumber\\&& 
     - 2 B_D{}^D \K - \half \lambda^2 (E_A E^A + B^\varphi{}^2) \\&&
      - \half \kappa (\rho_H - S_C{}^C + \tau) \,,\nonumber\\
  \S_{E^A} &=&  (4 D^{\mathrm{II}\,A} - 2 A^A) Z^\varphi 
    + (\chi + 3 \K - 2 \theta) E^A \nonumber\\&&
    - 2 B_B{}^A E^B - 2 B_D{}^D E^A \\&&
    + \epsilon^{AB} B^\varphi (3 L_B - 2 Z_B + D^\mathrm{I}{}_B) 
    - 2 \kappa S^A  \,,\nonumber\\
  \S_{B^\varphi} &=& \chi B^\varphi + \epsilon^{AB} E_A D^\mathrm{I}{}_B 
     - 2 B_D{}^D B^\varphi \,,\\
  \S_\theta &=& A_A (D^{\mathrm{I}\,A} - D^{\mathrm{II}\,A} + L^A - 2 Z^A) 
    \nonumber\\&& + (L_A + D^\mathrm{I}{}_A) (Z^A - L^A) 
    - \half D^\mathrm{I}{}_A D^{\mathrm{I}\,A}  \nonumber\\&&
    + D_{ABC} D^{ABC} - \half \Gamma_{ABC} \Gamma^{ABC} - 2 B_D{}^D \theta \\&&
    + \half (\chi^2 - \chi_{AB} \chi^{AB}) + \chi \K 
    - (\chi + \K) \theta \nonumber\\&&
    - \textstyle \frac{1}{4} \lambda^2 (E_A E^A + B^\varphi{}^2)
    - \kappa \rho_H \,,\nonumber\\
  \S_{Z_A} &=& 2 B_A{}^B Z_B - 2 B_D{}^D Z_A + A_A (\chi + \K - 2 \theta) 
     \nonumber\\&& 
     - L_A \K + \chi_{AB} ( D^{\mathrm{I}\,B} + L^B - 2 Z^B - A^B)  \\&&
     - \Gamma_{CAB} \chi^{BC} - \half \lambda^2 B^\varphi
     \epsilon_{AB} E^B - \kappa J_A  \,,\nonumber\\
  \S_{Z^\varphi} &=& \half E^A (D^\mathrm{I}{}_A + 3 L_A - 2 Z_A - A_A)
     - 2 B_D{}^D Z^\varphi - \kappa J^\varphi \,.
\end{eqnarray}}

Note that $H_{AB}, \lambda, \alpha$ and $\beta^A$ have vanishing fluxes and
thus trivially propagate along the normal lines. 
The twist variables $E^A, B^\varphi$ and $Z^\varphi$ form a decoupled 
subsystem on the level of principal parts (i.e., fluxes). 
In linearized theory, it completely decouples because the twist
variables enter the
source terms of the remaining equations only quadratically.


   \section{Hyperbolicity}
\label{sec:hyperbolicity}

\subsection{Generalities, well-posedness of the IVP}

To investigate the hyperbolicity of the Z(2+1)+1 system, we pick a unit
covector $\mu_A$ and define an orthogonal covector 
\begin{equation}
  \label{eq:pdef}
  \pi_A \equiv \epsilon_{AB} \mu^B \,,
\end{equation}
so that
\begin{equation*}
  \mu_A \mu^A = \pi_A \pi^A = 1 \,, \qquad 
  \mu_A \pi^A = 0 \,.
\end{equation*}
Thus $(\mu^A, \pi^A)$ form an orthonormal basis for the tangent space
of the slice $\Sigma(t)$.
Projection along $\mu$ and $\pi$ is denoted as\footnote{Here we use
  the opposite notation to \cite{RinneStewart05} because later
  (chapter \ref{sec:outerbcs}), 
  $\mu$ will be the normal ($\n$) to the boundary and $\pi$ will be
  parallel ($\p$) to it.}
\begin{equation}
  \label{eq:projnotation}
  V^\n \equiv V^A \mu_A \,, \qquad V^\p \equiv V^A \pi_A \,.
\end{equation}

Consider the Jacobian matrix of the flux in the $\mu$-direction,
\begin{equation}
  J \equiv \frac{\partial\F^\n}{\partial \vec u} \,.
\end{equation}
A vector $\vec r$ is a \emph{right eigenvector} of $J$ with 
\emph{eigenvalue} or \emph{characteristic speed} $\lambda$ if
\begin{equation}
  J \vec r = \lambda \vec r \,.
\end{equation}
A vector $\vec l$ is a \emph{left eigenvector} if
\begin{equation}
  J^T \vec l = \lambda \vec l \,.
\end{equation}
Note that $J^T$ has the same eigenvalues as $J$.
The \emph{characteristic variable} $l$ corresponding to a left eigenvector
$\vec l$ is defined to be $l = \vec l^T \vec u$.

The system is said to be \emph{weakly hyperbolic} if all the
eigenvalues are real, independently of the direction $\mu$.
It is \emph{strongly hyperbolic} if in addition
there exist complete sets of left and right eigenvectors (i.e., they
span the space), independently of $\mu$. 
Finally, it is \emph{symmetric hyperbolic} if $J$ is symmetrizable 
(i.e., there exists a symmetric positive-definite matrix $H$ such that 
$H J$ is symmetric) with a symmetrizer $H$ that is independent of $\mu$.
Clearly, symmetric hyperbolicity implies strong hyperbolicity, which in
turn implies weak hyperbolicity, but not the other way around.  

The significance of \emph{strongly} hyperbolic systems as opposed to weakly
hyperbolic ones is that at least for the case that the fluxes and
sources are linear and homogeneous in the unknowns $\vec u$, 
they admit a well-posed Cauchy or initial value problem (IVP) in the
following sense \cite{GKO, Stewart98}: 
for every initial data 
$\vec f \in C^\infty(x^A)$, 
$\vec u(0, x^A) = f(x^A)$, there exists a unique solution 
$\vec u(t, x^A) \in C^\infty(t, x^A)$ such that
\begin{equation}
  \label{eq:wellposedestimate}
  \lVert \vec u(t, \cdot) \rVert \leqslant K \e^{\alpha t} \lVert
  \vec f(\cdot) \rVert \,,
\end{equation}
where the constants $K$ and $\alpha$ are independent of $f$, and we are using
$L^2$ norms\footnote{Technically, one requires the additional condition that
  the matrix of eigenvectors and its inverse are uniformly bounded.}.
For nonlinear systems such as the one being considered in
this chapter, one can only hope for the estimate
\eqref{eq:wellposedestimate} to hold for a finite time. This is because 
in the nonlinear case, characteristics might cross to form shocks (as
is well-known in hydrodynamics) so
that a regular solution exists only for a finite time, 
or the nonlinear source terms might lead to an even more severe blow-up.  

The significance of \emph{symmetric} hyperbolicity is that it implies the
existence of a positive-definite \emph{energy} (which has no physical meaning
in general),
\begin{equation}
  \mathcal{E}(t) = \int_{\Sigma(t)} \vec u^T H \vec u \, \diff^2 x  > 0\,,
\end{equation}
where $H$ is the symmetrizer of the system. 
Suppose now that the slices $\Sigma(t)$ have a timelike boundary 
$\partial \Sigma$. Consider a simple linear constant-coefficient problem
\begin{equation}
  \vec u_{,t} = A^A \partial_A \vec u \,.
\end{equation}
Using the fact that the matrices $H A^A$ are symmetric, and Gauss'
theorem, we have
\begin{eqnarray}
  \label{eq:energyestimate}
  \partial_t \mathcal{E}(t) &=& 
  \int_{\Sigma(t)} 2 \vec u^T H A^A \vec u_{,A} \, \diff^2 x = 
  \int_{\Sigma(t)} \partial_A(\vec u^T H A^A \vec u) \, \diff^2 x 
  \nonumber\\&=& \int_{\partial \Sigma(t)} \vec u^T H A^\n \vec u \, 
  \diff x \,, 
\end{eqnarray}
where $A^\n = A^A \mu_A$ denotes the contraction of $A$ with the
normal $\mu$ to the boundary.
If the boundary conditions are chosen such that the last integral in
\eqref{eq:energyestimate} is always non-positive, it follows that
$0 \leqslant \mathcal{E}(t) \leqslant \mathcal{E}(0)$ for all $t \geq 0$.
Such energy estimates are the key ingredient of well-posedness proofs
for the initial boundary value problem \cite{Rauch85, Secchi96}.

\subsection{The dynamical shift case}
\label{sec:dynshifthyp}

We first deal with the general case in which the dynamical shift
condition \eqref{eq:d0beta} is included.

The system is found to be strongly hyperbolic provided that $f > 0$,
$\mu > 0$ and $d > 0$. The parameters $m$ and $a$ are generally
unconstrained. However, the following degenerate cases require more care:
\begin{itemize}
  \item $\underline{f = 1}$: $m = 2$ is needed for strong hyperbolicity.
  \item $\underline{d = 1}$: Here we must also set $\mu = 1$ and $a = 1$.   
  \item $\underline{d = f}$: $a = 1$ is required. 
\end{itemize}  
For $\mu \in \{ 1, f, d \}$ without any further degeneracies, 
the remaining parameters need not be adapted.

The characteristic speeds $\lambda$ and their multiplicities are
\begin{equation}
  \begin{array}{ll}
    \lambda_0 = 0 & (7)\\ 
    \lambda_1^\pm = \pm 1 & (2\times 6)\\ 
    \lambda_f^\pm = \pm \sqrt{f} & (2\times 1)\\ 
    \lambda_\mu^\pm = \pm \sqrt{\mu} & (2\times 1)\\
    \lambda_d^\pm = \pm \sqrt{d} & (2\times 1)
  \end{array}
\end{equation}
Note that because of the advection term and the factor of $\alpha$ 
in the fluxes in \eqref{eq:consform}, the actual characteristic speeds are 
$-\beta^\n + \alpha \lambda$.
For $f \leqslant 1$, $\mu \leqslant 1$ and $d \leqslant 1$, the
characteristic speeds are all causal. If the equality holds, they are
all ``physical'' (i.e., either zero or equal to the speed of light).

The characteristic variables are given by

\paragraph{Normal modes ($\lambda = 0)$:}
{\allowdisplaybreaks \begin{eqnarray}
  l_{0,1} &=& D_{\p\n\n} \,,\\
  l_{0,2} &=& D_{\p\p\n} \,,\\
  l_{0,3} &=& D_{\p\p\p} \,,\\
  l_{0,4} &=& L_\p - D_{\p\p\p} \,,\\
  l_{0,5} &=& A_\p \,,\\
  l_{0,6} &=& B_{\p\n} \,,\\
  l_{0,7} &=& B_{\p\p} \,,
\end{eqnarray}}
along with the zeroth-order variables $H_{AB}$, $\lambda$, $\alpha$, 
$\beta^\n$ and $\beta^\p$. 

\paragraph{Light cone modes ($\lambda = \pm 1$):}
{\allowdisplaybreaks \begin{eqnarray}
  l_{1,1}^\pm &=& \K - \chi_{\p\p} + 2 B_{\p\p} \pm (L_\n - D_{\n\p\p}) \,,\\
  l_{1,2}^\pm &=& E^\p \mp B^\varphi \,,\\
  l_{1,3}^\pm &=& \theta - 2 B_{\p\p} 
    \pm (D_{\n\p\p} + L_\n - D_{\p\p\n} - Z_\n) \,,\\
  l_{1,4}^\pm &=& \K + \chi_{\p\p} - \theta \pm (D_{\p\p\n} + Z_\n) \,,\\
  l_{1,5}^\pm &=& \chi_{\n\p} \pm \half ( A_\p + D_{\p\n\n} -
    D_{\p\p\p} + L_\p - 2 Z_\p) \,,\\
  l_{1,6}^\pm &=& E^\n \mp 2 Z^\varphi \,. 
\end{eqnarray}}

\paragraph{Lapse cone modes ($\lambda = \pm \sqrt{f}$):}
\begin{eqnarray}
  l_f^\pm &=& A_\n - f c_1 (D_{\n\p\p} + L_\n -
    D_{\p\p\n} - Z_\n) \nonumber\\ &&
    \pm \sqrt{f} \left[ \chi_{\n\n} + \chi_{\p\p} + \K 
    - \left( f c_1 + 2 \right) \theta + 2 c_1 B_{\p\p} \right] \,,
\end{eqnarray}
where we have set 
\begin{equation}
  \label{eq:c1}
  c_1 \equiv \frac{m-2}{f-1} \,.
\end{equation}
For $(f = 1\,,m = 2)$, the undefined expression $c_1$ is to be
replaced with an arbitrary fixed constant (e.g., $0$ for simplicity).

\paragraph{Transverse shift cone modes ($\lambda = \pm \sqrt{\mu}$):}
\begin{eqnarray}
  l_\mu^\pm &=& a A_\p + 2\mu (L_\p + D_{\p\n\n} - D_{\n\n\p} - Z_\p )
    \nonumber\\&& - d (D_{\p\n\n} + D_{\p\p\p} + L_\p) 
    \pm 2 \sqrt{\mu} (B_{\n\p} + B_{\p\n}) \,.
\end{eqnarray}

\paragraph{Longitudinal shift cone modes ($\lambda = \pm \sqrt{d}$):}
\begin{eqnarray}
  l_d^\pm &=& \left( f c_2 + 1 \right) (\chi_{\n\n} +
    \chi_{\p\p} + \K)  \nonumber\\&&
    + (f c_2 c_3 + 2 c_4) (2 B_{\p\p} - \theta)
    - f m c_2 \theta - 2 (B_{\n\n} + B_{\p\p}) \nonumber\\&&
    \pm \sqrt{d} \left[ D_{\n\n\n} + D_{\n\p\p} + L_\n +
    c_2 A_\n \right. \\&&\left.\qquad\quad
    - (f c_2 c_3 + 2 c_4) (L_\n + D_{\n\p\p} - D_{\p\p\n} - Z_\n)
    \right] \,,\nonumber
\end{eqnarray}
where we have set
\begin{equation}
  \label{eq:c234}
  c_2 \equiv \frac{a-1}{f-d} \,, \qquad
  c_3 \equiv \frac{m-2}{d-1} \,, \qquad
  c_4 \equiv \frac{\mu-1}{d-1} \,.
\end{equation}
As stated above, if $f = d$ then we must have $a = 1$, and $c_2$ is to
be replaced with an arbitrary constant. If $d = 1$, we also need $m =
2$ and $\mu = 1$, and both $c_3$ and $c_4$ are to be replaced with
arbitrary constants.

The inverse transformation from characteristic to conserved variables
is given by
{\allowdisplaybreaks \begin{eqnarray}
  D_{\n\n\n} &=& \half(f c_1 c_5 + 2 c_4) (l_{1,3}^+ - l_{1,3}^-) 
    - \half (l_{1,3}^+ - l_{1,3}^- + l_{1,4}^+ - l_{1,4}^-)
     \nonumber\\&& - \half c_2 (l_f^+ + l_f^-)
    + \textstyle \frac{1}{2 \sqrt{d}} (l_d^+ - l_d^-) \,,\\
  D_{\n\n\p} &=& \textstyle \frac{a-\mu}{2 \mu} l_{0,5}
    + \frac{\mu-d}{2 \mu} (l_{0,1} + 2 l_{0,3} + l_{0,4})
    + \half (l_{1,5}^+ - l_{1,5}^-) \nonumber\\&&\textstyle
    - \frac{1}{4 \mu} (l_\mu^+ + l_\mu^-) \,,\\
  D_{\n\p\p} &=& \textstyle
    \frac{1}{4} (-l_{1,1}^+ + l_{1,1}^- + l_{1,3}^+ - l_{1,3}^-
    + l_{1,4}^+ - l_{1,4}^- ) \,,\\
  D_{\p\n\n} &=& l_{0,1} \,,\\
  D_{\p\p\n} &=& l_{0,2} \,,\\
  D_{\p\p\p} &=& l_{0,3} \,,\\
  L_\n &=& \textstyle 
    \frac{1}{4} (l_{1,1}^+ - l_{1,1}^- + l_{1,3}^+ - l_{1,3}^-
    + l_{1,4}^+ - l_{1,4}^- ) \,,\\
  L_\p &=& l_{0,3} + l_{0,4} \,,\\
  A_\n &=& \half \left( l_f^+ + l_f^- \right)
    + \half f c_1 (l_{1,3}^+ - l_{1,3}^-)  \,,\\
  A_\p &=& l_{0,5} \,,\\
  B_{\n\n} &=& \textstyle -\frac{1}{4}(l_d^+ + l_d^-)
    + \frac{1}{4 \sqrt{f}} (f c_2 + 1) (l_f^+ - l_f^-) 
    + (m-1) l_{0,7}\nonumber\\&&
    \textstyle - \frac{1}{4} \left[ f c_1 (c_5 - 1) + 2 (c_4 - 1) \right] 
    (l_{1,3}^+ + l_{1,3}^-)  \,,\\
  B_{\n\p} &=& \textstyle \frac{1}{4 \sqrt{\mu}} (l_\mu^+ - l_\mu^-)
    - l_{0,6} \,,\\
  B_{\p\n} &=& l_{0,6} \,,\\
  B_{\p\p} &=& l_{0,7} \,,\\
  \chi_{\n\n} &=& \textstyle \half (f c_1 + 2) (l_{1,3}^+ + l_{1,3}^-)  
     + \frac{1}{2\sqrt{f}} \left( l_f^+ - l_f^- \right) \nonumber\\&&
     - \half (l_{1,3}^+ + l_{1,3}^- + l_{1,4}^+ + l_{1,4}^-) 
     + 2 (m-1) l_{0,7} \,,\\
  \chi_{\n\p} &=& \half (l_{1,5}^+ + l_{1,5}^-) \,,\\
  \chi_{\p\p} &=& \textstyle
    \frac{1}{4} (-l_{1,1}^+ - l_{1,1}^- + l_{1,3}^+ + l_{1,3}^-
     + l_{1,4}^+ + l_{1,4}^- ) + 2 l_{0,7} \,,\\
  \K &=& \textstyle \frac{1}{4} (l_{1,1}^+ + l_{1,1}^- + l_{1,3}^+ + l_{1,3}^-
    + l_{1,4}^+ + l_{1,4}^- ) \,,\\
  E^\n &=& \half (l_{1,6}^+ + l_{1,6}^-) \,,\\
  E^\p &=& \half (l_{1,2}^+ + l_{1,2}^-) \,,\\\
  B^\varphi &=& -\half (l_{1,2}^+ - l_{1,2}^-) \,,\\
  \theta &=& \half (l_{1,3}^+ + l_{1,3}^-) + 2 l_{0,7} \,,\\
  Z_\n &=&  \half (l_{1,4}^+ - l_{1,4}^-) - l_{0,2} \,,\\
  Z_\p &=& \half (l_{0,1} + l_{0,4} + l_{0,5}) 
    - \half(l_{1,5}^+ - l_{1,5}^-) \,,\\
  Z^\varphi &=& \textstyle -\frac{1}{4} (l_{1,6}^+ - l_{1,6}^-) \,,
\end{eqnarray}}
where in addition we have defined
\begin{equation}
  c_5 \equiv \frac{a-1}{d-1} \,.
\end{equation}

Unfortunately, the system with a dynamical shift is never symmetric
hyperbolic, not even for harmonic gauge ($f = d = \mu = a = 1$, $m = 2$).
This is because the antisymmetric part of $B_{AB}$ does not enter the
fluxes (only the symmetric part appears in the flux of $D_{ABC}$, equation
\eqref{eq:Dflux}).
However, $B_{[AB]}$ itself has a nonzero flux \eqref{eq:Bflux}. 
Hence a direction-independent symmetrizer does not exist.

\subsection{The vanishing shift case}

Next, we deal with the choice $\beta^A = 0$ for the shift vector. 
This is the case we considered in \cite{RinneStewart05}\footnote{The 
definitions of the characteristic variables in 
\cite{RinneStewart05} differ from those presented here in the ordering 
and by linear combinations.}. The following analysis would be unchanged 
if we chose $\beta^A$ to be some nonzero fixed vector (except that the
eigenvalues would be shifted by 
$\lambda \rightarrow \lambda - \alpha^{-1} \beta^A$).

The system is found to be strongly hyperbolic for all $f > 0$.
The parameter $m$ is unconstrained, expect for $f = 1$, 
in which case we need $m = 2$ in order to maintain strong hyperbolicity
(and hence we recover harmonic slicing \eqref{eq:d0alphaharm})\,.

The characteristic speeds and multiplicities are
\begin{equation}
  \begin{array}{ll}
    \lambda_0 = 0 & (7) \\
    \lambda_1^\pm = \pm 1 & (2\times 6) \\
    \lambda_f^\pm = \pm \sqrt{f} & (2\times 1)
  \end{array}  
\end{equation}

The characteristic variables can readily be obtained from the dynamical
shift case with the following modifications:

\begin{itemize}

  \item 
  Replace the normal modes $l_{0,6}$ and $l_{0,7}$ with 
  {\allowdisplaybreaks \begin{eqnarray}
    l_{0,6} &=& f m (D_{\n\p\p} + L_\n - D_{\p\p\n} - Z_\n) \nonumber\\&&
      - f (D_{\n\n\n} + D_{\n\p\p} + L_\n) + A_\n \,,\\
    l_{0,7} &=& f m (D_{\p\n\n} + L_\p - D_{\n\n\p} - Z_\p) \nonumber\\&&
      - f (D_{\p\n\n} + D_{\p\p\p} + L_\p) + A_\p \,.    
  \end{eqnarray}}

  Here we see very clearly how even though two normal modes ($B_{\p\n}$
  and $B_{\p\p}$) are lost, the system manages to remain strongly
  hyperbolic because two new normal modes appear.

  \item
  Set $B_{\p\p} = 0$ in $l_{1,1}^\pm$ and $l_f^\pm$.

  \item
  Clearly, there are no transverse and longitudinal shift modes in this case.

\end{itemize}

The inverse transformation is obtained from the dynamical shift case
by making the following changes:
\begin{itemize}
  
  \item 
  Replace the expressions for $D_{\n\n\n}$ and $D_{\n\n\p}$ with

  \begin{eqnarray} 
    D_{\n\n\n} &=& -\textstyle \frac{1}{f} l_{0,6} 
      + \half (f c_1 + 2) (l_{1,3}^+ - l_{1,3}^-) \nonumber\\&&  
      - \half (l_{1,3}^+ - l_{1,3}^- + l_{1,4}^+ - l_{1,4}^-) 
      + \frac{1}{2f} \left( l_f^+ + l_f^- \right) \,,\\
    D_{\n\n\p} &=& -\textstyle \frac{1}{fm} l_{0,7} 
      + \frac{(m-2)}{2m} (l_{0,1} + 2 l_{0,3} + l_{0,4}) \nonumber\\&&
      \textstyle - \frac{(fm-2)}{2fm} l_{0,2} 
      + \half (l_{1,5}^+ - l_{1,5}^-) \,.
  \end{eqnarray}

  \item
  Discard the equations for $B_{\n\n}$, $B_{\n\p}$, $B_{\p\p}$ and $B_{\p\p}$.

  \item
  Set $l_{0,7} = 0$ in the expressions for 
  $\chi_{\n\n}$, $\chi_{\p\p}$, $\K$ and $\theta$.

\end{itemize}

The case $(f=1,\, m=2)$ corresponding to harmonic slicing is special
in that it is the only choice of parameters for which the system is
symmetric hyperbolic. An explicit expression for a positive definite
energy is 

\begin{eqnarray}
  \label{eq:numenergy}
  \mathcal{E} &=& \chi_{AB} \chi^{AB} + \lambda_{CAB} \lambda^{CAB} 
  \nonumber\\&&
  + (\K + \chi - 2 \theta)^2 + A_A A^A \nonumber\\&& 
  + V_A V^A \\&&
  + {\K}^2 + L_A L^A \nonumber\\&&
  + E_A E^A + {B^\varphi}^2 + 4 {Z^\varphi}^2 \,,\nonumber
\end{eqnarray}
where
\begin{eqnarray}
  V_A &\equiv& A_A + D^\mathrm{I}{}_A + L_A - 2 D^\mathrm{II}{}_A 
  - 2 Z_A \,,\nonumber\\
  \lambda^C{}_{AB} &\equiv& D^C{}_{AB} + \delta_{(A}{}^C V_{B)} \,.
\end{eqnarray}
When computing the principal part of $\partial_t \mathcal{E}$, 
the terms in each individual line of \eqref{eq:numenergy} combine to
form a total divergence as in \eqref{eq:energyestimate}.


   \section{Regularity on axis}
\label{sec:regularity}

The Z(2+1)+1 equations presented so far in section
\ref{sec:firstorder} turn out to be singular on the axis $r = 0$ and
are thus unsuitable for numerical simulations. For instance, the term
$L_r = \lambda^{-1} \lambda_{,r}$ appearing several times in the fluxes
and sources is $O(r^{-1})$ for small $r$ because
$\lambda = O(r)$. We will see in this section how the regularity
conditions for axisymmetric tensor fields (chapter \ref{sec:impl}) can
be used to write the equations in a manifestly regular form.

\subsection{The main regularization procedure}

Let us first deal with one of the regularity conditions for
2-tensors $M_{\alpha\beta}$, which follows from \eqref{eq:regcond4},
\begin{equation}
  \frac{M_{\varphi\varphi}}{r^2 M_{rr}} = 1 + O(r^2) 
\end{equation}
near the axis.
For the metric $g_{\alpha\beta}$ this implies
\begin{equation}
  \label{eq:lambdacond}
  \frac{\lambda^2}{r^2 H_{rr}} = \frac{g_{\varphi\varphi}}{r^2 g_{rr}} 
  = 1 + O(r^2) \, .
\end{equation}
We enforce this condition by replacing $\lambda$ with a new variable
$s$ defined by
\begin{equation}
  \label{eq:sdef}
  \lambda = r \e^{r s} \sqrt{H_{rr}} \, ,
\end{equation}
where $s = O(r)$ near the axis.
Also, the logarithmic derivatives $L_A$ of $\lambda$ \eqref{eq:Ldef}
are replaced by the ordinary partial derivatives $s_A$ of $s$.
To satisfy the corresponding regularity condition for the extrinsic 
curvature, we introduce a new variable $Y$ via
\begin{equation}
  \label{eq:Ydef}
  \K = \frac{\chi_{rr}}{H_{rr}} + r Y 
\end{equation}
(note that $K_{\varphi\varphi} = \lambda^2 K_\varphi{}^\varphi$)
with $Y = O(r)$ on axis.
Similary for the energy-momentum tensor, we set
\begin{equation}
  \label{tautildedef}
  \tau = \frac{S_{rr}}{H_{rr}} + r \tilde \tau \, ,
\end{equation}
where $\tilde \tau = O(r)$ on axis.
We remark that the definitions of the variables $s$, $Y$ and $\tilde \tau$
can be viewed as generalizations of those in section \ref{sec:hypellreg}.

The second step of the regularization procedure is concerned with the
first $r$-derivatives of those variables $u$ that are $O(r)$ on the axis.
Consider the combination
\begin{equation}
  \label{eq:derivreg}
  (r^{-1} u)_{,r} = r^{-1} u_{r} - r^{-2} u \,.
\end{equation}
While each term on the right-hand-side is singular on the axis,
the left-hand-side shows that their difference is perfectly regular
(it is $O(r)$ on the axis). If we evolve the variables $u$ and $u_r$ 
separately in a numerical code, this subtle relationship will fail to hold 
because of
numerical errors, and the right-hand-side of \eqref{eq:derivreg} will blow up
on the axis. Such combinations do occur in the equations and so it is
essential to address this problem. We enforce regularity of \eqref{eq:derivreg}
by evolving instead of $u_r$ the variable
\begin{equation}
  \tilde u_r \equiv (r^{-1} u)_{,r} \,.
\end{equation}
This implies the following redefinitions:
\begin{eqnarray}
  D_{rrz} &\rightarrow& \tilde D_{rrz} 
    \equiv \half (r^{-1} H_{rz})_{,r} 
    = r^{-1} D_{rrz} - \half r^{-2} H_{rz} \,,\nonumber\\
  B_r{}^r &\rightarrow& \tilde B_r{}^r 
    \equiv \half \alpha^{-1} ( r^{-1} \beta^r)_{,r}
    = r^{-1} B_r{}^r - \half \alpha^{-1} r^{-2} \beta^r \,, \\
  s_r &\rightarrow& \tilde s_r 
    \equiv (r^{-1} s)_{,r} = r^{-1} s_r - r^{-2} s \,.\nonumber
\end{eqnarray}

\subsection{Choice of gauge source functions}
\label{sec:gaugesourcereg}

As pointed out in section \ref{sec:dyngauge}, the right-hand-side of
the evolution equations for the shift vector \eqref{eq:d0beta} is
singular on the axis, unless the gauge source functions $G^A$ in
\eqref{eq:dyngaugewithsource} are chosen appropriately. 
The offending term in \eqref{eq:d0beta} is 
\begin{equation}
  (2\mu - d) \partial^A \ln \lambda = (2 \mu - d) L^A \,.
\end{equation}
In terms of regularized variables,
\begin{equation}
  L_A = r s_A + D_{Arr} + \delta_A{}^r (s + r^{-1}) \,.
\end{equation}
We can cancel the irregular term by subtracting $r^{-1}$ from $L_r$.
This corresponds to setting
\begin{equation}
  \label{eq:Gr}
  G^r = - (2\mu - d) r^{-1} H^{rr}
\end{equation}
in \eqref{eq:dyngaugewithsource}. For the remaining gauge source
functions we choose
\begin{equation}
  G^0 = G^z = 0 \,.
\end{equation}
Different choices of gauge source functions are of course possible. The one
presented here is minimal in the sense that it precisely cancels the
singular term in \eqref{eq:d0beta}. We will see another application of
gauge source functions in section \ref{sec:TTgauge} in the context of
linearized theory.

A reasonable check for any gauge condition we choose
is that Minkowski space in standard cylindrical polar coordinates is a 
solution. In this chart, it is given by
\begin{equation}
 H_{AB} = \delta_{AB} \,,\qquad \lambda = r \, \Rightarrow \, s = 0
  \,,\qquad \alpha = 1\,, \qquad \beta^A = 0
\end{equation}
and of course $Z^A = 0$.
It is easy to check that (\ref{eq:dyngaugewithsource}) is satisfied for the
choice \eqref{eq:Gr}, but not for $G^r = 0$. Hence it it essential to
include a gauge source function in order to be able to evolve flat space
in standard coordinates!

\subsection{Regularized conservation forms}
\label{sec:regconsform}

It can now be verified with the help of a computer algebra programme
(see appendix \ref{sec:regconsformdetails}) that in terms of the 
\emph{regularized conserved variables} $\vec {\tilde u}$ 
(table \ref{tab:consrpar}), 
\begin{table}
  \begin{eqnarray*}
   H_{rr}, \underline{H_{rz}}, H_{zz}, \underline{s}, \alpha, 
    \underline{\beta^r}, \beta^z, \\
    \underline{D_{rrr}}, \underline{\tilde D_{rrz}}, \underline{D_{rzz}},
    D_{zrr}, \underline{D_{zrz}}, D_{zzz}, \underline{\tilde s_r}, 
    \underline{s_z}, \underline{A_r}, A_z, \underline{\tilde B_r{}^r}, 
    \underline{B_r{}^z}, \underline{B_z{}^r}, B_z{}^z, \\
    \chi_{rr}, \underline{\chi_{rz}}, \chi_{zz}, \underline{Y}, 
    \underline{E^r}, E^z, \underline{B^\varphi}, 
    \theta, \underline{Z_r}, Z_z, Z^\varphi, \\
    \rho_H, \sigma, J^\varphi, \underline{J_r}, J_z, \\
    \underline{\tilde \tau}, \underline{S_r}, S_z,
    \underline{\Sigma_r}, \Sigma_z, S_{rr}, 
    \underline{S_{rz}}, S_{zz} .
  \end{eqnarray*}
  \caption[Z(2+1)+1 system: $r$-parity of regularized conserved variables]
  {\label{tab:consrpar} \footnotesize
    The regularized conserved variables $\vec {\tilde u}$ 
    and their small-$r$ behaviour. Underlined variables are $O(r)$ on 
    the axis, all others are $O(1)$.}
\end{table}
the Z(2+1)+1 equations can again be written in conservation form 
\begin{equation}
  \label{eq:regconsform1}
  \vec {\tilde u}_{,t} + \left[ - \beta^D \vec {\tilde u}
   + \alpha \vec {\tilde \F}^D (\vec {\tilde u}) \right]_{,D} 
  = \alpha \vec {\tilde \S} (\vec {\tilde u}) \,,
\end{equation}
where now the fluxes $\vec {\tilde \F}^D$ and sources 
$\vec {\tilde \S}$ are regular on the axis, provided
that the appropriate boundary conditions are enforced.
For a variable $\tilde {\underline{u}}$ that is $O(r)$ on axis (the underlined 
variables in table \ref{tab:consrpar}), a Dirichlet condition 
\begin{equation}
  \tilde {\underline{u}} \vert_{r = 0} = 0
\end{equation}
is needed, and for a variable $\tilde u$ that is $O(1)$ on axis (the
remaining variables in table \ref{tab:consrpar}), we enforce a Neumann
condition 
\begin{equation}
  \partial_r \tilde u \vert_{r = 0} = 0 \,.
\end{equation}
This ensures that terms such as $r^{-1} \tilde {\underline{u}}$, 
$r^{-1} \partial_r \tilde u$, etc.~are
well-behaved on the axis. Numerically, this procedure works as long as
we do not evaluate the fluxes and sources at $r = 0$. This is one of
the reasons why we use a cell-centred grid (section \ref{sec:FDgrid}), 
in which the centre of the innermost cell is half a
grid spacing away from the axis.

In \cite{RinneStewart05}, the variables are further redefined by
taking out the leading order of $r$, i.e. 
\begin{equation}
  \underline{\tilde{\tilde u}} \equiv r^{-1} \underline{\tilde u}
\end{equation}
for the variables that are $O(r)$ on the axis (the remaining ones are
unchanged).
The equations can then be written in the form
\begin{equation}
  \label{eq:regconsform2}
  \partial_t \vec {\tilde {\tilde u}} 
   + \left[ - 2 r^2 \tilde {\tilde \beta}^r \vec {\tilde{\tilde u}}
   + \alpha \vec {\tilde {\tilde \F}}^{(r^2)} (\vec {\tilde
  {\tilde u}}) \right]_{,r^2}
  + \left[ - \beta^z  \vec {\tilde{\tilde u}} +
    \alpha \vec {\tilde {\tilde \F}}^{z} (\vec {\tilde {\tilde u}}) 
  \right]_{,z}  
  = \alpha \vec {\tilde {\tilde \S}} (\vec {\tilde {\tilde u}}) \,,
\end{equation}
where now the fluxes $\vec {\tilde {\tilde \F}}^D$ and the sources 
$\vec {\tilde {\tilde \S}}$ are \emph{manifestly} regular on the axis,
i.e., no negative powers of $r$ appear and they are even functions of $r$.

One might wonder whether one should discretize \eqref{eq:regconsform2} on a
grid that is uniform in $r$ or on one that is uniform in $r^2$, since the
derivatives are now taken with respect to $r^2$. On the former grid, one
can enforce Neumann conditions for all the modified variables $\vec
{\tilde {\tilde u}}$. On the latter grid, however, it is not so
clear what the boundary conditions should be. One might derive
boundary conditions by restricting the evolution equations to $r = 0$,
but those would include both time derivatives and spatial derivatives
tangential and normal to the $r = 0$ boundary. 
An earlier attempt of Nakamura et al.\cite{Nakamura87} for a similar
set of equations on an $r^2$ grid led to numerical instabilities on
the axis, which could only be controlled by adding a large amount of 
artificial viscosity.

Another problem with the $r^2$ grid is that the characteristic speeds 
are non-uniform (proportional to $r$) because
\begin{equation}
  \frac{\partial}{\partial (r^2)} = \frac{1}{2r}
 \frac{\partial}{\partial r} \,,
\end{equation}
which means that a factor of $2 r$ had to be taken out of
the flux $\vec{\tilde \F}^r$ in \eqref{eq:regconsform1} in order to
arrive at \eqref{eq:regconsform2}.

For these reasons, we choose to work on a grid that is uniform in $r$.
Both regularized versions of the equations 
(\ref{eq:regconsform1}, \ref{eq:regconsform2}) have been implemented,
but at some stage we decided to focus on the first version,
mainly for simplicity and because the ubiquitous factors of $r^2$ in
the second version led to instabilities caused by the outer boundary
conditions (chapter \ref{sec:outerbcs}).

\subsection{Hyperbolicity and the characteristic transformation}
\label{sec:charregularity}

The question arises whether the regularization procedure outlined
above affects the hyperbolicity of the system. This is not the case
because we have merely performed a \emph{linear} (position-dependent)
transformation $\vec {\tilde u} = T \vec u$ of those variables 
that occur in the fluxes, namely

\begin{eqnarray}
  \label{eq:regtrafoDrrz}
  \tilde D_{rrz} &=& r^{-1} D_{rrz} + \ldots \,,\\
  \tilde s_r &=& r^{-2} ( L_r - D_{rrr}) + \ldots \,,\\
  s_z &=& r^{-1} (L_z - D_{zrr}) \,,\\
  \tilde B_r{}^r &=& r^{-1} B_r{}^r + \ldots \,,\\
  \label{eq:regtrafoY}
  Y &=& r^{-1} (\K - \frac{\chi_{rr}}{H_{rr}}) \,.
\end{eqnarray}  
where the ellipses denote terms that have zero fluxes (and so has
$H_{rr}$ in \eqref{eq:regtrafoY}).
Hence the characteristic structure is unchanged.

To compute the characteristic variables, one starts from the
regularized variables $\vec{\tilde u}$,
computes the original conserved variables $\vec u$ and evaluates the
characteristic variables given in section \ref{sec:hyperbolicity}.
While this transformation is perfectly regular, the
inverse transformation contains factors of $r^{-1}$, which might cause
problems on the axis. 

The transformation from characteristic variables to regularized
conserved variables in the \emph{$z$-direction} (i.e., $\mu_A \propto
\delta_A{}^z$) turns out to be 
well-behaved at $r = 0$ provided that the characteristic variables have the
correct leading order in $r$ as summarized in table 
\ref{tab:charrpar}.\footnote{The basis of left eigenvectors in section 
\ref{sec:hyperbolicity} was chosen such that the regularity conditions 
on the characteristic variables have this simple form. For a different 
basis, they would involve linear combinations of characteristic variables.}
In turn, this small-$r$ behaviour is manifest when expressing the 
characteristic variables in terms of the regularized conserved variables
(using the conversions (\ref{eq:regtrafoDrrz}--\ref{eq:regtrafoY}), 
again leaving out the lower-order terms). This is worked out
explicitly in linearized theory in section \ref{sec:linchar}.
 
\begin{table}
  \begin{eqnarray*}
    \underline{l_{0,1}}, \overline{l_{0,2}}, \underline{l_{0,3}},
    \underline{l_{0,4}}, \underline{l_{0,5}}, \underline{l_{0,6}}, 
    \overline{l_{0,7}}, \nonumber\\
    \overline{l_{1,1}^\pm}, \underline{l_{1,2}^\pm}, l_{1,3}^\pm, 
    l_{1,4}^\pm, \underline{l_{1,5}^\pm}, l_{1,6}^\pm, \nonumber\\
    l_f^\pm, \underline{l_\mu^\pm}, l_d^\pm \, .
  \end{eqnarray*}
  \caption[Z(2+1)+1 system: $r$-parity of $z$-characteristic variables]
  {\label{tab:charrpar} \footnotesize
    Small-$r$ behaviour of the characteristic variables in the
    $z$-direction. Overlined variables are $O(r^2)$, underlined
    variables are $O(r)$ and the remaining variables are $O(1)$ on the axis.}
\end{table}

However, the transformation from the characteristic variables 
in the \emph{$r$-direction} (i.e., $\mu_A \propto \delta_A{}^r$) 
to regularized conserved variables is still
singular on the axis, and no simple regularity conditions on the
characteristic variables as the above can cure this problem.
To understand this, one should observe that unlike 
the characteristic variables in the $z$-direction, the characteristic
variables in the $r$-direction do not have a definite $r$-parity
(even and odd terms in $r$ are mixed). 

These results have two important numerical consequences.
Firstly, numerical methods that operate in the space of characteristic 
variables (typically ones based on the solution of the Riemann
problem\footnote{although there exist problems where the exact Riemann problem
  solution makes no reference to the characteristic structure, e.g., Euler's
  equations)} appear to be unusable near the axis because they require a
transformation between conserved and characteristic variables \emph{both}
in the $r$ \emph{and} the $z$ direction.
Secondly, suppose that the computational domain has outer boundaries
at $r = r_\mathrm{max}$ and $z = z_\mathrm{max}$. To set up boundary 
conditions, one typically only needs to transform between 
conserved and characteristic variables \emph{normal} to the boundary. 
The $r = r_\mathrm{max}$ boundary is unproblematic because all points on it 
are far away from the axis at $r = 0$.
At the $z = z_\mathrm{max}$ boundary, the characteristic transformation in the 
normal direction (i.e., the $z$ direction) is well-behaved even near the axis,
as pointed out above.
Hence it should be possible to impose outer boundary conditions that
respect regularity on axis. We shall see this explicitly in chapter
\ref{sec:outerbcs}.


   \section{Equation checks and code generation}
\label{sec:algebra}

We derived the regularized conservation forms \eqref{eq:regconsform1}
and \eqref{eq:regconsform2} of the Z(2+1)+1 equations using the computer
algebra language REDUCE \cite{REDUCE}. As can be appreciated from
appendix \ref{sec:regconsformdetails}, the resulting equations are
rather lengthy. It is indispensable to perform some sort of
consistency checks to make sure that they are correct.
Here, we verify that the equations are satisfied for a variety of
exact solutions of the field equations. 
We also generate C code implementing the equations directly from
within REDUCE using a source code optimization package.

\subsection{Checking the equations with exact solutions}
\label{sec:exactsolns}

The equations were checked with the following exact solutions,
also considered in \cite{BarnesPhD} for a different formulation. 
For all the solutions, we first
computed the (2+1)+1 variables as described in chapter \ref{sec:211}
and then the regularized conserved variables (section \ref{sec:regconsform}).
These were then inserted directly into the regularized conservation 
form of the equations (\eqref{eq:regconsform1} or alternatively 
\eqref{eq:regconsform2}).

\begin{itemize}

  \item
  A cylindrically symmetric Kasner metric \cite{Kasner21}
  \begin{equation}
    \diff s^2 = z^4 (\diff r^2 + r^2 \diff \varphi^2 + \diff z^2) 
    - z^{-2} \diff t^2 \,.
  \end{equation}
  This is a vacuum solution, it is static, and has zero twist.

  \item
  Another cylindrically symmetric Kasner metric \cite{Kasner21}
  \begin{equation}
    \diff s^2 = \textstyle t^\frac{4}{3} (\diff r^2 + r^2 \diff \varphi^2)
      + t^{-\frac{2}{3}} \diff z^2 - \diff t^2 \,.
  \end{equation}
  This is again matter- and twist-free, but not stationary.

  \item
  The JEKK metric \cite{JEK60, Kompaneets58}
  \begin{equation}
    \diff s^2 = - \e^{2(\gamma - \nu)} (\diff t^2 - \diff r^2)
      + \e^{2 \nu} (\diff z + \omega \diff \varphi)^2
      + r^2 \e^{-2\nu} \diff \varphi^2 \,,
  \end{equation}
  which is a cylindrically symmetric vacuum solution, with $\gamma$,
  $\nu$ and $\omega$ depending only on $t$ and $r$. It has nonzero
  twist for $\omega \neq 0$ (for $\omega = 0$, it reduces to the
  Einstein-Rosen waves \cite{WeberWheeler57}).
  It is a solution of the Einstein equations if and only if
  \begin{eqnarray}
    \nu_{,tt} - r^{-1} \nu_{,r} - \nu_{,rr} &=& \textstyle
      \half r^{-2} \e^{4\nu} \left( \omega_{,t}{}^2 - \omega_{,r}{}^2
      \right) \,, \\
    \omega_{,tt} + r^{-1} \omega_{,r} - \omega_{,rr} &=&
      4 (\omega_{,r} \nu_{,r} - \omega_{,t} \nu_{,t} \,,\\
    \gamma_{,r} &=& r \left( \nu_{,t}{}^2 + \nu_{,r}{}^2 \right)
      \nonumber\\&&\textstyle + \frac{1}{4} r^{-1} \e^{4\nu} 
      \left( \omega_{,t}{}^2 + \omega_{,r}{}^2 \right) \,,\\
    \gamma_{,t} &=& 2 r \nu_{,r} \nu_{,t} + \textstyle \half r^{-1} \e^{4 \nu}
      \omega_{,r} \omega_{,t} \,.
  \end{eqnarray}

  \item
  The Robertson-Walker metric \cite{Garfinkle97}
  \begin{equation} \textstyle
    \diff s^2 = - \diff t^2 + t^\frac{2}{3} (\diff r^2 + \diff z^2 
    + r^2 \diff \varphi^2 ) \,,
  \end{equation}
  a non-rotating perfect fluid solution for the equation of state
  \begin{equation}
    p = \rho = \textstyle \frac{1}{3} \kappa^{-1} t^{-2}
  \end{equation}
  four-velocity
  \begin{equation}
    u_\alpha = - \delta_\alpha{}^t 
  \end{equation}
  and number density $N^\alpha = n u^\alpha$ with
  \begin{equation}
    n \propto t^{-1} \,.
  \end{equation}

  \item
  The Kramer metric \cite{Kramer88}
  \begin{equation}
    \diff s^2 = \e^{a^2 r^2} \left( - \diff t^2 + \diff r^2 \right)
      + \diff z^2 + r^2 \diff \varphi^2 \,,
  \end{equation}
  a non-rotating static perfect fluid solution for
  \begin{eqnarray}
    p = \rho &=& \kappa^{-1} a^2  \e^{-a^2 r^2} \,,\\
    u_\alpha &=&  - \e^{\frac{1}{2} a^2 r^2} \delta_\alpha{}^t \,,\\
    n &=& \mathrm{const.} \,,
  \end{eqnarray}
  where $a$ is a constant.

  \item
  The Tabensky-Taub metric \cite{TabenskyTaub73}
  \begin{equation}
    \diff s^2 = V (-\diff t^2 + \diff z^2) 
      + z (\diff r^2 + r^2 \diff \varphi^2) \,,
  \end{equation}
  a non-rotating static perfect fluid solution for
  \begin{eqnarray}
    p = \rho &=& \half \kappa^{-1} a^2 V^{-1} \,,\\
    u_\alpha &=& - V^{\frac{1}{2}} \delta_\alpha{}^t \,,\\
    n &=& \mathrm{const.} \,,
  \end{eqnarray}
  where 
  \begin{equation}
     V = z^{-\frac{1}{2}} \e^{-\frac{1}{2} a^2 z^2}
  \end{equation}
  and $a$ is a constant.

  \item
  The solution given in equation (6.1) (taking $A = 1$) of Davidson
  \cite{Davidson97},
  \begin{eqnarray}
    \diff s^2 &=& \textstyle - (1 + r^2) \diff t^2 +
    (1 + r^2)^\frac{1}{3} \diff r^2 + (1 + r^2)^{-\frac{2}{3}}
    \diff z^2 \nonumber\\&&\textstyle
    + r^2 (1 - \frac{5}{3} r^2 - \frac{8}{3} r^4) \diff \varphi^2
    - 2 \sqrt{\frac{11}{3}} r^2 (1 + r^2) \diff t \diff \varphi \,,
  \end{eqnarray}
  a rotating perfect fluid solution for
  \begin{eqnarray}
    p = \textstyle \frac{3}{5} \rho &=& 4 \kappa^{-1} (1 + r^2)^{-\frac{4}{3}}
    \,,\\
    u^\alpha &=& (1 + r^2)^{-\frac{1}{2}} \delta_t{}^\alpha \,,\\
    n &=& \mathrm{const.} 
  \end{eqnarray} 

\end{itemize}

In addition, it was verified that the equations are satisfied by the
solutions presented in chapter \ref{sec:lin} in linearized theory.

\subsection{Code generation}

Since the equations that we would like to implement are very long,
it is highly desirable to produce code directly from the
computer algebra programme used to derive the equations.
This is provided for REDUCE by the Source Code Optimization PackagE
SCOPE \cite{SCOPE}, which in addition minimizes the number of
algebraic operations in the output. We used SCOPE's
straightforward {\tt OPTIMIZE} command. A combination with the
automatic code GENerator and TRANslator package GENTRAN \cite{GENTRAN},
also described in \cite{SCOPE}, failed for very long expressions.

To make sure that the implementation is correct, we chose random data
for all the variables and verified that the fluxes and sources
computed with the C code agree with those computed within REDUCE.


   \chapter{A test problem in linearized theory}
\label{sec:lin}

To check that the implementation of the Z(2+1)+1 system is correct, 
it is highly desirable to have an exact solution 
which the numerical approximation can be compared with.
In this thesis, we are mainly interested in asymptotically flat radiative
vacuum spacetimes. Not many exact solutions of the fully nonlinear 
Einstein equations with those properties are known. 
The cylindrically symmetric Einstein-Rosen
waves mentioned in section \ref{sec:algebra} as a special case of the 
JEKK solution are not asymptotically flat. In fact, as shown by Bi\v{c}\'ak 
and Schmidt \cite{BicakSchmidt83}, the only isometry in addition to
axisymmetry admitting gravitational radiation and asymptotical flatness
is boost symmetry. 
Examples of such boost-rotation-symmetric solutions can be found in 
\cite{BicakSchmidt89}.

Here, we take a different approach: we focus on axisymmetric gravitational 
wave solutions of the \emph{linearized} field equations.
We begin by writing out the linearized Z(2+1)+1 equations in terms of
the regularized variables (section \ref{sec:Z211lin}), 
which also serves as another illustration of regularity on axis.
Next we discuss the transverse-traceless gauge and its compatibility with the
dynamical gauge conditions used in the Z(2+1)+1 system (section
\ref{sec:TTgauge}).
The linearized quadrupole solution of Teukolsky \cite{Teukolsky82} is then
presented and the corresponding Z(2+1)+1 variables are computed (section
\ref{sec:teuko}).
In addition, we derive an even-parity twisting octupole solution (section
\ref{sec:octupole}).
Some features of the numerical implementation are described and convergence of
the numerical solution to the exact one is demonstrated,
both for vanishing and dynamical shift vector (section \ref{sec:linnum}).

\section{The linearized Z(2+1)+1 equations}
\label{sec:Z211lin}

We express the linearized Z(2+1)+1 equations in terms of the
regularized variables (table \ref{tab:consrpar}).
All variables $u$ are linearized about their flat-space values $u_0$,
\begin{equation}
  u = u_0 + \epsilon (\hat u - u_0) \,, \qquad \epsilon \ll 1 \,,
\end{equation}
and we shall omit the hats in the following.
For $u \in \{ H_{rr}, H_{zz}, \alpha \}$ we have $u_0 = 1$, for all
remaining variables $u_0 = 0$.
As a shorthand, we set
\begin{eqnarray}
  X_0 &\equiv& f( 2 \chi_{rr} + \chi_{zz} - m \theta + r Y) \,,\\
  X_1 &\equiv& a A_r - d (2 D_{rrr} + D_{rzz} + r^2 \tilde s_r) \nonumber\\&&
    + 2 \mu (D_{rrr} + D_{rzz} - D_{zrz} + r^2 \tilde s_r - Z_r) \,,\\
  X_2 &\equiv& a A_z - d (2 D_{zrr} + D_{zzz} + r s_z)\nonumber\\&&
    + 2 \mu (-r \tilde D_{rrz} + 2 D_{zrr} + r s_z - Z_z) \,.
\end{eqnarray}
Written in conservation form with sources, 
the linearized evolution equations are given by 
{\allowdisplaybreaks \begin{eqnarray}
  \partial_t H_{rr} &=& 2 (2 r \tilde B_r{}^r + r^{-1} \beta^r - \chi_{rr})
    \,,\\
  \partial_t H_{rz} &=& 2 (B_r{}^z + B_z{}^r - \chi_{rz})\,,\\
  \partial_t H_{zz} &=& 2 (2 B_z{}^z - \chi_{zz}) \,,\\
  \partial_t s &=& -2 \tilde B_r{}^r - Y \,,\\
  \label{eq:lind0alpha}
  \partial_t \alpha &=&  - X_0 \,,\\
  \label{eq:lind0betar}
  \partial_t \beta^r &=& - X_1 - 2 (2 \mu - d) s \,,\\
  \label{eq:lind0betaz}
  \partial_t \beta^z &=& - X_2 + (3 \mu - d) r^{-1} H_{rz} \,,\\
  \partial_t D_{rrr} &=& -\partial_r [ -2 r \tilde B_r{}^r + \chi_{rr} ] 
    + 2 \tilde B_r{}^r \,,\\
  \partial_t \tilde D_{rrz} &=& -\partial_r [ r^{-1} (-B_r{}^z -
    B_z{}^r + \chi_{rz}) ] \,,\\
  \partial_t D_{rzz} &=& -\partial_r [ -2 B_z{}^z + \chi_{zz} ] \,,\\
  \partial_t D_{zrr} &=& - \partial_z [ - 2 r \tilde B_r{}^r + \chi_{rr} ]
    + 2 r^{-1} B_z{}^r \,,\\
  \partial_t D_{zrz} &=& - \partial_z [ -B_r{}^z - B_z{}^r + \chi_{rz} ]  
    \,,\\
  \partial_t D_{zzz} &=& - \partial_z [ - 2 B_z{}^z + \chi_{zz} ] \,,\\ 
  \partial_t s_r &=& - \partial_r [r^{-1}(2 \tilde B_r{}^r + Y)]\,,\\
  \partial_t s_z &=& - \partial_z [2 \tilde B_r{}^r + Y]\,,\\
  \label{eq:lind0Ar}
  \partial_t A_r &=& - \partial_r X_0 \,,\\
  \label{eq:lind0Az}
  \partial_t A_z &=& - \partial_z X_0 \,,\\
  \label{eq:lind0Brr}
  \partial_t \tilde B_r{}^r &=& - \partial_r [\half r^{-1} X_1] 
    + (d - 2 \mu) \tilde s_r \,,\\
  \label{eq:lind0Brz}
  \partial_t B_r{}^z &=& - \partial_r [\half X_2] + (3 \mu - d) \tilde
    D_{rrz} \,,\\
  \label{eq:lind0Bzr}
  \partial_t B_z{}^r &=& - \partial_z [\half X_1] + (d - 2 \mu) s_z \,,\\
  \label{eq:lind0Bzz}
  \partial_t B_z{}^z &=& - \partial_z [\half X_2] + (3 \mu - d) r^{-1}
    D_{zrz} \,,\\
  \partial_t \chi_{rr} &=& - \partial_r [A_r + D_{rrr} + D_{rzz} - 2
    D_{zrz} + r^2 \tilde s_r - 2 Z_r] \nonumber\\&& - \partial_z D_{zrr} 
    + r^{-1} (-D_{rrr} - 4 r^2 \tilde s_r - 6 s) \,,\\
  \partial_t \chi_{rz} &=& - \partial_r [\half(A_z + 2 D_{zrr} -
    D_{zzz} + r s_z - 2 Z_z)] \nonumber\\&&
    - \partial_z [\half(A_r + D_{rzz} + r^2 \tilde s_r - 2 Z_r)]
    \textstyle - 2 s_z \,,\\
  \partial_t \chi_{zz} &=& - \partial_r D_{rzz} - \partial_z [A_z - 2
    r \tilde D_{rrz} + 2 D_{zrr} + r s_z - 2 Z_z] \nonumber\\&&
    + r^{-1} (-D_{rzz} + 4 D_{zrz}) \,,\\  
  \partial_t Y &=& -\partial_r[r^{-1}(-A_r - D_{rzz} + 2 D_{zrz} + 2 Z_r)]
    - \partial_z s_z \,,\\
  \label{eq:lind0Er}
  \partial_t E^r &=& - \partial_r[-2 Z^\varphi] - \partial_z[-
    B^\varphi]\,\\\  
  \label{eq:lind0Ez}
  \partial_t E^z &=& - \partial_r B^\varphi - \partial_z[-2 Z^\varphi]
    - 3 r^{-1} B^\varphi \,,\\
  \label{eq:lind0Bp}
  \partial_t B^\varphi &=& - \partial_r E^z - \partial_z [-E^r] \,,\\
  \partial_t \theta &=& - \partial_r[D_{rrr} + D_{rzz} - D_{zrz} + r^2 \tilde
    s_r - Z_r] \nonumber\\&&
    - \partial_z[-r \tilde D_{rrz} + 2 D_{zrr} + r s_z - Z_z] \nonumber\\&&
    + r^{-1} (-D_{rrr} - D_{rzz} + 3 D_{zrz} - 4 r^2 \tilde s_r
    - 6 s + Z_r)\,,\\
  \partial_t Z_r &=& - \partial_r[\chi_{rr} + \chi_{zz} + r Y - \theta] 
    - \partial_z[-\chi_{rz}] - Y \,,\\
  \partial_t Z_z &=& - \partial_r [-\chi_{rz}] - \partial_z[2
    \chi_{rr} + r Y - \theta] + r^{-1} \chi_{rz} \,,\\
  \label{eq:lind0Zp}
  \partial_t Z^\varphi &=& - \partial_r [-\half E^r] - \partial_z
    [-\half E^z] + \textstyle\frac{3}{2} r^{-1} E^r \,.
\end{eqnarray} }

We have used the minimal gauge source function (\ref{eq:Gr}) to cancel the
singular term in \eqref{eq:lind0betar}.
Note that all the above equations are regular on axis provided that
the appropriate boundary conditions (table \ref{tab:consrpar}) are enforced.
Another point to observe is that the evolution equations for the twist
variables (\ref{eq:lind0Er}--\ref{eq:lind0Bp}, \ref{eq:lind0Zp})
decouple completely from the remaining system, as already mentioned in
section \ref{sec:firstorder}.


   \section{Transverse-traceless gauge}
\label{sec:TTgauge}

All the linearized solutions presented in this chapter adopt the
\emph{transverse-traceless (TT) gauge}, which is described in the following,
stressing its relation to other familar gauges.

As usual in linearized theory, we write the metric as 
\begin{equation}
  g_{\alpha\beta} = \eta_{\alpha\beta} + h_{\alpha\beta} \,,
\end{equation}
where $\eta_{\alpha\beta}$ is the Minkowski metric and 
$h_{\alpha\beta}$ is a small perturbation.
$h_{\alpha\beta}$ is chosen to obey the \emph{Lorentz gauge} condition
\begin{equation}
  \label{eq:Lorentzgauge}
  h^{\alpha\beta}{}_{\vert\beta} - \half \partial^\alpha h  = 0 \,.
\end{equation}
Here a vertical bar denotes a covariant derivative in flat space (where we
will be using polar coordinates), indices are raised with 
$\eta^{\alpha\beta}$, and $h \equiv h_\gamma{}^\gamma$.
We recognize in \eqref{eq:Lorentzgauge} the linearized version of the harmonic
gauge condition (\ref{eq:harmgauge}). Hence it is not surprising that in this
gauge the linearized vacuum Einstein equations become a flat-space 
wave equation,
\begin{equation}
  \label{eq:linEinstein}
  h_{\alpha\beta\vert\gamma}{}^\gamma = 0 \,.
\end{equation}
The Lorentz gauge condition \eqref{eq:Lorentzgauge} is invariant under
infinitesimal coordinate transformations
\begin{equation}
  x^\alpha \rightarrow x^\alpha + \zeta^\alpha
\end{equation}
provided that
\begin{equation}
  \zeta^\alpha{}_{\vert \gamma}{}^\gamma = 0 \,.
\end{equation}
This remaining gauge freedom can be exploited to impose the additional
conditions
\begin{eqnarray}
  \label{eq:transverse}
  h_{0 \alpha} &=& 0 \,,\\
  \label{eq:traceless}
  h &=& 0 \,,
\end{eqnarray}
i.e., $h_{\alpha\beta}$ is \emph{transverse} to the time direction and 
\emph{traceless}.
Equations (\ref{eq:transverse}--\ref{eq:traceless})
are actually only four conditions because once \eqref{eq:transverse} is
enforced, the time component of \eqref{eq:Lorentzgauge} implies that $h$ is
constant in time, and we can choose the initial conditions such that $h = 0$.

The question arises whether TT gauge is compatible with
the dynamical gauge conditions (section \ref{sec:dyngauge})
used in the Z(2+1)+1 system.
In ADM language, \eqref{eq:transverse} implies that to linear order
\begin{equation}
  \label{eq:geodesicgauge}
  \alpha = 1 \,, \qquad \beta^A = 0 \,,
\end{equation}
which is also known as \emph{geodesic gauge}.
In terms of regularized Z(2+1)+1 variables, the $t$, $r$ and $z$ components 
of \eqref{eq:Lorentzgauge} read 
\begin{eqnarray}
  \label{eq:Lorentzt}
  -\half \partial_t (2 H_{rr} + H_{zz} + 2 r s) = 
    2 \chi_{rr} + \chi_{zz} + r Y &=& 0 \,,\\
  \label{eq:Lorentzr}
  -D_{rzz} + 2 D_{zrz} - r^2 \tilde s_r - 4 s &=& 0 \,,\\
  \label{eq:Lorentzz}
  -2 D_{zrr} + D_{zzz} - r s_z + 2 r \tilde D_{rrz} + 2 r^{-1} H_{rz}&=&0 \,.
\end{eqnarray}
Let us now compare these results with the linearized dynamical gauge
conditions (\ref{eq:lind0alpha}--\ref{eq:lind0betaz}).
The evolution equation for $\alpha$ (\ref{eq:lind0alpha}) is
satisfied for any choice of the parameter $f$.
The evolution equation for $\beta^z$ (\ref{eq:lind0betaz}) is consistent if
and only if we choose $\mu = d = 1$. This choice of parameters is also
necessary for the evolution equation for $\beta^r$
\eqref{eq:lind0betar} to be satisfied, but not sufficient: the right-hand-side 
of \eqref{eq:lind0betar} still fails to vanish by a term $2s$.
To cancel this term, we have to add a gauge source function
\begin{equation}
  \label{eq:lingaugesource}
  G^r = - 2s
\end{equation}
to the right-hand-side of \eqref{eq:lind0betar} (in addition to the minimal
one, equation (\ref{eq:Gr})). Note that this does not affect the principal
parts of the system.

We conclude that our dynamical gauge conditions are compatible with the TT
gauge used for the exact solutions if either the shift vector vanishes,
or it is dynamical with parameters $\mu = d = 1$ (i.e., harmonic shift), 
in which case we need to include a gauge source function 
\eqref{eq:lingaugesource}.


   \section{Teukolsky's quadrupole solution}
\label{sec:teuko}

General solutions of equations (\ref{eq:Lorentzgauge}--\ref{eq:linEinstein})
and (\ref{eq:transverse}--\ref{eq:traceless}) can be constructed as 
multipole expansions using tensor spherical harmonics \cite{Burke71} 
with ``quantum numbers'' $L$ and $M$. Teukolsky \cite{Teukolsky82} focuses 
on quadrupole radiation ($L = 2$), which is likely to be the strongest 
mode from realistic sources (see, for example, \cite{MTW}). 
Axisymmetry implies that the azimuthal quantum number is $M = 0$ in our case.

\subsection{The even-parity solution}
\label{sec:teukoeven}

First we consider the solution that is symmetric under 
$\theta \rightarrow \pi - \theta$, or equivalently $z \rightarrow -z$.
The line element can be written in spherical polar coordinates $(t, R, \theta,
\varphi)$ as
\begin{eqnarray}
  \label{eq:evenlineelement}
  \diff s^2 &=& - \diff t^2 + (1 + A f_{RR}) \diff R^2
   + (2 B f_{R\theta}) R \, \diff R \, \diff \theta \nonumber\\&&
   + \left(1 + C \,f^{(1)}_{\theta\theta} + A\,f^{(2)}_{\theta\theta}\right) 
     R^2 \diff \theta^2 \nonumber\\&&
   + \left(1 + C\,f^{(1)}_{\varphi\varphi} + A\,f^{(2)}_{\varphi\varphi}\right)
      R^2 \sin^2 \theta\, \diff \varphi^2\,.
\end{eqnarray}
The functions $f$ only depend on the polar angle $\theta$ and are given
by
\begin{eqnarray}
  f_{RR} &=& 2 - 3 \sin^2 \theta \,,\\
  f_{R\theta} &=& - 3 \sin \theta \cos \theta \,,\\
  f^{(1)}_{\theta\theta} &=& 3 \sin^2 \theta \,,\\
  f^{(2)}_{\theta\theta} &=& -1 \,,\\
  f^{(1)}_{\varphi\varphi} &=& - 3 \sin^2 \theta \,,\\
  f^{(2)}_{\varphi\varphi} &=& 3 \sin^2 \theta - 1\,.
\end{eqnarray}
The functions $A$, $B$ and $C$ only depend on $t$ and $R$ and can be
expressed as
\begin{eqnarray}
  \label{eq:teukoABC}
  A &=& 3 \left( \frac{F^{(2)}}{R^3} \pm \frac{3 F^{(1)}}{R^4} +
    \frac{3 F}{R^5} \right) \,,\\
  B &=& - \left( \pm \frac{F^{(3)}}{R^2} + \frac{3 F^{(2)}}{R^3}
    \pm \frac{6 F^{(1)}}{R^4} + \frac{6 F}{R^5} \right)\,,\\
  C &=& \frac{1}{4} \left( \frac{F^{(4)}}{R} \pm \frac{2 F^{(3)}}{R^2}
    + \frac{9 F^{(2)}}{R^3} \pm \frac{21 F^{(1)}}{R^4} 
    + \frac{21 F}{R^5} \right) \,,
\end{eqnarray}
where 
\begin{equation}
  \label{eq:teukoF}
  F = F(t \mp R) \,, \qquad F^{(n)} \equiv
    \frac{\diff^n F(x)}{\diff x^n}{\Big \vert}_{x = t \mp R} \,.
\end{equation}
The mode function $F$ can be freely specified. The upper sign in
(\ref{eq:teukoABC}--\ref{eq:teukoF}) corresponds to an outgoing
solution, the lower sign to an ingoing one. Clearly, linear combinations
of outgoing and ingoing solutions are also solutions.
Using a Taylor expansion of $F$ about $R = 0$, one can show that the
only linear combination that is regular at $R = 0$ is (up to an overall factor)
\begin{equation}
  \vec u_\mathrm{reg} = \vec u_\mathrm{out} - \vec u_\mathrm{in} \,,
\end{equation}
where $\vec u_\mathrm{out}$ and $\vec u_\mathrm{in}$ are out- and ingoing
solutions with the \emph{same} mode function $F$.

Given the line element, we can now compute the regularized Z(2+1)+1
variables. To obtain the 2-metric $H_{AB}$, the metric tensor has to be 
transformed to cylindrical polar components $r, z$ given by 
\begin{equation}
  \label{eq:polsphertrafo}
  r = R \sin \theta \,, \quad z = R \cos \theta \quad \Leftrightarrow \quad
  R = \sqrt{r^2 + z^2} \,, \quad \theta = \tan^{-1} \frac{r}{z} \,.
\end{equation}
We find
\begin{eqnarray}
  H_{rr} &=& 1 - A + 3 \sin^2\theta \cos^2\theta (A - 2 B + C) \,,\\
  H_{rz} &=& 3 \sin\theta \cos\theta [ \cos^2\theta (A - B) +
    \sin^2\theta (B - C) ] \,,\\
  H_{zz} &=& 1 + 2 A + 3 \sin^2\theta(C - A) - 3 \sin^2\theta
    \cos^2\theta (A - 2 B + C) 
\end{eqnarray}
To compute the variable $s$, we use its definition
\begin{equation}
  s = r^{-1} \ln \left( \frac{\lambda}{r \sqrt{H_{rr}}} \right) \,,
\end{equation}
first linearize the right-hand-side and then insert the results for
$H_{rr}$ and $\lambda^2 = g_{\varphi\varphi}$, obtaining
\begin{equation}
  s = \textstyle\frac{3}{2} R^{-1} \sin\theta [ \sin^2\theta(A - C) +
  2 \cos^2 \theta (B - C) ] \,.
\end{equation}
The spatial derivatives of the 2-metric and the
extrinsic curvature variables can be computed from their definitions
and using
\begin{eqnarray}
  \chi_{AB} &=& - \half \partial_t H_{AB} \,,\\
  Y &=& - \partial_t s \,.
\end{eqnarray}
As explained in section \ref{sec:TTgauge}, the gauge variables are
\begin{equation}
  \alpha = 1 \,, \qquad \beta^r = \beta^z = 0 \,,
\end{equation}
and clearly
\begin{equation}
  \theta = Z^r = Z^z = Z^\varphi = 0
\end{equation} 
for an exact solution.

The important point to observe is that the twist variables 
vanish for the even-parity solution,
\begin{equation}
  E^r = E^z = B^\varphi = 0 \,,
\end{equation}
because there are no $(R\varphi)$ and $(\theta\varphi)$ components in
the line element \eqref{eq:evenlineelement}.

As with all the solutions presented in this chapter, it has been
checked with REDUCE that the above solution obeys equations 
(\ref{eq:Lorentzgauge}--\ref{eq:linEinstein}) and 
(\ref{eq:transverse}--\ref{eq:traceless}) 
as well as the linearized Z(2+1)+1 equations (section \ref{sec:Z211lin}).

\subsection{The odd-parity solution}
\label{sec:teukoodd}

Next we consider the solution that is antisymmetric under 
$\theta \rightarrow \pi - \theta$. Its line element is
\begin{eqnarray}
  \label{eq:oddlineelement}
  \diff s^2 &=& - \diff t^2 + \diff R^2 + R^2 \diff \theta^2
    + R^2 \sin^2\theta \diff \varphi^2 \nonumber\\&&
    + 2 K d_{R\varphi} R \sin\theta \, \diff R \, \diff \varphi 
    + 2 L d_{\theta\varphi} R^2 \sin\theta \, \diff \theta \, \diff
    \varphi \,.
\end{eqnarray}
The angular functions are
\begin{eqnarray}
  d_{R \varphi} &=& - 4 \cos\theta \sin\theta \,,\\
  d_{\theta\varphi} &=& - \sin^2 \theta \,.
\end{eqnarray}
The functions $K$ and $L$ are given by
\begin{eqnarray}
  K &=& \frac{G^{(2)}}{R^2} \pm \frac{3 G^{(1)}}{R^3} + \frac{3
    G}{R^4} \,,\\
  L &=& \pm \frac{G^{(3)}}{R} + \frac{2 G^{(2)}}{R^2} \pm \frac{3
    G^{(1)}}{R^3} + \frac{3 G}{R^4} \,,
\end{eqnarray}  
where 
\begin{equation}
  G = G(t \mp R) \,, \qquad G^{(n)} \equiv
    \frac{\diff^n G(x)}{\diff x^n}{\Big \vert}_{x = t \mp R} \,.
\end{equation}

The mode function $G$ can be freely specified. 
Again, the upper sign corresponds to an outgoing solution and the
lower sign to an ingoing one, and superpositions of the two are
also solutions. To obtain a regular solution at $R = 0$, one has to
form the combination
\begin{equation}
  \vec u_\mathrm{reg} = \vec u_\mathrm{out} - \vec u_\mathrm{in} \,,
\end{equation}
where $\vec u_\mathrm{out}$ and $\vec u_\mathrm{in}$ are out- and ingoing
solutions with the same mode function $G$.

In the odd-parity case, the twist variables do not vanish:
\begin{eqnarray}
  \label{eq:teukoddEr}
  E^r &=& R^{-1} \sin\theta \cos\theta \, \partial_t(L + 4 K)  \,,\\
  E^z &=& R^{-1} \left[- \partial_t L 
    + \cos^2\theta \, \partial_t (L + 4 K) \right] \,,\\
  \label{eq:teukoddBp}
  B^\varphi &=& -R^{-2} \sin\theta (R \, \partial_R L + 4 K) \,.
\end{eqnarray}
To obtain these, one first computes the twist vector \eqref{eq:twist}
and then uses definitions (\ref{eq:E}--\ref{eq:B}).
We have checked that when expressed in cylindrical polar coordinates
$(t, r, z, \varphi)$, the variables (\ref{eq:teukoddEr}--\ref{eq:teukoddBp}) 
are manifestly regular on the axis $r = 0$.

The remaining Z(2+1)+1 variables are found to be trivial:
\begin{equation}
  H_{AB} = \delta_{AB} \,, \qquad s = 0 
\end{equation}
and thus 
\begin{equation}
  D_{CAB} = s_A = \chi_{AB} = Y = 0 \,.
\end{equation}

Hence the odd-parity solution only involves the twist geometry,
whereas the even-parity solution involves the remaining variables: 
the two polarization states can be understood as a twisting
state and a non-twisting one. This reflects a similar decoupling of
the evolution equations, section \ref{sec:Z211lin}.


   \section{An even-parity twisting octupole solution}
\label{sec:octupole}

The twisting solution presented in section \ref{sec:teukoodd} is
antisymmetric under reflection about the $z = 0$ plane. However,
we would like to impose reflection symmetry about $z = 0$ so that we
only need to evolve the upper half of the $(r,z)$-plane.
It would thus be interesting to find an \emph{even}-parity solution
that is purely twisting. It turns out that the even polarization state
of the octupole solution ($L = 3$) has that property.

The line element can again be written in the form \eqref{eq:oddlineelement}.
However, the angular functions are different:
\begin{eqnarray}
  d_{R\varphi} &=& \sin\theta (4 - 5 \sin^2\theta) \,,\\
  d_{\theta\varphi} &=& \sin^2\theta \cos\theta \,.
\end{eqnarray}
The radial functions are also modified:
\begin{eqnarray}
  K &=& \pm \frac{G^{(3)}}{R^2} + \frac{6 G^{(2)}}{R^3} 
    \pm \frac{15 G^{(1)}}{R^4} + \frac{15 G}{R^5} \,,\\
  L &=& \frac{G^{(4)}}{R} \pm \frac{5 G^{(3)}}{R^2} + \frac{15 G^{(2)}}{R^3}
    \pm \frac{30 G^{(1)}}{R^4} + \frac{30 G}{R^5} \,,
\end{eqnarray}
where as before
\begin{equation}
  G = G(t \mp R) \,, \qquad G^{(n)} \equiv
    \frac{\diff^n G(x)}{\diff x^n}{\Big \vert}_{x = t \mp R} \,,
\end{equation}
and a regular solution can be obtained by forming
\begin{equation}
  \vec u_\mathrm{reg} = \vec u_\mathrm{out} - \vec u_\mathrm{in} \,,
\end{equation}
where $\vec u_\mathrm{out}$ and $\vec u_\mathrm{in}$ are out- and ingoing
solutions with the same mode function $G$.

The twist variables are found to be
\begin{eqnarray}
  \label{eq:octupoleEr}
  E^r &=& R^{-1} \sin\theta (\cos^2\theta \, \partial_t(L + 4 K) 
    - \sin^2\theta \, \partial_t K) \,,\\
  \label{eq:octupoleEz}
  E^z &=& R^{-1} \cos\theta (4 \cos^2\theta \, \partial_t K -
    \sin^2\theta \, \partial_t (L + K) \,,\\
  \label{eq:octupoleBp}
  B^\varphi &=& R^{-2} \sin\theta \cos\theta (R \, \partial_R L + 10 K) \,.
\end{eqnarray}
Noting the transformation \eqref{eq:polsphertrafo}, these have both the
desired $r$ and $z$ parities (tables \ref{tab:consrpar}, and 
\ref{tab:conszpar} in the following section).
The remaining Z(2+1)+1 variables are trivial, as in section 
\ref{sec:teukoodd}.

It is worth explaining how we derived this solution. Instead of using
tensor spherical harmonics as in \cite{Teukolsky82}, it is 
easier to work directly with the twist subsystem of the linearized Z(2+1)+1
equations.
One can begin by postulating the desired angular behaviour, i.e.,
\begin{equation}
  E^r \propto \sin\theta\,,\qquad 
  E^z \propto \cos\theta\,,\qquad
  B^\varphi \propto \sin\theta \cos\theta \,.
\end{equation}
For the spherical polar components, this means
\begin{equation}
  E^R \propto (4 - 5 \sin^2 \theta) \,,\qquad
  E^\theta \propto \sin\theta \cos\theta \,,\qquad
  B^\varphi \propto \sin\theta \cos\theta \,.
\end{equation}
One now makes an ansatz of a form similar to the previously found solutions,
\begin{eqnarray}
  E^R &=& (4 - 5 \sin^2 \theta) \, R^{-p} \sum_{n = 0}^{N} a_n^\pm R^n
    G^{(n)}(t \pm R) \,,\\
  E^\theta &=& \sin\theta \cos\theta \, R^{-(p+1)} \sum_{n = 0}^{N}
    b_n^\pm R^n G^{(n)}(t \pm R) \,,\\
  B^\varphi &=& \sin\theta \cos\theta \, R^{-p} \sum_{n = 0}^{N}
    c_n^\pm R^n G^{(n)}(t \pm R) 
\end{eqnarray}
and inserts it into the spherical polar version of equations
(\ref{eq:lind0Er}--\ref{eq:lind0Bp}, \ref{eq:lind0Zp}) (with
$Z^\varphi = 0$),
\begin{eqnarray}
  0 &=& E^R{}_{,t} + R^{-1} B^\varphi{}_{,\theta} + 3 R^{-1}
    \cot\theta \, B^\varphi \,,\\
  0 &=& E^\theta{}_{,t} - R^{-1} B^\varphi{}_{,R} - 3 R^{-2} B^\varphi
    \,,\\
  0 &=& E^\theta{}_{,t} - R^{-1} B^\varphi{}_{,R} - 3 R^{-2} B^\varphi \,,\\
  0 &=& E^R{}_{,R} + R^{-1} E^R + E^\theta{}_{,\theta} 
    + 3(R^{-1} E^R + \cot \theta \, E^\theta) \,.
\end{eqnarray}
After some experimentation one finds that $p = 6$ and $N = 4$ are
required and that the only nontrivial solution for the constants
$a_n^\pm, b_n^\pm, c_n^\pm$ is (up to an overall factor)
\begin{eqnarray}
  a_n^\pm &=& (15, \mp 15, 6, \mp 1, 0) \,,\\
  b_n^\pm &=& (30, \mp 30, 15, \mp 5 , 1) \,,\\
  c_n^\pm &=& (0, -15, \pm 15, -6, \pm 1) \,.
\end{eqnarray}
Transforming back to cylindrical polar coordinates, one arrives at 
(\ref{eq:octupoleEr}--\ref{eq:octupoleBp}).


   \section{Numerical evolutions}
\label{sec:linnum}

We are now ready to perform numerical evolutions and compare them with
the exact solutions. 

The initial data is taken to be that of the exact
solutions at $t = 0$. The even-parity non-twisting quadrupole solution
(section \ref{sec:teukoeven}) and the even-parity twisting octupole
solution (section \ref{sec:octupole}) are considered separately. The mode
functions are taken to be 
\begin{equation}
  \label{eq:modefunctions}
  F(x) = F_0 x \e^{-x^2} \,, \qquad
  G(x) = G_0 x \e^{-x^2} \,,
\end{equation}
and in both cases we form a regular combination of outgoing and
ingoing solutions as described in the preceding sections.
Although the exact solutions are only valid in linearized theory, we
evolve them using the fully nonlinear Z(2+1)+1 system. This is
consistent if the amplitudes $F_0, G_0 \ll 1$ in \eqref{eq:modefunctions}.
The amplitudes we choose are $F_0 = G_0 = 10^{-4}$.

The gauge parameters are taken to be those of harmonic gauge, $f = d =
\mu = a = 1$, $m = 2$. Both vanishing and dynamical shift are considered.

We impose the appropriate Dirichlet or Neumann conditions on the axis
$r = 0$ (table \ref{tab:consrpar}).
Because the exact solutions we consider are reflection-symmetric about 
$z = 0$, we only evolve the upper half of the $(r, z)$-plane and impose
either a Dirichlet or a Neumann condition at $z = 0$, depending on the
$z$-parity of the variables (table \ref{tab:conszpar}).
The outer boundaries are placed at $r_\mathrm{max} = z_\mathrm{max} = 5$.
In this chapter, we impose the exact solution at the outer boundaries
(chapter \ref{sec:outerbcs} is devoted entirely to general outer boundary
conditions).

\begin{table}
  \begin{eqnarray*}   
    H_{rr}, \underline{H_{rz}}, H_{zz}, s, \alpha, 
    \beta^r, \underline{\beta^z}, \nonumber\\
    D_{rrr}, \underline{\tilde D_{rrz}}, D_{rzz},
    \underline{D_{zrr}}, D_{zrz}, \underline{D_{zzz}}, \tilde s_r, 
    \underline{s_z}, A_r, \underline{A_z}, \tilde B_r{}^r, \underline{B_r{}^z},
    \underline{B_z{}^r}, B_z{}^z, \nonumber\\
    \chi_{rr}, \underline{\chi_{rz}}, \chi_{zz}, Y, 
    E^r, \underline{E^z}, \underline{B^\varphi}, 
    \theta, Z_r, \underline{Z_z}, Z^\varphi \,, \\
    J^\varphi, J_r, \underline{J_z}, \rho_H, 
    \tilde \tau, S_r, \underline{S_z}, S_{rr}, 
    \underline{S_{rz}}, S_{zz} \,. \nonumber
  \end{eqnarray*}
  \caption[Z(2+1)+1 system: $z$-parity of regularized conserved variables]
  {\label{tab:conszpar}  \footnotesize
   $z$-parity of the regularized Z(2+1)+1 variables
   if reflection symmetry is assumed. Underlined variables are odd functions
   of $z$, the remaining ones are even.}
\end{table}

\subsection{Numerical method}

The equations are discretized using second-order accurate finite
differencing on a single cell-centred grid that is uniform in $r$ and $z$
(section \ref{sec:FD}). 
The conservative form of the equations is retained on the discrete level,
i.e.,
\begin{equation}
  \partial_t \vec u = - \partial_r \F^r(\vec u) - \partial_z \F^z(\vec u) 
  + \S (\vec u)
\end{equation}
is discretized as
\begin{eqnarray}
  \partial_t \vec u_{ij} &=& -\tfrac{1}{2h} \big[ \F^r(\vec u)_{i+1,j} 
  - \F^r(\vec u)_{i-1,j} + \F^z (\vec u)_{i,j+1} -
  \F^z (\vec u)_{i, j-1} \big] \nonumber\\&&+ \S(\vec u)_{ij} \,.
\end{eqnarray}
The numerical solution is advanced in time
using the method of lines with the third-order Runge-Kutta scheme
(\ref{eq:RK3}b). The Courant number is taken to be $\Delta t / h = 0.8$.
Fourth-order Kreiss-Oliger dissipation \eqref{eq:diss4} with
amplitude $\epsilon_D = 0.5$ is added at all interior points. 
The boundary conditions are implemented using the method of ghost cells
(section \ref{sec:ghosts}). The ghosts at the outer boundaries are filled with
the exact solution.

\subsection{Snapshots of the evolution}

As an example, figure \ref{fig:lins} shows the variable $s$ of the
even-parity quadrupole solution (section \ref{sec:teukoeven}) with
vanishing shift at a number of consecutive times. 
The numerical approximation and the exact solution are overlaid.
The resolution is very coarse (32 points) -- for higher resolutions, the
difference between the exact solution and the numerical approximation
is hardly visible.
Note that the numerical evolution is perfectly regular on the axis.

\begin{figure}[p]
  \centering
  \begin{minipage}[t]{0.49\textwidth}
    \includegraphics[scale = 0.49]{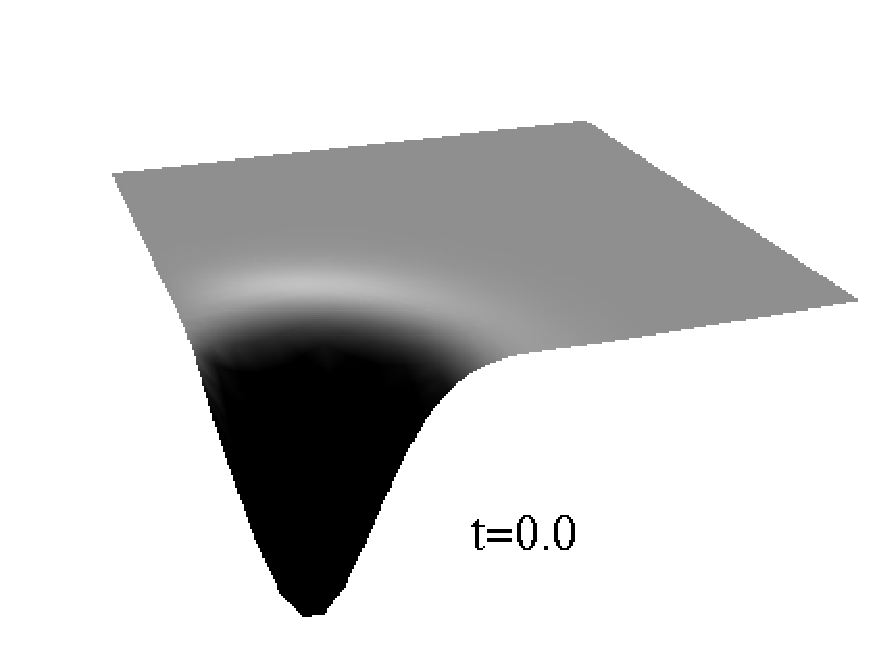}
  \end{minipage}
  \hfill  
  \begin{minipage}[t]{0.49\textwidth}
    \includegraphics[scale = 0.49]{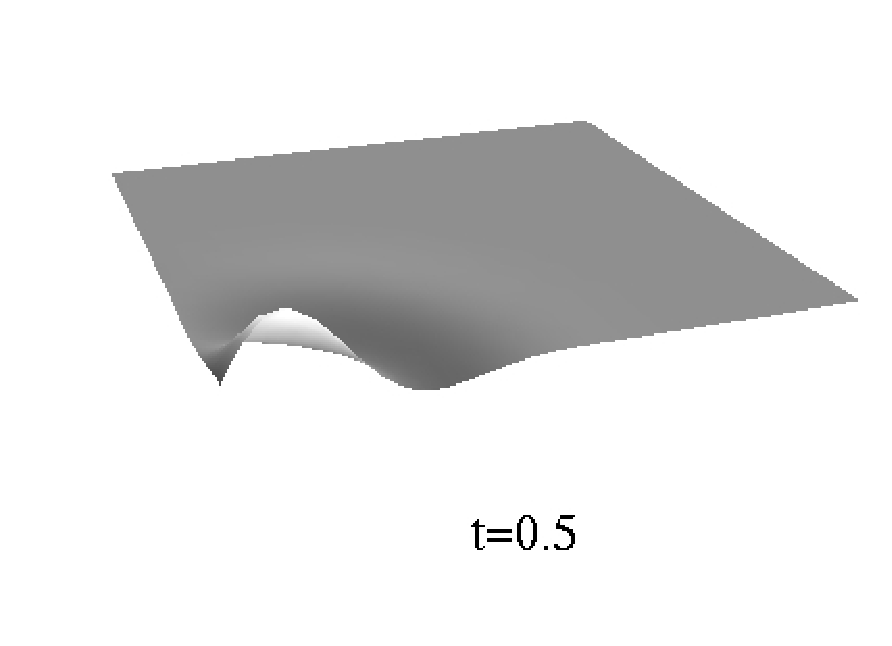}
  \end{minipage}
  \hfill  
  \begin{minipage}[t]{0.49\textwidth}
    \includegraphics[scale = 0.49]{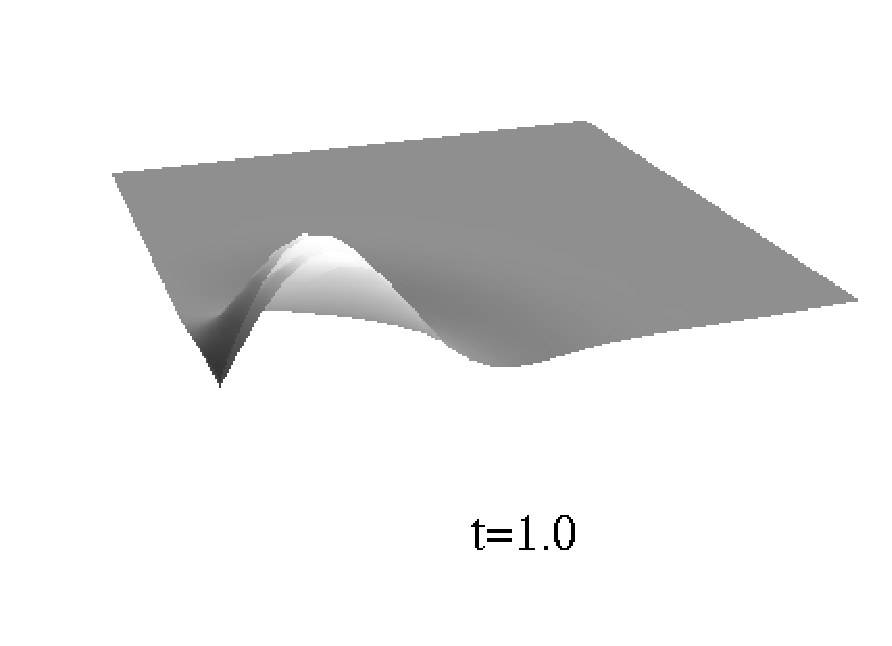}
  \end{minipage}
  \hfill  
  \begin{minipage}[t]{0.49\textwidth}
    \includegraphics[scale = 0.49]{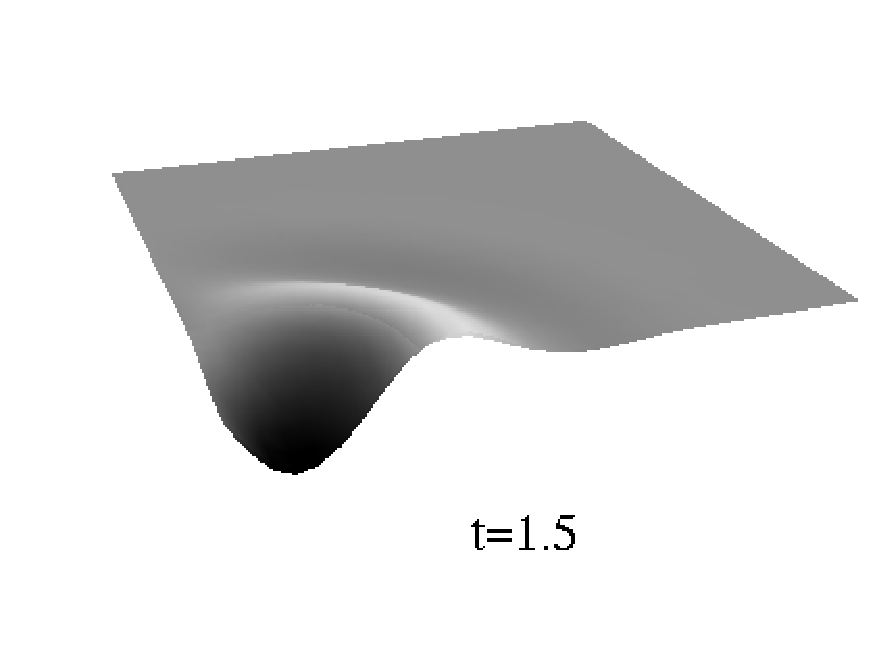}
  \end{minipage}
  \hfill  
  \begin{minipage}[t]{0.49\textwidth}
    \includegraphics[scale = 0.49]{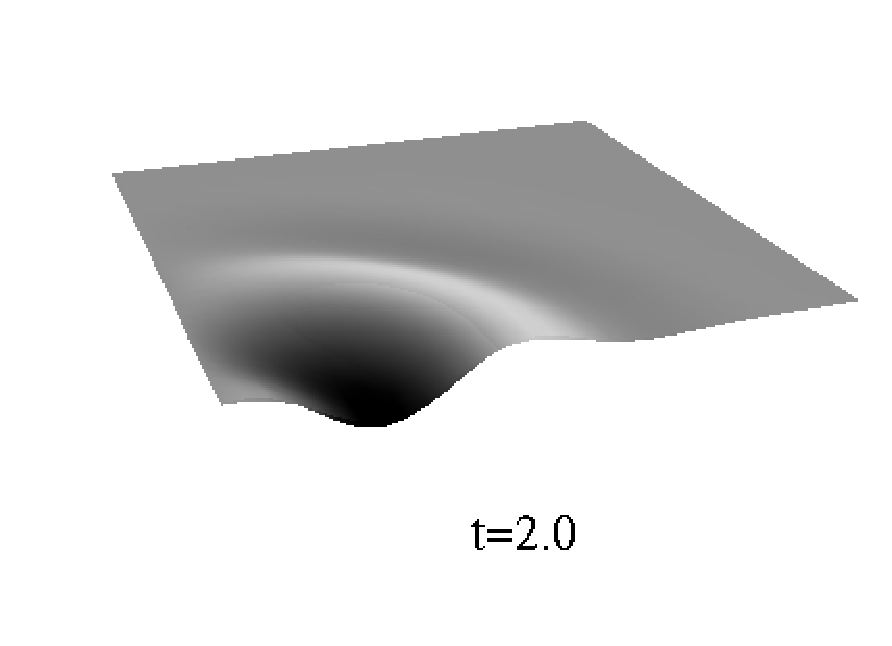}
  \end{minipage}
  \hfill  
  \begin{minipage}[t]{0.49\textwidth}
    \includegraphics[scale = 0.49]{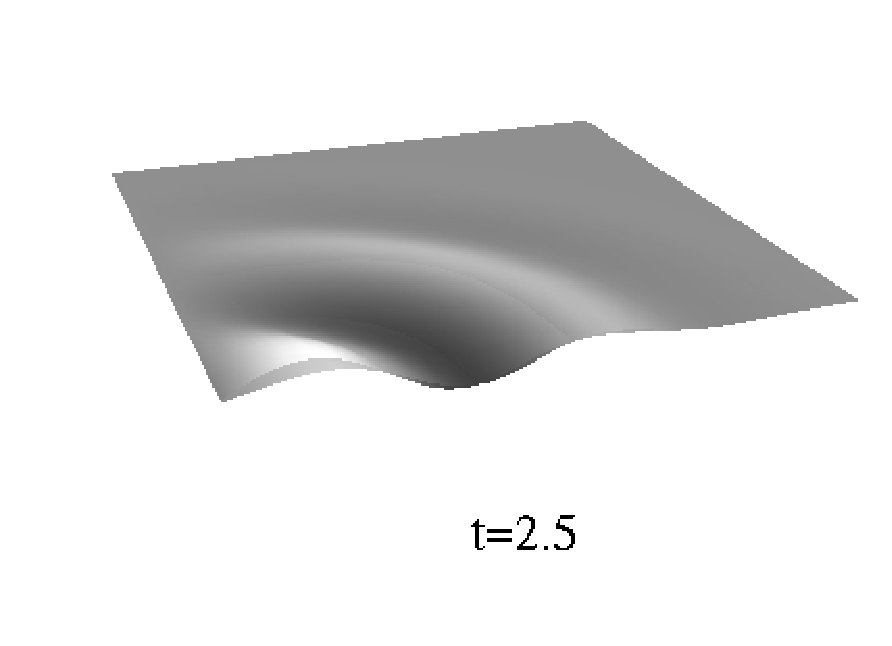}
  \end{minipage}
  \hfill  
  \begin{minipage}[t]{0.49\textwidth}
    \includegraphics[scale = 0.49]{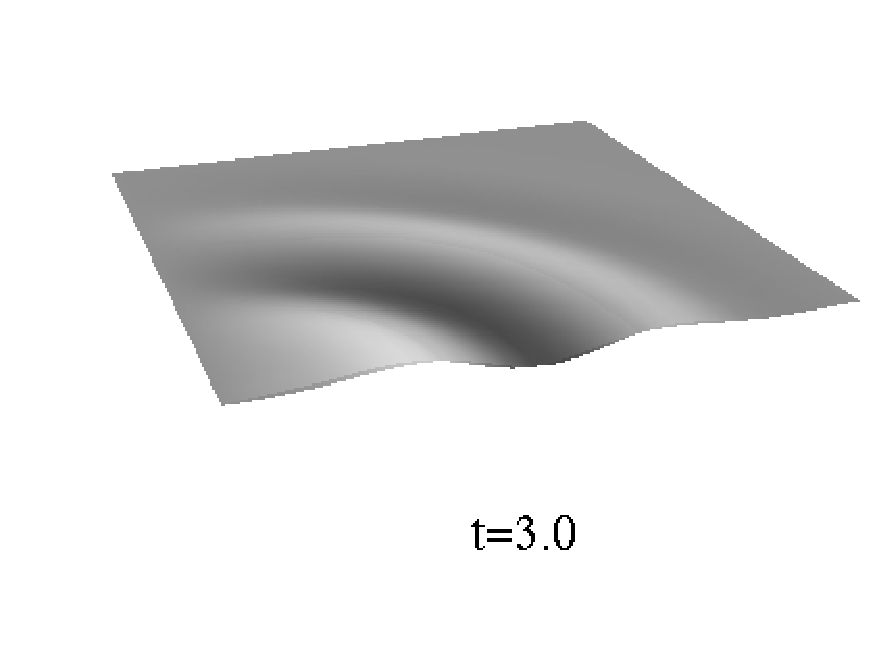}
  \end{minipage}
  \hfill  
  \begin{minipage}[t]{0.49\textwidth}
    \includegraphics[scale = 0.49]{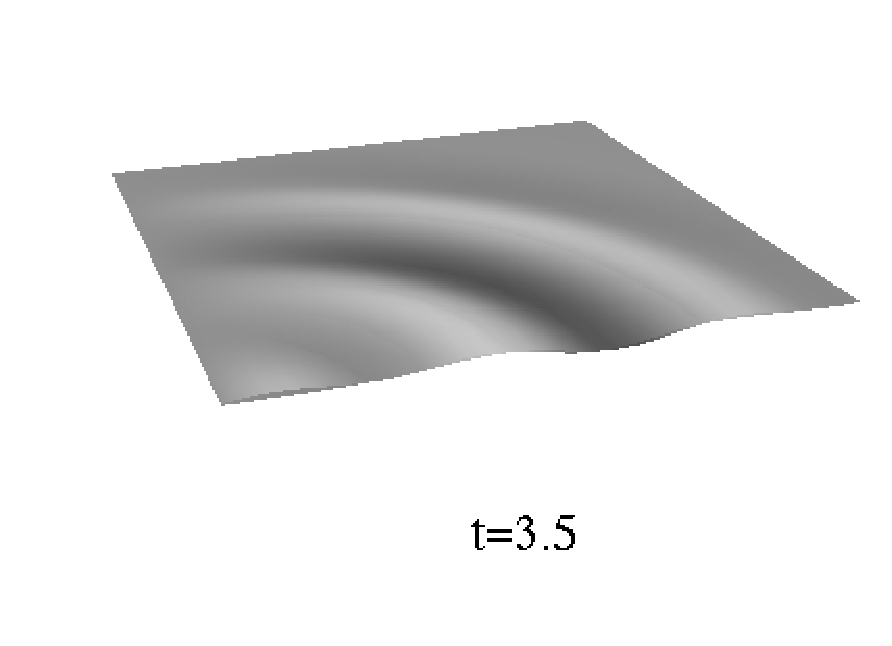}
  \end{minipage}
  \hfill  
  \caption[Snapshots of $s$: even quadrupole, vanishing shift]
  {\label{fig:lins}  \footnotesize
  Evolution of the variable $s$ for the even-parity quadrupole
  solution with vanishing shift. The numerical approximation and the
  exact solution are overlaid.} 
\end{figure}

\subsection{Convergence tests}

An important touchstone for a numerical code is a convergence test.
Because our implementation uses second-order accurate
finite-differencing and at least a second-order accurate time
integrator, the numerical error with respect to the exact solution
should behave like $\sim h^2$, where $h$ is the grid spacing.
That is, if we double the number of points per spatial dimension, the
error should decrease by a factor of $4$. 
The following plots show for three different resolutions (32, 64 and
128 points) the total (discrete) $L^2$ norm of the error $\vec e =
\vec u - \vec u_\mathrm{exact}$ as a function of time,
\begin{equation}
  \label{eq:L2norm}
  e_{L^2}(t) \equiv h \left( \sum_{ijn} {e_{ij}^n(t)}^2 \right)^{1/2} \,,
\end{equation}
where the index $n$ labels the component of the solution (all the
Z(2+1)+1 variables are included), and the indices $i,j$ refer to the 
grid point.
Alternatively, we have tried the supremum norm
\begin{equation}
  e_\textrm{sup}(t) \equiv \max_{ijn} \lvert e_{ij}^n(t) \rvert \,,
\end{equation}
which leads to the same qualitative results.

Consider first the even-parity non-twisting quadrupole solution 
(section \ref{sec:teukoeven}). This was evolved both with vanishing shift
(figure \ref{fig:linconv1}) and with dynamical shift (figure 
\ref{fig:linconv2}). Both evolutions show approximate (not perfect) 
second-order convergence. The average convergence factor of the vanishing
shift evolution is shown in figure \ref{fig:linconv1bdry}. 
In order to study the influence of the boundary location, 
we have performed the same run with twice the domain size. 
Discrepancies between the convergence factors can be observed
from $t \approx 4$ onwards, when the wave is about to arrive at the 
boundary of the smaller domain. However, this does not lead to a 
significant loss of convergence at later times.

\begin{figure}[h]
  \centering
  \begin{minipage}[t]{0.49\textwidth}
    \includegraphics[scale = 1]{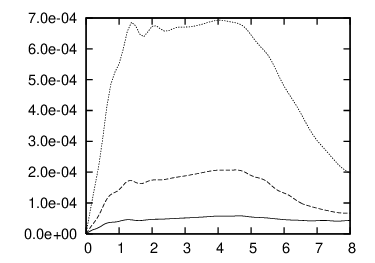}
    \centerline{total}
  \end{minipage}
  \hfill  
  \begin{minipage}[t]{0.49\textwidth}
    \includegraphics[scale = 1]{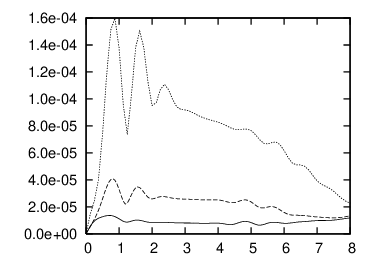}
    \centerline{$\theta$}
  \end{minipage}
  \hfill  
  \bigskip
  \begin{minipage}[t]{0.49\textwidth}
    \includegraphics[scale = 1]{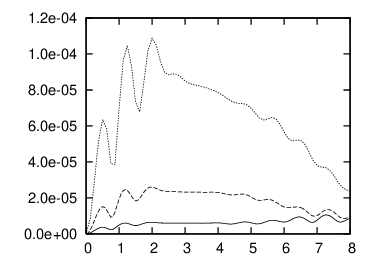}
    \centerline{$Z^r$}
  \end{minipage}
  \hfill  
  \begin{minipage}[t]{0.49\textwidth}
    \includegraphics[scale = 1]{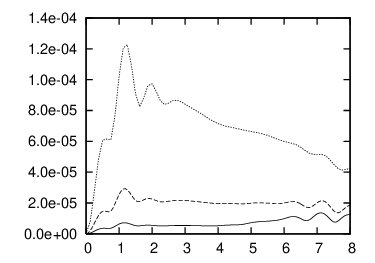}
    \centerline{$Z^z$}
  \end{minipage}
  \hfill  
  \bigskip
  \caption[Convergence test: even quadrupole, vanishing shift]
  {\label{fig:linconv1} \footnotesize
  Convergence test for the even-parity quadrupole solution 
  using vanishing shift.
  $L^2$ norm of the error as a function of time for 32 (dotted), 64
  (dashed) and 128 (solid) points per dimension.
  The total error of all the Z(2+1)+1 variables and the components
  of the $Z$ vector are shown separately.}
\end{figure}

\begin{figure}[h]
  \centering
  \includegraphics[scale = 1.0]{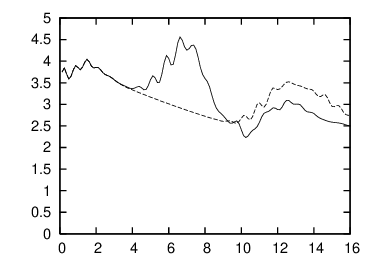}
  \caption[Convergence factor, dependence on boundary location]
  {\label{fig:linconv1bdry} \footnotesize
  Average convergence factor of the Z(2+1)+1 variables as a function of time
  for the even-parity quadrupole solution using vanishing shift.
  Solid line: $r_\mathrm{max} = z_\mathrm{max} = 5$ as in figure
  \ref{fig:linconv1},
  dashed line: $r_\mathrm{max} = z_\mathrm{max} = 10$.}
\end{figure}

Note that in the vanishing shift case, the error decays with time and
all the variables assume (very nearly) their flat-space values after
the wave has left the computational domain. In contrast, the dynamical
shift evolution suffers from a growth of the error, which also affects
the constraints. We have verified that the growth rates are virtually
independent of the boundary location (the results for twice the domain size
are visually indistinguishable from the plots in figure \ref{fig:linconv2}).
It would be interesting to investigate the origin of
this growth in future work.

\begin{figure}[t]
  \centering
  \begin{minipage}[t]{0.49\textwidth}
    \includegraphics[scale = 1]{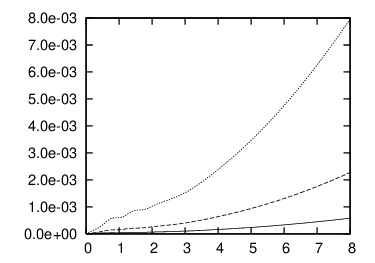}
    \centerline{total}
  \end{minipage}
  \hfill  
  \begin{minipage}[t]{0.49\textwidth}
    \includegraphics[scale = 1]{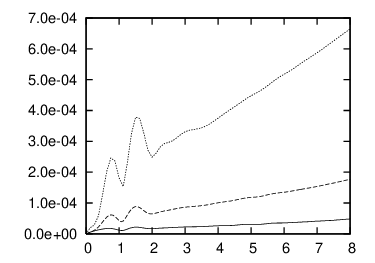}
    \centerline{$\theta$}
  \end{minipage}
  \hfill  
  \bigskip
  \begin{minipage}[t]{0.49\textwidth}
    \includegraphics[scale = 1]{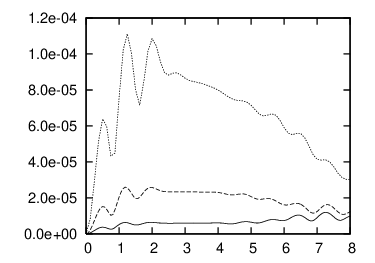}
    \centerline{$Z^r$}
  \end{minipage}
  \hfill  
  \begin{minipage}[t]{0.49\textwidth}
    \includegraphics[scale = 1]{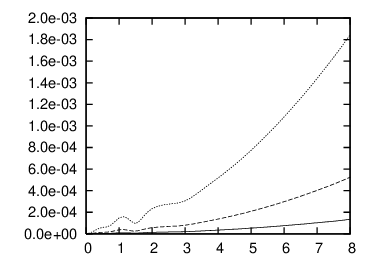}
    \centerline{$Z^z$}
  \end{minipage}
  \hfill  
  \bigskip
  \caption[Convergence test: even quadrupole, dynamical shift]
  {\label{fig:linconv2}  \footnotesize
  Convergence test for the even-parity quadrupole solution 
  using dynamical shift.
  $L^2$ norm of the error as a function of time for 32 (dotted), 64
  (dashed) and 128 (solid) points per dimension.
  The total error of all the Z(2+1)+1 variables and the components
  of the $Z$ vector are shown separately.}
\end{figure}

\begin{figure}[t]
  \centering
  \begin{minipage}[t]{0.49\textwidth}
    \includegraphics[scale = 1]{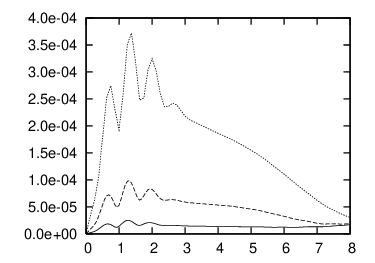}
    \centerline{total}
  \end{minipage}
  \hfill  
  \begin{minipage}[t]{0.49\textwidth}
    \includegraphics[scale = 1]{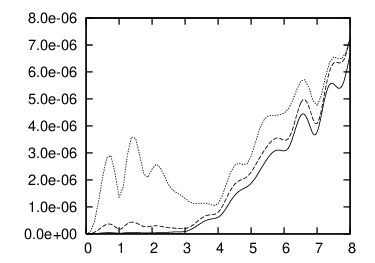}
    \centerline{$Z^\varphi$}
  \end{minipage}
  \bigskip
  \caption[Convergence test: even octupole]
  {\label{fig:linconv3}  \footnotesize
  Convergence test for the even-parity octupole solution.
  $L^2$ norm of the error as a function of time for 32 (dotted), 64
  (dashed) and 128 (solid) points per dimension.
  Total error of all the Z(2+1)+1 variables and the constraint 
  $Z^\varphi$.}
\end{figure}

\begin{figure}[h]
  \centering
  \includegraphics[scale = 1]{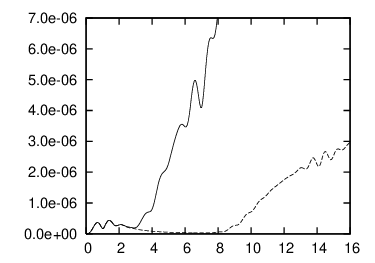}
  \bigskip
  \caption[Dependence of growth of $Z^\varphi$ on boundary location]
  {\label{fig:linconv3bdry} \footnotesize
  Dependence of the constraint growth on the boundary location for the
  even-parity octupole solution. Shown is the variable $Z^\varphi$
  as a function of time.
  Solid line: $r_\mathrm{max} = z_\mathrm{max} = 5$ as in figure
  \ref{fig:linconv3},
  dashed line: $r_\mathrm{max} = z_\mathrm{max} = 10$.
  The resolution is taken to be 64 points per dimension.}
\end{figure}

Figure \ref{fig:linconv3} shows a similar convergence test for the
even-parity twisting octupole solution (section \ref{sec:octupole}).
Here it does not matter whether we use vanishing or dynamical shift 
because the right-hand-side of the evolution equations for the shift
(\ref{eq:lind0betar}--\ref{eq:lind0betaz}) is zero anyway.
We see approximate second-order convergence up to $t \approx 3$. 
After this, the constraint $Z^\varphi$ begins to grow.
In this particular case, the onset and growth rate of the instability 
depend on the location of the outer boundary, as demonstrated by figure
\ref{fig:linconv3bdry}.

\subsection{Conclusions}

To summarize, we have demonstrated second-order convergence of the
code (at least at early times) for two different exact solutions of 
linearized theory and two different shift conditions, which strongly 
indicates that the implementation is correct. An unexpected growth of
the error occurs in the dynamical shift case.


   \emptypage

   \chapter{Outer boundary conditions}
\label{sec:outerbcs}

Having derived the appropriate boundary conditions on the axis $r = 0$ 
(table \ref{tab:consrpar}) and at $z = 0$ if reflection symmetry 
is assumed (table \ref{tab:conszpar}), we now turn to the question of how to
impose boundary conditions at the outer boundaries of the computational
domain.

Outer boundary conditions are currently a very active field of research in
numerical relativity, particularly since hyperbolic formulations of the
Einstein equations were introduced. In such formulations, one knows the
characteristic variables and their propagation speeds with respect to the
boundary, and one can use this information to construct boundary conditions.

Throughout this thesis we assume that spacetime is asymptotically flat, by
which we mean in a broad sense that the metric approaches the Minkowski 
metric towards spacelike infinity (no precise definition of asymptotic
flatness is required for our purposes). By placing the outer boundary
sufficiently far out, we may assume that linearized theory is valid near the
boundary. Hence all the calculations in this chapter are performed in
linearized theory. 

We begin by writing down the characteristic variables in terms of the
regularized Z(2+1)+1 variables (section \ref{sec:linchar}), 
which we shall need throughout the following.
As a first class of boundary conditions, we discuss dissipative boundary 
conditions (section \ref{sec:dissbcs}), 
for which well-posedness theorems of the initial boundary value 
problem are known.
In an alternative approach to the problem, we begin with certain physical
considerations, which are then translated into a prescription for the 
characteristic variables. These considerations fall into three different
groups: outgoing radiation boundary conditions based on the Newman-Penrose
scalars (section \ref{sec:radbcs}), 
constraint-preserving boundary conditions based on the so-called 
subsidiary system (section \ref{sec:cpbcs}), and gauge boundary conditions
(section \ref{sec:gaugebcs}).
Stability of the thus derived boundary conditions is analysed using the
Fourier-Laplace transform technique (section \ref{sec:fourierlaplace}). 
We perform numerical evolutions of the exact linearized solutions of chapter
\ref{sec:lin} to compare the various boundary conditions with regard to their
stability and efficiency in avoiding spurious reflections from the outer 
boundaries (section \ref{sec:bcnum}).

\section{Linearized characteristic variables}
\label{sec:linchar}

The linearized Z(2+1)+1 equations have already been expressed in terms
of regularized variables in section \ref{sec:Z211lin}. Here we derive
the characteristic variables of that system. One starts from the 
results in section \ref{sec:hyperbolicity} and makes the replacements 
(\ref{eq:regtrafoDrrz}--\ref{eq:regtrafoY}). Note that the terms
with zero fluxes (the ellipses in 
(\ref{eq:regtrafoDrrz}--\ref{eq:regtrafoY})) are not to be included
because they do not enter the principal parts of the regularized
evolution equations. Finally we linearize the characteristic variables
about flat space.
We have checked explicitly with REDUCE that the results below are indeed the
characteristic variables of the system stated in section \ref{sec:Z211lin}.
\pagebreak

For the characteristic variables in the $r$-direction we find in the
dynamical shift case
{\allowdisplaybreaks \begin{eqnarray}
    l_{0,1} &=& D_{zrr} \,,\\
    l_{0,2} &=& D_{zrz} \,,\\
    l_{0,3} &=& D_{zzz} \,,\\
    l_{0,4} &=& r s_z + D_{zrr} - D_{zzz} \,,\\
    l_{0,5} &=& A_z \,,\\
    l_{0,6} &=& B_z{}^r \,,\\
    l_{0,7} &=& B_z{}^z \,,\\
    \label{eq:l11mr}
    l_{1,1}^\pm &=& r Y + \chi_{rr} - \chi_{zz} + 2 B_z{}^z
      \pm (r^2 \tilde s_r + D_{rrr} - D_{rzz}) \,,\\
    \label{eq:l12mr}
    l_{1,2}^\pm &=& E^z \pm B^\varphi \,,\\
    l_{1,3}^\pm &=& \theta - 2 B_z{}^z \pm (D_{rzz} + r^2 \tilde s_r +
      D_{rrr} - D_{zrz} - Z_r) \,,\\
    l_{1,4}^\pm &=& r Y + \chi_{rr} + \chi_{zz} - \theta \pm (D_{zrz}
      + Z_r) \,,\\
    l_{1,5}^\pm &=& \chi_{rz} \pm \half (A_z + 2 D_{zrr} - D_{zzz} + r
       s_z - 2 Z_z) \,,\\
    l_{1,6}^\pm &=& E^r \mp 2 Z^\varphi \,,\\
    l_f^\pm &=& A_r - f c_1(D_{rzz} + r^2 \tilde s_r + D_{rrr} -
       D_{zrz} - Z_r) \nonumber\\&&
       \pm \sqrt{f} (2 \chi_{rr} + \chi_{zz} + r Y - (f c_1 + 2)
       \theta + 2 c_1 B_z{}^z) \,,\\
    l_\mu^\pm &=& a A_z + 2 \mu(r s_z + 2 D_{zrr} - r \tilde D_{rrz} -
       Z_z) \nonumber\\&&
       - d (2 D_{zrr} + D_{zzz} + r s_z) \pm 2 \sqrt{\mu} (B_r{}^z +
       B_z{}^r) \,,\\
    l_d^\pm &=& (f c_2 + 1) (2 \chi_{rr} + \chi_{zz} + r Y) + (f c_2
       c_3 + 2 c_4) (2 B_z{}^z - \theta) \nonumber\\&&
       - f m c_2 \theta - 2 (r \tilde B_r{}^r + B_z{}^z) \nonumber\\&&
       \pm \sqrt{d} [2 D_{rrr} + D_{rzz} + r^2 \tilde s_r + c_2 A_r)
       \\&& \qquad
       - (f c_2 c_3 + 2 c_4)(r^2 \tilde s_r + D_{rrr} + D_{rzz} - D_{zrz} -
       Z_r)] \,.\nonumber
\end{eqnarray}}
In the vanishing shift case, $l_{0,6}$ and $l_{0,7}$ are replaced with
\begin{eqnarray}
  l_{0,6} &=& f m (D_{rzz} + r^2 \tilde  s_r + D_{rrr} - D_{zrz} - Z_r)
    \nonumber\\&&
    - f (2 D_{rrr} + D_{rzz} + r^2 \tilde s_r) + A_r \,,\\
  l_{0,7} &=& f m (2 D_{zrr} + r s_z - r \tilde D_{rrz} - Z_z)\nonumber\\&&
    - f (2 D_{zrr} + D_{zzz} + r s_z) + A_z \,.
\end{eqnarray}

The characteristic variables in the $z$-direction for dynamical shift
are found to be
{\allowdisplaybreaks \begin{eqnarray}
    l_{0,1} &=& D_{rzz} \,,\\
    l_{0,2} &=& r \tilde D_{rrz} \,,\\
    l_{0,3} &=& D_{rrr} \,,\\
    l_{0,4} &=& r^2 \tilde s_r \,,\\
    l_{0,5} &=& A_r \,,\\
    l_{0,6} &=& B_r{}^z \,,\\
    l_{0,7} &=& r \tilde B_r{}^r \,,\\
    \label{eq:l11mz}
    l_{1,1}^\pm &=& r (Y + 2 \tilde B_r{}^r \pm s_z) \,,\\
    \label{eq:l12mz}
    l_{1,2}^\pm &=& E^r \mp B^\varphi \,,\\
    l_{1,3}^\pm &=& \theta - 2 r \tilde B_r{}^r \pm (2 D_{zrr} + r s_z
      - r \tilde D_{rrz} - Z_z) \,,\\
    l_{1,4}^\pm &=& r Y + 2 \chi_{rr} - \theta \pm (r \tilde D_{rrz} +
      Z_z) \,,\\
    l_{1,5}^\pm &=& \chi_{rz} \pm \half (A_r + D_{rzz} + r^2 \tilde
      s_r - 2 Z_r)\,,\\
    l_{1,6}^\pm &=& E^z \mp 2 Z^\varphi \,,\\
    l_f^\pm &=& A_z - f c_1 (2 D_{zrr} + r s_z - r \tilde D_{rrz} -
      Z_z) \nonumber\\&&
      \pm \sqrt{f} (\chi_{zz} + 2 \chi_{rr} + r Y - (f c_1 + 2) \theta
      + 2 c_1 r \tilde B_r{}^r) \,,\\
    l_\mu^\pm &=& a A_r + 2 \mu (r^2 \tilde s_r + D_{rrr} + D_{rzz} -
      D_{zrz} - Z_r) \nonumber\\&&
      - d (D_{rzz} + 2 D_{rrr} + r^2 \tilde s_r) \pm 2 \sqrt{\mu}
      (B_z{}^r + B_r{}^z) \,,\\
    l_d^\pm &=& (f c_2 + 1)(\chi_{zz} + 2 \chi_{rr} + r Y) + (f c_2
      c_3 + 2 c_4)(2 r \tilde B_r{}^r - \theta) \nonumber\\&&
      - f m c_2 \theta - 2(B_z{}^z + r \tilde B_r{}^r) \nonumber\\&&
      \pm \sqrt{d}(D_{zzz} + 2 D_{zrr} + r s_z + c_2 A_z) \\&&\qquad
      - (f c_2 c_3 + 2 c_4)(r s_z + 2 D_{zrr} - r \tilde D_{rrz} -
      Z_z) \,.\nonumber
\end{eqnarray}}
In the vanishing shift case, $l_{0,6}$ and $l_{0,7}$ are replaced with
\begin{eqnarray}
  l_{0,6} &=& f m (2 D_{zrr} + r s_z - r \tilde D_{rrz} - Z_z)\nonumber\\&&
    - f (2 D_{zrr} + D_{zzz} + r s_z) + A_z \,,\\
  l_{0,7} &=& f m (D_{rzz} + r^2 \tilde s_r + D_{rrr} - D_{zrz} - Z_r)
    \nonumber\\&& 
    - f (2 D_{rrr} + D_{rzz} + r^2 \tilde s_r) + A_r \,.
\end{eqnarray}
For the numerical implementation of the characteristic transformation in the
$z$-direction, we replace $l_{0,2}$, $l_{0,4}$, the
dynamical-shift version of $l_{0,7}$, and $l_{1,1}^\pm$ with 
\begin{eqnarray}
  \tilde l_{0,2} &\equiv& r^{-1} l_{0,2} = \tilde D_{rrz} \,,\\
  \tilde l_{0,4} &\equiv& r^{-2} l_{0,4} = \tilde s_r \,,\\
  \tilde l_{0,7} &\equiv& r^{-1} l_{0,7} = \tilde B_r{}^r \,,\\
  \label{eq:l11tilde}
  \tilde l_{1,1}^\pm &\equiv& r^{-1} l_{1,1}^\pm 
    = Y + 2 \tilde B_r{}^r \pm s_z 
\end{eqnarray}
because then the inverse transformation from $z$-characteristic to
regularized conserved variables does not involve any negative powers of $r$.


   \section{Dissipative boundary conditions}
\label{sec:dissbcs}

Let $\vec v^-$ be the vector of ingoing characteristic variables in
the direction normal to the boundary under consideration and 
$\vec v^+$ the vector of the remaining (outgoing and zero-speed)
characteristic variables. In this section, we consider 
\emph{(homogeneous) dissipative boundary conditions} of the form
\begin{equation}
  \label{eq:dissbcs}
  \vec v^- \doteq M \vec v^+ \,,
\end{equation}
where $M$ is a constant real matrix and here and in the following $\doteq$
denotes equality on the boundary.

Under certain additional assumptions, there exist theorems that
guarantee the well-posedness of the initial boundary value problem
(IBVP) for such boundary conditions, 
which we shall outline in the following.

\subsection{Well-posedness of the IBVP}
\label{sec:IBVP}

We focus on a single boundary here and so we take the spatial domain 
$\Omega \subset \mathbb{R}^2$
to be a bounded open set lying on one side of its boundary $\Gamma$.
Suppose we have a linear symmetric hyperbolic system 
\begin{equation}
  \partial_t \vec u + A^A \partial_A \vec u + B \vec u = 0  \,,
\end{equation}
in $[0, \infty) \times \Omega$, i.e., the matrices $A^A$ can be
assumed to be symmetric (after a suitable symmetrization).
The boundary conditions are taken to be of the form \eqref{eq:dissbcs} on 
$[0, \infty) \times \Gamma$.
Assume first that the boundary is \emph{non-characteristic}, i.e. that 
$A^\n \equiv A^A \mu_A$ is invertible on the entire boundary $\Gamma$, 
where $\mu_A$ is the boundary normal. 
One can prove \cite{GKO} that the above initial boundary-value problem is
\emph{strongly well-posed} in the following sense: 
for every initial data  $\vec f \in C^\infty(x^A)$, $\vec u(0, x^A) = f(x^A)$, 
there exists a unique solution $\vec u(t, x^A) \in C^\infty(t, x^A)$ such that
\begin{equation}
  \label{eq:wellposedestimate2}
  \lVert \vec u(t, \cdot) \rVert \leqslant K \e^{\alpha t} \lVert
  \vec f(\cdot) \rVert \,,
\end{equation}
where the constants $K$ and $\alpha$ are independent of $f$, and we are using
$L^2$ norms (this is the same type of estimate as in section 
\ref{sec:hyperbolicity}).

This result has been generalized by Majda and Osher
\cite{MajdaOsher75} to the case of a \emph{uniformly characteristic
  boundary}, which means that $A^\n$ has constant
(not necessarily maximal) rank across the whole boundary $\Gamma$.
This wider class includes many important examples in physics such as Maxwell's
equations and the linearized Einstein equations.
Finally, Rauch \cite{Rauch85} and Secchi \cite{Secchi96} generalized
the well-posedness theorem to quasilinear systems (such as the fully nonlinear
Z(2+1)+1 system). In this case,
however, the estimate \eqref{eq:wellposedestimate2} only holds for a
finite time in general and the solution cannot be taken to be $C^\infty$
(instead, it lies in an appropriate Sobolev space, see \cite{Secchi96}
for details).

Since the above theorems all require a symmetric hyperbolic system,
they are only applicable to the zero-shift version of the Z(2+1)+1
equations with parameters $(f = 1, m = 2)$.
In practice, however, dissipative boundary conditions work well for
most strongly hyperbolic systems, too. 
We would like to stress that in the dynamical shift case, the boundary
is uniformly characteristic only in linearized theory 
because otherwise the speed of the normal modes
depends on the shift vector, which may change sign along the boundary.

\subsection{Absorbing boundary conditions}
\label{sec:absbcs}

The simplest example of dissipative boundary conditions are
\emph{absorbing boundary conditions}
\begin{equation}
  \label{eq:absbcs}
  \vec v^- \doteq 0 \,,
\end{equation}
i.e., the incoming modes are set to zero at the outer boundaries.
These boundary conditions have proven to be very stable in numerical
experiments (section \ref{sec:bcnum}).

One might ask to what extent absorbing boundary conditions
are satisfied by the exact linearized solutions of chapter \ref{sec:lin}.
Even if they are not satisfied identically,
one can still evaluate the residuals of the boundary conditions
(i.e., the incoming modes) and expand them into a series in inverse 
powers of $r$ (for the outer boundary at $r = r_\mathrm{max}$) or $z$ 
(for the outer boundary at $z = z_\mathrm{max}$). One finds that 
the incoming modes vanish to leading order except for $l_{1,3}^\pm,
l_{1,4}^\pm, l_{1,5}^\pm$ and $l_{1,6}^\pm$.

Can we construct ``better'' boundary conditions of dissipative type?

\subsection{Zero-$Z$ boundary conditions}
\label{sec:cpdissbcs}

A different boundary condition one might impose is the vanishing of
the algebraic constraints
\begin{equation}
  \label{eq:zeroZ}
  \theta \doteq Z_r \doteq Z_Z \doteq Z^\varphi \doteq 0 \,,
\end{equation}
which is clearly satisfied by a solution of Einstein's equations.
Using the expressions for the characteristic variables 
(section \ref{sec:linchar}), we can rewrite \eqref{eq:zeroZ} in
dissipative form: 
\begin{eqnarray} 
  \label{eq:zerozbcsfirst}
  l_{1,3}^- &\doteq& -l_{1,3}^+ - 4 l_{0,7} \,,\\
  l_{1,4}^- &\doteq& l_{1,4}^+ - 2 l_{0,2} \,,\\
  l_{1,5}^- &\doteq& l_{1,5}^+ - l_{0,1} - l_{0,4} - l_{0,5} \,,\\
  \label{eq:zerozbcslast}
  l_{1,6}^- &\doteq& l_{1,6}^+ 
\end{eqnarray}
(the equations are the same in the $r$- and in the $z$-direction).
Supplemented with absorbing boundary conditions for the remaining
incoming modes, we obtain a set of boundary conditions that are all
satisfied by the exact solutions at least to leading order in the respective
inverse coordinate. In this sense, they would appear to be superior to
pure absorbing boundary conditions. 
However, we shall see in section \ref{sec:cpbcs} that the $Z$ vector
obeys a wave equation (\ref{eq:BoxZ}). The boundary conditions 
(\ref{eq:zerozbcsfirst}--\ref{eq:zerozbcslast}) are Dirichlet boundary
conditions and hence any violations of the $Z_\alpha = 0$ constraints 
hitting the outer boundaries will be reflected. 
This is confirmed by our numerical experiments in section \ref{sec:bcnum}.


   \section{Outgoing-radiation boundary conditions}
\label{sec:radbcs}

We now start to look for a different set of boundary conditions based
on more physical considerations.
A reasonable requirement for an isolated system is that no
gravitational radiation should enter the computational domain from the
outside.
If we were dealing with a scalar wave equation 
\begin{equation}
  \Box u = 0 \,,
\end{equation}
the appropriate outgoing-radiation boundary condition would be a
\emph{Sommerfeld condition}
\begin{equation}
  u = \frac{f(t - R)}{R} \Rightarrow (\partial_t + \partial_R) (R u) = 0 \,,
\end{equation}
where $R = \sqrt{r^2 + z^2}$. 
This is the condition we imposed componentwise on all the variables
evolved with the hyperbolic-elliptic system (section \ref{sec:hypellnum}).
However, the gravitational field has only two
degrees of freedom (the two polarization states, cf. chapter \ref{sec:lin})
and hence we are only allowed to impose two conditions.

\subsection{Newman-Penrose scalars and the peeling theorem}

More insight can be obtained by looking at the asymptotic behaviour of
the \emph{Weyl tensor}
\begin{eqnarray}
  C_{\alpha\beta\gamma\delta} &=& R_{\alpha\beta\gamma\delta}
  - \half (g_{\alpha\gamma} R_{\delta\beta} + g_{\beta\delta} R_{\gamma\alpha}
         - g_{\alpha\delta} R_{\gamma\beta} - g_{\beta\gamma} R_{\delta\alpha})
  \nonumber\\&&+ \tfrac{1}{6} R (g_{\alpha\gamma} g_{\delta\beta} 
    - g_{\alpha\delta} g_{\gamma\beta}) \,.
\end{eqnarray}       
The Weyl tensor is the tracefree part of the Riemann tensor, i.e., the part
that is not determined by the matter sources via Einstein's equations. 
It thus contains the gravitational-wave information, 
and it reduces to the Riemann tensor in vacuum.

In the \emph{Newman-Penrose (NP) formalism} \cite{PenroseRindler, 
StewartAdvGR}, 
one forms a complex null tetrad $(l, k, m, \bar m)$ consisting of 
two real null vectors $l$ and $k$, a complex null vector $m$ and 
its complex conjugate $\bar m$ satisfying
\begin{equation}
  \label{eq:NPtetradrel}
  l \cdot k = -1 \,, \qquad m \cdot \bar m = 1 \,.
\end{equation}
Here we adapt the null tetrad to the boundary under consideration in
the following way: first we choose an orthonormal basis $\{ e_0,
e_1, e_2, e_3 \}$ with the properties
\begin{eqnarray}
  \label{eq:onb1}
  e_0 \propto n &\,& (\textrm{future-directed timelike normal}) \,,\\
  e_1 \propto \mu &\,& (\textrm{spacelike outward-pointing 
    normal to the boundary}) \,,\\
  e_2 \propto \pi &\,& (\textrm{tangent to the boundary},
    \mu_A\pi^A = 0) \,,\\
  \label{eq:onb4}
  e_3 \propto \xi &\,& (\textrm{Killing vector}) \,.
\end{eqnarray}
Then we define the NP tetrad by
\begin{eqnarray}
  \label{eq:NPtetradfirst}
  l &\equiv& \textstyle \frac{1}{\sqrt{2}} (e_0 + e_1) \,,\\
  k &\equiv& \textstyle \frac{1}{\sqrt{2}} (e_0 - e_1) \,,\\
  \label{eq:NPtetradlast}
  m &\equiv& \textstyle \frac{1}{\sqrt{2}} (e_2 - i e_3) \,,
\end{eqnarray}
which satisfies the relations \eqref{eq:NPtetradrel}.

One now forms the five independent complex projections of the Weyl tensor
with respect to the NP tetrad,
{\allowdisplaybreaks \begin{eqnarray}
  \label{eq:Psi0}
  \Psi_0 &\equiv& C_{\alpha\beta\gamma\delta} 
     l^\alpha m^\beta l^\gamma m^\delta \,,\\
  \Psi_1 &\equiv& C_{\alpha\beta\gamma\delta} 
     l^\alpha k^\beta l^\gamma m^\delta \,,\\
  \Psi_2 &\equiv& C_{\alpha\beta\gamma\delta} 
     l^\alpha m^\beta \bar m^\gamma k^\delta \,,\\
  \Psi_3 &\equiv& C_{\alpha\beta\gamma\delta} 
     l^\alpha k^\beta \bar m^\gamma k^\delta \,,\\
  \label{eq:Psi4}
  \Psi_4 &\equiv& C_{\alpha\beta\gamma\delta} 
     k^\alpha \bar m^\beta k^\gamma \bar m^\delta \,.
\end{eqnarray}}
The \emph{peeling theorem} implies that 
\begin{equation}
  \label{eq:peeling}
  \Psi_i \sim x^{-5+i} \,, \quad i = 0, 1, 2, 3, 4 
\end{equation}
as future null infinity is approached along the outgoing null
geodesics with tangent $l$ and affine parameter $x$. 
For the $r = r_\mathrm{max}$ ($z = z_\mathrm{max}$) boundary, 
$x$ may be taken to be the coordinate $r$ ($z$).
Strictly speaking, \eqref{eq:peeling} is only valid for solutions of
the Einstein equations that are algebraically general. We have
checked that all the linearized solutions in chapter \ref{sec:lin}
have the peeling behaviour \eqref{eq:peeling}, as expected for generic 
gravitational radiation.

One can use the peeling theorem to derive an
outgoing radiation condition at the outer boundaries: because $x$
is large there, $\Psi_0 \sim x^{-5}$ is suppressed as compared with
the other Weyl scalars and so it is reasonable to impose
\begin{equation}
  \label{eq:Psi0bc}
  \Psi_0 \doteq 0 
\end{equation}
at the outer boundaries. 
This condition was used by Friedrich and Nagy \cite{Friedrich98} in their
well-posed initial boundary value formulation of the Einstein equations, 
and similar conditions have recently been applied to numerical
relativity (e.g., \cite{Kidder05, SarbachTiglio05}).
Because $\Psi_0$ is complex, \eqref{eq:Psi0bc} constitute two real conditions, 
as desired. They correspond to the two gravitational degrees of freedom,
as we will see explicitly in the following.

\subsection{Construction of the NP tetrad}
\label{sec:NPtetrad}

We begin by setting up the orthormal basis (\ref{eq:onb1}--\ref{eq:onb4}).
For this we need the full four metric $g_{\alpha\beta}$ and its
inverse $g^{\alpha\beta}$. In linearized theory, we may set
\begin{eqnarray}
  \alpha &=& 1 + \delta \alpha \,,\\
  H_{AB} &=& \delta_{AB} + \delta H_{AB} \,,\\
  \lambda &=& r (1 + \delta \lambda) \,,
\end{eqnarray}
where $\delta \alpha$, $\delta H_{AB}$ and $\delta \lambda$ as well as
the shift vector $\beta^A$ and the components $\xi_t, \xi_A$ of the
Killing vector are small quantities.
In $(t, r, z, \varphi)$ coordinates, the linearized 4-metric then 
takes the form  
\begin{equation}
  g_{\alpha\beta} = \left( \begin{array}{cccc}
      -1 - 2 \delta \alpha & \beta^r & \beta^z & \xi_t \\
      \beta^r & 1 + \delta H_{rr} & \delta H_{rz} & \xi_r \\
      \beta^z & \delta H_{rz} & 1 + \delta H_{zz} & \xi_z \\
      \xi_t & \xi_r & \xi_z & r^2 (1 + 2 \delta \lambda) 
    \end{array} \right) 
\end{equation}
and its inverse is
\begin{equation}
  g^{\alpha\beta} = \left( \begin{array}{cccc}
      -1 + 2 \delta \alpha & \beta^r & \beta^z & r^{-2} \xi_t \\
      \beta^r & 1 - \delta H_{rr} & -\delta H_{rz} & -r^{-2} \xi_r \\
      \beta^z & -\delta H_{rz} & 1 - \delta H_{zz} & -r^{-2} \xi_z \\
      r^{-2} \xi_t & -r^{-2} \xi_r & -r^{-2} \xi_z & 
        r^{-2} (1 - 2 \delta \lambda) 
    \end{array} \right) \, .
\end{equation}
Note that $\xi_t$, $\xi_r$ and $\xi_z$ are \emph{not} in the Geroch
space $\N$ (section \ref{sec:Geroch}).

The Killing vector is $\xi^\alpha = \delta_\varphi{}^\alpha$ and so we have 
\begin{equation}
  e_3{}^\alpha = (0, 0, 0, r^{-1} (1 - \delta \lambda)) \,,
\end{equation}
normalized such that $e_3{}^\alpha e_{3\,\alpha} = 1$.
Lowering indices with $g_{\alpha\beta}$, the covariant version becomes
to linear order
\begin{equation}
  e_{3\,\alpha} = (r^{-1} \xi_t, r^{-1} \xi_r, r^{-1} \xi_z,
    r(1 + \delta\lambda)) \,.
\end{equation}
The timelike normal $n$ satisfies $n_a = - \alpha \delta_a{}^t$ and
$n_\alpha \xi^\alpha = 0$, which yields 
\begin{equation}
  e_{0 \, \alpha} = (-1 - \delta \alpha, 0, 0, 0) \,,
\end{equation}
and raising indices with $g^{\alpha\beta}$,
\begin{equation}
  e_0{}^\alpha = (-1 + \delta \alpha, \beta^r, \beta^z, r^{-2} \xi_t) \,.
\end{equation}
Consider first the $r = r_\mathrm{max}$ boundary. Its spacelike normal 
satisfies $\mu_A \propto \delta_A{}^r$, 
$\mu_a n^a = 0$ and $\mu_\alpha \xi^\alpha = 0$, which implies
\begin{equation}
  e_{1 \, \alpha} = (\beta^r, 1 + \half \delta H_{rr}, 0, 0) 
\end{equation}
and so
\begin{equation}
  e_1{}^\alpha = (0, 1 - \half \delta H_{rr}, -\delta H_{rz}, -r^{-2}
  \xi_r) \,.
\end{equation}
The tangent to the boundary, $\pi^A$ satisfies $\pi^A \mu_A = 0$,
$\pi^a n_a = 0$ and $\pi^\alpha \xi_\alpha = 0$, which together with
the normalization fixes
\begin{equation}
  e_2{}^\alpha = (0, 0, 1 - \half \delta H_{zz}, -r^{-2} \xi_z)
\end{equation}
and thus finally
\begin{equation}
  e_{2 \, \alpha} = (\beta^z, \delta H_{rz}, 1 + \half \delta H_{zz},
  0) \,.
\end{equation}
For the $z = z_\mathrm{max}$ boundary we have instead
\begin{eqnarray}
  e_{1 \,\alpha} &=& ( \beta^z, 0, 1 + \half \delta H_{zz}, 0) \,,\\
  e_1{}^\alpha &=& (0, -\delta H_{rz}, 1 - \half H_{zz}, -r^{-2} \xi_z )
\end{eqnarray}
and
\begin{eqnarray}
  e_2{}^\alpha &=& (0, 1 - \half \delta H_{rr}, 0, -r^{-2} \xi_r) \,,\\
  e_{2\,\alpha} &=& (\beta^r, 1 + \half \delta H_{rr}, \delta H_{rz}, 0) \,.
\end{eqnarray}

From the orthonormal basis, we finally form the NP tetrad as defined by 
equations (\ref{eq:NPtetradfirst}--\ref{eq:NPtetradlast}).

\subsection{Computation of $\Psi_0$}

Having constructed the NP tetrad in covariant and
contravariant form, we can compute the Weyl scalars 
(\ref{eq:Psi0}--\ref{eq:Psi4}) using, for example, the algorithm of
Campbell and Wainwright \cite{Campbell77}.\footnote{A REDUCE
  version of this algorithm was written by John Stewart.}

The resulting expressions contain various (up to second-order) 
time and spatial derivatives of $\delta H_{AB}$, $\delta \lambda$, 
$\delta \alpha$, $\beta^A$ and $\xi_t, \xi_A$, 
which we need to translate into (2+1)+1 language.
In particular, one might worry about the components of the Killing
vector $\xi$ because they are not in $\N$. Fortunately, they only
appear in the following combinations:
\begin{eqnarray}
  2 r^{-2} (\xi_{[r,t]} + r^{-1} \xi_t) &=& E^r \,,\\
  2 r^{-2} \xi_{[z,t]} &=& E^z \,,\\
  2 r^{-2} (\xi_{[r,z]} + r^{-1} \xi_z) &=& B^\varphi \,,
\end{eqnarray}
so that we recover the twist variables, which are in $\N$ (we have used
definitions \eqref{eq:twist} and (\ref{eq:E}--\ref{eq:B})).
This was to be expected, of course, because $\Psi_0$ is a spacetime
scalar and as such is in $\N$.
The remaining time derivatives in the expressions for the NP scalars 
are eliminated using the linearized evolution
equations (section \ref{sec:Z211lin}). The spatial derivatives of 
$\delta H_{AB}, \delta \lambda, \delta \alpha$ and $\beta^A$
are substituted using the definitions of the first-order variables
(\ref{eq:Ddef}--\ref{eq:Bdef}).\footnote{The ordering of the second
  spatial derivatives must be chosen carefully so that the results
  below obtain.}
Everything is expressed in terms of regularized variables; 
e.g., one should note that
\begin{equation}
  \delta \lambda = r s + \half \delta H_{rr} 
\end{equation}
in linearized theory.

For the $r = r_\mathrm{max}$ boundary, we find
\begin{eqnarray}
  \label{eq:Psi0rRe}
  \mathrm{Re} \Psi_0 &=& \half \{ - \partial_r [ r Y + \chi_{rr} -
    \chi_{zz} - r^2 \tilde s_r - D_{rrr} + D_{rzz} ]  \nonumber\\&& 
    \quad + \partial_z [-\chi_{rz} + 2 r \tilde D_{rrz} - D_{zrr} ]\\
    && \quad + r^{-1} D_{rrr} + 2 r^{-1} D_{zrz} - 2 Y 
    + 4 r \tilde s_r + 6 r^{-1} s \} \,,\nonumber\\
  \label{eq:Psi0rIm}
  \mathrm{Im} \Psi_0 &=& \textstyle\frac{r}{4} \{ 
    2 \partial_r [E^z - B^\varphi] - \partial_z E^r
    - 6 r^{-1} B^\varphi + 3 r^{-1} E^z \} \,.
\end{eqnarray}

Thus the imaginary part of $\Psi_0$ involves only the twist variables and
the real part only the remaining variables. Recall from chapter \ref{sec:lin}
that these two subsystems correspond to the two polarization states of
the gravitational field so that we obtain one separate boundary
condition for each polarization, as desired.
Note also that (\ref{eq:Psi0rRe}--\ref{eq:Psi0rIm}) are manifestly 
gauge-independent (they do not contain the lapse, shift or derivatives
thereof).

Next observe that the $r$-fluxes in (\ref{eq:Psi0rRe}--\ref{eq:Psi0rIm}) 
are proportional to the incoming modes $l_{1,1}^-$ \eqref{eq:l11mr} 
and $l_{1,2}^-$ \eqref{eq:l12mr}! 
Hence we can express the boundary conditions $\Psi_0 \doteq 0$ as 
conditions on the normal derivatives of two of the incoming modes:
\begin{eqnarray}
  \label{eq:radbc1r}
  \partial_r \, l_{1,1}^- &\doteq& \partial_z [ - \chi_{rz} + 2 r \tilde
    D_{rrz} - D_{zrr} + 2 B_r{}^z ] \nonumber\\&&
    + r^{-1} D_{rrr} + 2 r^{-1} D_{zrz} - Y + 4 r \tilde s_r 
    + 6 r^{-1} s \,,\\
  \label{eq:radbc2r}
  \partial_r \, l_{1,2}^- &\doteq& \half \partial_z E^r + 3 r^{-1} B^\varphi 
    - \textstyle\frac{3}{2} r^{-1} E^z \,.
\end{eqnarray}

Similarly, we find for the $z = z_\mathrm{max}$ boundary
\begin{eqnarray}
  \label{eq:Psi0zRe}
  \mathrm{Re} \Psi_0 &=& \half \{ -\partial_z [ r(Y - s_z)]
    + \partial_r [r^{-1} (-\chi_{rz} - D_{rzz} + 2 D_{zrz})] \}\,,\\
  \label{eq:Psi0zIm}
  \mathrm{Im} \Psi_0 &=& \textstyle\frac{r}{4} \{
    2 \partial_z [E^r + B^\varphi] - \partial_r E^z \} \,.
\end{eqnarray}
We identify $l_{1,1}^-$ \eqref{eq:l11mz} and $l_{1,2}^-$
\eqref{eq:l12mz} in the $z$-fluxes so that $\Psi_0 \doteq 0$ can be written as
\begin{eqnarray}
  \label{eq:radbc1z}
  \partial_z \tilde l_{1,1}^- &\doteq& 
    \partial_r[ r^{-1} (-\chi_{rz} - D_{rzz} + 2 D_{zrz} + 2 B_z{}^r)] \,,\\
  \label{eq:radbc2z}
  \partial_z \, l_{1,2}^- &\doteq& \half \partial_r E^z \,,
\end{eqnarray}
where $\tilde l_{1,1}^-$ was defined in \eqref{eq:l11tilde}, and here
we see very clearly the reason for that definition: written in the form
\eqref{eq:radbc1z}, the boundary condition is regular at $r = 0$
because of table \ref{tab:consrpar}.

We have checked that the residuals of the boundary conditions
(\ref{eq:radbc1r}--\ref{eq:radbc2r}) 
are indeed of the order $r^{-5}$ when evaluated for the
exact linearized solutions of chapter \ref{sec:lin}.
However, this fall-off is not uniform in $z$: the leading-order
coefficient of the expansion of the residual in inverse powers of $r$ has a 
$z$-dependence of $\sim z^4$ or lower, depending on which solution we choose. 
A similar statement with $r$ and $z$ interchanged 
holds for equations (\ref{eq:radbc1z}--\ref{eq:radbc2z}).
This means that if the size of the grid is doubled both in the $r$ and 
in the $z$ direction, the supremum of the residual evaluated along the 
entire boundary will not decrease by a factor of $2^5 = 32$ but only
$2$ (in the worst case). 

In order to improve on this, one could consider an NP tetrad that
points in the $R$-direction, where $R = \sqrt{r^2 + z^2}$, for this
coordinate is large everywhere on the outer boundary. The residual then
falls off like $R^{-5}$. However, in this case we cannot translate the
boundary conditions into a prescription for the normal derivatives of
the incoming modes as done above.


   \section{Constraint-preserving boundary conditions}
\label{sec:cpbcs}

Further boundary conditions can be obtained by requiring that no
violations of the constraints enter the computational domain from the
outside. The basic strategy for deriving such \emph{constraint-preserving 
boundary conditions} was first developed by Stewart \cite{Stewart98},
and there has been much recent work on this subject 
(e.g., \cite{Calabrese02, Calabrese03, Bona05}).

In the Z4 formalism, the standard Einstein or ADM constraints are replaced
with the algebraic constraints
\begin{equation}
  Z_\alpha = 0 \,.
\end{equation}
If those hold at all times, the Einstein constraints are automatically
satisfied by virtue of the evolution equations for the $Z$ vector 
(\ref{eq:d0theta}--\ref{eq:d0Zphi}).
In order to set up constraint-preserving boundary conditions, we need
to understand how the constraints propagate. By applying the
contracted Bianchi identities 
\begin{equation}
  \nabla_\beta G^{\alpha\beta} = 0
\end{equation}
to the Z4-Einstein equations (\ref{eq:Z4}), we obtain a homogeneous
wave equation for the $Z$ vector:
\begin{equation}
   \label{eq:BoxZ}
   \nabla_\beta \nabla^\beta Z_\alpha + R_{\alpha\beta} Z^\beta = 0 \,.
\end{equation}
This equation forms the \emph{constraint propagation} or 
\emph{subsidiary system}.

The (2+1)+1 reduction of \eqref{eq:BoxZ} can be obtained by simply 
taking a second time derivative of the evolution equations for the 
$Z$ vector (\ref{eq:d0theta}--\ref{eq:d0Zphi}).
As we would like to write the resulting system in first-order form, we
have to introduce new variables for the first-order space and time
derivatives of the $Z$ vector,
\begin{eqnarray}
  \theta_A \equiv \partial_A \theta\,, \quad
  &Z_{BA} \equiv \partial_B Z_A \,, \quad
  &Z_A{}^\varphi \equiv \partial_A Z^\varphi \,, \\
  \label{eq:Ztimedervs}
  \theta_0 \equiv  \eth_t \theta \,, \quad
  &Z_{0A} \equiv  \eth_t Z_A \,, \quad
  &Z_0{}^\varphi \equiv  \eth_t Z^\varphi \,,
\end{eqnarray}
where 
\begin{equation}
  \eth_t \equiv \alpha^{-1} (\partial_t - \beta^A \partial_A) \,.
\end{equation}
We also need the \emph{ordering constraints}
\begin{eqnarray}
  \mathcal{D}_{ABCD} &\equiv& 2 \partial_{[A} D_{B]CD} \,,\\
  \mathcal{L}_{AB} &\equiv& 2 \partial_{[A} L_{B]} \,,\\
  \mathcal{A}_{AB} &\equiv& 2 \partial_{[A} A_{B]} \,,\\
  \mathcal{B}_{AB}{}^C &\equiv& 2 \partial_{[A} B_{B]}{}^C \,,
\end{eqnarray}
which vanish because of the definitions of the first-order variables 
(\ref{eq:Ddef}--\ref{eq:Bdef}).

The subsidiary system now takes the form
\begin{eqnarray}
  \label{eq:substheta}
  \eth_t \theta_0 -\partial_B \theta^B &\simeq& 
    -2 \partial_B \mathcal{B}^B{}_A{}^A \,,\\
  \label{eq:subsZA}
  \eth_t Z_{0A} - \partial_B Z^B{}_A &\simeq&
    \half \partial_B \left( \mathcal{A}_A{}^B + \mathcal{L}_A{}^B 
    + \mathcal{D}_A{}^{BC}{}_C -2\mathcal{D}_C{}^{BC}{}_A \right)\,,\\
  \label{eq:subsZphi}
  \eth_t Z_0{}^\varphi - \partial_B Z^{B\varphi} 
    &\simeq& 0 \,,
\end{eqnarray}    
to principal parts ($\simeq$).
We clearly recognize the wave operator of equation \eqref{eq:BoxZ} on the
left-hand-sides of (\ref{eq:substheta}--\ref{eq:subsZphi}).
The ordering constraints on the right-hand-sides of
(\ref{eq:substheta}--\ref{eq:subsZA}) are not present in \eqref{eq:BoxZ};
they are a consequence of the first-order reduction we used to derive
the Z(2+1)+1 system. Analytically of course, the ordering constraints
vanish, but this may not obtain numerically and so we have to include
them in the subsidiary system. Fortunately, the ordering constraints
propagate along the normal lines:
\begin{equation}
  \label{eq:subsorder}
  \eth_t \mathcal{D}_{ABCD} = \eth_t \mathcal{L}_{AB} 
   = \eth_t \mathcal{A}_{AB} = \eth_t \mathcal{B}_{AB}{}^C = 0 \,.
\end{equation}

Constraint-preserving boundary conditions are obtained by requiring
the incoming modes of the subsidiary system to vanish at the boundary.
If $\mu_A$ is the unit outward-pointing normal to the boundary and
$\n$ denotes contraction with $\mu_A$ (cf. section \ref{sec:hyperbolicity}), 
the conditions read
\begin{eqnarray}
  \theta_0 + \theta_\n - 2 \mathcal{B}_{\n A}{}^A \doteq 0 \,,\\
  Z_{0A} + Z_{\n A} + \half (\mathcal{A}_{A\n} + \mathcal{L}_{A\n}
    + \mathcal{D}_{A\n C}{}^C - 2 \mathcal{D}_{C\n}{}^C{}_A ) \doteq 0 \,,\\
  Z_0{}^\varphi + Z_\n{}^\varphi \doteq 0 \,.
\end{eqnarray}  

We now express these conditions in terms of regularized Z(2+1)+1
variables and in linearized theory. The time derivatives of the $Z$
vector \eqref{eq:Ztimedervs} are substituted using the evolution
equations (\ref{eq:d0theta}--\ref{eq:d0Zphi}). 
The boundary conditions at $r = r_\mathrm{max}$ can be written as
{\allowdisplaybreaks\begin{eqnarray}
  \label{eq:cpbc3r}
  \partial_r l_{1,3}^- &=& \partial_r [\theta - 2 B_z{}^z - D_{rzz} -
    r^2 \tilde s_r - D_{rrr} + D_{zrz} + Z_r] \nonumber\\
    &\doteq& \partial_z [-2 B_r{}^z{} - r \tilde D_{rrz} + 2 D_{zrr} 
    + r s_z - Z_z]
    \\&& + r^{-1} (D_{rrr} + D_{rzz} - 3 D_{zrz} - Z_r + 6 s) 
    + 4 r \tilde s_r \,,\nonumber\\
  \label{eq:cpbc4r}
  \partial_r l_{1,4}^- &=& \partial_r [r Y + \chi_{rr} + \chi_{zz} -
    \theta - D_{zrz} - Z_r]\nonumber\\
    &\doteq& \partial_z [\chi_{rz} - r \tilde D_{rrz}] - Y 
    - r^{-1} D_{zrz} \,,\\
  \label{eq:cpbc5r}
  \partial_r l_{1,5}^- &=& \partial_r [\chi_{rz} - \half (A_z + 2
    D_{zrr} - D_{zzz} + r s_z - 2 Z_z)] \nonumber\\
    &\doteq& \partial_z [ 2 \chi_{rr} + r Y - \theta - \half(A_r - D_{rzz} +
    2 D_{rrr} + r^2 \tilde s_r) ] \\&&
    - r^{-1} \chi_{rz} - s_z \,,\nonumber\\
  \label{eq:cpbc6r}
  \partial_r l_{1,6}^- &=& \partial_r [ E^r + 2 Z^\varphi]
    \doteq - \partial_z E^z - 3 r^{-1} E^r \,. 
\end{eqnarray}}
For the $z = z_\mathrm{max}$ boundary we find
{\allowdisplaybreaks\begin{eqnarray}
  \label{eq:cpbc3z}
  \partial_z l_{1,3}^- &=& \partial_z [\theta - 2 r \tilde
    B_r{}^r - 2 D_{zrr} - r s_z + r \tilde D_{rrz} + Z_z] \nonumber\\
    &\doteq& \partial_r [-2 B_z{}^r + D_{rrr} + D_{rzz} - D_{zrz} - Z_r 
    + r^2 \tilde s_r] \\&&
    + r^{-1} (B_z{}^r + D_{rrr} + D_{rzz} - 3 D_{zrz} - Z_r + 6 s) 
    + 4 r \tilde s_r \,,\nonumber\\
  \label{eq:cpbc4z}
  \partial_z l_{1,4}^- &=& \partial_z [r Y + 2 \chi_{rr} - \theta
    - r \tilde D_{rrz} - Z_z] \nonumber\\
    &\doteq& \partial_r [ \chi_{rz} - D_{zrz}] + r^{-1} (\chi_{rz} +
    D_{zrz}) \,,\\
  \label{eq:cpbc5z}
  \partial_z l_{1,5}^- &=& \partial_z [\chi_{rz} - \half (A_r +
    D_{rzz} + r^2 \tilde s_r - 2 Z_r)] \nonumber\\
    &\doteq& \partial_r [ r Y + \chi_{rr} + \chi_{zz} - \theta 
    - \half (A_z + D_{zzz} + r s_z) ] \\&& + Y + s_z \,,\nonumber\\
  \label{eq:cpbc6z}
  \partial_z l_{1,6}^- &=& \partial_z [E^z + 2 Z^\varphi] 
    \doteq - \partial_r E^r - 3 r^{-1} E^r \,.
\end{eqnarray}}

As in section \ref{sec:radbcs}, these boundary conditions are again 
prescriptions for the normal derivatives of the incoming modes.
The conditions at the $z = z_\mathrm{max}$ boundary are regular at 
$r = 0$ provided that the on-axis conditions hold 
(table \ref{tab:consrpar}).
We have verified that equations (\ref{eq:cpbc3r}--\ref{eq:cpbc6r})
and (\ref{eq:cpbc3z}--\ref{eq:cpbc6z}) are satisfied identically by
the exact solutions of chapter \ref{sec:lin}. This is as it should be
because the constraints vanish for an exact solution and so do the
incoming modes of the subsidiary system.


   \section{Gauge boundary conditions and summary}
\label{sec:gaugebcs}

To complete the boundary conditions derived so far in sections
\ref{sec:radbcs} and \ref{sec:cpbcs}, we have to prescribe boundary
conditions for the gauge variables $\alpha$ and $\beta^A$.
Because we are free to specify the gauge in any way we like, this
procedure is essentially arbitrary. 
The simplest choice would be absorbing boundary conditions for the
gauge modes,
\begin{eqnarray}
  \label{eq:lfabs}
  l_f^- \doteq 0 \,,\\
  \label{eq:lmudabs}
  l_\mu^- \doteq l_d^- \doteq 0 \,.
\end{eqnarray}
Provided that the gauge parameters are chosen to be those of harmonic gauge 
($f = \mu = d = a = 1, m = 2$), equation \eqref{eq:lfabs} is satisfied
identically by the exact solutions of chapter \ref{sec:lin}, 
but equations \eqref{eq:lmudabs} only hold to leading order in 
$r^{-1}$ ($z^{-1}$).

Can we construct gauge boundary conditions that are all satisfied
identically by the exact solutions? 
In harmonic gauge, the lapse $\alpha$ and
the components of the shift $\beta^A$ each satisfy a wave equation to
principal parts: in linearized theory,
\begin{eqnarray}
  \label{eq:boxalpha}
  \partial_t^2 \alpha - \partial_B \partial^B \alpha &\simeq& 0 \,,\\
  \label{eq:boxbeta}
  \partial_t^2 \beta^A - \partial_B \partial^B \beta^A &\simeq& 0 \,.
\end{eqnarray}
This is clear from the harmonic gauge condition (\ref{eq:harmgauge}) 
but can also be verified directly from the
linearized Z(2+1)+1 equations (section \ref{sec:Z211lin}).
It is important to note that harmonic gauge is the only choice of
gauge parameters for which one obtains a \emph{closed} evolution system for
the gauge variables.

We can now construct boundary conditions for
(\ref{eq:boxalpha}--\ref{eq:boxbeta}) in a similar way as we did
for the constraint evolution system (\ref{eq:BoxZ}), which 
also formed a wave equation.
In order that the waves in the gauge variables leave the computational
domain without causing reflections, it is reasonable to set the 
incoming modes of (\ref{eq:boxalpha}--\ref{eq:boxbeta}) to zero 
at the outer boundaries, i.e.
\begin{eqnarray}
  \partial_t \alpha + \partial_\n \alpha &\doteq& 0 \,,\\
  \partial_t \beta^A + \partial_\n \beta^A &\doteq& 0 \,,
\end{eqnarray}
where as usual $\n$ denotes a derivative normal to the boundary.

Substituting the time derivatives using the evolution equations
(\ref{eq:lind0alpha}--\ref{eq:lind0betaz}) we obtain in the $r$-direction
\begin{eqnarray}
  \label{eq:gaugebc01r}
  0 &\doteq& A_r - 2 \chi_{rr} - \chi_{zz} - r Y + 2 \theta \nonumber\\
    &=& l_f^- \,,\\
  \label{eq:gaugebc02r}
  0 &\doteq& 2 r \tilde B_r{}^r + r^{-1} \beta^r - A_r - D_{rzz} + 2 D_{zrz} -
    r^2 \tilde s_r - 4 s + 2 Z_r \nonumber\\
    &=& -l_d^- - l_f^- + 2 l_{1,3}^- + 2 l_{0,7} + r^{-1} \beta^r - 4 s \,,\\
  \label{eq:gaugebc03r}
  0 &\doteq& 2 B_r{}^z - A_z + 2 r \tilde D_{rrz} - 2 D_{zrr} + D_{zzz} 
    + 2 r^{-1} H_{rz}  \nonumber\\
    &&- r s_z + 2 Z_z \\
    &=& -l_\mu^- - 2 l_{0,6} + 2 r^{-1} H_{rz} \,.\nonumber
\end{eqnarray}
Equation \eqref{eq:gaugebc01r} is simply an absorbing boundary 
condition for $l_f^-$.
However, (\ref{eq:gaugebc02r}) is more problematic because it
specifies $l_d^-$ in terms of $l_{1,3}^-$, of which only the
normal derivative is known if we 
impose constraint-preserving boundary conditions (\ref{eq:cpbc3r}).
What happens though if we take a time derivative of
(\ref{eq:gaugebc01r}--\ref{eq:gaugebc03r})? 
Equation \eqref{eq:gaugebc01r} becomes
\begin{equation}
  \label{eq:gaugebc1r}
  \partial_r l_f^- = - \partial_z A_z - r^{-1} A_r \,.
\end{equation}
As for (\ref{eq:gaugebc02r}), terms $\partial_r l_f^-$ and
$\partial_r l_{1,3}^-$ will appear (recall that $l_f^-$ and
$l_{1,3}^-$ are eigenfields in the $r$-direction both with speed 1 in the
harmonic case). These can be eliminated using 
\eqref{eq:gaugebc1r} and \eqref{eq:cpbc3r}. The result is
\begin{eqnarray}
  \label{eq:gaugebc2r}
  \partial_r l_d^- &\doteq& \partial_z [ A_z + 2 (-B_r{}^z + B_z{}^r 
    - r \tilde D_{rrz} + 2 D_{zrr} + r s_z - Z_z )] \nonumber\\&&
    + r^{-1} [ A_r + 2(2 r \tilde B_r{}^r + D_{rrr} + D_{rzz} - 3 D_{zrz} -
    Z_r \\&&\qquad + 2 r^2 \tilde s_r + 4s)] \,.\nonumber
\end{eqnarray}
Thus we have managed to obtain a prescription for the normal
derivative of $l_d^-$ in terms of tangential derivatives and source terms.
The time derivative of \eqref{eq:gaugebc03r} becomes
\begin{equation}
  \label{eq:gaugebc3r}
  \partial_r l_\mu^- = \partial_z [2(B_z{}^z - r \tilde B_r{}^r)]
    + 4 \tilde D_{rrz} + 2 r^{-1} (B_r{}^z - B_z{}^r) \,.
\end{equation}

Applying the same procedure in the $z$-direction yields
\begin{eqnarray}
  \label{eq:gaugebc1z}
  \partial_z l_f^- &\doteq& -\partial_r A_r - r^{-1} A_r \,,\\
  \label{eq:gaugebc2z}
  \partial_z l_d^- &\doteq& \partial_r[A_r + 2 (B_r{}^z + D_{rrr} + D_{rzz}
    - D_{zrz} + r^2 \tilde s_r - Z_r)] \nonumber\\&&
    + r^{-1} [A_r + 2 ( B_r{}^z + 2 D_{zrz} + D_{rrr} + D_{rzz} - 3
  D_{zrz} \\&&\qquad + 4 r^2 \tilde s_r + 6 s - Z_r)] \,,\nonumber\\
  \label{eq:gaugebc3z}
  \partial_z l_\mu^- &\doteq& \partial_z[ 2(r \tilde B_r{}^r - B_z{}^z)]
    + 4 (\tilde B_r{}^r - s_z) \,.
\end{eqnarray}
These equations are regular on axis because of table \ref{tab:consrpar}.

As expected, the boundary conditions (\ref{eq:gaugebc1r}--\ref{eq:gaugebc3r})
and (\ref{eq:gaugebc1z}--\ref{eq:gaugebc3z}) are now satisfied
identically by the exact solutions.

\paragraph{Summary.}
Using information about the Newman-Penrose scalar $\Psi_0$ (section
\ref{sec:radbcs}), the constraint propagation system (section
\ref{sec:cpbcs}) and the gauge propagation system (section
\ref{sec:gaugebcs}), we have obtained a total of 9 boundary conditions 
for each outer boundary. 
The conditions all specify the normal derivatives of the incoming
modes in terms of tangential derivatives and source terms.

We have checked that the boundary conditions are satisfied
identically by the exact linearized solutions of chapter \ref{sec:lin},
except for the two outgoing-radiation boundary conditions, whose
residuals are of the optimal order of $r^{-5}$ ($z^{-5}$).

Since the linearized Z(2+1)+1 system with dynamical shift 
has 9 incoming modes, we have a complete set of boundary conditions
specifying all the incoming modes of the system.
In the vanishing shift case, one deletes the boundary conditions for
$l_d^-$ and $l_\mu^-$ and obtains a total of 7 conditions, which is
again the required number.

It is a fortunate coincidence of axisymmetry that the numbers work out 
in such a convenient way. In the case without symmetries, there are
more characteristic variables and hence more boundary conditions are required.
For instance, there are 6 additional pairs of 
light cone modes \cite{Bona03a}.
This problem can be addressed by considering instead of $\Psi_0$ the
evolution system of the electric and magnetic parts of the Weyl tensor, 
which leads to the right number of boundary conditions for the 
light cone modes \cite{Kidder05}.


   \section{Fourier-Laplace analysis}
\label{sec:fourierlaplace}

Having derived a set of boundary conditions in sections
\ref{sec:radbcs}--\ref{sec:gaugebcs}, the question arises whether
the associated initial boundary value problem is well-posed.
Because the boundary conditions are not of the dissipative type 
(\ref{eq:dissbcs}), the standard theorems mentioned in section
\ref{sec:IBVP} do not apply. 

We now prove a necessary condition for well-posedness of the IBVP in
the high-frequency limit using the Fourier-Laplace technique \cite{GKO}. 
This limit is also known as the WKB approximation, or geometrical optics.
It implies that we may neglect the source terms in the linearized 
Z(2+1)+1 equations against the flux terms.
Harmonic slicing is used ($f = 1, m = 2$), and we consider first the
case of a vanishing shift vector.

To begin with, it is convenient to rewrite the linearized Z(2+1)+1 equations
(section \ref{sec:Z211lin}) in characteristic space, i.e., 
as evolution equations for the
characteristic variables. This can easily be done with the help of a
computer algebra programme using the transformation between conserved
and characteristic variables (section \ref{sec:linchar}).
The characteristic form of the equations has the advantage that it is
the same both for characteristic variables in the $r$-direction and in
the $z$-direction (this has been checked explicitly). Hence we use
general indices $\n$ and $\p$, where either $(x^\n, x^\p) = (r,z)$
or $(x^\n, x^\p) = (z, r)$.
The coordinate normal to the boundary under consideration is $x^\n$
and the one parallel to it is $x^\p$.  
The characteristic variables refer to the $x^\n$-direction.
Neglecting the source terms, we find

{\allowdisplaybreaks \begin{eqnarray}
  \label{eq:charevolfirst0}
  \partial_t l_{0,1} &=& -\half \partial_\p 
    (l_{1,3}^- + l_{1,3}^+ - l_{1,4}^- - l_{1,4}^+ - l_f^- + l_f^+) \,,\\
  \partial_t l_{0,2} &=& -\half \partial_\p 
    (l_{1,5}^- + l_{1,5}^+) \,,\\
  \partial_t l_{0,3} &=& -\quarter \partial_\p 
    (-l_{1,1}^- - l_{1,1}^+ + l_{1,3}^- + l_{1,3}^+ + l_{1,4}^- + l_{1,4}^+) 
    \,,\\
  \partial_t l_{0,4} &=& -\half \partial_\p 
    (l_{1,1}^- + l_{1,1}^+) \,,\\
  \partial_t l_{0,5} &=& -\half \partial_\p 
    (-l_f^- + l_f^+) \,,\\
  \partial_t l_{0,6} &=& 0 \,,\\
  \label{eq:charevollast0}
  \partial_t l_{0,7} &=& 0 \,,\\
  \label{eq:charevolfirst1}
  \partial_t l_{1,1}^\pm &=& \mp \partial_\n l_{1,1}^\pm 
    - \partial_\p (l_{0,4} - l_{0,7}) \,,\\
  \partial_t l_{1,2}^\pm &=& \mp \partial_\n l_{1,2}^\pm
    \pm \partial_\p l_{1,6}^\mp \,,\\
  \partial_t l_{1,3}^\pm &=& \mp \partial_\n l_{1,3}^\pm
    - \half \partial_\p (l_{0,1} + 2 l_{0,3} + l_{0,4} - l_{0,5} +
    l_{0,7}) \,,\\
  \partial_t l_{1,4}^\pm &=& \mp \partial_\n l_{1,4}^\pm
    - \half \partial_\p (-l_{0,1} + 2 l_{0,3} + l_{0,4} + l_{0,5} +
    l_{0,7} )\,,\\
  \partial_t l_{1,5}^\pm &=& \mp \partial_\n l_{1,5}^\pm
    - \half \partial_\p (2 l_{0,2} + l_{0,6})\,,\\
  \partial_t l_{1,6}^\pm &=& \mp \partial_\n l_{1,6}^\pm 
    \mp \partial_\p l_{1,2}^\mp \,,\\
  \label{eq:charevollast1}
  \partial_t l_f^\pm &=& \mp \partial_\n l_f^\pm
    \mp \partial_\p l_{0,5} \,.
\end{eqnarray}}

We can solve these equations by means of a Laplace transformation in
time and a Fourier transformation in the $x^\p$ direction. That is, we
write the solution as a superposition of modes of the form
\begin{equation}
  \vec u (t, x^\n, x^\p) = 
  \vec {\hat u} (x^\n) \e^{s t + i \omega x^\p} \,,
\end{equation}
where $s \in \mathbb{C}$, $\mathrm{Re}(s) > 0$.
Substituting this into (\ref{eq:charevolfirst0}--\ref{eq:charevollast1}),
we obtain a set of ordinary differential equations in the coordinate $x^\n$
coupled to algebraic conditions. 
For simplicity we set
\begin{equation}
  \label{eq:fourlap}
  \xi \equiv \omega x^\n \,, \qquad \zeta \equiv \frac{s}{\omega} \,,
\end{equation}
(we may assume $\omega > 0$ because we are only interested in high
frequencies) and we leave out the hats on the transformed variables.

Substituting \eqref{eq:fourlap} into 
(\ref{eq:charevolfirst0}--\ref{eq:charevollast0}) yields the
following algebraic conditions for the zero-speed modes:
{\allowdisplaybreaks \begin{eqnarray}
  \label{eq:algfirst}
  l_{0,1} &=& -\tfrac{i}{2\zeta} (l_{1,3}^- + l_{1,3}^+ -
    l_{1,4}^- - l_{1,4}^+ - l_f^- + l_f^+) \,,\\ 
  l_{0,2} &=& -\tfrac{i}{2\zeta}(l_{1,5}^- + l_{1,5}^+) \,,\\
  l_{0,3} &=& -\tfrac{i}{4\zeta}(-l_{1,1}^- - l_{1,1}^+ +
    l_{1,3}^- + l_{1,3}^+ + l_{1,4}^- + l_{1,4}^+) \,,\\
  l_{0,4} &=& -\tfrac{i}{2\zeta} (l_{1,1}^- + l_{1,1}^+) \,,\\
  l_{0,5} &=& -\tfrac{i}{2\zeta} (-l_f^- + l_f^+) \,,\\
  l_{0,6} &=& 0 \,,\\
  \label{eq:alglast}
  l_{0,7} &=& 0 \,.
\end{eqnarray}}
These can now be used to eliminate the zero-speed modes in the Fourier-Laplace 
transform of (\ref{eq:charevolfirst1}--\ref{eq:charevollast1}):
{\allowdisplaybreaks \begin{eqnarray}
  \partial_\xi l_{1,1}^\pm 
     &=& \mp \zeta l_{1,1}^\pm \mp i (l_{0,4} - l_{0,7}) \nonumber\\
     &=& \mp \zeta l_{1,1}^\pm \mp \tfrac{1}{2\zeta}
       (l_{1,1}^- + l_{1,1}^+) \,,\\
  \partial_\xi l_{1,2}^\pm    
     &=& \mp \zeta l_{1,2}^\pm + i l_{1,6}^\mp \,,\\
  \partial_\xi l_{1,3}^\pm
     &=& \mp \zeta l_{1,3}^\pm \mp \tfrac{i}{2} (l_{0,1} + 2
       l_{0,3} + l_{0,4} - l_{0,5} + l_{0,7}) \nonumber\\
     &=& \mp \zeta l_{1,3}^\pm \mp \tfrac{1}{2\zeta} 
       (l_{1,3}^- + l_{1,3}^+) \,,\\
  \partial_\xi l_{1,4}^\pm
     &=& \mp \zeta l_{1,4}^\pm \mp \tfrac{i}{2} (-l_{0,1} + 2
       l_{0,3} + l_{0,4} + l_{0,5} + l_{0,7}) \nonumber\\
     &=& \mp \zeta l_{1,4}^\pm \mp \tfrac{1}{2\zeta} 
       (l_{1,4}^- + l_{1,4}^+) \,,\\
  \partial_\xi l_{1,5}^\pm
     &=& \mp \zeta l_{1,5}^\pm \mp \tfrac{i}{2} (2 l_{0,2}
       + l_{0,6}) \nonumber\\
     &=& \mp \zeta l_{1,5}^\pm \mp \tfrac{1}{2\zeta} 
       (l_{1,5}^- + l_{1,5}^+) \,,\\
  \partial_\xi l_{1,6}^\pm
     &=& \mp \zeta l_{1,6}^\pm - i l_{1,2}^\mp \,,\\
  \partial_\xi l_f^\pm 
     &=& \mp \zeta l_f^\pm - i l_{0,5} \nonumber\\
     &=& \mp \zeta l_f^\pm - \tfrac{1}{2\zeta} 
       (-l_f^- + l_f^+) \,.
\end{eqnarray}}

This system of ODEs decouples into the following $2\times 2$ blocks:
{\allowdisplaybreaks\begin{eqnarray}
  \label{eq:ODEblockfirst}
  \partial_\xi (l_{1,1}^-, l_{1,1}^+)^T &=& M_1 (l_{1,1}^-, l_{1,1}^+)^T \,,\\
  \partial_\xi (l_{1,3}^-, l_{1,3}^+)^T &=& M_1 (l_{1,3}^-, l_{1,3}^+)^T \,,\\
  \partial_\xi (l_{1,4}^-, l_{1,4}^+)^T &=& M_1 (l_{1,4}^-, l_{1,4}^+)^T \,,\\
  \partial_\xi (l_{1,5}^-, l_{1,5}^+)^T &=& M_1 (l_{1,5}^-, l_{1,5}^+)^T \,,\\
  \partial_\xi (l_f^-, l_f^+)^T &=& M_2 (l_f^-, l_f^+)^T \,,\\
  \partial_\xi (l_{1,2}^-, l_{1,6}^+)^T &=& M_3 (l_{1,2}^-, l_{1,6}^+)^T \,,\\
  \label{eq:ODEblocklast}
  \partial_\xi (l_{1,2}^+, l_{1,6}^-)^T &=& M_4 (l_{1,2}^+, l_{1,6}^-)^T \,,
\end{eqnarray}}
where the matrices $M_i$ are given by
\begin{eqnarray}
  M_1 = \left( \begin{array}{cc} 
      \zeta + \tfrac{1}{2\zeta} & \tfrac{1}{2\zeta} \\
      -\tfrac{1}{2\zeta} & -\zeta - \tfrac{1}{2\zeta} 
      \end{array} \right) \,, &\,&
  M_2 = \left( \begin{array}{cc} 
      \zeta + \tfrac{1}{2\zeta} & -\tfrac{1}{2 \zeta} \\
      \tfrac{1}{2\zeta} & - \zeta - \tfrac{1}{2 \zeta}
      \end{array} \right) \,, \nonumber\\
  M_3 = \left( \begin{array}{cc} 
      \zeta & i \\ -i & - \zeta 
      \end{array} \right) \,, &\,&
  M_4 = \left( \begin{array}{cc} 
      -\zeta & i \\ -i & \zeta 
      \end{array} \right) \,.
\end{eqnarray}
Each matrix $M_i$ has eigenvalues $\pm \lambda$ where
\begin{equation}
  \label{eq:fltev}
  \lambda = \sqrt{1 + \zeta^2} \,,
\end{equation}
with the sign of the square root chosen such that
$\mathrm{Re}(\lambda) > 0$ for $\mathrm{Re}(\zeta) > 0$.
The corresponding solutions of the ODEs have a $\xi$-dependence of
$\exp(\pm \lambda \xi)$. We only admit solutions that are $L^2$
in the interior and so we have to exclude the $\exp(-\lambda \xi)$
solutions\footnote{It is more common in the
  literature to have the boundary at $\xi = 0$, with $\xi > 0$ being
  in the interior, in which case the opposite sign of $\lambda$ has to
  be chosen here.} 
because they blow up as $\xi \rightarrow -\infty$.
The (right) eigenvectors of the $M_i$ with eigenvalue $+ \lambda$ and
hence the admissible solutions of the ODEs are found to be
{\allowdisplaybreaks\begin{eqnarray}
  \label{eq:decsolnfirst}
  (l_{1,1}^-, l_{1,1}^+)^T &=& a_1 \e^{\lambda \xi} 
     (-1, (\zeta - \lambda)^2)^T \,,\\
  (l_{1,3}^-, l_{1,3}^+)^T &=& a_2 \e^{\lambda \xi} 
     (-1, (\zeta - \lambda)^2)^T \,,\\
  (l_{1,4}^-, l_{1,4}^+)^T &=& a_3 \e^{\lambda \xi} 
     (-1, (\zeta - \lambda)^2)^T \,,\\
  (l_{1,5}^-, l_{1,5}^+)^T &=& a_4 \e^{\lambda \xi} 
     (-1, (\zeta - \lambda)^2)^T \,,\\
  (l_f^-, l_f^+)^T &=& a_5 \e^{\lambda \xi} 
     (1, (\zeta - \lambda)^2)^T \,,\\
  (l_{1,2}^-, l_{1,6}^+)^T  &=& a_6 \e^{\lambda \xi} 
     (-i, \zeta - \lambda)^T \,,\\
  \label{eq:decsolnlast}
  (l_{1,2}^+, l_{1,6}^-)^T  &=& a_7 \e^{\lambda \xi} 
     (i, \zeta + \lambda)^T \,,
\end{eqnarray}}
and the general admissible solution is a superposition of these with
arbitrary complex constants $a_i$.

In order for the IBVP to be well-posed, the constants $a_i$ have to be 
uniquely determined by the boundary conditions, for all $\zeta \in
\mathbb{C}$ with $\mathrm{Re}(\zeta) > 0$. Otherwise, there exists a
nontrivial solution for some $\zeta$ with $\mathrm{Re}(\zeta) > 0$, 
which after reversing the Fourier-Laplace transformation takes the form
\begin{equation}
  \vec u (t, x^\n, x^\p) = \vec {\hat u} (x^\n) 
  \e^{\omega(\zeta t + i x^\p)} \,.
\end{equation}
Here $\omega$ can be arbitrarily large. Hence no estimate
of the form \eqref{eq:wellposedestimate2} holds, and the IBVP is ill-posed.

The boundary conditions we want to impose consist of 
the outgoing-radiation conditions 
((\ref{eq:radbc1r}--\ref{eq:radbc2r}) and 
(\ref{eq:radbc1z}--\ref{eq:radbc2z})),
the constraint-preserving conditions
((\ref{eq:cpbc3r}--\ref{eq:cpbc6r}) and (\ref{eq:cpbc3z}--\ref{eq:cpbc6z}))
and the gauge boundary condition for the lapse function
(\eqref{eq:gaugebc1r} and \eqref{eq:gaugebc1z}).
In characteristic space and after performing the Fourier-Laplace
transformation, these boundary conditions can be written as
\begin{eqnarray}
  \label{eq:fltbc11}
  \partial_\xi l_{1,1}^- 
     &\doteq& -\tfrac{i}{2} (2 l_{0,1} + 2 l_{0,7} + 3 l_{1,5}^- - l_{1,5}^+) 
     \nonumber\\
     &=& -\tfrac{1}{2\zeta}(l_{1,3}^- + l_{1,3}^+ - l_{1,4}^- -
     l_{1,4}^+ - l_f^- + l_f^+) \\&&
     - \tfrac{i}{2} (3 l_{1,5}^- - l_{1,5}^+) \,,\nonumber\\
  \label{eq:fltbc12}
  \partial_\xi l_{1,2}^- 
     &\doteq& \tfrac{i}{4} (l_{1,6}^+ + l_{1,6}^-) \,,\\
  \label{eq:fltbc13}
  \partial_\xi l_{1,3}^-
     &\doteq& -\tfrac{i}{2} (-l_{0,1} - 2 l_{0,3} - l_{0,4} + l_{0,5} 
     - l_{0,7}) \nonumber\\
     &=& \tfrac{1}{2\zeta}(l_{1,3}^- + l_{1,3}^+) \,,
\end{eqnarray}
\begin{eqnarray}
  \label{eq:fltbc14}
  \partial_\xi l_{1,4}^- 
     &\doteq& -\tfrac{i}{2} (-l_{0,7} - 2 l_{1,5}^-) \nonumber\\
     &=& i l_{1,5}^- \,,\\
  \label{eq:fltbc15}
  \partial_\xi l_{1,5}^- 
     &\doteq& -\tfrac{i}{2} (-l_{0,6} - l_{1,1}^- - l_{1,3}^- + l_{1,4}^- +
     2 l_f^-) \nonumber\\
     &=& -\tfrac{i}{2} (-l_{1,1}^- - l_{1,3}^- + l_{1,4}^- + 2 l_f^-) \,,\\
  \label{eq:fltbc16}
  \partial_\xi l_{1,6}^- 
     &\doteq& -\tfrac{i}{2} (l_{1,2}^- + l_{1,2}^+) \,,\\
  \label{eq:fltbcf}
  \partial_\xi l_f^- 
     &\doteq& -i l_{0,7} \nonumber\\
     &=& -\tfrac{1}{2 \zeta}(-l_f^- + l_f^+) \,,
\end{eqnarray}
where we have again used the algebraic conditions
(\ref{eq:algfirst}--\ref{eq:alglast}) to eliminate the incoming modes.
(It does not matter whether we choose $(\n, \p) = (r, z)$ or 
$(\n, \p) = (z, r)$, the result is the same.)
Inserting the superposition of (\ref{eq:decsolnfirst}--\ref{eq:decsolnlast})
into (\ref{eq:fltbc11}--\ref{eq:fltbcf}), we obtain the
following relations for the coefficients $a_i$.
Equation \eqref{eq:fltbc13} yields
\begin{equation}
  0 = - \lambda a_2 + \tfrac{1}{2 \zeta} [ 1 - (\zeta - \lambda)^2 ] a_2 
    = \zeta a_2 \,,    
\end{equation}
where we have used \eqref{eq:fltev}. Hence $a_2 = 0$ because
$\mathrm{Re}(\zeta) > 0$.
Similarly, \eqref{eq:fltbcf} becomes
\begin{equation}
  0 = \lambda a_5 + \tfrac{1}{2 \zeta} [- 1 + (\zeta - \lambda)^2 ] a_5 
    = -\zeta a_5  
\end{equation}
so that $a_5 = 0$.
Equation \eqref{eq:fltbc15} tells us that
\begin{equation}
  a_4 = -i \lambda a_3 \,.
\end{equation}
Inserting the results for $a_2, a_4$ and $a_5$ 
into equations \eqref{eq:fltbc11} and
\eqref{eq:fltbc15} leads to the following linear system for the
coefficients $a_1$ and $a_3$:
\begin{equation}
  \left( \begin{array}{cc}
      - \lambda & - \lambda + \zeta^2 (\zeta - \lambda) \\
      \tfrac{i}{2} & \tfrac{i}{2} (2 \zeta^2 + 1) \end{array} \right) 
  \left( \begin{array}{c} a_1 \\ a_3 \end{array} \right) = 0 \,.
\end{equation}
Its determinant is found to be 
\begin{equation}
  \label{eq:det1}
  D_1 (\zeta) = -\tfrac{i}{2} \zeta^2 (\zeta + \lambda) \,.
\end{equation}
$D_1(\zeta) \neq 0$ 
because $\mathrm{Re} (\zeta) > 0$ and $\mathrm{Re} (\lambda) > 0$.
Hence $a_1 = a_3 = 0$ is the only solution.
The twist subsystem (\eqref{eq:fltbc12} and \eqref{eq:fltbc16})
implies that 
\begin{equation}
  \left( \begin{array}{cc}
      \lambda + \tfrac{1}{4} (\zeta - \lambda) & \tfrac{1}{4} (\zeta +
      \lambda) \\ \half & \lambda (\zeta + \lambda) - \half
      \end{array} \right) \left( \begin{array}{c} a_6 \\ a_7
      \end{array} \right) = 0 \,.
\end{equation}
Its determinant is
\begin{equation}
  \label{eq:det2}
  D_2 (\zeta) = \tfrac{1}{4} [ \zeta (4 \zeta^2 + 3) + \lambda (4 \zeta^2 +
  1) ] \,.
\end{equation}
Let us multiply this with 
\begin{equation}
  \hat D_2 (\zeta) \equiv \tfrac{1}{4} [ \zeta (4 \zeta^2 + 3) 
  - \lambda (4 \zeta^2 + 1)] \,,
\end{equation}
obtaining 
\begin{equation}
  D_2(\zeta) \hat D_2 (\zeta) = \tfrac{1}{16} [
  \zeta^2 (4 \zeta^2 + 3)^2 - (1 + \zeta^2)(4 \zeta^2 + 1)^2 ]
  = - \tfrac{1}{16} \neq 0 \,.
\end{equation}
Hence also $D_2 (\zeta) \neq 0$ for all $\zeta \in \mathbb{C}$.

We conclude that the only solution that satisfies the
boundary conditions is the trivial one ($a_i = 0 \, \forall i$), 
and thus we have proven a necessary condition for well-posedness of
the IBVP in the high-frequency limit.

It should be stressed that the above \emph{determinant condition} is
necessary but not sufficient for well-posedness. If the boundary
conditions are algebraic, a stronger condition called the \emph{Kreiss
  condition} \cite{Kreiss70} is sufficient: 
the IBVP is well-posed if the determinant 
of the coefficient matrix is uniformly bounded away from zero for
$\mathrm{Re}(\zeta) \geqslant 0$. This condition is not satisfied in
our example, as equation \eqref{eq:det1} clearly shows. 
Even if the Kreiss condition is satisfied, this does not guarantee
well-posedness of the IBVP if the boundary conditions are differential
as in our case. A counter-example can be found in \cite{ReulaSarbach04}.
Furthermore, we have only investigated the limit of high-frequencies.
There could well be an instability for low frequencies where the
source terms may not be neglected.

In the above analysis, we have assumed that the shift vector
vanishes. The algebra in the dynamical shift case turns out to be
considerably more complicated -- the system of ODEs no longer decouples
into simple $2 \times 2$ blocks as in equations
(\ref{eq:ODEblockfirst}--\ref{eq:ODEblocklast}).
A REDUCE programme has been written to carry out the calculation in
this case. Of course, the twist subsystem is unchanged. 
Remarkably, the determinant of the coefficient matrix for the
remaining system turns out to be (up to a constant factor) again  
$D_1$ defined above \eqref{eq:det1}! 
Hence the same stability result applies in the dynamical shift case.


   \section{Numerical experiments}
\label{sec:bcnum}

In the preceding sections, we have developed three different sets of
boundary conditions:
\begin{itemize}
  \item absorbing boundary conditions (section \ref{sec:absbcs})
  \item dissipative boundary conditions with zero $Z$ vector (section
    \ref{sec:cpdissbcs})
  \item differential boundary conditions (the combination of sections
  \ref{sec:radbcs}, \ref{sec:cpbcs} and \ref{sec:gaugebcs})
\end{itemize}

We now perform some numerical experiments to assess the performance of
these boundary conditions in practice. 
The ideal boundary conditions would be numerically stable (i.e., no
``blow-up'' of the numerical solution occurs) and
they would minimize spurious reflections.
These originate when the wave arrives at the outer boundaries. The
reflections then propagate in and grow in amplitude as a consequence
of cylindrical polar coordinates (the time-reverse, an outgoing wave,
loses amplitude as it travels out in these coordinates).
The maximum of the reflections occurs when they reach the origin,
i.e., approximately two light-crossing times after the pulse is
emitted from the origin.

To measure the reflections, we use the exact linearized solutions of
chapter \ref{sec:lin} as a test problem and monitor the numerical error.
The three test problems we look at are
\begin{itemize}
  \item the even-parity non-twisting quadrupole solution (section
    \ref{sec:teukoeven}) evolved with vanishing shift
  \item the same evolved with dynamical shift
  \item the even-parity twisting octupole solution (section
    \ref{sec:octupole}; here the shift does not enter the linearized evolution
    equations)
\end{itemize}
For all the runs, the resolution is taken to be 32 points in both spatial
dimensions and the outer boundaries are placed at $r_\textrm{max} =
z_\textrm{max} = 5$. The mode functions are of the form
(\ref{eq:modefunctions}) with amplitudes $F_0 = G_0 = 10^{-4}$.
The outer boundaries are not very far out compared with the width of
the initial pulse ($\approx 1$) and the resolution is rather low, but
this suffices for our purposes here. 
The amplitude of the reflections decreases approximately linearly as
$r_\textrm{max}$ and $z_\textrm{max}$ are increased and it is nearly
independent of the resolution. The qualitative comparisons we draw
between the different boundary conditions below are robust under
changes of these parameters.

\subsection{Numerical method}

The interior grid points are evolved in the same way as in section
\ref{sec:linnum}. In particular, fourth-order numerical dissipation is
added at all interior points unless otherwise stated.
This implies that we need two layers of ghost cells at each boundary.

In all cases, the outgoing and zero-speed characteristic variables
$\vec v^+$ are computed from the conserved variables $\vec u$
at the outermost interior cells
and are linearly extrapolated to the ghost cells.

For the dissipative (absorbing and zero-$Z$) boundary conditions,
the incoming modes at the ghost cells are then set in terms of the
outgoing and zero-speed modes there using the dissipative boundary 
conditions (\eqref{eq:absbcs} or
(\ref{eq:zerozbcsfirst}--\ref{eq:zerozbcslast}), respectively).
Finally the ghost cells are transformed back to conserved variables.

The differential boundary conditions have the form (we write out the
$r$-direction, the $z$-direction follows by symmetry)
\begin{equation}
  \partial_r v^- = \partial_z f + s
\end{equation}
and are discretized as described in section \ref{sec:ghosts}, equation
\eqref{eq:modeldiffbcdiscr}. We have also implemented the second-order
discretization \eqref{eq:modeldiffbcdiscr2} but in the cases in which the
boundary conditions were unstable, the instability showed up earlier when
using that discretization.

\subsection{Numerical results}

\paragraph{The quadrupole solution with vanishing shift.}

Consider first the non-twisting quadrupole solution (section
\ref{sec:teukoeven}) evolved with vanishing shift vector.
Figure \ref{fig:bc1} shows the $L^2$ norm of the error for the 
variable $s$ and the constraint $\theta$ 
(the qualitative results are similar for the remaining variables). 

\begin{figure}[p]
  \centering
  \begin{minipage}[t]{0.49\textwidth}
    \includegraphics[scale = 1]{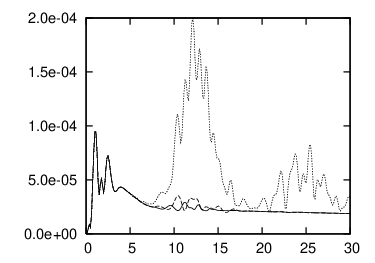}
    \centerline{$s$, dissipative}
  \end{minipage}
  \hfill  
  \begin{minipage}[t]{0.49\textwidth}
    \includegraphics[scale = 1]{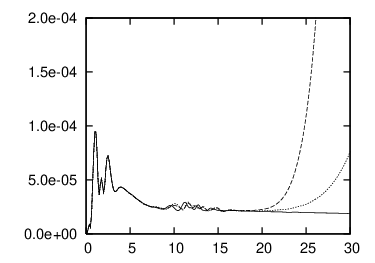}
    \centerline{$s$, differential}
  \end{minipage}
  \hfill  
  \begin{minipage}[t]{0.49\textwidth}
    \includegraphics[scale = 1]{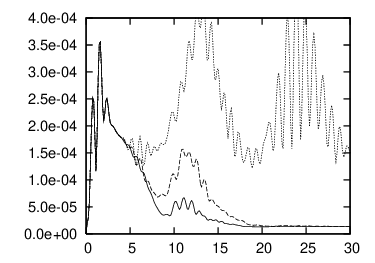}
    \centerline{$\theta$, dissipative}
  \end{minipage}
  \hfill  
  \begin{minipage}[t]{0.49\textwidth}
    \includegraphics[scale = 1]{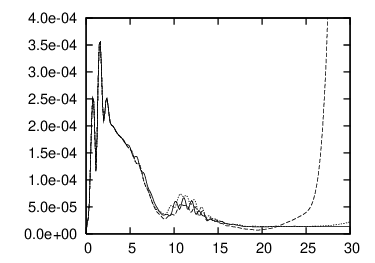}
    \centerline{$\theta$, differential}
  \end{minipage}
  \bigskip
  \caption[Test of boundary conditions: even quadrupole, vanishing shift] 
  {\label{fig:bc1} \footnotesize
  Test of boundary conditions for the quadrupole solution with
  vanishing shift. $L^2$ norm of the error as a function of time for the
  variables $s$ (top panels) and $\theta$ (bottom panels).
  Left panels: absorbing (dashed) and zero-$Z$ (dotted) boundary conditions,
  right panels: differential boundary conditions without (dashed) and
  with (dotted) modified dissipation near the boundaries.
  For comparison, the solid lines show the error if the exact 
  solution is imposed at the boundaries.}
\end{figure}

As expected, the reflection-induced peak of the error occurs after two
light-crossing times, at $t \approx 10$.
The zero-$Z$ dissipative boundary
conditions cause much stronger reflections than the absorbing
boundary conditions, particularly for the constraints. 
This is somewhat surprising because we argued in section 
\ref{sec:cpdissbcs} that the zero-$Z$ boundary conditions are
satisfied by the exact solution, in contrast to absorbing ones. 
However, the zero-$Z$ boundary conditions are not constraint-preserving 
and are indeed highly reflective because they form Dirichlet conditions for 
the wave equation (\ref{eq:BoxZ}).
Both types of dissipative boundary conditions (absorbing and zero-$Z$)
are stable.

The differential boundary conditions perform best in minimizing the 
reflections. However, at late times ($t \gtrsim 20$) the error begins
to grow exponentially. In an attempt to cure this instability,
we tried applying second-order dissipation \eqref{eq:diss2} 
instead of fourth-order dissipation \eqref{eq:diss4} near
the boundaries (at the outermost interior cells and along the first ghost
layer). This postponed the blow-up to a later time but could not
eliminate it completely (see the right half of figure \ref{fig:bc1}).

\begin{figure}[t]
  \centering
  \begin{minipage}[t]{0.49\textwidth}
    \includegraphics[angle = -90, scale = 0.5]{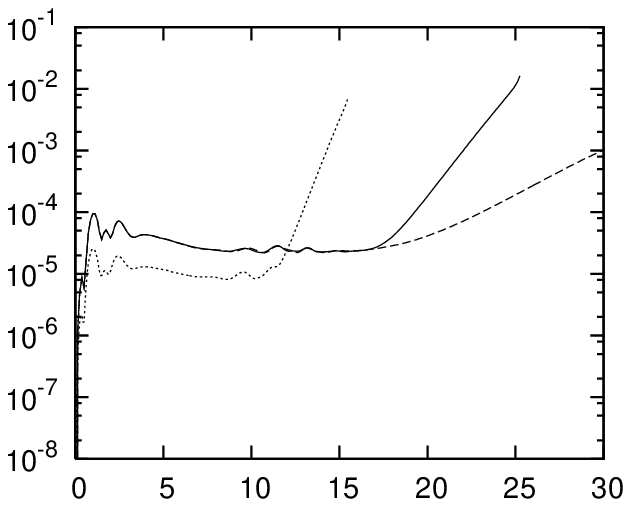}
    \centerline{\footnotesize (a)}
  \end{minipage}
  \hfill  
  \begin{minipage}[t]{0.49\textwidth}
    \includegraphics[scale = 0.5]{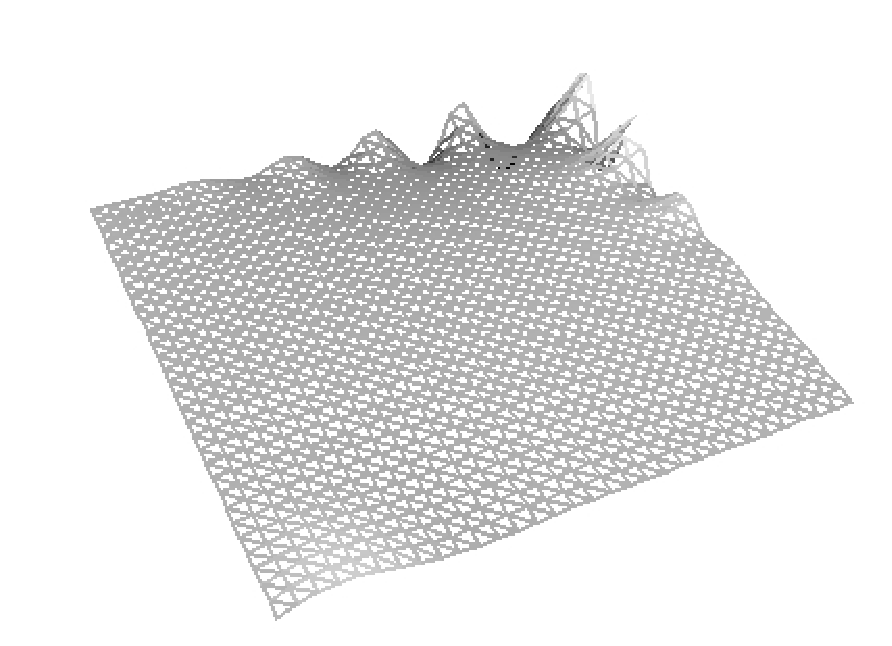}
    \centerline{\footnotesize (b)}
  \end{minipage}
  \bigskip
  \caption[Instability of differential boundary conditions] 
  {\label{fig:bcinstab}  \footnotesize
  Instability of the differential boundary conditions (quadrupole
  solution with vanishing shift). 
  (a) $L^2$ norm of the error as a function of time for the variable $s$.
  Solid line: without modified dissipation near the boundaries, 
  dashed line: with modified dissipation near the boundaries,
  dotted line: same as solid line but with twice the resolution.
  (b) The variable $s$ at $t = 25$.}
\end{figure}

In order to determine the nature of the instability, we performed the
same evolution again but with twice the resolution. Figure
\ref{fig:bcinstab}a shows that the instability sets in earlier and
with a higher exponential growth rate. The ratio of the growth rates
is found to be $2.00 \pm 0.02$. This suggests that the numerical solution
behaves like $\sim \exp(a t / h)$ at late times. The dependence on the
grid spacing $h$ means that the instability is not present in the
continuum problem but is caused by the finite-differencing used.
Modified second-order dissipation near the boundaries reduces the
growth rate but cannot eliminate the instability (for no value of 
$0 \leqslant \epsilon_D \leqslant 1$). 
So far we have not found a stable discretization. Figure
\ref{fig:bcinstab}b indicates that the instability might emanate
from the outer corner. A more careful treatment of the discretization at
the corner will be required (one-sided differences are used in
(\ref{eq:modeldiffbcdiscr})).

\paragraph{The quadrupole solution with dynamical shift.}

The results for the evolution with a dynamical shift vector are
similar (figure \ref{fig:bc2}). 
With regard to the avoidance of reflections, the differential
boundary conditions perform better than the absorbing ones, which in
turn are better than the zero-$Z$ dissipative boundary conditions.
(We do not display the constraint $\theta$ here because it is affected by
the linear drift of the error (section \ref{sec:linnum})
and differences between the boundary conditions are hardly visible.)
Again, a late-time instability occurs for the differential boundary
conditions, which can be postponed but not eliminated by modifying the
numerical dissipation near the boundaries.

\begin{figure}[t]
  \centering
  \begin{minipage}[t]{0.49\textwidth}
    \includegraphics[scale = 1]{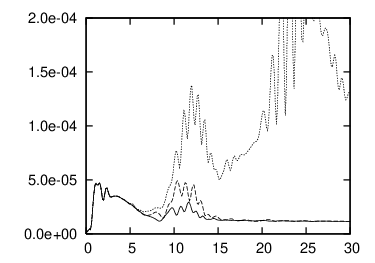}
    \centerline{$s$, dissipative}
  \end{minipage}
  \hfill  
  \begin{minipage}[t]{0.49\textwidth}
    \includegraphics[scale = 1]{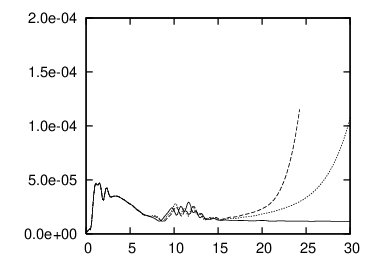}
    \centerline{$s$, differential}
  \end{minipage}
  \bigskip
  \caption[Test of boundary conditions: even quadrupole, dynamical shift] 
  {\label{fig:bc2}  \footnotesize
  Test of boundary conditions for the quadrupole solution with
  dynamical shift. $L^2$ norm of the error as a function of time for the
  variable $s$.
  Left panel: absorbing (dashed) and zero-$Z$ (dotted) boundary conditions.
  Right panel: differential boundary conditions without (dashed) and
  with (dotted) modified dissipation near the boundaries.
  The solid lines show the error if the exact solution is imposed at 
  the boundaries.}
\end{figure}

\paragraph{The twisting octupole solution.}

The twisting octupole solution (section \ref{sec:octupole}) is the
only test problem for which the zero-$Z$ dissipative boundary
conditions outperform the absorbing ones (figure \ref{fig:bc3}). 
The differential boundary conditions are about as good,
and in this case they are also stable.

\begin{figure}[t]
  \centering
  \begin{minipage}[t]{0.49\textwidth}
    \includegraphics[scale = 1]{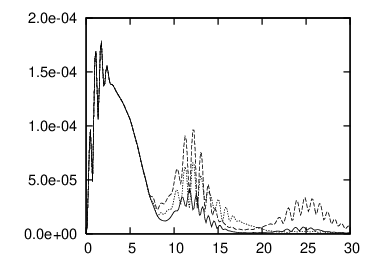} 
    \centerline{$B^\varphi$, dissipative}
  \end{minipage}
  \hfill  
  \begin{minipage}[t]{0.49\textwidth}
    \includegraphics[scale = 1]{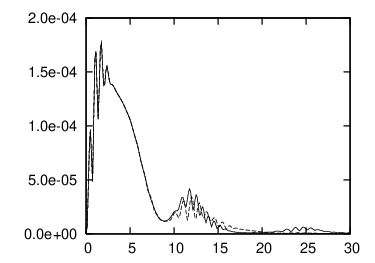}
    \centerline{$B^\varphi$, differential}
  \end{minipage}
  \hfill  
  \begin{minipage}[t]{0.49\textwidth}
    \includegraphics[scale = 1]{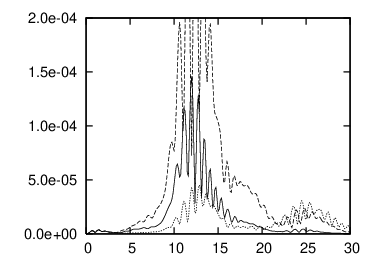}
    \centerline{$Z^\varphi$, dissipative}
  \end{minipage}
  \hfill  
  \begin{minipage}[t]{0.49\textwidth}
    \includegraphics[scale = 1]{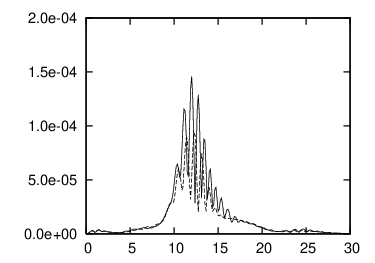}
    \centerline{$Z^\varphi$, differential}
  \end{minipage}
  \bigskip
  \caption[Test of boundary conditions: even octupole] 
  {\label{fig:bc3}  \footnotesize
  Test of boundary conditions for the octupole solution.
  $L^2$ norm of the error as a function of time for the
  variables $B^\varphi$ (top panels) and $Z^\varphi$ (bottom panels).
  Left panels: absorbing (dashed) and zero-Z (dotted) boundary conditions.
  Right panels: differential boundary conditions (dashed).
  The solid lines show the error if the exact solution is imposed at 
  the boundaries.}
\end{figure}


\subsection{Conclusions}

We conclude that in all cases, the differential boundary conditions do
the best job in reducing reflections from the outer boundaries.
However, their non-twisting part suffers from a late-time
instability, which appears to be caused by the particular
discretization that is currently used.
For the time being, the differential boundary conditions cannot be
used in numerical simulations,
at least not for a long time (one might switch to more stable boundary
conditions at late times, see section \ref{sec:brillconvtest}).

As expected, the boundary conditions of dissipative type are stable in
all cases. With regard to reflections, absorbing boundary conditions
clearly outperform the zero-$Z$ ones for the nontwisting subsystem.
For the twist subsystem, the zero-$Z$ boundary conditions are
marginally better.

In section \ref{sec:brillconvtest}, the boundary conditions are further
tested from the point of view of mass conservation.


  \chapter[Evolutions of nonlinear Brill waves]
{Adaptive evolutions of nonlinear generalized Brill waves}
\label{sec:brill}

The numerical evolutions with the Z(2+1)+1 system presented so far
were all linear. In this final chapter, we turn to nonlinear
evolutions of axisymmetric gravitational waves in vacuum.
A nonzero twist is included and hence we refer to
the problem investigated here as \emph{generalized Brill waves}.

We begin by explaining our choices of initial data and gauge in
section \ref{sec:brillini}. The pros and cons of the different gauge 
conditions are discussed and the need for adaptive mesh refinement (AMR) 
is demonstrated.
A 3-grid convergence test is performed in section \ref{sec:brillconvtest},
indicating the accuracy of our implementation.
Once we form a black hole in supercritical Brill wave evolutions, we
would like to detect it, and so we describe our method of finding
apparent horizons in section \ref{sec:brillapphor}.
Our results on adaptive evolutions of both sub- and supercritical
generalized Brill waves are presented in section \ref{sec:brillcollapse}.

\section{Initial data and gauge choices}
\label{sec:brillini}

Our method of generating initial data is similar to the formalism used
in chapter \ref{sec:hypell}. 
We take the initial 2-metric to be conformally flat,
\begin{equation}
  H_{AB} = \psi^4 \delta_{AB} \,.
\end{equation}
Free data is prescribed for the Z(2+1)+1 variables\footnote{Note that
  the Z(2+1)+1 definition of $s$ differs from the one in chapter 
  \ref{sec:hypell} by a minus sign.} $s$, $\alpha$ and 
\begin{equation}
  \label{eq:confbp}
  \hat B^\varphi \equiv \psi^{9/2} B^\varphi \,. 
\end{equation}
We choose Gaussian profiles
\begin{eqnarray}
  s &=& - A_s \, r \, \exp \left[ -(\tfrac{r}{\sigma_{r,s}})^2 
    -(\tfrac{z}{\sigma_{z,s}})^2 \right] \,,\\
  \hat B^\varphi &=& A_B \, r z \, \exp \left[ -(\tfrac{r}{\sigma_{r,B}})^2 
    -(\tfrac{z}{\sigma_{z,B}})^2 \right] \,,\\
  \label{eq:inialpha}
  \alpha &=& 1 - A_\alpha \exp \left[ -(\tfrac{r}{\sigma_{r,\alpha}})^2 
    -(\tfrac{z}{\sigma_{z,\alpha}})^2 \right] \,.
\end{eqnarray}
Note that the factors of $r$ and $z$ have to be included for the
correct behaviour at small $r$ and $z$ (as before, we impose reflection 
symmetry about $z = 0$).
Unless otherwise stated, we take all the widths $\sigma$ to be 1 and
$A_\alpha = 0$.
The variables $\chi_{AB}$, $Y$ and $E^A$ are chosen to vanish initially.
As already mentioned in section \ref{sec:hypellnum}, this initial data
is more general than Brill's orginial one \cite{Brill59} (which has zero
twist, $A_B = 0$) but is still time-symmetric, so that the term 
\emph{generalized Brill waves} is justified.

The momentum constraints (\ref{eq:momcons}) and the Geroch constraint
(\ref{eq:gercons}) are automatically satisfied for this choice of
initial data.
The Hamiltonian constraint (\ref{eq:hamcons}) takes the form
\begin{eqnarray}
  \label{eq:brillhamcons}
  0 &=& \psi_{,rr} + \psi_{,zz} + (s + r s_{,r} + r^{-1}) \psi_{,r} 
  + r s_{,z} \psi_{,z} \nonumber\\
  && + \quarter [r s_{,rr} + 4 s_{,r} + 2 r^{-1} s + (s + r s_{,r})^2
  + r s_{,zz} + r^2 s_{,z}{}^2] \psi \nonumber\\
  && + \tfrac{1}{16} r^2 \e^{2rs} \hat {B^\varphi}^2 \,.
\end{eqnarray}
This elliptic equation is solved for the conformal factor $\psi$ using
Multigrid (section \ref{sec:MG}). Note that the variable 
$B^\varphi$ has been conformally rescaled \eqref{eq:confbp}. Otherwise
$B^\varphi$ would be multiplied with a positive power of $\psi$ in
\eqref{eq:brillhamcons}, the equation would not be linearization-stable, 
and Multigrid would fail to converge (cf.~section \ref{sec:ellsolv}).
After solving the Hamiltonian constraint, the original variable
$B^\varphi$ is formed and the derivatives of the 2-metric are computed
numerically.

Next the question arises which gauge in the family of generalized
harmonic gauge conditions (section \ref{sec:dyngauge}) one should use 
in order to evolve this initial data. The vanishing shift case and the 
dynamical shift case turn out to behave in a completely different way,
as illustrated by figure \ref{fig:dynvsvan}: whereas the variables
clearly show a wavelike behaviour and eventually assume their flat-space
values in the dynamical shift case, they settle down to a non-trivial
static solution in the vanishing shift case (we shall see below 
that this is also Minkowski space, but in non-standard 
coordinates). A similar residual ``lump of gauge'' for harmonic
slicing with zero shift has also been reported by Eppley \cite{Eppley79}.
The explanation for this lies in the fact that in pure harmonic gauge
(i.e., including a dynamical shift), all the variables obey the wave
equation to principal parts (section \ref{sec:dyngauge}), which does
not hold in the vanishing shift case.
The linear evolutions in chapter \ref{sec:lin} were wavelike even for
vanishing shift only because the initial data was taken to be that of
the exact solution. It is not clear which restrictions one has to
impose on general initial data such that this property extends to
the nonlinear case.
These difficulties were our main motivation for adding dynamical shift 
conditions to our original paper \cite{RinneStewart05}.

\begin{figure}[t]
  \centering
    \begin{minipage}[t]{0.49\textwidth}
      \includegraphics[scale = 0.49]{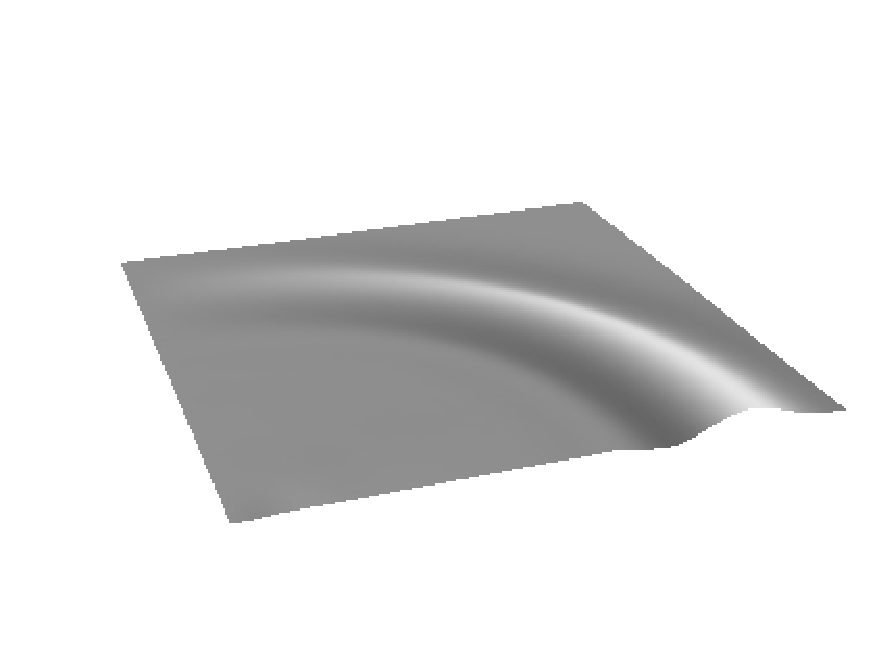}
      \centerline{\footnotesize (a) dynamical shift}
    \end{minipage}
    \hfill  
    \begin{minipage}[t]{0.49\textwidth}
      \includegraphics[scale = 0.49]{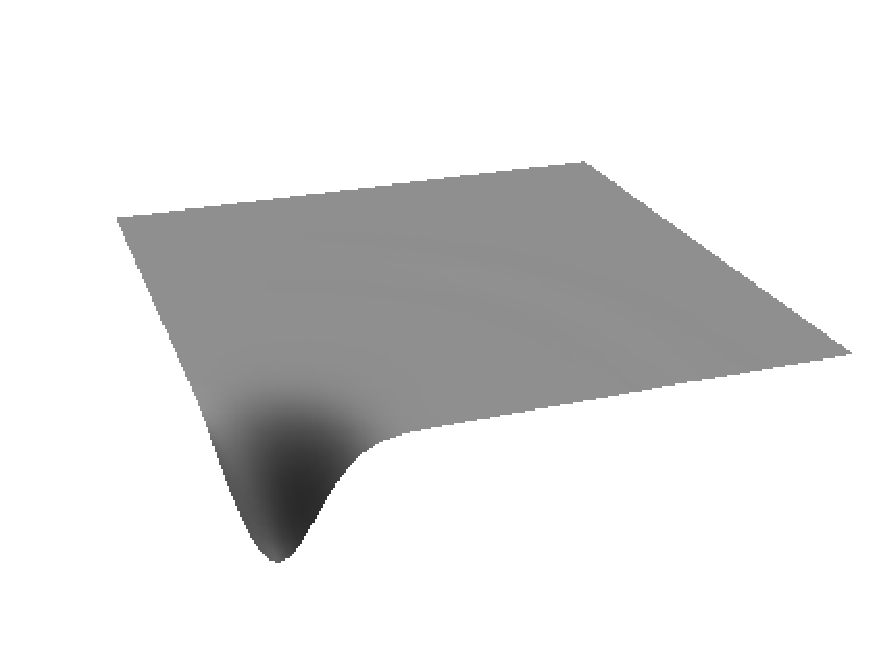}
      \centerline{\footnotesize (b) vanishing shift}
    \end{minipage}
    \hfill  
  \bigskip
  \caption[Dynamical vs. vanishing shift]
  {\label{fig:dynvsvan}  \footnotesize
  The variable $s$ at time $t = 8.75$ for a Brill wave with amplitudes
  $A_s = 1$ and $A_B = 0$ using harmonic gauge with 
  (a) dynamical and (b) vanishing shift vector (gauge parameters $f = d = \mu
  = a = 1$, $m = 2$). 
  Single grid with 64 points in each dimension, $r_\mathrm{max} =
  z_\mathrm{max} = 5$.}
\end{figure}

In order to convince ourselves that the final state of the
vanishing-shift evolution is indeed Minkowski space, we need to
compute the curvature.
A useful quantity to look at is the \emph{Kretschmann scalar}
\begin{equation}
  \label{eq:kretschmann}
  I \equiv \four{R_{\alpha\beta\gamma\delta}} 
  \four{R^{\alpha\beta\gamma\delta}} 
\end{equation}
evaluated at the origin $r = z = 0$, where we expect the curvature to
be maximal. To simplify the calculation, one can first note that when 
computing the Riemann tensor for a general metric of the form 
\eqref{eq:regcond4} and finally setting $r = 0$, the $\varphi t$, 
$\varphi r$ and $\varphi z$ components do not contribute.
We also assume that the shift vanishes. Hence we consider the metric
\begin{equation}
  g_{\alpha\beta} = \left( \begin{array}{cccc}
      -\alpha^2 & 0 & 0 & 0 \\ 0 & H_{rr} & H_{rz} & 0 \\
      0 & H_{rz} & H_{zz} & 0 \\ 0 & 0 & 0 & r^2 H_{rr} \e^{2rs} 
      \end{array} \right) \,.
\end{equation}
For this we compute the Riemann tensor directly using REDUCE,
substituting the evolution equations for the time derivatives.
The resulting expression is manifestly regular.
It contains up to second-order spatial derivatives of the metric, i.e., 
first-order derivatives of the first-order variables 
(\ref{eq:Ddef}--\ref{eq:Adef}). 
To second order in the grid spacing $h$, it is consistent to evaluate
$I$ at the innermost grid point $r = z = h/2$ and to set
$u_{,r} = 0$ for a variable that is even in $r$ and $u_{,r} = 2u/h$ 
for a variable that is odd in $r$ (and similarly for $z$).
Figure \ref{fig:kretschmann} shows that $I$ for the above evolution 
starts off at a large value of $\approx 200$, 
then drops rapidly and after a few bounces
settles down at a constant small value of $\approx 2$ (this is mainly
determined by the time step and decays as it is decreased).
This suggests that the curvature of the final state indeed vanishes.

\begin{figure}[t]
  \centering
  \includegraphics[scale = 1]{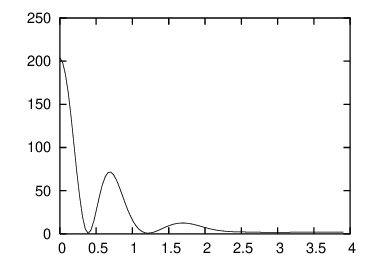}
  \caption[Kretschmann scalar]
  {\label{fig:kretschmann}  \footnotesize
  The Kretschmann scalar $I$ as a function of time for a Brill wave
  with amplitudes $A_s = 1$ and $A_B = 0$ using harmonic slicing with 
  vanishing shift.}
\end{figure}

Unexpected difficulties occur for strong waves (amplitudes
$A_s \gtrsim 3$, irrespective of $A_B$). In the dynamical shift case, the
solution blows up exponentially as the waves travel out. The first
variables to grow are the constraints $\theta, Z_r, Z_z$, and soon
after all the remaining variables are affected. The blow-up occurs at
$z = 0$ (where most of the variables have their extrema), well away
from the axis $r = 0$. We have checked that the location and growth
rate are essentially unchanged as the resolution is increased. 
Hence it is likely that we are faced with a continuum instability. 

Motivated by work of Brodbeck et al.~\cite{Brodbeck99}, Gundlach et 
al.~\cite{Gundlach05} have recently proposed the addition of
\emph{constraint-damping} terms to the Z4 equations \eqref{eq:Z4},
\begin{equation}  
  R_{\alpha\beta} + 2 \nabla_{(\alpha} Z_{\beta)} 
  - \kappa_\mathrm{CD} (2 n_{(\alpha} Z_{\beta)} 
  - g_{\alpha\beta} n^\gamma Z_\gamma) = 0 \,,
 \end{equation}
where $n_\alpha$ is the unit timelike normal to the foliation and
$\kappa_\mathrm{CD} > 0$ is a constant. 
After performing the Geroch and ADM reductions, this implies that we
should add the following terms to the right-hand-sides of the Z(2+1)+1
equations:
\begin{eqnarray}
  \Lie{n} \theta &=& \ldots - 2 \kappa_\mathrm{CD} \, \theta \,,\\
  \Lie{n} Z_A &=& \ldots - \kappa_\mathrm{CD \,} Z_A \,,\\
  \Lie{n} Z^\varphi &=& \ldots - \kappa_\mathrm{CD} \, Z^\varphi \,,\\
  \Lie{n} \chi_{AB} &=& \ldots - \kappa_\mathrm{CD} \, \theta H_{AB} \,.
\end{eqnarray}
The authors of \cite{Gundlach05} showed that in the high-frequency (or
geometrical optics) approximation, all constraints are damped 
exponentially if the damping terms are included, except modes that are 
constant in space. 
Recently, constraint damping has been used successfully 
in binary black hole simulations \cite{Pretorius05}.
However, we found that it could not eliminate the blow-up in the 
gravitational wave evolutions with dynamical shift considered here,
for any value of $\kappa_\mathrm{CD}$.
It should be stressed that the analysis in \cite{Gundlach05} is only valid for
high-frequency constraint violations. It is unclear if the inclusion of 
such damping terms renders the constraint manifold $Z_\alpha = 0$ an 
attractor if the wavelength of the constraint violations becomes comparable 
with the curvature scale, as is expected in nonlinear gravitational wave
evolutions.
We have also tried eliminating all nonlinear couplings with the $Z$ 
vector in the source terms of the Z(2+1)+1 equations, again without 
any improvements.
Hence harmonic gauge with dynamical shift appears to be unusable for
the problem at hand.

One is faced with a different obstruction when using a vanishing shift
vector: the occurrence of steep gradients and highly distorted waves.
This is not surprising if we recall that the \emph{coordinate}
characteristic speeds depend on the spatial metric:
the \emph{physical} speeds in section \ref{sec:hyperbolicity}
were computed for the projection of the flux vector along the 
\emph{unit} normal to the boundary, $\F^\n = \F^A \mu_A \,$.
For the $r = r_\mathrm{max}$ boundary, for example, 
$\mu_A = \delta_A{}^r / \sqrt{H^{rr}}$.
Hence the coordinate speeds in the $r$-direction (corresponding to the
$r$-component of the flux vector, $\F^r$) are obtained from the
physical speeds by multiplying with $\sqrt{H^{rr}}$.
If the metric is wavelike as in the dynamical shift case, then it is 
essentially constant when averaged in time, and the characteristic 
speeds should be uniform across the grid on average. If however the 
metric is essentially static and non-constant as in the vanishing shift case, 
then there can well be regions in which the characteristics converge 
(signals to the left travel faster than those to the right for a right-moving
wave), and steep gradients can build up. 

It has been claimed by Alcubierre \cite{Alcubierre97, AlcubierreMasso98,
Alcubierre03} that under certain circumstances, true discontinuities
can develop when using hyperbolic gauge conditions, so-called
\emph{gauge shocks}. 
Our results strongly suggest that this is not the case in our problem,
provided that we choose the gauge parameter to be $f = 1$.
We do observe a steep gradient in the profile of the lapse function
$\alpha$, or equivalently, a sharp peak in the variable $A_r =
\alpha^{-1} \alpha_{,r}$, which travels out at the speed of light. 
However, if we switch on the adaptive mesh refinement
(section \ref{sec:AMR}) and sufficiently refine the region around the 
gradient, we can show that the peak is completely smooth 
(figure \ref{fig:gaugeshock}). Its height increases as one increases 
the amplitude $A_s$. We have checked that the peak is well-resolved
during the entire evolution even for for the largest amplitudes 
considered in section \ref{sec:brillcollapse}.
This would not be possible on a single coarse grid: there the feature
\emph{looks} like a ``delta-function'' and the finite-difference code would
crash because of the Gibbs oscillations that this causes.
Hence adaptive mesh refinement appears to be crucial in order to be
able to evolve radiative spacetimes with harmonic slicing and zero shift.
Preliminary results indicate that for parameters $f \neq 1$, the 
peak in the gradient of the lapse becomes narrower and narrower during
the evolution, which ultimately crashes the code. 
This is in agreement with the work of Alcubierre, who showed that
generalized harmonic slicing will always develop gauge shocks for $f < 1$.

\begin{figure}[t]
  \centering
   \begin{minipage}[t]{0.49\textwidth}
     \includegraphics[scale = 0.49]{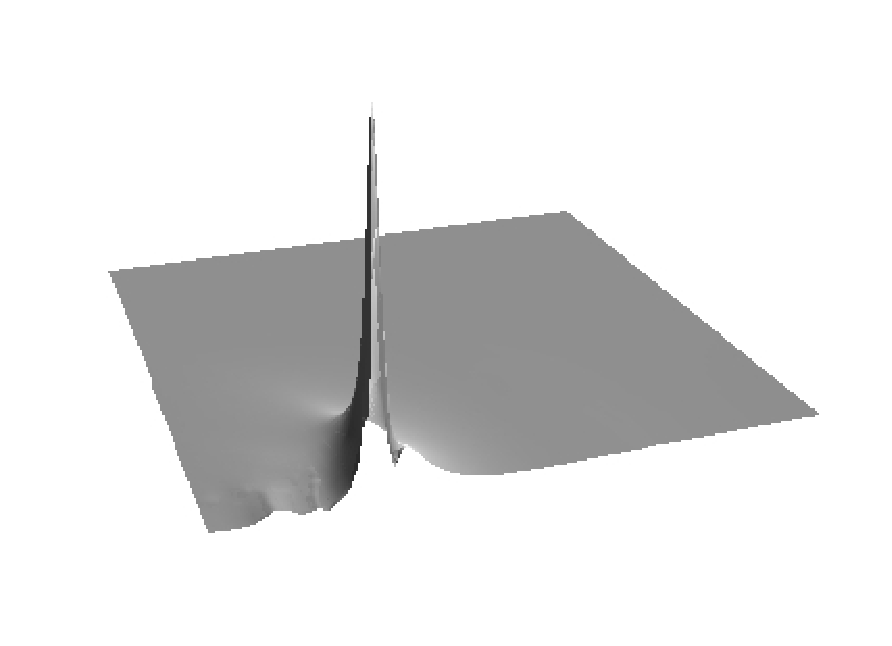}
     \centerline{\footnotesize (a)}
   \end{minipage}
   \hfill  
   \begin{minipage}[t]{0.49\textwidth}
     \includegraphics[scale = 0.4]{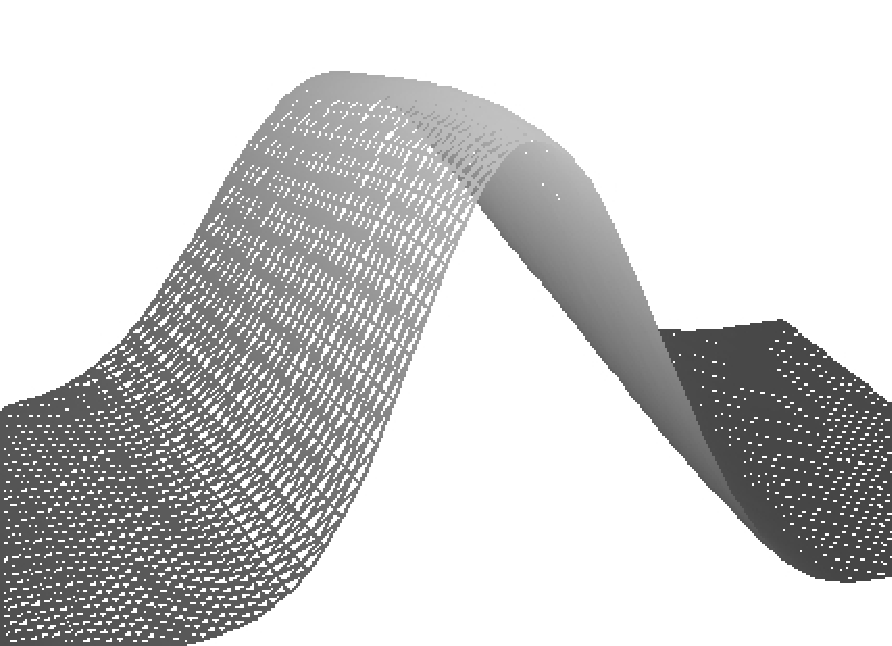}
     \centerline{\footnotesize (b)}
   \end{minipage}
   \hfill  
  \bigskip
  \caption[The need for AMR]
  {\label{fig:gaugeshock}  \footnotesize 
  The variable $A_r = \alpha^{-1} \alpha_{,r}$ at time $t = 1.72$ 
  when the spike reaches its maximum height, for a Brill wave with
  amplitudes $A_s = 4$ and $A_B = 0$, using harmonic slicing with
  vanishing shift. 
  (a) shows the base grid, (b) the finest grid containing the spike.
  The base grid resolution is 64 points and 4 levels of refinement 
  are added.  
}
\end{figure}


  \section{Convergence test}
\label{sec:brillconvtest}

Unlike in linearized theory, no exact solutions are known for
nonlinear Brill waves and so the method of testing the convergence of
the code used in section \ref{sec:linnum} is not applicable.
However, we can perform a 3-grid convergence test based on Richardson
extrapolation.
As explained in section \ref{sec:AMR}, we estimate the error $\vec e^h$ 
on a grid $G^h$ with grid spacing $h$ by
\begin{equation}
  \vec e^h \approx \third (\vec u^{2h} - \vec u^h) \,,
\end{equation}
where $\vec u^h$ denotes the numerical approximation on $G^h$.
This involves interpolating the approximation from $G^{2h}$ to $G^h$,
and it is important to use an interpolation scheme that is more than
second-order accurate in order not to affect the leading order of the
error estimate (we use biquadratic interpolation).
Similarly, we can estimate the error on $G^{2h}$ by 
\begin{equation}
  \vec e^{2h} \approx \third (\vec u^{4h} - \vec u^{2h}) \,.
\end{equation}
For a second-order accurate code, the ratio of the errors should be
\begin{equation}
  \lVert \vec e^{2h} \rVert / \lVert \vec e^h \rVert \approx 4 \,,
\end{equation}
where we use the discrete $L^2$ norm and all variables are summed over. 

We evolve a twisting Brill wave with amplitudes $A_s = A_B = 1$ using
harmonic slicing ($f = 1$, $m = 2$) with vanishing shift.
The coarsest grid has a resolution of 32 points, with the outer
boundaries placed at $r_{max} = z_{max} = 5$. The Courant number is
taken to be $0.5$.
First we can check that the initial data solver works correctly.
The estimated errors on the two finest grids obtained via Richardson
extrapolation are $9.91\times 10^{-3}$ and $2.25\times 10^{-3}$, yielding
a ratio of $4.40$. The constraint residuals (evaluated independently from the
Multigrid solver) are $5.11\times 10^{-2}$, $1.28 \times 10^{-2}$ and
$3.23 \times 10^{-3}$, with ratios $3.99$ and $3.96$. This is
perfectly second-order convergent.
Next we look at the evolution of the errors and constraints.
Figure \ref{fig:brillconvtest} shows the estimated errors as well as the
residuals of the Einstein constraints (\ref{eq:hamcons}--\ref{eq:gercons}),
the $Z$ constraints $\theta = Z_r = Z_z = Z^\varphi = 0$ and the 
differential constraints associated with the definitions of the
first-order variables (\ref{eq:Ddef}--\ref{eq:Adef}).
The results indicate not perfect, but approximate second-order convergence
up to one light-crossing time ($t = 5$). At this point we switched from
differential boundary conditions to absorbing ones (chapter
\ref{sec:outerbcs}) in order to avoid the instability of the former at
late times. This leads to a loss of convergence and an increase particularly 
of the constraint residuals -- this is what we expect because the
absorbing boundary conditions are not constraint-preserving.
However, we do achieve a stable evolution in this way.

\begin{figure}[t]
  \centering
    \begin{minipage}[t]{0.49\textwidth}
      \includegraphics[scale = 1]{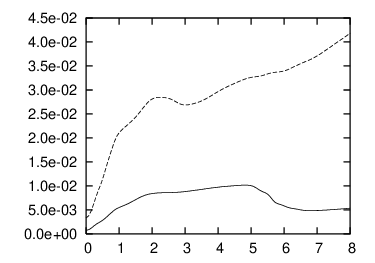}
      \centerline{\footnotesize (a) error}
    \end{minipage}
    \hfill  
    \begin{minipage}[t]{0.49\textwidth}
      \includegraphics[scale = 1]{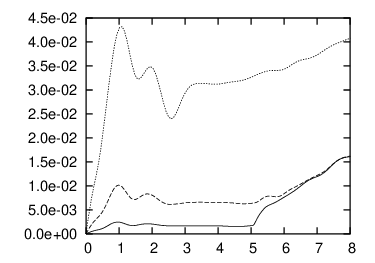}
      \centerline{\footnotesize (b) Einstein constraints}
    \end{minipage}
    \hfill  
    \begin{minipage}[t]{0.49\textwidth}
      \includegraphics[scale = 1]{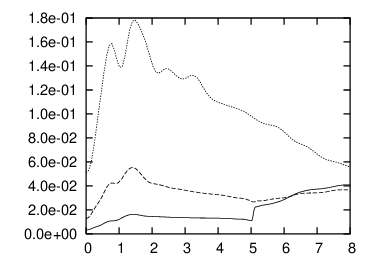}
      \centerline{\footnotesize (c) $Z$ constraints}
    \end{minipage}
    \hfill  
    \begin{minipage}[t]{0.49\textwidth}
      \includegraphics[scale = 1]{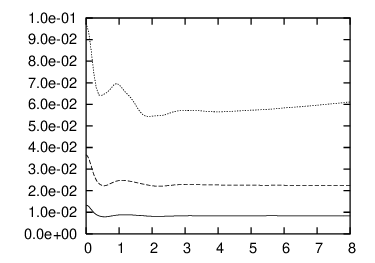}
      \centerline{\footnotesize (c) differential constraints}
    \end{minipage}
    \hfill  
  \bigskip
  \caption[Convergence test]
  {\label{fig:brillconvtest}  \footnotesize
  3-grid convergence test for a Brill wave with $A_s = A_B = 1$.
  The resolutions are 32 points (dotted), 64 points (dashed) and 128
  points (solid) per dimension.
  Shown are as functions of time the $L^2$ norms of 
  (a) the estimated error on the two finest grids; 
  (b) the Einstein constraints, 
  (c) the $Z$ constraints and 
  (d) the differential constraints on all three grids.}
\end{figure}

A useful quantity to monitor during the evolution is the \emph{ADM
  mass}, the mass of an asymptotically flat spacetime measured at
spacelike infinity. This can be derived by writing the Hamiltonian
constraint \eqref{eq:hamcons} in linearized theory in conservation form
\begin{equation}
  \kappa \rho = \tilde \nabla_A J^A \,,
\end{equation}
where $\tilde \nabla$ denotes the flat-space connection and the
current $J^A$ is given in our variables by
\begin{eqnarray}
  J^r &=& - D_{rrr} - D_{rzz} + D_{zrz} - r^2 \tilde s_r - 3 s \,,\\
  J^z &=& r \tilde D_{rrz} - 2 D_{zrr} + r^{-1} H_{rz} - r s_z \,.
\end{eqnarray}
The ADM mass is then defined by an integral over a 2-surface $S_\infty$
at spacelike infinity,
\begin{equation}
  M_\textrm{ADM} = \kappa^{-1} \int_{S_\infty} J^A \diff^2 S_A \,,
\end{equation}
$\diff^2 S_A$ denoting the area element on $S_\infty$.
We choose the $S_\infty$ to be aligned with the grid boundaries,
\begin{equation}
  \label{eq:ADMmass}
  M_\textrm{ADM} = \lim_{r_0, z_0 \rightarrow \infty}
  \half \left( \int_0^{r_0} r J^z \diff r \big \rvert_{z = z_0} 
    + r_0 \int_0^{z_0} J^r \diff z \big \rvert_{r = r_0} \right) \,.
\end{equation}
We have verified that we obtain the same result for the ADM mass
if we start from the standard expression found in the literature \cite{MTW},
which is only valid in \emph{Cartesian} coordinates, and carefully
transform it to cylindrical polar coordinates.

In the numerical implementation, we evaluate the integrals in
\eqref{eq:ADMmass} at $r_0 = 0.9 r_\textrm{max}$, $z_0 = 0.9 z_\textrm{max}$
(a few grid points away from the outer boundaries in order to reduce
the influence of possible reflections). One would expect the ADM mass
to be constant until the wave reaches the outer boundaries of the
computational domain and to drop afterwards. 
This is confirmed by figure \ref{fig:masscons}a, which shows the ADM
mass as a function of time on the three grids. The higher the
resolution, the longer the mass is conserved.
In order to study the influence of the boundary, we perform another run with
twice the domain size, leaving the surface of integration (the ``detector'')
at the same location (figure \ref{fig:massbdry}). This shows that the initial
drop of the numerically evaluated ADM mass at $t \approx 4$ is indeed caused
by the wave passing through the detector rather than by reflections from the
boundary. However, the (unphysical) negative value of the ADM mass at $t
\gtrsim 6$ appears to be a boundary effect -- this is less severe in the run
with twice the domain size.

\begin{figure}[t]
  \centering
    \begin{minipage}[t]{0.49\textwidth}
      \includegraphics[scale = 1]{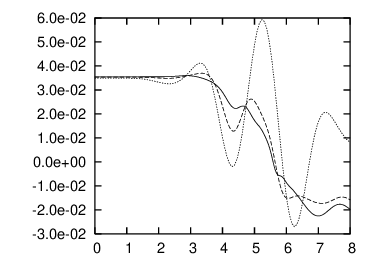}
      \centerline{\footnotesize (a)}
    \end{minipage}
    \hfill  
    \begin{minipage}[t]{0.49\textwidth}
      \includegraphics[scale = 1]{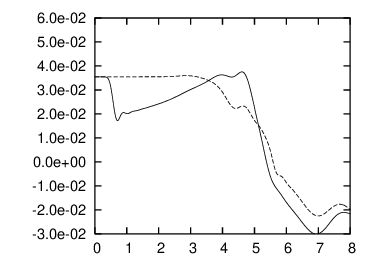}
      \centerline{\footnotesize (b)}
    \end{minipage}
    \hfill  
  \bigskip
  \caption[Mass conservation]
  {\label{fig:masscons}  \footnotesize
  The ADM mass as a function of time for a Brill wave with $A_s = A_B = 1$.
  (a) 3-grid convergence test as in figure \ref{fig:brillconvtest},
  (b) absorbing (solid) vs.~differential (dashed) boundary conditions
  on the finest grid.}
\end{figure}

\begin{figure}[t]
  \centering
  \includegraphics[scale = 1]{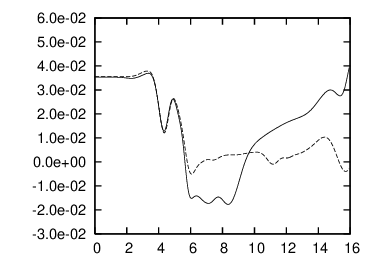}
  \bigskip
  \caption[Dependence of ADM mass on boundary location]
  {\label{fig:massbdry}  \footnotesize
  Dependence of the numerically evaluated ADM mass on the location of the
  outer boundary for a Brill wave with $A_s = A_B = 1$.
  Solid line: $r_\mathrm{max} = z_\mathrm{max} = 5$, 
  dashed line: $r_\mathrm{max} = z_\mathrm{max} = 10$.
  In both cases, the integral \eqref{eq:ADMmass} is evaluated at 
  $r_0 = z_0 = 4.5$.
  The resolution is 64 points per dimension and differential boundary
  conditions are used.}
\end{figure}

It turns out that the ADM mass is extremely sensitive to the outer
boundary conditions we impose, and we can use this to assess the
various choices of boundary conditions that are available.
Figure \ref{fig:masscons}b shows that the differential boundary
conditions of chapter \ref{sec:outerbcs} perform very well in this
respect but that absorbing boundary conditions are not very
mass-conserving at all. This is what we expect because the
differential boundary conditions are designed such that no incoming
radiation enters the domain from the outside, whereas this does not
hold for absorbing boundary conditions.
For this reason, we henceforth use differential boundary conditions as
long as we can and switch to absorbing ones at late times just before the 
instability sets in.


  \section{Apparent horizon finder}
\label{sec:brillapphor}

Once a black hole has formed in our numerical spacetime, we would like
to be able to detect it. An indication of black hole formation is the
existence of \emph{trapped surfaces}, i.e., closed two-surfaces whose
outgoing null geodesics have zero expansion (``light cannot
escape''). The outermost\footnote{It can be very difficult to verify that a
  trapped surface is the outermost one. Often the terms ``apparent
  horizon'' and ``trapped surface'' are used synonymously.}
trapped surface in a given spacelike slice is called the \emph{apparent
  horizon}. Since only the data on a given spacelike slice is
required, apparent horizons can be determined at each time step during
a numerical simulation.

The \emph{(future) event horizon} is defined to be the boundary of 
the causal past of future null infinity. This is a global property of 
spacetime: in contrast to an apparent horizon, the event horizon can
only be determined if the entire future development of the given slice is
known, i.e., at the end of a numerical simulation.

Under certain technical assumptions, the existence of an apparent
horizon implies the existence of an event horizon containing the
apparent horizon in its interior \cite{HawkingEllis}. Unfortunately, the
converse is not true:
one can construct slicings of Schwarzschild spacetime such that there
is no apparent horizon \cite{Wald91}, although an event horizon does 
of course exist.
However, generally an apparent horizon is a good approximation to
the event horizon. In particular, the two coincide in stationary spacetimes.

In the following, we shall focus on apparent horizons and derive an equation
determining the horizon in the $(2+1)+1$ formalism.

Suppose we are given an apparent horizon $\mathcal{H}$ on a
\emph{three}-dimensional spacelike hypersurface $\three{\Sigma}$. 
Let $s^\alpha$ be the outward-pointing unit normal to the horizon, 
$\three{g}_{\alpha\beta}$ the metric on $\three{\Sigma}$, 
\begin{equation}
  \label{eq:g3}
  \three{g}_{\alpha\beta} = g_{\alpha\beta} + n_\alpha n_\beta \, ,
\end{equation}
$\three{\nabla}$ the covariant derivative of that metric, 
and $\three{K}_{\alpha\beta}$ the second fundamental form, 
\begin{equation}
  \label{eq:K3}
  \three{K}_{\alpha\beta} = -\three{g}_\alpha{}^\gamma \,
  \three{g}_\beta{}^\delta \, n_{(\gamma ; \delta)} \, .
\end{equation}

From the unit spatial normal $s^\alpha$ to the horizon and the unit
timelike normal $n^\alpha$ to $\three{\Sigma}$,
we can construct the future-pointing outgoing null vector field
\begin{equation}
  \label{eq:nullvectors}
  l^\alpha = n^\alpha + s^\alpha \, .
\end{equation}
The expansion $\Theta$ of the null vectors is given by the 
(four-)divergence of $l^\alpha$ projected into the hypersurface $\mathcal{H}$,
\begin{equation}
  \label{eq:expansion1}
  \Theta = (\three{g^{\alpha\beta}} - s^\alpha s^\beta) \nabla_\beta 
  l_\alpha \, .
\end{equation}
Substituting (\ref{eq:nullvectors}) into (\ref{eq:expansion1}) and
using the definition of the second fundamental form (\ref{eq:K3}), 
we obtain \cite{York89}
\begin{equation}
  \label{eq:expansion}
  \Theta = \three{\nabla}_\alpha s^\alpha + \three{K}_{\alpha\beta} 
  s^\alpha s^\beta - \three{K} \, .
\end{equation}
For an apparent horizon, this quantity has to vanish.

We would like to write equation (\ref{eq:expansion}) in $(2+1)+1$ form. 
The first term on the right-hand-side can be rewritten as
\begin{equation}
  \label{eq:divs}
  \three{\nabla}_\alpha s^\alpha = \diff_A s^A 
  + \lambda^{-1} s^A \partial_A \lambda \, .
\end{equation}
Using the definitions of $\three{K}_{\alpha\beta}$ (\ref{eq:K3}) and 
$\chi_{\alpha\beta}$ (\ref{eq:chi2}), we can express 
$\three{K}_{\alpha\beta}$ in terms of $(2+1)+1$ quantities:
\begin{equation}
  \label{eq:K3tochi2}
  \three{K}_{\alpha\beta} = \chi_{\alpha\beta} + \lambda^{-3} \xi_{(\alpha} 
  \epsilon_{\beta ) \gamma} \omega^\gamma + \lambda^{-2} \xi_\alpha \xi_\beta
  \K \, .
\end{equation}
Hence we obtain the \emph{apparent horizon equation} in $(2+1)+1$ form,
\begin{equation}
  \label{eq:apphoreqn}
  \diff_A s^A + \lambda^{-1} s^A \partial_A \lambda + \chi_{AB} s^A s^B 
  - \chi - \K = 0 \, ,
\end{equation}
which is clearly an equation in $\N$. 

Since the horizon is a closed \emph{curve} in $\three{\Sigma} \cap \N$, 
we can parametrize its coordinates as $x^A = x^A(\tau)$. 
The horizon normal is then given by
\begin{equation}
  s^A = N H^{AB} \epsilon_{BC} \frac{\diff x^C}{\diff \tau} \, , \quad
  N \equiv \left( H_{AB} \frac{\diff x^A}{\diff \tau} 
    \frac{\diff x^B}{\diff \tau} \right)^{-1/2} \, .
\end{equation}
Let us also introduce the unit tangent to the horizon,
\begin{equation}
  t^A = N \frac{\diff x^A}{\diff \tau} \, .
\end{equation}
Clearly, $t^A s_A = 0$. Differentiating that relation, we obtain
\begin{eqnarray}
  0 &=& \diff_B (t^A s_A) = s_A \diff_B t^A + t^A \diff_B s_A \nonumber\\
  \Rightarrow 0 &=& t^B s_A \diff_B t^A + t^B t^A \diff_B s_A = 
  t^B s_A \diff_B t^A + H^{AB} \diff_B s_A  \, ,
\end{eqnarray}
where in the last step we have used that the two-metric can be written
as $H^{AB} = s^A s^B + t^A t^B$ and that $s_A s^A = 1$.
Hence we find
\begin{eqnarray}
  d_A s^A &=& -t^B s_A \diff_B t^A = -t^B s_A \left( \partial_B t^A +
  \Gamma^A{}_{BC} t^C \right) \nonumber\\
  &=& -N^3 \left[ \epsilon_{AB} \frac{\diff^2 x^A}{\diff \tau^2}
    \frac{\diff x^B}{\diff \tau} + \epsilon_{AB} 
    \frac{\diff x^B}{\diff \tau} \frac{\diff x^C}{\diff \tau}  
    \frac{\diff x^D}{\diff \tau} \Gamma^A{}_{CD} \right] \,,  
\end{eqnarray}
which agrees with \cite[eqn. $(2 \cdot 23)$]{Nakamura87}.

In practice, we choose the parameter $\tau$ to be the spherical polar 
angle $\theta$.\footnote{This parametrization only works when the apparent
  horizon forms a star-shaped domain with respect to the centre.}
The cylindrical polar coordinates of the horizon are 
\begin{equation}
  \quad x^1 \equiv r = R(\theta) \sin \theta \, , \quad 
  x^2 \equiv z = R(\theta) \cos \theta \,,
\end{equation}
where the spherical polar radius $R(\theta)$ is the unknown function
we need to determine.

With these definitions, \eqref{eq:apphoreqn} becomes a nonlinear
second-order ODE for $R(\theta)$,
\begin{equation}
  \label{eq:apphortbvp1}
  f \big ( \theta, R(\theta), R'(\theta), R''(\theta), \vec u(R(\theta),
    \theta) \big ) = 0 \,.
\end{equation}
Here $\vec u$ denotes the vector of (2+1)+1 variables.
We require that the horizon be smooth on the axes, which implies
Neumann boundary conditions
\begin{equation}
  \label{eq:apphortbvp2}
  R'(0) = R'(\tfrac{\pi}{2}) = 0 
\end{equation}
(note again that we impose reflection symmetry about $z = 0
\Leftrightarrow \theta = \tfrac{\pi}{2}$).

We cover the interval $[0, \tfrac{\pi}{2}]$ with a uniform cell-centred grid
consisting of $N_H$ points with grid spacing $h_H = \tfrac{\pi}{2 N_H}$. 
Second-order accurate centred finite differences
are used to discretize the derivatives of $R(\theta)$, 
\begin{eqnarray}
  \label{eq:ahdiscr1}
  R'_{i} &\rightarrow& \tfrac{1}{2h_H} (R_{i+1} - R_{i-1}) \,, \nonumber\\
  R''_{i} &\rightarrow& \tfrac{1}{h_H{}^2} (R_{i+1} - 2 R_i + R_{i-1}) \,,
\end{eqnarray}
where $1 \leqslant i \leqslant N_H$, and ghost cells are employed to
implement the boundary conditions \eqref{eq:apphortbvp2},
\begin{equation}
  \label{eq:ahdiscr2}
  R_0 = R_1 \,, \qquad R_{N_H + 1} = R_{N_H} \,.
\end{equation}
The nonlinear two-point boundary value problem 
(\ref{eq:apphortbvp1}--\ref{eq:apphortbvp2}) is solved
using the Newton-Raphson method. At each step of the iteration, we
approximate the Jacobian matrix 
\begin{equation}
  J_{ij} = \frac{\partial f_i}{\partial R_j} 
\end{equation}
numerically by a difference quotient
\begin{equation}
  \label{eq:numjac}
  J_{ij} \approx \tfrac{1}{2 \Delta R} \left[ f_i (R_j + \Delta R) - f_i (R_j
    - \Delta R) \right] \,.
\end{equation}
Fortunately, the discretization (\ref{eq:ahdiscr1}--\ref{eq:ahdiscr2}) yields a
tridiagonal Jacobian matrix, which can be solved exactly in $O(N_H)$
operations using the Thomas algorithm \cite{NR}.

We have tested our apparent horizon finder for a Schwarzschild black 
hole of mass $M$ in isotropic coordinates,
\begin{equation}
  \diff s^2 = - \left(\frac{M - 2 R}{M + 2 R}\right)^2 \diff t^2
  + \left(1 + \frac{M}{2R} \right)^4 (\diff r^2 + \diff z^2 + r^2 \diff
  \varphi^2) \,.
\end{equation}
Its horizon is a sphere of radius $R = M/2$.

\begin{figure}[t]
  \centering
    \begin{minipage}[t]{0.49\textwidth}
      \includegraphics[scale = 1]{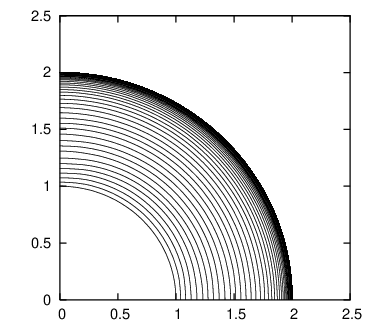}
      \centerline{\footnotesize (a)}
    \end{minipage}
    \hfill  
    \begin{minipage}[t]{0.49\textwidth}
      \includegraphics[scale = 1]{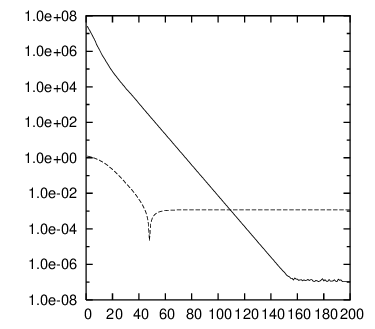}
      \centerline{\footnotesize (b)}
    \end{minipage}
    \hfill  
  \bigskip
  \caption[Test of the apparent horizon finder]
  {\label{fig:apphor}  \footnotesize
  Test of the apparent horizon finder for a Schwarzschild black hole
  of mass $M = 1$ and initial guess $R = 1$ for the horizon radius.
  The horizon grid resolution is 64 points, the data grid resolution
  128 points in each dimension, with $r_\mathrm{max} = z_\mathrm{max} = 5$.
  Shown are 
  (a) the approximation to the horizon shape after each
  Newton iteration and
  (b) the $L^2$ norms of the residual of the apparent horizon equation (solid) 
  and the error with respect to the exact solution (dashed) as functions of 
  the iteration number.}
\end{figure}

Figure \ref{fig:apphor} shows the convergence of the Newton iteration.
Here we have chosen the mass to be $M = 4$ and the initial guess for the
horizon to be a sphere of radius $R_0 = 1$. The algorithm converged for
initial radii $R_0 \in [0.05, 4.5]$. One could probably enlarge the
radius of convergence by including a line search in Newton's method
\cite{NR}.
The radius and rate of convergence turn out to depend 
crucially on the displacement $\Delta R$ used to evaluate the Jacobian 
matrix \eqref{eq:numjac}. The best performance was achieved by
choosing $\Delta R \approx 0.5 N_H{}^{-1} R$.
As seen in figure \ref{fig:apphor}, the final error and residual 
settle down at a constant level after a few iterations. This is mainly
determined by the resolution of the grid holding the (2+1)+1 variables, 
from which the data is interpolated.

If a spacelike slice does not contain an apparent horizon, the approximation
typically shrinks to a point and we stop the iteration once its radius 
is smaller than one grid spacing.

If an apparent horizon is found, one can compute the mass $M$ of the
black hole via the (proper) horizon area $A_H$:
the area radius is defined as
\begin{equation}
  R_A = \sqrt{\frac{A_H}{4 \pi}} 
\end{equation}
and the horizon mass is then given by
\begin{equation}
  \label{eq:horizonmass}
  M = \half R_A \,.
\end{equation}
The area can be calculated as
\begin{equation}
  A_H = 2 \pi \int_{\theta = 0}^{\pi} \lambda \, \diff s \,,
\end{equation}
where $\lambda$ is the norm of the Killing vector 
and $\diff s$ is the line element of the horizon curve,
\begin{equation}
  \diff s^2 = H_{AB} \frac{\diff x^A}{\diff \theta} 
  \frac{\diff x^B}{\diff \theta} \diff \theta^2 \,.
\end{equation}
For our test problem above, the algorithm determined the mass to be
$M = 4.00014$, corresponding to a relative error of $3 \times 10^{-5}$.

One should remark here that formula \eqref{eq:horizonmass} only holds for
non-rotating black holes. For rotating black holes, it has to be
replaced with
\begin{equation}
  M = \frac{1}{2 R_A} \sqrt{R_A^4 + 4 J^4} \,,
\end{equation}
where $J$ is the angular momentum of the black hole (see 
e.g.~\cite{Dreyer03} for a discussion in the isolated horizon framework).
However, axisymmetric initial data on a spacelike slice $\Sigma$ that 
does not contain any trapped surfaces has zero angular momentum in vacuum. 
(Angular momentum in axisymmetry can be defined in an unambiguous way 
by the Komar integral \cite{Komar59} associated with the Killing
vector $\xi$,
\begin{equation}
  J = \kappa^{-1} \oint_{\partial \Sigma} \diff S_{\alpha\beta}
  \nabla^\alpha \xi^\beta \,,
\end{equation}
and a little calculation shows that this vanishes in vacuum by virtue 
of the angular momentum or Geroch constraint \eqref{eq:gercons}).
Angular momentum conservation implies that if a black hole forms 
when evolving such initial data, it must also have zero angular momentum.

With regard to the relation between angular momentum and twist, one should
note that a nonzero angular momentum implies a nonzero twist, but not the
other way around.


  \section{Adaptive collapse simulations}
\label{sec:brillcollapse}

In this final section, we present some evolutions of strong Brill
waves close to the threshold of black hole formation.
The initial data is taken to be that of section \ref{sec:brillini}.
We focus on non-twisting waves here ($A_B = 0$). Amplitudes $A_s$
in the range $4 \leqslant A_s \leqslant 6$ are considered. 
The width of the pulse is taken to be $\sigma_{r,s} = \sigma_{z,s} = 1$.
The same initial data was chosen (in a 3D code) by Alcubierre 
et al.~\cite{Alcubierre00} and (in an axisymmetric code) 
by Garfinkle and Duncan \cite{Garfinkle00}.
The former authors determined the critical amplitude of black hole
formation to be $A^\ast_s = 4.85 \pm 0.15$ and the latter reported 
$4 \leqslant A^\ast_s \leqslant 6$.

The two codes used different gauge conditions (maximal slicing with 
zero shift vs.~maximal slicing with Wilson shift (cf.~section 
\ref{sec:ellgauge})) but the critical amplitude should of course be 
independent of the gauge. Our gauge condition is again different: we
use harmonic slicing ($f = 1, m = 2$) with zero shift.
It is found empirically that by choosing the initial lapse function to have a
slight dip at the origin, $A_\alpha \sim 0.5$ in \eqref{eq:inialpha},
the initial rise of the peak in the gradient $A_r$ of the lapse function is
less drastic (about half the growth rate), making it easier for the
code to cope with this feature.

Adaptive mesh refinement is used with a refinement criterion based on
truncation error estimation as described in section \ref{sec:AMRregrid}.
A reasonable value for the threshold of the $L^2$ norm of the error
appears to be $0.1$. The largest values attained by the variables
during the evolutions are of the order of $10^2$ so that the threshold 
corresponds to a relative error of $\sim 10^{-3}$. 
We have also experimented with ``ad hoc'' refinement indicators such 
as a combination of the quantities $h^2 A_r$ and $h^2 \theta$ 
(some power of the grid spacing $h$ has to be included here so that 
the algorithm does not refine indefinitely).
The first quantity ensures that the gauge peak is tracked during the
evolution, the second one takes highly oscillatory features close to
the origin into account that typically lead to constraint violations,
particularly of the variable $\theta$. 
This refinement criterion gave similar results as the one based on
truncation error estimation. We decided to use the latter in order not to
lose track of features that cannot be controlled with the ``ad hoc'' 
criterion.
Figure \ref{fig:AMRinaction} shows a typical AMR hierarchy.
The resolution of the base grid is taken to be 128 points
in each dimension and up to three levels of refinement are added. This is
the minimum number of levels needed in order to keep all features well 
resolved.
Two levels are used {\it ab initio} in order to keep the
residual of the Hamiltonian constraint at a tolerable level close to
the origin.

\begin{figure}[t]
  \centering
  \includegraphics[scale = 0.45]{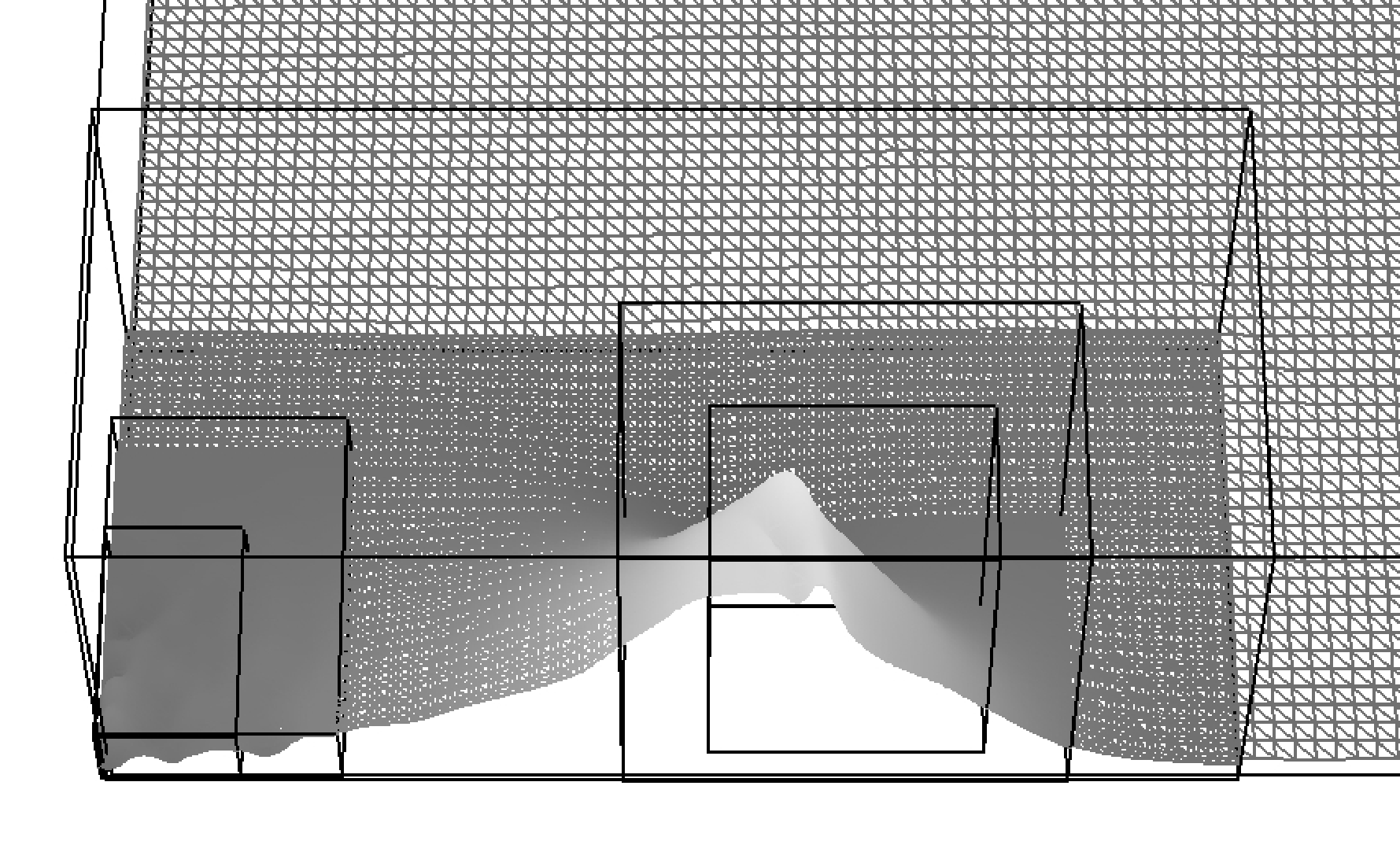}
  \bigskip
  \caption[Example of the AMR hierarchy] 
  {\label{fig:AMRinaction}  \footnotesize
  Typical AMR hierarchy in strong Brill wave evolutions: the
  variable $Y$ at time $t \approx 3$ for the amplitude $A_s = 5$.
  Only a quarter of the base grid is shown. The high resolution region
  on the right coincides with the position of the gauge peak.}
\end{figure}

The outer boundaries are placed at $r_\mathrm{max} = z_\mathrm{max} = 5$.
This is sufficient for supercritical evolutions $A_s \gtrsim 5$,
which do not produce much gravitational radiation because the wave
essentially collapses. For the dispersing waves with $A_s \lesssim 5$,
the results should only be trusted until times $t \sim 10$, after
which the solution becomes dominated by reflections from the outer
boundaries. As explained in section \ref{sec:brillconvtest}, we start with
differential boundary conditions and switch to absorbing ones at $t = 3.5$.

All the parameters -- resolution, location of the outer boundaries,
evolution time -- should be enlarged considerably in the future if more
powerful computer equipment is available. 
The runtime for the strongest wave ($A_s = 6$) presented here was 
$\approx 6$ hours on a 3 GHz single-processor machine, and the code is 
still in the testing phase. The code would have to be parallelized in
order to make efficient use of multi-processor architectures.

The following plots refer to Brill waves with amplitudes $A_s = 4$,
5 and 6. The corresponding ADM masses are $M_\mathrm{ADM} = 0.48$,
$0.67$ and $0.94$.
Figure \ref{fig:clapsekretsch}a shows the logarithm of the lapse
function at the origin as a function of time.
Whereas the lapse eventually returns to its flat-space value for the $A_s = 4$
evolution as the wave disperses, it continues to collapse for the $A_s = 6$
evolution. 
The code could not be run long enough (for reasons discussed below) 
to determine the final fate of the $A_s = 5$ wave.
This qualitative behaviour of the lapse in the three evolutions 
is consistent with the results of section \ref{sec:hypellnum} 
(figure \ref{fig:axistrongclapse}), where a very different
formulation was used. It also agrees with \cite{Alcubierre00} and
\cite{Garfinkle00}.
The claim that the $A_s = 4$ wave disperses and the $A_s = 6$ wave collapses
is further substantiated by figure \ref{fig:clapsekretsch}b,
which shows the evolution of the Kretschmann scalar $I$ 
\eqref{eq:kretschmann} evaluated at the origin. 
This quantity decays at late times 
in the $A_s = 4$ case and grows exponentially in the $A_s = 6$ case, which
indicates that a singularity is approached. The Kretschmann scalar is still
highly oscillatory at the end of the runtime of the near-critical $A_s = 5$
evolution.

\begin{figure}[t]
  \centering
  \begin{minipage}[t]{0.49\textwidth}
    \includegraphics[scale = 1]{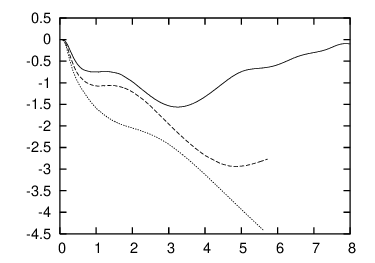}
    \centerline{\footnotesize (a)}
  \end{minipage}
  \hfill  
  \begin{minipage}[t]{0.49\textwidth}
    \includegraphics[scale = 1]{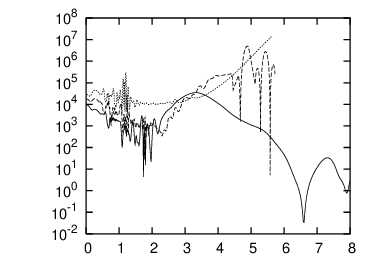}
    \centerline{\footnotesize (b)}
  \end{minipage}
  \bigskip
  \caption[Central lapse and Kretschmann scalar] 
  {\label{fig:clapsekretsch}  \footnotesize
  (a) Logarithm of the lapse function at the origin and 
  (b) Kretschmann scalar at the origin 
  as functions of time for amplitudes $A_s = 4$ (solid), 5 (dashed) and 
  6 (dotted).}
\end{figure}

No apparent horizon was found in the supercritical $A_s = 6$ 
evolution during the runtime of the simulation (until $t \approx 5.7$).
To make sure that this is not caused by bad convergence of the apparent
horizon finder, we used a sequence of circular trial curves spanning 
the entire domain of interest as initial guesses. There appeared to be a 
trend for the average expansion of curves with radius $\approx 1$  to 
decrease but we would have to wait a little longer for it to pass through
zero.
Alcubierre et al.~\cite{Alcubierre00} report the formation of the apparent
horizon at $t = 7.7$ in their coordinates.

In order to see why the simulations crashed, we display the $L^2$ norm of the
$Z$ vector as a function of time in figure \ref{fig:strongz}a (the remaining
constraints behave in a similar way).
For near- and supercritical evolutions, the constraints begin to grow 
exponentially fast at late times. This growth
then affects all the other variables and ultimately leads to a breakdown of
the numerical evolution. 
The growth mainly occurs close to the origin across a rather large spatial
scale (again, this is not a high-frequency instability).
The growth rate is robust under variations of the Courant number (we used
$\Delta t / h = 0.5$ for the results presented here) and of the resolution,
which indicates that we are faced with a continuum instability.
Figure \ref{fig:strongzbdry} demonstrates that the onset and growth rate of
the instability depend only weakly on the location of the outer boundary.
This suggests that the predominant source of the constraint growth is the 
formulation of the equations in the bulk, not the boundary conditions.

\begin{figure}[t]
  \centering
  \begin{minipage}[t]{0.49\textwidth}
    \includegraphics[scale = 1]{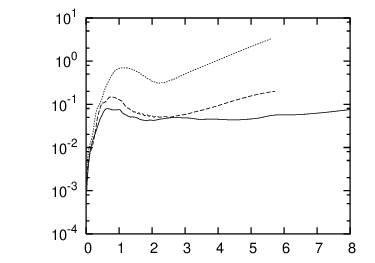}
    \centerline{\footnotesize (a)}
  \end{minipage}
  \hfill  
  \begin{minipage}[t]{0.49\textwidth}
    \includegraphics[scale = 1]{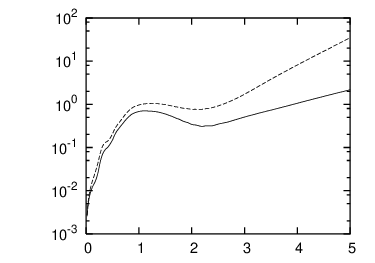}
    \centerline{\footnotesize (b)}
  \end{minipage}
  \bigskip
  \caption[$Z$ constraints and constraint-damping] 
  {\label{fig:strongz}  \footnotesize
  $L^2$ norm of the $Z$ vector as a function of time.
  (a) The $A_s = 4$ (solid), 5 (dashed) and 6 (dotted) evolutions
  with constraint-damping constant $\kappa_\mathrm{CD} = 4$,
  (b) the $A_s = 6$ evolution with $\kappa_\mathrm{CD} = 4$ (solid) and
  0 (dashed).}
\end{figure}

\begin{figure}[t]
  \centering
  \includegraphics[scale = 1]{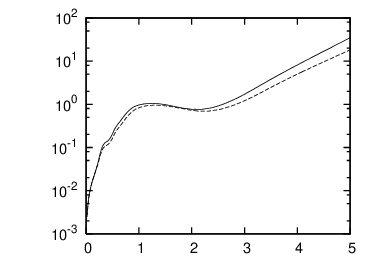}
  \bigskip
  \caption[Dependence of constraint growth on boundary location] 
  {\label{fig:strongzbdry}  \footnotesize
  Dependence of the constraint growth on the location of the outer boundary.
  Shown is the $L^2$ norm of the $Z$ vector as a function of time for a 
  Brill wave with amplitude $A_s = 6$. 
  Solid line: $r_\mathrm{max} = z_\mathrm{max} = 5$,
  dashed line: $r_\mathrm{max} = z_\mathrm{max} = 7.5$. 
  (No constraint damping is included here.)}
\end{figure}

We included constraint-damping terms in the evolution equations as described 
in section \ref{sec:brillini}. 
Figure \ref{fig:strongz}b shows that this does have a positive effect: 
the growth rate of the constraints is smaller if a nonzero
$\kappa_\mathrm{CD} > 0$ is chosen. For large values $\kappa_\mathrm{CD}
\gtrsim 10$, however, instabilities at the outer boundaries quickly developed.
A good compromise appeared to be $\kappa_{CD} \approx 4$.
For no value of $\kappa_\mathrm{CD}$ could constraint-damping eliminate the
exponential growth completely.
We also tried setting $Z^\alpha = 0$ every few timesteps (a simple example
of ``constraint projection'' \cite{Holst04}). However, the increase in the
constraint variables became increasingly rapid after the projections,
again ultimately leading to a blow-up of the numerical solution. 

The development of more sophisticated methods for controlling the growth of 
the constraints in this formulation of the Einstein equations will be crucial 
in order to be able to evolve long enough such that interesting physical
phenomena can be studied.
The work of Abrahams and Evans \cite{Abrahams93} suggests that critical
behaviour will not set in before $t \gtrsim 20$ (although this will be
gauge-dependent).

To close on a more positive note, we demonstrate that we can evolve 
twisting spacetimes as well. Figure \ref{fig:s4b2snaps} shows
a few snapshots of the variable $B^\varphi$ for an evolution with 
amplitudes $A_s = 4$ and $A_B = 2$.
As pointed out in section \ref{sec:ADM}, the evolution equations 
for the twist variables are essentially
Maxwell's equations and as expected, we see a wavelike behaviour,
although a rather complicated one because the twist system 
is now coupled to the remaining evolution equations in a nonlinear way.

\begin{figure}
  \centering
  \begin{minipage}[t]{0.49\textwidth}
    \includegraphics[scale = 0.45]{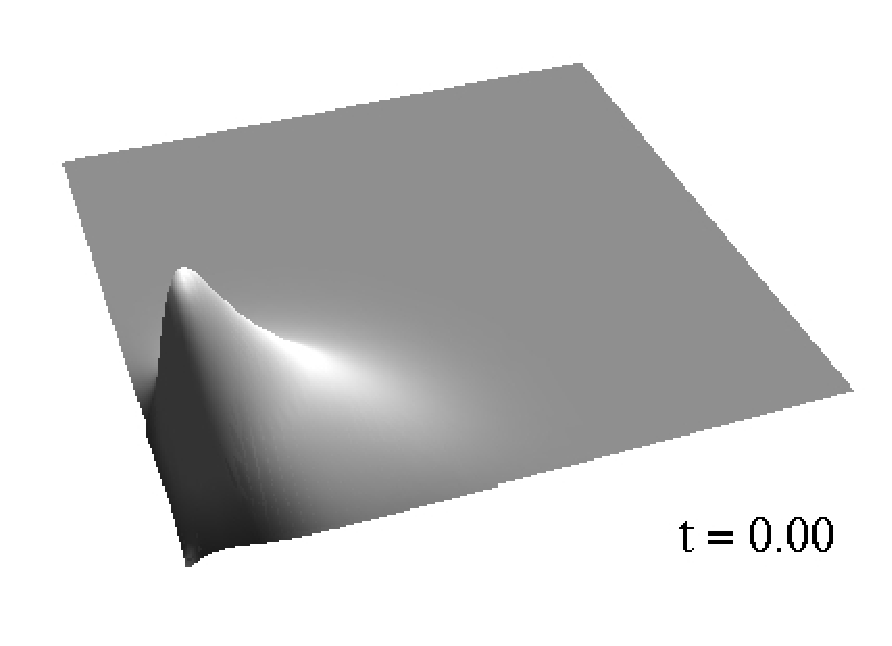}
  \end{minipage}
  \hfill  
  \begin{minipage}[t]{0.49\textwidth}
    \includegraphics[scale = 0.45]{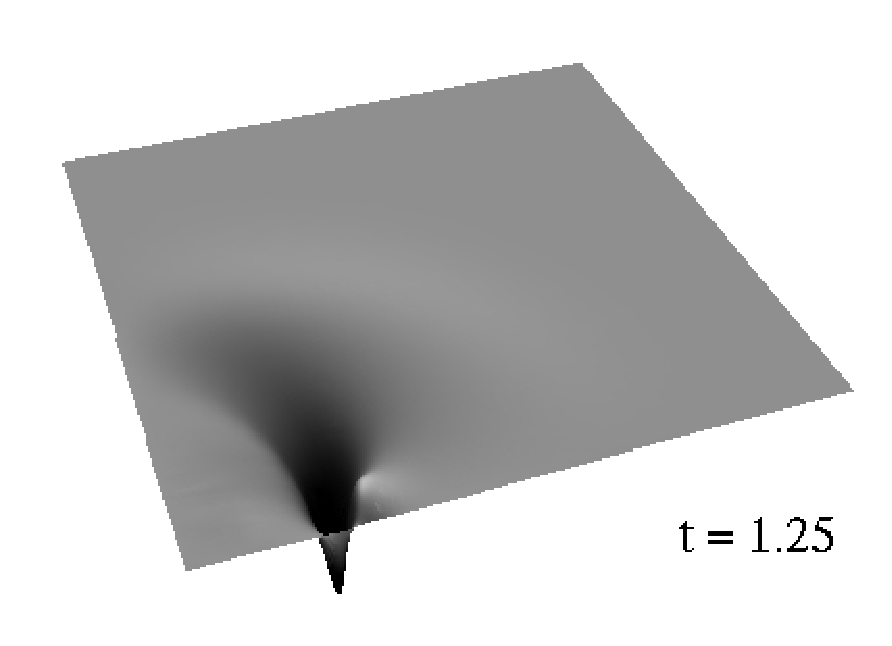}
  \end{minipage}
  \hfill  
  \begin{minipage}[t]{0.49\textwidth}
    \includegraphics[scale = 0.45]{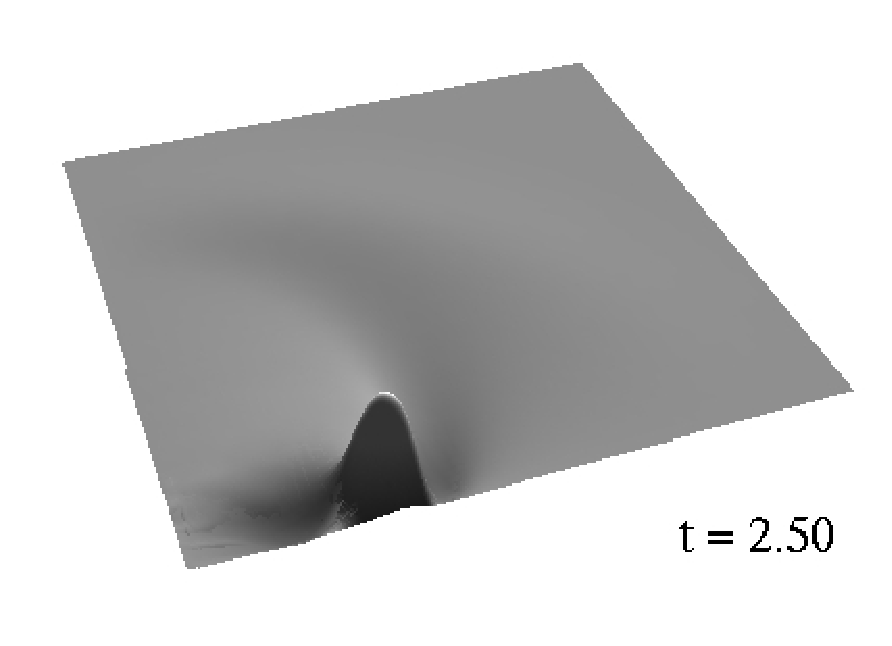}
  \end{minipage}
  \hfill  
  \begin{minipage}[t]{0.49\textwidth}
    \includegraphics[scale = 0.45]{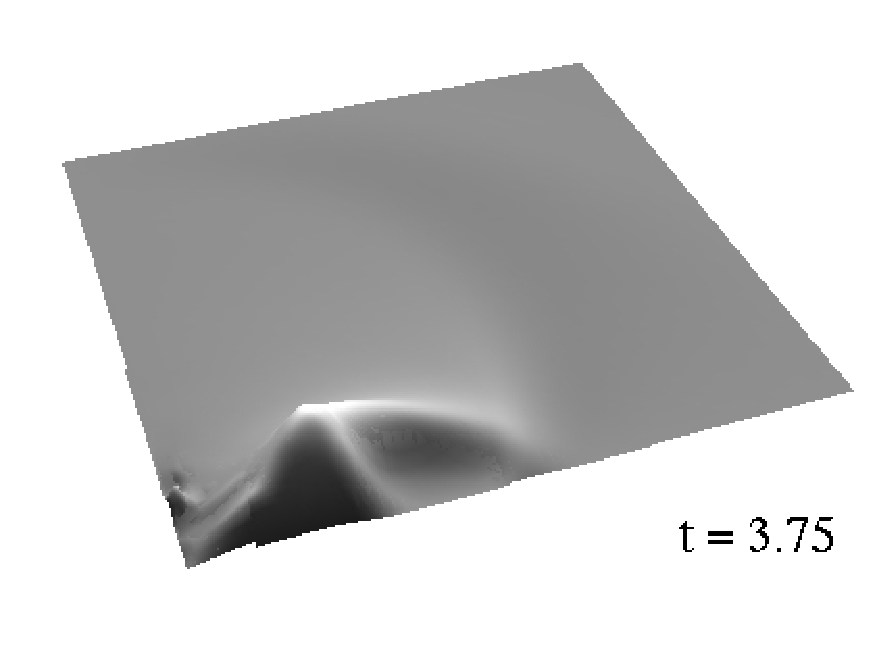}
  \end{minipage}
  \hfill  
  \begin{minipage}[t]{0.49\textwidth}
    \includegraphics[scale = 0.45]{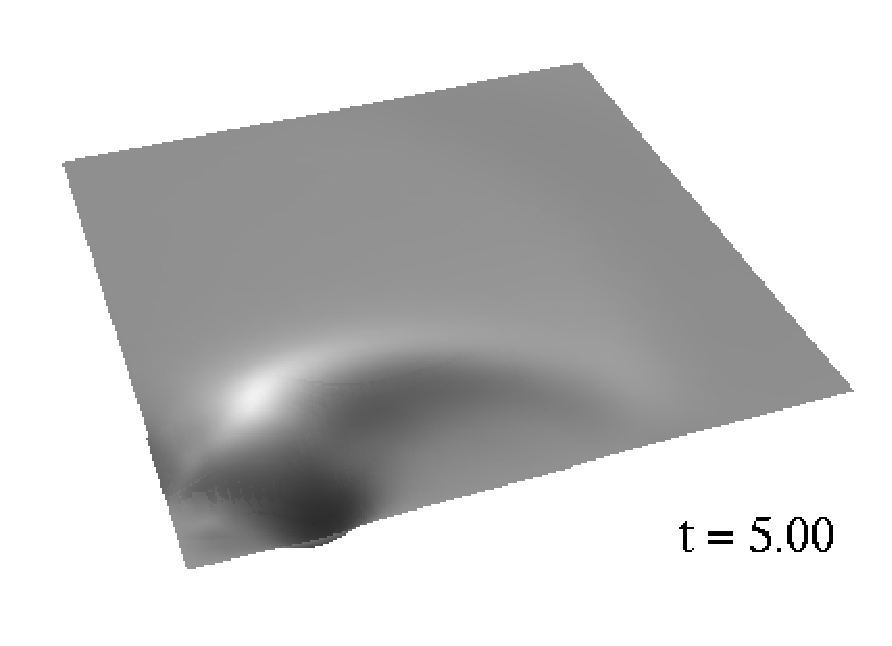}
  \end{minipage}
  \hfill  
  \begin{minipage}[t]{0.49\textwidth}
    \includegraphics[scale = 0.45]{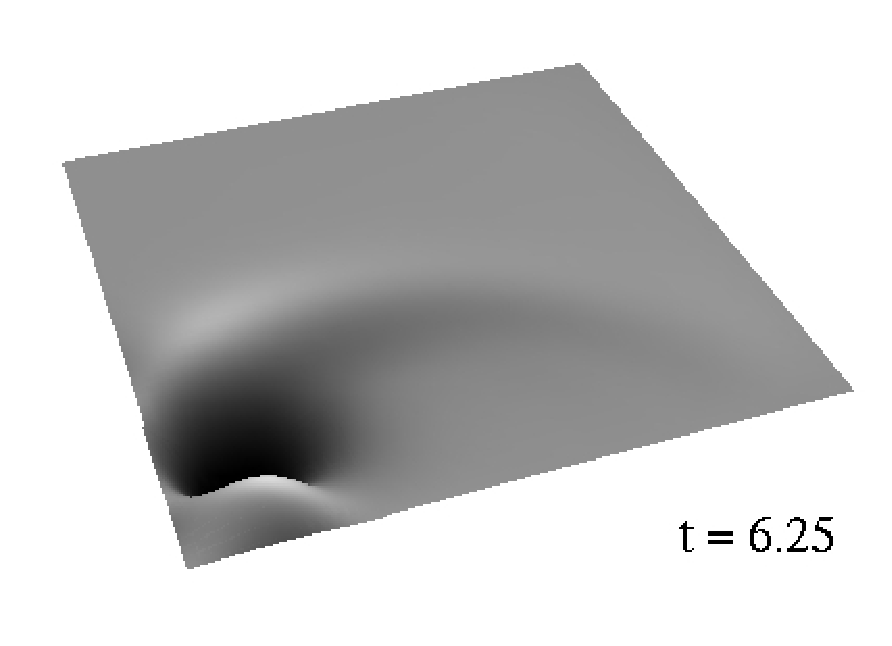}
  \end{minipage}
  \hfill  
  \begin{minipage}[t]{0.49\textwidth}
    \includegraphics[scale = 0.45]{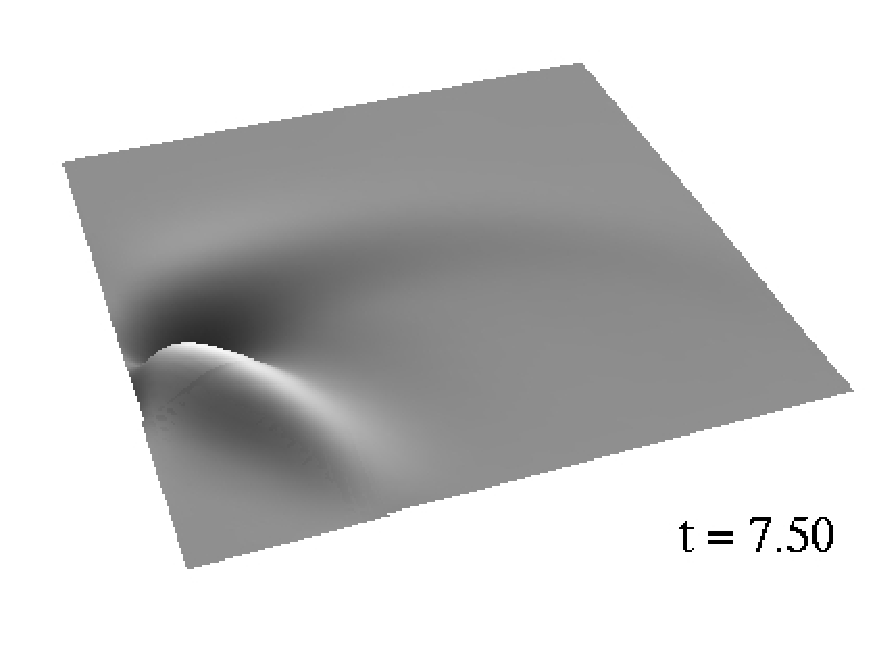}
  \end{minipage}
  \hfill  
  \begin{minipage}[t]{0.49\textwidth}
    \includegraphics[scale = 0.45]{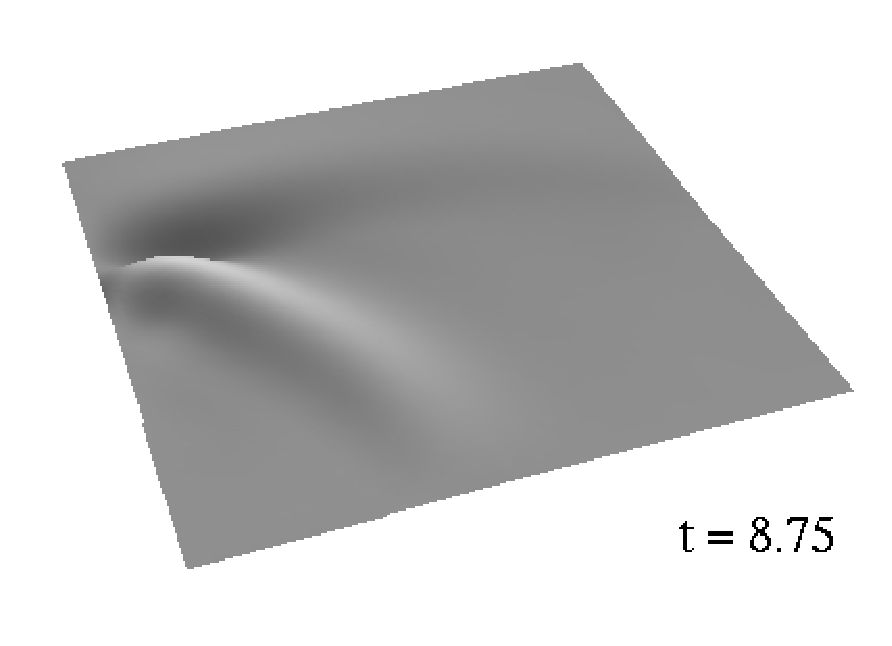}
  \end{minipage}
  \hfill  
  \caption[Snapshots of a twisting Brill wave]
  {\label{fig:s4b2snaps}  \footnotesize
  Evolution of the twist variable $B^\varphi$ for a strong generalized Brill
  wave with amplitudes $A_s = 4$ and $A_B = 2$.} 
\end{figure}


  \chapter{Conclusions and outlook}
\label{sec:concl}

\section{Conclusions}

This thesis has been concerned with formulations of the Einstein equations
suitable for the numerical evolution of axisymmetric spacetimes,
mainly focusing on the vacuum geometry.

We started out by trying to understand why many previous attempts to evolve
these spacetimes failed because of instabilities on the axis.
This led to a detailed study of the behaviour of the components of
axisymmetric tensor fields with respect to cylindrical polar coordinates,
given that the components with respect to Cartesian coordinates were regular
in a neighbourhood of the axis.

In order to exploit the axisymmetry and simplify the system of equations as
much as possible, we first performed a dimensional reduction to the Lorentzian
three-manifold formed by the trajectories of the Killing vector. 
This manifold was then foliated into spacelike hypersurfaces by an ADM 
decomposition, arriving at what is known as the (2+1)+1 formalism
\cite{Maeda80}. We included general matter sources and rotational degrees of
freedom, which have been neglected in previous numerical studies.

The first evolution system we presented adopted elliptic gauge conditions 
arising from maximal slicing and conformal flatness of the two-metric, 
as previously considered in \cite{Eppley79,Garfinkle00,Choptuik03a}. 
The hyperbolic evolution equations were integrated using the method of lines
with second-order finite differencing, and the elliptic equations were solved
using an efficient Multigrid algorithm. 
In strong field situations the Multigrid solver failed when trying solve the
Hamiltonian constraint and the slicing condition during the evolution.
This was explained in terms of a lack of diagonal dominance of the
discretization matrix. In addition, an analytical investigation indicated that
the equations concerned might actually be ill-posed in the sense that they are
not linearization-stable.
If on the other hand we used free evolution, the constraints suffered from a
severe numerical violation. We showed that the constraint evolution system was
in fact ill-posed in this case.
These observations led us to consider a partially constrained evolution 
scheme, in which only the momentum constraints were solved but not the 
Hamiltonian constraint and for which the elliptic equations were well-posed.
Using this modified scheme, we were able to evolve strong Brill waves and
estimated the critical amplitude of black hole formation by looking at
the collapse of the lapse function. For the first time, a nonzero
twist was included.
The runtime of the code for near-critical evolutions is at present 
limited by the resolution. Adaptive mesh refinement would be needed to explore
the critical behaviour more closely.

The problems we experienced with this mixed hyperbolic-elliptic system
motivated the search for a completely hyperbolic formulation of the Einstein
equations. We used the Z4 formalism developed by Bona et al.~\cite{Bona03a}
but applied it to the (2+1)+1 formalism, arriving at what we called the
Z(2+1)+1 system. Generalized harmonic gauge conditions were included, both
with vanishing and dynamical shift vector.
We wrote the equations in first-order form as conservation
laws with sources. The system was shown to be strongly hyperbolic and, for one
choice of parameters, symmetric hyperbolic.
By a judicious choice of dependent variables based on our study of the
behavior of axisymmetric tensor fields, we were able to write the equations in
a form that was well-behaved on the axis and suitable for numerical
evolutions. The incompatibility of the harmonic shift conditions with
axisymmetry was addressed by adding a suitable gauge source function.
 
As a first test problem for our implementation,
we considered exact solutions of linearized theory. The quadrupole waves of
Teukolsky \cite{Teukolsky82} were expressed in terms of Z(2+1)+1 variables
and the two polarization states were understood in terms of twisting and
non-twisting solutions.
In addition, we derived a new even-parity twisting solution with octupolar
angular dependence.
The solutions were shown to satisfy the Z(2+1)+1 equations provided that the
gauge parameters and gauge source functions were chosen such that
transverse-traceless gauge and (generalized) harmonic gauge are compatible.
Second-order convergence of our code to the exact solutions was demonstrated
up to the point when the waves interacted with the boundary.
Whereas the error decayed with time in the vanishing shift case, 
it grew linearly if a dynamical shift was used, the cause of which
would need further investigation.

Next we discussed various choices of outer boundary conditions.
The dissipative boundary conditions we considered included absorbing boundary
conditions and boundary conditions with vanishing $Z$ vector.
A study of the Newman-Penrose scalars and the constraint and gauge propagation 
systems led to a set of differential boundary conditions, where the normal
derivatives of the incoming modes were prescribed.
Whereas dissipative boundary conditions have been proven to be stable subject
to certain restrictions on the hyperbolic system \cite{Rauch85, Secchi96},
those theorems do not apply to the differential boundary conditions.
Hence we analyzed the latter using the Fourier-Laplace technique.
This suggested that our differential boundary conditions were stable in the 
high-frequency limit, although it could not rule out a low-frequency
instability.
Numerical evolutions of the linearized solutions showed that the dissipative
boundary conditions were stable as expected but that (the non-twisting
part of) the differential boundary conditions suffered from a late-time 
instability, which appeared to be a finite-difference instability
rather than a continuum one.
In minimizing spurious reflections from the outer boundaries, the differential
boundary conditions performed better than the absorbing ones, which in turn
outperformed the zero-$Z$ boundary conditions in most cases.

Finally, we turned to nonlinear evolutions of generalized Brill waves
(including twist). Initial data was generated in the same way as for the
hyper\-bolic-elliptic system by requiring that the 2-metric be conformally flat
and that the extrinsic curvature be zero initially. When using harmonic
slicing with zero shift, subcritical initial data of this type evolved to a
nontrivial representation of Minkowski space. Adaptive mesh refinement 
turned out to be essential in order to resolve the highly distorted waveforms 
that occurred as a consequence of this. We showed that for pure harmonic
slicing ($f = 1$), no gauge shocks appeared.
A 3-grid convergence test was carried out for a moderately strong Brill wave,
and a study of the numerical conservation of the ADM mass showed that
differential boundary conditions are superior to those of 
dissipative type in this respect.
Adaptive evolutions of near-critical Brill waves were then performed. 
We obtained bounds on the critical amplitude by looking at the evolution 
of the lapse function and the Kretschmann scalar at the origin. 
These are consistent with the results obtained with our hyperbolic-elliptic 
system and by other authors.
The simulations could not be run long enough yet for an apparent horizon to
form in the supercritical case. The main limitation to the runtime is
currently an exponential growth of the constraints. The inclusion of
constraint-damping terms \cite{Gundlach05} in the evolution equations 
decreased the growth rate but could not eliminate the blow-up completely.

In a sense, the situation is more complicated in Z4-like formulations than
in different approaches because extra constraint variables (the $Z$ vector)
are introduced. Only solutions with $Z = 0$ are solutions of the Einstein
equations, but it is not at all clear whether the constraint manifold $Z = 0$
is an attractor in the fully nonlinear case. On the other hand, terms
homogeneous in the $Z$ vector can easily be added to the evolution equations
without affecting the characteristic structure (in general, this is not 
possible in more conventional approaches). It is quite possible that constraint
additions will be found in the future that eliminate the constraint blow-up
completely.

\section{Outlook on future work}

Once the growth of the constraints is under control, we will hopefully be able
to tune closer to the critical point of black hole formation and to evolve 
long enough so that the potentially interesting physics that occurs at the 
threshold can be studied.
Apart from comparing with the results of Abrahams and Evans \cite{Abrahams93,
  Abrahams94}, we would like to find out whether a nonzero twist might 
modify the critical behaviour. Another question that should be addressed is
whether highly distorted initial data can lead to the formation of naked
singularities in Brill wave collapse \cite{Abrahams92,Garfinkle00}.

Our long-term goal is to include matter in the form of a perfect fluid.
This is much more interesting physically because perfect fluid spacetimes
can carry angular momentum. When studying the gravitational collapse of a
rotating fluid, the question arises what happens with the angular momentum at
the threshold of black hole formation.
A perturbation analysis by Gundlach \cite{Gundlach98} predicts that for a
slightly non-spherical and slowly rotating fluid, the critical solution 
will be the spherically symmetric one, which for the ultrarelativistic 
equation of state was found by Evans and Coleman \cite{Evans94}.
According to Gundlach, the angular momentum in supercritical evolutions 
will obey a similar power-law as the black hole mass \eqref{eq:scaling},
and an expression for the dependence of the angular momentum exponent 
on the mass exponent and the equation of state has been derived.
It would be very interesting to probe those results numerically.


   \appendix

   \chapter{Perfect fluid}
\label{sec:fluid}

In section \ref{sec:generalmatter}, we derived the evolution equations
for general matter in the (2+1)+1 formalism. Here, we specify the
matter model to be a perfect fluid. We write the equations in
conservation form and work out their characteristic decomposition.
The transformation from conserved to primitive variables is cast in a
form that helps avoid superluminal speeds in numerical simulations.

\section{Conservation form}
\label{sec:matter_cons}

The matter evolution equations 
(\ref{eq:d0rhoH}--\ref{eq:d0Jp}, \ref{eq:d0sigma}) can clearly be written 
in conservation form (with sources and a common advection term),
\begin{equation}
  \label{eq:matter_consform}  
  \partial_t \vec u 
  + \left[ - \beta^A \vec u + \alpha \vec \F^D (\vec u) \right]_{,D} 
  = \alpha \vec \S (\vec u) \,.
\end{equation}
Following \cite{Banyuls97}, 
we replace $\rho_H$ with  $\rho_K = \rho_H - \sigma$ 
(kinetic energy) and regard as the set of conserved variables
\begin{equation}
  \label{eq:matterconsvars}
  \vec u = (\rho_K, J_A, J^\varphi, \sigma)^T \,.
\end{equation}
The fluxes are
\begin{eqnarray}
  \F^D{}_{\rho_K} &=& J^D - \Sigma^D \,,\\
  \F^D{}_{J_A} &=& S_A{}^D \,,\\
  \F^D{}_{J^\varphi} &=& S^D \,,\\
  \F^D{}_{\sigma} &=& \Sigma^D \,,
\end{eqnarray}
and the source terms are
\begin{eqnarray}
  \S_{\rho_K} &=& (\Sigma^A - J^A)(D^\mathrm{I}{}_A + L_A)  
    + \K (\tau + \rho_K) + \chi_{AB} S^{AB}  \nonumber\\ &&
    - J^A A_A + \chi \rho_K + \lambda^2 E^A S_A \,,\\
  \S_{J_A} &=&  - S_{AB} (A^B + L^B) + J_A (\chi + \K) + 2 B_A{}^B J_B
    \nonumber\\&& - A_A \rho_H + L_A \tau 
    + \lambda^2 (E_A J^\varphi + \epsilon_{AB} S^B B^\varphi) \,,\\
  \S_{J^\varphi} &=& - (D^\mathrm{I}{}_A + 3 L_A) S^A + J^\varphi (\chi + 3 \K)
    \,,\\
  \S_{\sigma} &=& - (D^\mathrm{I}{}_A + L_A) \Sigma^A + \sigma (\chi + \K) \,. 
\end{eqnarray}
We use the notation of section \ref{sec:firstorder} for the first-order
derivatives of the metric.

\section{Matter model}
\label{sec:matter_fluid}

To evaluate the characteristic structure, we need to specify the matter model.
Here, we consider a perfect fluid with four-velocity $u^\alpha$,
normalized such that 
\begin{equation}
  u_\alpha u^\alpha = -1 \,,
\end{equation}
rest mass density $\rho$, pressure $p$ and internal energy $\epsilon$. 
The dependence of the pressure on the density and the internal energy is
given by the \emph{equation of state} 
\begin{equation}
  \label{eq:EOS}
  p = p(\rho, \epsilon) \,.
\end{equation}
With those definitions, the number density is
\begin{equation}
  N^\alpha = \rho u^\alpha
\end{equation}
and the energy-momentum tensor is given by
\begin{equation}
  T^{\alpha\beta} = \rho h u^\alpha u^\beta + p g^{\alpha\beta} \, ,
\end{equation}
where $h$ is the specific enthalpy,
\begin{equation}
  \label{eq:enthalpy}
  h = 1 + \epsilon + \frac{p}{\rho} \, .
\end{equation}
The Lorentz factor is defined as
\begin{equation}
  \label{eq:Lorentz}
  W \equiv - u^\alpha n_\alpha \, .
\end{equation}
Observers who are at rest in a slice $\Sigma(t)$ (i.e., who have
four-velocity $n^\alpha$) measure a coordinate velocity
\begin{equation}
  v^A = W^{-1}  h_\alpha{}^A u^\alpha \, ,
\end{equation}
and the angular velocity is
\begin{equation}
  v^\varphi =  W^{-1} \lambda^{-2} \xi_\alpha u^\alpha \,.
\end{equation}
Hence we obtain the familiar relation
\begin{equation}
  \label{eq:Lorentzrelation}
  W = (1 - v^2)^{-1/2} \,,
\end{equation}
where
\begin{equation}
  v^2 =  v_A v^A + \lambda^2 {v^\varphi}^2 \,.
\end{equation}

The variables
\begin{equation}
  \vec w = (v_A, v^\varphi, \rho, p, \epsilon, h, W)^T
\end{equation}
are called \emph{primitive variables}. Note only five of
these are independent because of \eqref{eq:EOS}, \eqref{eq:enthalpy}
and \eqref{eq:Lorentzrelation}.
The conserved variables can be expressed in terms of the primitive 
variables as
\begin{eqnarray}
  \label{eq:prim2cons}
  \rho_K &=& \rho h W^2 - p - \rho W \,,\nonumber\\
  J_A &=& \rho h W^2 v_A \,,\nonumber\\
  J^\varphi &=& \rho h W^2 v^\varphi \,,\\
  \sigma &=& \rho W \,,\nonumber
\end{eqnarray}
and the remaining matter variables are
\begin{eqnarray}
  \label{eq:prim2noncons}
  \tau &=& \rho h W^2 \lambda^2 {v^\varphi}^2 + p \,,\nonumber\\
  S_A &=& \rho h W^2 v^\varphi v_A \,,\nonumber\\
  S_{AB} &=& \rho h W^2 v_A v_B + p H_{AB} \,,\\
  \Sigma_A &=& \rho W v_A \,.\nonumber
\end{eqnarray}

\section{Characteristic decomposition}
\label{sec:matter_char}

The characteristic decomposition for 3+1 general relativistic hydrodynamics
was first derived by the Valencia group \cite{Banyuls97}. 
The application to our (2+1)+1 system is straightforward.
Note however the additional source terms that occur in our case.
Our method differs slightly from \cite{Banyuls97} in that we choose a 
general orthonormal basis $(\mu^A, \pi^A)$ in two-space as in section 
\ref{sec:hyperbolicity} and project vectors along $\mu$ (index $\n$) 
and $\pi$ (index $\p$). 

Following the notation of \cite{Font00}, we introduce a few abbreviations. 
From the equation of state \eqref{eq:EOS}, we form
\begin{equation}
  \chi \equiv \frac{\partial p}{\partial \rho} \, , \qquad \kappa \equiv
  \frac{\partial p}{\partial \epsilon} \, , \qquad h c_s^2 \equiv \chi +
  \frac{p}{\rho^2} \kappa \, , 
\end{equation}
where $c_s$ is known as the \emph{sound speed}. 
Also set
\begin{equation*}
  \Kf^{-1} = 1 - \frac{c_s^2 \rho}{\kappa} \,, \qquad
  \V^\pm = \frac{v_\n - \lambda_s^\pm}{1 - v_\n \lambda_s^\pm} 
    \,, \qquad
  \A^\pm = \frac{1 - v_\n^2}{1 - v_\n \lambda_s^\pm} \,,
\end{equation*}
\begin{equation}
  \C^\pm = v_\n - \V^\pm \,, \qquad
  \xi = 1 - v_\n^2 \,, 
\end{equation}
\begin{equation*}
  \Delta = h^3 W (1 - \Kf^{-1})(\C^+ - \C^-) \xi \,.\nonumber
\end{equation*}
Our definitions of $\xi$
and $\Delta$ differ from those in \cite{Font00} by a factor of
$\lambda^2$ to ensure regularity on axis.
We have defined $\Kf^{-1}$ instead of $\Kf$ to allow for the special
case of the ultrarelativistic equation of state (\ref{eq:EOS_spec}), 
for which $\Kf^{-1} = 0$.
As a consequence, $\Delta$ above has been multiplied by $\Kf^{-1}$ and the 
characteristic variable $l_{0,1}$ has been divided by $\Kf^{-1}$.

The system is found to be strongly hyperbolic.
The characteristic speeds in the $\mu$-direction are 
\begin{eqnarray}
  \lambda_0 &=& v_\n \, , \nonumber\\
  \lambda_s^\pm &=& \frac{1}{1 - v^2 c_s^2 } \Big\{ v_\n (1-c_s^2) 
     \nonumber\\&&\qquad\qquad \pm c_s \sqrt{ (1 - v^2)
     \left[ (1 - v^2 c_s^2) - v_\n^2 (1 - c_s^2) \right] } \Big\} \, .
\end{eqnarray}
The characteristic variables $l$ (corresponding to the left eigenvectors) are
{\allowdisplaybreaks \begin{eqnarray}
  l_{0,1} &=& \frac{W}{1 - \Kf^{-1}} \Big\{ h \sigma - W (\sigma + \rho_K)
    \nonumber\\&& \qquad \qquad
    + W (v_\n J_\n + v_\p J_\p + \lambda^2 v^\varphi J^\varphi) \Big\} \,,\\
  l_{0,2} &=& \frac{1}{h \xi} \left\{ - v_\p (\sigma + \rho_K) 
    + v_\n v_\p J_\n + (1 - v_\n^2) J_\p \right\} \,, \\
  l_{0,3} &=& \frac{1}{h \xi} \left\{ - v^\varphi (\sigma + \rho_K)
    + v^\varphi v_\n J_\n + (1 - v_\n^2) J^\varphi \right\} \,,
\end{eqnarray}}
\begin{eqnarray}
  l_s^\mp &=& \frac{h^2}{\Delta} \Big\{ \Kf^{-1} h W \V^\pm \xi \sigma
    + \left[ \Kf^{-1} - \A^\pm - (2 - \Kf^{-1}) v_\n \right] J_\n 
    \nonumber\\ && \qquad
    + (2 - \Kf^{-1}) \V^\pm W^2 \xi (v_\n J_\n + v_\p J_\p 
    + \lambda^2 v^\varphi J^\varphi) \\ && \qquad
    + \left[ (\Kf^{-1} - 1) \left( - v_\n + \V^\pm (W^2 \xi - 1) \right) 
    \right. \nonumber\\&&\qquad \qquad \qquad \qquad \qquad \qquad
    \left. - W^2 \V^\pm \xi \right] (\sigma + \rho_K) \Big\} \,.
\end{eqnarray}
The inverse transformation (corresponding to the right eigenvectors)
is given by
{\allowdisplaybreaks\begin{eqnarray}
  \sigma &=& \frac{1}{h W} l_{0,1} + W(v_\p l_{0,2} + \lambda^2
    v^\varphi l_{0,3}) + l_s^+ + l_s^- \,,\\
  J_\n &=& \Kf^{-1} v_\n l_{0,1} + 2 h W^2 v_\n (v_\p l_{0,2} + \lambda^2
    v^\varphi l_{0,3}) \nonumber\\&&
    + h W (\C^+ l_s^+ + \C^- l_s^- ) \,,\\
  J_\p &=& \Kf^{-1} v_\p l_{0,1} + h l_{0,2} + 2 h W^2 v_\p (v_\p l_{0,2} +
    \lambda^2 v^\varphi l_{0,3}) \nonumber\\&&
    + h W v_\p (l_s^+ + l_s^-) \,,\\
  J^\varphi &=& \Kf^{-1} v^\varphi l_{0,1} + h l_{0,3} + 2 h W^2 v^\varphi 
    (v_\p l_{0,2} + \lambda^2 v^\varphi l_{0,3}) \nonumber\\&&
    + h W v^\varphi (l_s^+ + l_s^-) \,,\\
  \rho_K &=& \left( \Kf^{-1} - \frac{1}{h W} \right) l_{0,1} + W
    (2 h W - 1) (v_\p l_{0,2} + \lambda^2 v^\varphi l_{0,3}) \nonumber\\&&
    + h W (\A^+ l_s^+ + \A^- l_s^-) - l_s^+ - l_s^- \,.
\end{eqnarray}}

\section[From conserved to primitive variables]
{Transformation from conserved to primitive variables}
\label{sec:matter_trafo}

The conserved matter variables \eqref{eq:matterconsvars} are the ones
that are evolved in a numerical algorithm. To compute the remaining
matter variables \eqref{eq:prim2noncons} and the eigenvectors, 
the primitive variables have to be calculated from the conserved
variables in an intermediate step.
This transformation is much more involved than the opposite
direction \eqref{eq:prim2cons}.
To make it explicit, we have to specify an equation of state. 
Here, we consider the ultrarelativistic equation of state,
\begin{equation}
  \label{eq:EOS_spec}
  p = (\Gamma - 1) \rho_{tot} = (\Gamma - 1) \rho (\epsilon + 1)
    = \frac{\Gamma - 1}{\Gamma} \rho h \,,
\end{equation}
where $\rho_{tot}$ is the total energy density.

Suppose we are given the conserved variables, and also form $\rho_H =
\rho_K + \sigma$. Consider the quantity
\begin{equation}
  J^2 \equiv J_A J^A + \lambda^2 {J^\varphi}^2 \, .
\end{equation}
Using \eqref{eq:prim2cons}, \eqref{eq:EOS_spec} and
\eqref{eq:Lorentzrelation},
we can express $J^2$ and $\rho_H$ in terms of the primitive variables as
\begin{eqnarray}
  J^2 &=& \left( \frac{\Gamma}{\Gamma-1} \right)^2 p^2 W^2 (W^2 -1)
    \, , \nonumber\\
  \rho_H &=& p \, \left(\frac{\Gamma}{\Gamma-1} W^2 - 1\right) \, .
\end{eqnarray}
Eliminating $W$ yields an equation for the pressure in terms of
conserved variables:
\begin{equation}
  p = - 2 \beta \rho_H + \sqrt{4 \beta^2 \rho_H^2 +
  (\Gamma-1)(\rho_H^2 - J^2)} \, ,
\end{equation}
where $\beta \equiv (2 - \Gamma)/4$.
Next define
\begin{equation}
  \chi_A \equiv \frac{(\Gamma - 1) J_A}{\Gamma p} \, , \qquad
  \chi^\varphi \equiv \frac{(\Gamma - 1) J^\varphi}{\Gamma p} \, , \qquad
  \chi^2 \equiv \chi_A \chi^A + \lambda^2 {\chi^\varphi}^2 \, .
\end{equation}
We identify $\chi^A = W^2 v^A$ and $\chi^\varphi = W^2 v^\varphi$ and
hence with \eqref{eq:Lorentzrelation} we obtain
\begin{equation}
  \label{eq:Wfromcons}
  W^{-2} = \frac{1}{2 \chi^2} \left( \sqrt{1 + 4 \chi^2}
  - 1 \right) \, .
\end{equation}
This now enables us to calculate the velocities,
\begin{equation}
  v_A = W^{-2} \chi_A \, , \qquad v^\varphi = W^{-2} \chi^\varphi \, .
\end{equation}
The form of $W^{-2}$ in \eqref{eq:Wfromcons} guarantees that $|v_A|,
\, |v^\varphi| \leqslant 1$. This is most important since evolved speeds
greater than unity (i.e., greater than the speed of light) can easily
cause the numerical code to crash.
  
Finally, we can calculate the specific enthalpy and rest mass
energy density from \eqref{eq:prim2cons} and \eqref{eq:EOS_spec},
\begin{equation}
  h = \frac{J^\varphi}{\sigma v^\varphi W} \, , \qquad 
  \rho = \frac{\Gamma p}{(\Gamma-1)h} \, .
\end{equation}

A similar method of calculating the primitive variables to the one
described here is used by 
Choptuik and Neilsen \cite{Neilsen00a, Neilsen00b} 
and Hawke \cite{HawkePhD}.


   \chapter{Regularized conservation form}
\label{sec:regconsformdetails}

In this appendix we write out the fluxes and sources of the
regularized conservation form \eqref{eq:regconsform1} 
of the Z(2+1)+1 equations. The equations were
generated with the computer algebra language REDUCE \cite{REDUCE},
from which we created LaTeX code using the TeX-REDUCE Interface TRI
\cite{TRI}.

To demonstrate regularity on axis, we define
\begin{equation}
  \underline{\hat u} \equiv r^{-1} \underline{u}
\end{equation}
for a variable $\underline{u}$ that is $O(r)$ on the axis
(see table \ref{tab:axizpar}).
In terms of the hatted variables, the fluxes and sources are
manifestly regular on axis. They are either even or odd functions of $r$.
As a shorthand, we introduce
\begin{equation}
  H \equiv H_{rr} H_{zz} - H_{rz}{}^2
\end{equation}
for the determinant of the 2-metric.
We use the minimal gauge source function (\ref{eq:Gr}).

\section{Fluxes in the $r$ direction}
$$\displaylines{\qdd
\tilde \F^r 
\(H_{rr} 
\)
=
\[0
\]
\cr}$$
$$\displaylines{\qdd
\tilde \F^r 
\(H_{rz} 
\)
=
\[0
\]
\cr}$$
$$\displaylines{\qdd
\tilde \F^r 
\(H_{zz} 
\)
=
\[0
\]
\cr}$$
$$\displaylines{\qdd
\tilde \F^r 
\(s 
\)
=
\[0
\]
\cr}$$
$$\displaylines{\qdd
\tilde \F^r 
\(\alpha 
\)
=
\[0
\]
\cr}$$
$$\displaylines{\qdd
\tilde \F^r 
\({\beta^r} 
\)
=
\[0
\]
\cr}$$
$$\displaylines{\qdd
\tilde \F^r 
\({\beta^z} 
\)
=
\[0
\]
\cr}$$
$$\displaylines{\qdd
\tilde \F^r 
\(D_{rrr} 
\)
=
\[\alpha \chi_{rr} 
  -2\alpha r^{2}
  \({\hat {\tilde B}_r{}^r} H_{rr} 
    +{\hat B_r{}^z} \hat H_{rz} 
  \)
  -\hat D_{rrr} r^{2}\hat {\beta^r} 
\]
\cr}$$
$$\displaylines{\qdd
\tilde \F^r 
\(\tilde D_{rrz} 
\)
=
\[\alpha 
  \(-r^{2}{\hat {\tilde B}_r{}^r} \hat H_{rz} 
    -{\hat B_r{}^z} H_{zz} \nl 
    \off{5396347}
    -{\hat B_z{}^r} H_{rr} 
    -{B_z{}^z} \hat H_{rz} 
    +\hat \chi_{rz} 
  \)
  -\hat {\tilde D}_{rrz} r^{2}\hat {\beta^r} 
\]
\Nl}$$
$$\displaylines{\qdd
\tilde \F^r 
\(D_{rzz} 
\)
=
\[\alpha \chi_{zz} 
  -2\alpha 
  \(r^{2}{\hat B_z{}^r} \hat H_{rz} 
    +{B_z{}^z} H_{zz} 
  \)
  -\hat D_{rzz} r^{2}\hat {\beta^r} 
\]
\cr}$$
$$\displaylines{\qdd
\tilde \F^r 
\(D_{zrr} 
\)
=
\[-D_{zrr} r\hat {\beta^r} 
\]
\cr}$$
$$\displaylines{\qdd
\tilde \F^r 
\(D_{zrz} 
\)
=
\[-\hat D_{zrz} r^{2}\hat {\beta^r} 
\]
\cr}$$
$$\displaylines{\qdd
\tilde \F^r 
\(D_{zzz} 
\)
=
\[-D_{zzz} r\hat {\beta^r} 
\]
\cr}$$
$$\displaylines{\qdd
\tilde \F^r 
\(\tilde s_r 
\)
=
\[\alpha 
  \(2{\hat {\tilde B}_r{}^r} 
    +2{\hat B_r{}^z} H_{rr} ^{-1}\hat H_{rz} 
    +\hat Y 
  \)
  -r^{2}\hat {\beta^r} \hat {\tilde s}_r 
\]
\cr}$$
$$\displaylines{\qdd
\tilde \F^r 
\(s_z 
\)
=
\[-r^{2}\hat {\beta^r} \hat s_z 
\]
\cr}$$
$$\displaylines{\qdd
\tilde \F^r 
\(A_r 
\)
=
\[2\alpha \chi_{rr} fH_{rr} ^{
  -1}
  +H^{-1}\alpha \chi_{rr} r^{2}fH_{rr} ^{
  -1}\hat H_{rz} ^{2}
  +H^{-1}\alpha \chi_{zz} fH_{rr} \nl 
  \off{4069301}
  +\alpha f
  \(r^{2}\hat Y 
    -m\theta 
  \)
  -2H^{-1}\alpha r^{2}\hat \chi_{rz} f\hat H_{rz} 
  -r^{2}\hat A_r \hat {\beta^r} 
\]
\Nl}$$
$$\displaylines{\qdd
\tilde \F^r 
\(A_z 
\)
=
\[-rA_z \hat {\beta^r} 
\]
\cr}$$
$$\displaylines{\qdd
\tilde \F^r 
\({\tilde B_r{}^r} 
\)
=
\[H^{-1}\alpha \hat D_{rrr} H_{rr} ^{
  -1}H_{zz} 
  \(-d
    +\mu 
  \)
  \nl 
  \off{4245275}
  -
  \frac{1}{
        2}H^{-2}\alpha \hat D_{rrr} r^{2}dH_{rr} ^{
  -1}\hat H_{rz} ^{2}H_{zz} 
  +H^{-1}\alpha \hat {\tilde D}_{rrz} r^{2}dH_{rr} ^{
  -1}\hat H_{rz} \nl 
  \off{4245275}
  +H^{-2}\alpha \hat {\tilde D}_{rrz} r^{4}dH_{rr} ^{
  -1}\hat H_{rz} ^{3}
  +H^{-1}\alpha \hat D_{rzz} 
  \(-
    \frac{1}{
          2}d
    +\mu 
  \)
  \nl 
  \off{4245275}
  -
  \frac{1}{
        2}H^{-2}\alpha \hat D_{rzz} r^{2}d\hat H_{rz} ^{2}
  +H^{-1}\alpha D_{zrr} H_{rr} ^{
  -1}\hat H_{rz} 
  \(d
    -\mu 
  \)
  \nl 
  \off{4245275}
  +
  \frac{1}{
        2}H^{-2}\alpha D_{zrr} r^{2}dH_{rr} ^{
  -1}\hat H_{rz} ^{3}
  -H^{-1}\alpha \hat D_{zrz} \mu \nl 
  \off{4245275}
  -H^{-2}\alpha \hat D_{zrz} r^{2}d\hat H_{rz} ^{2}
  +
  \frac{1}{
        2}H^{-2}\alpha D_{zzz} dH_{rr} \hat H_{rz} \nl 
  \off{4245275}
  +\alpha H_{rr} ^{-1}
  \(-
    \frac{1}{
          2}r^{2}d\hat {\tilde s}_r 
    +r^{2}\mu \hat {\tilde s}_r 
    +
    \frac{1}{
          2}a\hat A_r 
    -\mu \hat Z_r 
  \)
  \nl 
  \off{4245275}
  +H^{-1}\alpha \hat H_{rz} 
  \(-
    \frac{1}{
          2}r^{4}dH_{rr} ^{-1}\hat H_{rz} 
    \hat {\tilde s}_r 
    +r^{4}H_{rr} ^{-1}\hat H_{rz} \mu \hat {\tilde s}_r \nl 
    \off{7852337}
    +
    \frac{1}{
          2}r^{2}a\hat A_r H_{rr} ^{
    -1}\hat H_{rz} 
    +
    \frac{1}{
          2}r^{2}d\hat s_z 
    -r^{2}H_{rr} ^{-1}\hat H_{rz} \mu \hat Z_r \nl 
    \off{7852337}
    -r^{2}\mu \hat s_z 
    -
    \frac{1}{
          2}aA_z 
    +\mu Z_z 
  \)
  -r^{2}\hat {\beta^r} {\hat {\tilde B}_r{}^r} 
\]
\Nl}$$
$$\displaylines{\qdd
\tilde \F^r 
\({B_r{}^z} 
\)
=
\[H^{-1}\alpha \hat D_{rrr} r^{2}H_{rr} ^{
  -1}\hat H_{rz} 
  \(d
    -\mu 
  \)
  \nl 
  \off{4257064}
  +
  \frac{1}{
        2}H^{-2}\alpha \hat D_{rrr} r^{4}dH_{rr} ^{
  -1}\hat H_{rz} ^{3}
  -H^{-1}\alpha \hat {\tilde D}_{rrz} r^{2}\mu \nl 
  \off{4257064}
  -H^{-2}\alpha \hat {\tilde D}_{rrz} r^{4}d\hat H_{rz} ^{2
  }
  +
  \frac{1}{
        2}H^{-2}\alpha \hat D_{rzz} r^{2}dH_{rr} 
  \hat H_{rz} \nl 
  \off{4257064}
  +H^{-1}\alpha D_{zrr} 
  \(-d
    +2\mu 
  \)
  -
  \frac{1}{
        2}H^{-2}\alpha D_{zrr} r^{2}d\hat H_{rz} ^{2}\nl 
  \off{4257064}
  +H^{-2}\alpha \hat D_{zrz} r^{2}dH_{rr} \hat H_{rz} 
  -
  \frac{1}{
        2}H^{-2}\alpha D_{zzz} dH_{rr} ^{2}\nl 
  \off{4257064}
  +H^{-1}\alpha 
  \(\frac{1}{
          2}r^{4}d\hat H_{rz} \hat {\tilde s}_r 
    -r^{4}\hat H_{rz} \mu \hat {\tilde s}_r 
    -
    \frac{1}{
          2}r^{2}a\hat A_r \hat H_{rz} \nl 
    \off{6759947}
    -
    \frac{1}{
          2}r^{2}dH_{rr} \hat s_z 
    +r^{2}H_{rr} \mu \hat s_z 
    +r^{2}\hat H_{rz} \mu \hat Z_r \nl 
    \off{6759947}
    +
    \frac{1}{
          2}aA_z H_{rr} 
    -H_{rr} \mu Z_z 
  \)
  -r^{2}\hat {\beta^r} {\hat B_r{}^z} 
\]
\Nl}$$
$$\displaylines{\qdd
\tilde \F^r 
\({B_z{}^r} 
\)
=
\[-r^{2}\hat {\beta^r} {\hat B_z{}^r} 
\]
\cr}$$
$$\displaylines{\qdd
\tilde \F^r 
\({B_z{}^z} 
\)
=
\[-r\hat {\beta^r} {B_z{}^z} 
\]
\cr}$$
$$\displaylines{\qdd
\tilde \F^r 
\(\chi_{rr} 
\)
=
\[\alpha \hat D_{rrr} rH_{rr} ^{
  -1}
  +H^{-1}\alpha \hat D_{rzz} rH_{rr} 
  +H^{-1}\alpha D_{zrr} r\hat H_{rz} \nl 
  \off{4245275}
  -2H^{-1}\alpha \hat D_{zrz} rH_{rr} 
  +\alpha r
  \(r^{2}\hat {\tilde s}_r 
    +\hat A_r 
    -2\hat Z_r 
  \)
  -\chi_{rr} r\hat {\beta^r} 
\]
\Nl}$$
$$\displaylines{\qdd
\tilde \F^r 
\(\chi_{rz} 
\)
=
\[H^{-1}\alpha \hat D_{rzz} r^{2}\hat H_{rz} 
  +\alpha D_{zrr} H_{rr} ^{-1}\nl 
  \off{4257064}
  +
  \frac{1}{
        2}H^{-1}\alpha D_{zrr} r^{2}H_{rr} ^{
  -1}\hat H_{rz} ^{2}
  -H^{-1}\alpha \hat D_{zrz} r^{2}\hat H_{rz} \nl 
  \off{4257064}
  -
  \frac{1}{
        2}H^{-1}\alpha D_{zzz} H_{rr} 
  +\alpha 
  \(\frac{1}{
          2}r^{2}\hat s_z 
    +
    \frac{1}{
          2}A_z 
    -Z_z 
  \)
  -r^{2}\hat {\beta^r} \hat \chi_{rz} 
\]
\Nl}$$
$$\displaylines{\qdd
\tilde \F^r 
\(\chi_{zz} 
\)
=
\[H^{-1}\alpha \hat D_{rzz} rH_{zz} 
  -H^{-1}\alpha D_{zzz} r\hat H_{rz} 
  -\chi_{zz} r\hat {\beta^r} 
\]
\cr}$$
$$\displaylines{\qdd
\tilde \F^r 
\(Y 
\)
=
\[H^{-1}\alpha \hat D_{rrr} r^{2}H_{rr} ^{
  -2}\hat H_{rz} ^{2}
  -H^{-1}\alpha \hat D_{rzz} 
  -2H^{-1}\alpha D_{zrr} H_{rr} ^{
  -1}\hat H_{rz} \nl 
  \off{3813679}
  +2H^{-1}\alpha \hat D_{zrz} 
  +\alpha H_{rr} ^{-1}
  \(-\hat A_r 
    +2\hat Z_r 
  \)
  \nl 
  \off{3813679}
  +H^{-1}\alpha r^{2}\hat H_{rz} 
  \(r^{2}H_{rr} ^{-1}\hat H_{rz} \hat {\tilde s}_r 
    -\hat s_z 
  \)
  -r^{2}\hat {\beta^r} \hat Y 
\]
\Nl}$$
$$\displaylines{\qdd
\tilde \F^r 
\({E^r} 
\)
=
\[-2H^{-1}\alpha H_{zz} {Z^\varphi} 
  -r^{2}\hat {\beta^r} \hat {E^r} 
\]
\cr}$$
$$\displaylines{\qdd
\tilde \F^r 
\({E^z} 
\)
=
\[2H^{-1}\alpha r\hat H_{rz} {Z^\varphi} 
  -r\hat {\beta^r} {E^z} 
  +
  \(\sqrt{H}
  \)
  ^{-1}\alpha r\hat {B^\varphi} 
\]
\cr}$$
$$\displaylines{\qdd
\tilde \F^r 
\({B^\varphi} 
\)
=
\[-r^{2}\hat {\beta^r} \hat {B^\varphi} 
  +
  \(\sqrt{H}
  \)
  ^{-1}\alpha 
  \(r^{2}\hat {E^r} \hat H_{rz} 
    +{E^z} H_{zz} 
  \)
\]
\cr}$$
$$\displaylines{\qdd
\tilde \F^r 
\(\theta 
\)
=
\[H^{-1}\alpha \hat D_{rrr} rH_{rr} ^{
  -1}H_{zz} 
  +H^{-1}\alpha \hat D_{rzz} r
  -H^{-1}\alpha D_{zrr} rH_{rr} ^{
  -1}\hat H_{rz} \nl 
  \off{3613430}
  -H^{-1}\alpha \hat D_{zrz} r
  +\alpha rH_{rr} ^{-1}
  \(r^{2}\hat {\tilde s}_r 
    -\hat Z_r 
  \)
  \nl 
  \off{3613430}
  +H^{-1}\alpha r\hat H_{rz} 
  \(r^{4}H_{rr} ^{-1}\hat H_{rz} \hat {\tilde s}_r 
    -r^{2}H_{rr} ^{-1}\hat H_{rz} \hat Z_r \nl 
    \off{7497958}
    -r^{2}\hat s_z 
    +Z_z 
  \)
  -r\hat {\beta^r} \theta 
\]
\Nl}$$
$$\displaylines{\qdd
\tilde \F^r 
\(Z_r 
\)
=
\[\alpha \chi_{rr} H_{rr} ^{-1}
  +H^{-1}\alpha \chi_{zz} H_{rr} \nl 
  \off{4025155}
  +\alpha 
  \(r^{2}\hat Y 
    -\theta 
  \)
  -H^{-1}\alpha r^{2}\hat \chi_{rz} \hat H_{rz} 
  -r^{2}\hat {\beta^r} \hat Z_r 
\]
\Nl}$$
$$\displaylines{\qdd
\tilde \F^r 
\(Z_z 
\)
=
\[H^{-1}\alpha \chi_{zz} r\hat H_{rz} 
  -H^{-1}\alpha r\hat \chi_{rz} H_{zz} 
  -r\hat {\beta^r} Z_z 
\]
\cr}$$
$$\displaylines{\qdd
\tilde \F^r 
\({Z^\varphi} 
\)
=
\[-
  \frac{1}{
        2}\alpha r\hat {E^r} 
  -r\hat {\beta^r} {Z^\varphi} 
\]
\cr}$$

\section{Fluxes in the $z$ direction}
$$\displaylines{\qdd
\tilde \F^z 
\(H_{rr} 
\)
=
\[0
\]
\cr}$$
$$\displaylines{\qdd
\tilde \F^z 
\(H_{rz} 
\)
=
\[0
\]
\cr}$$
$$\displaylines{\qdd
\tilde \F^z 
\(H_{zz} 
\)
=
\[0
\]
\cr}$$
$$\displaylines{\qdd
\tilde \F^z 
\(s 
\)
=
\[0
\]
\cr}$$
$$\displaylines{\qdd
\tilde \F^z 
\(\alpha 
\)
=
\[0
\]
\cr}$$
$$\displaylines{\qdd
\tilde \F^z 
\({\beta^r} 
\)
=
\[0
\]
\cr}$$
$$\displaylines{\qdd
\tilde \F^z 
\({\beta^z} 
\)
=
\[0
\]
\cr}$$
$$\displaylines{\qdd
\tilde \F^z 
\(D_{rrr} 
\)
=
\[-\hat D_{rrr} r{\beta^z} 
\]
\cr}$$
$$\displaylines{\qdd
\tilde \F^z 
\(\tilde D_{rrz} 
\)
=
\[-\hat {\tilde D}_{rrz} r{\beta^z} 
\]
\cr}$$
$$\displaylines{\qdd
\tilde \F^z 
\(D_{rzz} 
\)
=
\[-\hat D_{rzz} r{\beta^z} 
\]
\cr}$$
$$\displaylines{\qdd
\tilde \F^z 
\(D_{zrr} 
\)
=
\[\alpha \chi_{rr} 
  -2\alpha r^{2}
  \({\hat {\tilde B}_r{}^r} H_{rr} 
    +{\hat B_r{}^z} \hat H_{rz} 
  \)
  -D_{zrr} {\beta^z} 
\]
\cr}$$
$$\displaylines{\qdd
\tilde \F^z 
\(D_{zrz} 
\)
=
\[\alpha r
  \(-r^{2}{\hat {\tilde B}_r{}^r} \hat H_{rz} 
    -{\hat B_r{}^z} H_{zz} \nl 
    \off{5697331}
    -{\hat B_z{}^r} H_{rr} 
    -{B_z{}^z} \hat H_{rz} 
    +\hat \chi_{rz} 
  \)
  -\hat D_{zrz} r{\beta^z} 
\]
\Nl}$$
$$\displaylines{\qdd
\tilde \F^z 
\(D_{zzz} 
\)
=
\[\alpha \chi_{zz} 
  -2\alpha 
  \(r^{2}{\hat B_z{}^r} \hat H_{rz} 
    +{B_z{}^z} H_{zz} 
  \)
  -D_{zzz} {\beta^z} 
\]
\cr}$$
$$\displaylines{\qdd
\tilde \F^z 
\(\tilde s_r 
\)
=
\[-r{\beta^z} \hat {\tilde s}_r 
\]
\cr}$$
$$\displaylines{\qdd
\tilde \F^z 
\(s_z 
\)
=
\[\alpha r
  \(2{\hat {\tilde B}_r{}^r} 
    +2{\hat B_r{}^z} H_{rr} ^{-1}\hat H_{rz} 
    +\hat Y 
  \)
  -r{\beta^z} \hat s_z 
\]
\cr}$$
$$\displaylines{\qdd
\tilde \F^z 
\(A_r 
\)
=
\[-r\hat A_r {\beta^z} 
\]
\cr}$$
$$\displaylines{\qdd
\tilde \F^z 
\(A_z 
\)
=
\[2\alpha \chi_{rr} fH_{rr} ^{
  -1}
  +H^{-1}\alpha \chi_{rr} r^{2}fH_{rr} ^{
  -1}\hat H_{rz} ^{2}
  +H^{-1}\alpha \chi_{zz} fH_{rr} \nl 
  \off{4092819}
  +\alpha f
  \(r^{2}\hat Y 
    -m\theta 
  \)
  -2H^{-1}\alpha r^{2}\hat \chi_{rz} f\hat H_{rz} 
  -A_z {\beta^z} 
\]
\Nl}$$
$$\displaylines{\qdd
\tilde \F^z 
\({\tilde B_r{}^r} 
\)
=
\[-r{\beta^z} {\hat {\tilde B}_r{}^r} 
\]
\cr}$$
$$\displaylines{\qdd
\tilde \F^z 
\({B_r{}^z} 
\)
=
\[-r{\beta^z} {\hat B_r{}^z} 
\]
\cr}$$
$$\displaylines{\qdd
\tilde \F^z 
\({B_z{}^r} 
\)
=
\[H^{-1}\alpha \hat D_{rrr} rH_{rr} ^{
  -1}H_{zz} 
  \(-d
    +\mu 
  \)
  \nl 
  \off{4268823}
  -
  \frac{1}{
        2}H^{-2}\alpha \hat D_{rrr} r^{3}dH_{rr} ^{
  -1}\hat H_{rz} ^{2}H_{zz} \nl 
  \off{4268823}
  +H^{-1}\alpha \hat {\tilde D}_{rrz} r^{3}dH_{rr} ^{
  -1}\hat H_{rz} 
  +H^{-2}\alpha \hat {\tilde D}_{rrz} r^{5}dH_{rr} ^{
  -1}\hat H_{rz} ^{3}\nl 
  \off{4268823}
  +H^{-1}\alpha \hat D_{rzz} r
  \(-
    \frac{1}{
          2}d
    +\mu 
  \)
  -
  \frac{1}{
        2}H^{-2}\alpha \hat D_{rzz} r^{3}d\hat H_{rz} ^{2}
  \nl 
  \off{4268823}
  +H^{-1}\alpha D_{zrr} rH_{rr} ^{
  -1}\hat H_{rz} 
  \(d
    -\mu 
  \)
  \nl 
  \off{4268823}
  +
  \frac{1}{
        2}H^{-2}\alpha D_{zrr} r^{3}dH_{rr} ^{
  -1}\hat H_{rz} ^{3}
  -H^{-1}\alpha \hat D_{zrz} r\mu \nl 
  \off{4268823}
  -H^{-2}\alpha \hat D_{zrz} r^{3}d\hat H_{rz} ^{2}
  +
  \frac{1}{
        2}H^{-2}\alpha D_{zzz} rdH_{rr} \hat H_{rz} \nl 
  \off{4268823}
  +\alpha rH_{rr} ^{-1}
  \(-
    \frac{1}{
          2}r^{2}d\hat {\tilde s}_r 
    +r^{2}\mu \hat {\tilde s}_r 
    +
    \frac{1}{
          2}a\hat A_r 
    -\mu \hat Z_r 
  \)
  \nl 
  \off{4268823}
  +H^{-1}\alpha r\hat H_{rz} 
  \(-
    \frac{1}{
          2}r^{4}dH_{rr} ^{-1}\hat H_{rz} 
    \hat {\tilde s}_r 
    +r^{4}H_{rr} ^{-1}\hat H_{rz} \mu \hat {\tilde s}_r \nl 
    \off{8153351}
    +
    \frac{1}{
          2}r^{2}a\hat A_r H_{rr} ^{
    -1}\hat H_{rz} 
    +
    \frac{1}{
          2}r^{2}d\hat s_z 
    -r^{2}H_{rr} ^{-1}\hat H_{rz} \mu \hat Z_r \nl 
    \off{8153351}
    -r^{2}\mu \hat s_z 
    -
    \frac{1}{
          2}aA_z 
    +\mu Z_z 
  \)
  -r{\beta^z} {\hat B_z{}^r} 
\]
\Nl}$$
$$\displaylines{\qdd
\tilde \F^z 
\({B_z{}^z} 
\)
=
\[H^{-1}\alpha \hat D_{rrr} r^{2}H_{rr} ^{
  -1}\hat H_{rz} 
  \(d
    -\mu 
  \)
  \nl 
  \off{4280552}
  +
  \frac{1}{
        2}H^{-2}\alpha \hat D_{rrr} r^{4}dH_{rr} ^{
  -1}\hat H_{rz} ^{3}
  -H^{-1}\alpha \hat {\tilde D}_{rrz} r^{2}\mu \nl 
  \off{4280552}
  -H^{-2}\alpha \hat {\tilde D}_{rrz} r^{4}d\hat H_{rz} ^{2
  }
  +
  \frac{1}{
        2}H^{-2}\alpha \hat D_{rzz} r^{2}dH_{rr} 
  \hat H_{rz} \nl 
  \off{4280552}
  +H^{-1}\alpha D_{zrr} 
  \(-d
    +2\mu 
  \)
  -
  \frac{1}{
        2}H^{-2}\alpha D_{zrr} r^{2}d\hat H_{rz} ^{2}\nl 
  \off{4280552}
  +H^{-2}\alpha \hat D_{zrz} r^{2}dH_{rr} \hat H_{rz} 
  -
  \frac{1}{
        2}H^{-2}\alpha D_{zzz} dH_{rr} ^{2}\nl 
  \off{4280552}
  +H^{-1}\alpha 
  \(\frac{1}{
          2}r^{4}d\hat H_{rz} \hat {\tilde s}_r 
    -r^{4}\hat H_{rz} \mu \hat {\tilde s}_r 
    -
    \frac{1}{
          2}r^{2}a\hat A_r \hat H_{rz} \nl 
    \off{6783435}
    -
    \frac{1}{
          2}r^{2}dH_{rr} \hat s_z 
    +r^{2}H_{rr} \mu \hat s_z 
    +r^{2}\hat H_{rz} \mu \hat Z_r \nl 
    \off{6783435}
    +
    \frac{1}{
          2}aA_z H_{rr} 
    -H_{rr} \mu Z_z 
  \)
  -{\beta^z} {B_z{}^z} 
\]
\Nl}$$
$$\displaylines{\qdd
\tilde \F^z 
\(\chi_{rr} 
\)
=
\[-H^{-1}\alpha \hat D_{rrr} r^{2}\hat H_{rz} 
  +H^{-1}\alpha D_{zrr} H_{rr} 
  -\chi_{rr} {\beta^z} 
\]
\cr}$$
$$\displaylines{\qdd
\tilde \F^z 
\(\chi_{rz} 
\)
=
\[-
  \frac{1}{
        2}H^{-1}\alpha \hat D_{rrr} r^{3}H_{rr} ^{
  -1}\hat H_{rz} ^{2}
  -H^{-1}\alpha \hat {\tilde D}_{rrz} r^{3}\hat H_{rz} \nl 
  \off{4268823}
  +
  \frac{1}{
        2}H^{-1}\alpha \hat D_{rzz} rH_{rr} 
  +H^{-1}\alpha D_{zrr} r\hat H_{rz} \nl 
  \off{4268823}
  +\alpha r
  \(\frac{1}{
          2}r^{2}\hat {\tilde s}_r 
    +
    \frac{1}{
          2}\hat A_r 
    -\hat Z_r 
  \)
  -r{\beta^z} \hat \chi_{rz} 
\]
\Nl}$$
$$\displaylines{\qdd
\tilde \F^z 
\(\chi_{zz} 
\)
=
\[-2\alpha \hat {\tilde D}_{rrz} r^{2}H_{rr} ^{
  -1}
  -2H^{-1}\alpha \hat {\tilde D}_{rrz} r^{4}H_{rr} ^{
  -1}\hat H_{rz} ^{2}\nl 
  \off{4280552}
  +H^{-1}\alpha \hat D_{rzz} r^{2}\hat H_{rz} 
  +2\alpha D_{zrr} H_{rr} ^{-1}\nl 
  \off{4280552}
  +H^{-1}\alpha D_{zrr} r^{2}H_{rr} ^{
  -1}\hat H_{rz} ^{2}
  +\alpha 
  \(r^{2}\hat s_z 
    +A_z 
    -2Z_z 
  \)
  -\chi_{zz} {\beta^z} 
\]
\Nl}$$
$$\displaylines{\qdd
\tilde \F^z 
\(Y 
\)
=
\[H^{-1}\alpha r
  \(-r^{2}\hat H_{rz} \hat {\tilde s}_r 
    +H_{rr} \hat s_z 
  \)
  -r{\beta^z} \hat Y 
\]
\cr}$$
$$\displaylines{\qdd
\tilde \F^z 
\({E^r} 
\)
=
\[2H^{-1}\alpha r\hat H_{rz} {Z^\varphi} 
  -r{\beta^z} \hat {E^r} 
  -
  \(\sqrt{H}
  \)
  ^{-1}\alpha r\hat {B^\varphi} 
\]
\cr}$$
$$\displaylines{\qdd
\tilde \F^z 
\({E^z} 
\)
=
\[-2H^{-1}\alpha H_{rr} {Z^\varphi} 
  -{\beta^z} {E^z} 
\]
\cr}$$
$$\displaylines{\qdd
\tilde \F^z 
\({B^\varphi} 
\)
=
\[-r{\beta^z} \hat {B^\varphi} 
  -
  \(\sqrt{H}
  \)
  ^{-1}\alpha r
  \(\hat {E^r} H_{rr} 
    +{E^z} \hat H_{rz} 
  \)
\]
\cr}$$
$$\displaylines{\qdd
\tilde \F^z 
\(\theta 
\)
=
\[-H^{-1}\alpha \hat D_{rrr} r^{2}H_{rr} ^{
  -1}\hat H_{rz} 
  -H^{-1}\alpha \hat {\tilde D}_{rrz} r^{2}
  +2H^{-1}\alpha D_{zrr} \nl 
  \off{3625189}
  +H^{-1}\alpha 
  \(-r^{4}\hat H_{rz} \hat {\tilde s}_r 
    +r^{2}H_{rr} \hat s_z 
    +r^{2}\hat H_{rz} \hat Z_r 
    -H_{rr} Z_z 
  \)
  -{\beta^z} \theta 
\]
\Nl}$$
$$\displaylines{\qdd
\tilde \F^z 
\(Z_r 
\)
=
\[H^{-1}\alpha \chi_{rr} r\hat H_{rz} 
  -H^{-1}\alpha r\hat \chi_{rz} H_{rr} 
  -r{\beta^z} \hat Z_r 
\]
\cr}$$
$$\displaylines{\qdd
\tilde \F^z 
\(Z_z 
\)
=
\[2\alpha \chi_{rr} H_{rr} ^{-1}
  +H^{-1}\alpha \chi_{rr} r^{2}H_{rr} ^{
  -1}\hat H_{rz} ^{2}\nl 
  \off{4048673}
  +\alpha 
  \(r^{2}\hat Y 
    -\theta 
  \)
  -H^{-1}\alpha r^{2}\hat \chi_{rz} \hat H_{rz} 
  -{\beta^z} Z_z 
\]
\Nl}$$
$$\displaylines{\qdd
\tilde \F^z 
\({Z^\varphi} 
\)
=
\[-
  \frac{1}{
        2}\alpha {E^z} 
  -{\beta^z} {Z^\varphi} 
\]
\cr}$$

\section{Sources}
$$\displaylines{\qdd
\tilde \S 
\(H_{rr} 
\)
=
\[-2\alpha \chi_{rr} 
  +4\alpha r^{2}
  \({\hat {\tilde B}_r{}^r} H_{rr} 
    +{\hat B_r{}^z} \hat H_{rz} 
  \)
  \nl 
  \off{4260297}
  +2\hat D_{rrr} r^{2}\hat {\beta^r} 
  +2D_{zrr} {\beta^z} 
  +2\hat {\beta^r} H_{rr} 
\]
\Nl}$$
$$\displaylines{\qdd
\tilde \S 
\(H_{rz} 
\)
=
\[2\alpha r
  \(r^{2}{\hat {\tilde B}_r{}^r} \hat H_{rz} 
    +{\hat B_r{}^z} H_{zz} 
    +{\hat B_z{}^r} H_{rr} 
    +{B_z{}^z} \hat H_{rz} 
    -\hat \chi_{rz} 
  \)
  \nl 
  \off{4272056}
  +2\hat {\tilde D}_{rrz} r^{3}\hat {\beta^r} 
  +2\hat D_{zrz} r{\beta^z} 
  +2r\hat {\beta^r} \hat H_{rz} 
\]
\Nl}$$
$$\displaylines{\qdd
\tilde \S 
\(H_{zz} 
\)
=
\[-2\alpha \chi_{zz} 
  +4\alpha 
  \(r^{2}{\hat B_z{}^r} \hat H_{rz} 
    +{B_z{}^z} H_{zz} 
  \)
  +2\hat D_{rzz} r^{2}\hat {\beta^r} 
  +2D_{zzz} {\beta^z} 
\]
\cr}$$
$$\displaylines{\qdd
\tilde \S 
\(s 
\)
=
\[\alpha r
  \(-2{\hat {\tilde B}_r{}^r} 
    -2{\hat B_r{}^z} H_{rr} ^{-1}\hat H_{rz} 
    -\hat Y 
  \)
  +r
  \(r^{2}\hat {\beta^r} \hat {\tilde s}_r 
    +2\hat {\beta^r} \hat s 
    +{\beta^z} \hat s_z 
  \)
\]
\cr}$$
$$\displaylines{\qdd
\tilde \S 
\(\alpha 
\)
=
\[-2\alpha ^{2}\chi_{rr} fH_{rr} ^{
  -1}
  -H^{-1}\alpha ^{2}\chi_{rr} r^{2}fH_{rr} ^{
  -1}\hat H_{rz} ^{2}\nl 
  \off{3589538}
  -H^{-1}\alpha ^{2}\chi_{zz} fH_{rr} 
  +\alpha ^{2}f
  \(-r^{2}\hat Y 
    +m\theta 
  \)
  \nl 
  \off{3589538}
  +2H^{-1}\alpha ^{2}r^{2}\hat \chi_{rz} f\hat H_{rz} 
  +\alpha 
  \(r^{2}\hat A_r \hat {\beta^r} 
    +A_z {\beta^z} 
  \)
\]
\Nl}$$
$$\displaylines{\qdd
\tilde \S 
\({\beta^r} 
\)
=
\[2H^{-1}\alpha ^{2}\hat D_{rrr} rH_{rr} ^{
  -1}H_{zz} 
  \(d
    -\mu 
  \)
  \nl 
  \off{3979643}
  +H^{-2}\alpha ^{2}\hat D_{rrr} r^{3}dH_{rr} ^{
  -1}\hat H_{rz} ^{2}H_{zz} \nl 
  \off{3979643}
  -2H^{-1}\alpha ^{2}\hat {\tilde D}_{rrz} r^{3}dH_{rr} ^{
  -1}\hat H_{rz} 
  -2H^{-2}\alpha ^{2}\hat {\tilde D}_{rrz} r^{5}dH_{rr} ^{
  -1}\hat H_{rz} ^{3}\nl 
  \off{3979643}
  +H^{-1}\alpha ^{2}\hat D_{rzz} r
  \(d
    -2\mu 
  \)
  +H^{-2}\alpha ^{2}\hat D_{rzz} r^{3}d\hat H_{rz} ^{2}\nl 
  \off{3979643}
  +2H^{-1}\alpha ^{2}D_{zrr} rH_{rr} ^{
  -1}\hat H_{rz} 
  \(-d
    +\mu 
  \)
  \nl 
  \off{3979643}
  -H^{-2}\alpha ^{2}D_{zrr} r^{3}dH_{rr} ^{
  -1}\hat H_{rz} ^{3}
  +2H^{-1}\alpha ^{2}\hat D_{zrz} r\mu \nl 
  \off{3979643}
  +2H^{-2}\alpha ^{2}\hat D_{zrz} r^{3}d\hat H_{rz} ^{2}
  -H^{-2}\alpha ^{2}D_{zzz} rdH_{rr} \hat H_{rz} \nl 
  \off{3979643}
  +\alpha ^{2}rH_{rr} ^{-1}
  \(r^{2}d\hat {\tilde s}_r 
    -2r^{2}\mu \hat {\tilde s}_r 
    -a\hat A_r 
    +2d\hat s 
    -4\mu \hat s 
    +2\mu \hat Z_r 
  \)
  \nl 
  \off{3979643}
  +H^{-1}\alpha ^{2}r\hat H_{rz} 
  \(r^{4}dH_{rr} ^{-1}\hat H_{rz} \hat {\tilde s}_r 
    -2r^{4}H_{rr} ^{-1}\hat H_{rz} \mu \hat {\tilde s}_r 
    \nl 
    \off{8140878}
    -r^{2}a\hat A_r H_{rr} ^{-1}\hat H_{rz} 
    +2r^{2}dH_{rr} ^{-1}\hat H_{rz} \hat s 
    -r^{2}d\hat s_z \nl 
    \off{8140878}
    -4r^{2}H_{rr} ^{-1}\hat H_{rz} \mu \hat s 
    +2r^{2}H_{rr} ^{-1}\hat H_{rz} \mu \hat Z_r \nl 
    \off{8140878}
    +2r^{2}\mu \hat s_z 
    +aA_z 
    -dH_{rr} ^{-1}\hat H_{rz} 
    -2\mu Z_z 
  \)
  \nl 
  \off{3979643}
  -H^{-2}\alpha ^{2}r^{3}dH_{rr} ^{
  -1}\hat H_{rz} ^{4}
  +2\alpha r
  \(r^{2}\hat {\beta^r} {\hat {\tilde B}_r{}^r} 
    +{\beta^z} {\hat B_z{}^r} 
  \)
  +r\hat {\beta^r} ^{2}
\]
\Nl}$$
$$\displaylines{\qdd
\tilde \S 
\({\beta^z} 
\)
=
\[2H^{-1}\alpha ^{2}\hat D_{rrr} r^{2}H_{rr} ^{
  -1}\hat H_{rz} 
  \(-d
    +\mu 
  \)
  \nl 
  \off{3991402}
  -H^{-2}\alpha ^{2}\hat D_{rrr} r^{4}dH_{rr} ^{
  -1}\hat H_{rz} ^{3}
  +2H^{-1}\alpha ^{2}\hat {\tilde D}_{rrz} r^{2}\mu \nl 
  \off{3991402}
  +2H^{-2}\alpha ^{2}\hat {\tilde D}_{rrz} r^{4}d
  \hat H_{rz} ^{2}
  -H^{-2}\alpha ^{2}\hat D_{rzz} r^{2}dH_{rr} \hat H_{rz} 
  \nl 
  \off{3991402}
  +2H^{-1}\alpha ^{2}D_{zrr} 
  \(d
    -2\mu 
  \)
  +H^{-2}\alpha ^{2}D_{zrr} r^{2}d\hat H_{rz} ^{2}\nl 
  \off{3991402}
  -2H^{-2}\alpha ^{2}\hat D_{zrz} r^{2}dH_{rr} 
  \hat H_{rz} 
  +H^{-2}\alpha ^{2}D_{zzz} dH_{rr} ^{2}\nl 
  \off{3991402}
  +H^{-1}\alpha ^{2}
  \(-r^{4}d\hat H_{rz} \hat {\tilde s}_r 
    +2r^{4}\hat H_{rz} \mu \hat {\tilde s}_r 
    +r^{2}a\hat A_r \hat H_{rz} 
    +r^{2}dH_{rr} \hat s_z \nl 
    \off{6770992}
    -2r^{2}d\hat H_{rz} \hat s 
    -2r^{2}H_{rr} \mu \hat s_z 
    +4r^{2}\hat H_{rz} \mu \hat s 
    -2r^{2}\hat H_{rz} \mu \hat Z_r \nl 
    \off{6770992}
    -aA_z H_{rr} 
    -d\hat H_{rz} 
    +2H_{rr} \mu Z_z 
    +3\hat H_{rz} \mu 
  \)
  \nl 
  \off{3991402}
  +H^{-2}\alpha ^{2}r^{2}d\hat H_{rz} ^{3}
  +2\alpha 
  \(r^{2}\hat {\beta^r} {\hat B_r{}^z} 
    +{\beta^z} {B_z{}^z} 
  \)
\]
\Nl}$$
$$\displaylines{\qdd
\tilde \S 
\(D_{rrr} 
\)
=
\[-2\alpha \hat D_{rrr} r{B_z{}^z} 
  +2\alpha D_{zrr} r{\hat B_r{}^z} 
  +2\alpha r{\hat {\tilde B}_r{}^r} H_{rr} 
  +2\hat D_{rrr} r\hat {\beta^r} 
\]
\cr}$$
$$\displaylines{\qdd
\tilde \S 
\(\tilde D_{rrz} 
\)
=
\[-2\alpha \hat {\tilde D}_{rrz} r{B_z{}^z} 
  +2\alpha \hat D_{zrz} r{\hat B_r{}^z} 
  +2\alpha r{\hat {\tilde B}_r{}^r} \hat H_{rz} 
  +2\hat {\tilde D}_{rrz} r\hat {\beta^r} 
\]
\cr}$$
$$\displaylines{\qdd
\tilde \S 
\(D_{rzz} 
\)
=
\[-2\alpha \hat D_{rzz} r{B_z{}^z} 
  +2\alpha D_{zzz} r{\hat B_r{}^z} 
\]
\cr}$$
$$\displaylines{\qdd
\tilde \S 
\(D_{zrr} 
\)
=
\[2\alpha \hat D_{rrr} r^{2}{\hat B_z{}^r} 
  -2\alpha D_{zrr} r^{2}{\hat {\tilde B}_r{}^r} 
  +2\alpha {\hat B_z{}^r} H_{rr} 
  +D_{zrr} \hat {\beta^r} 
\]
\cr}$$
$$\displaylines{\qdd
\tilde \S 
\(D_{zrz} 
\)
=
\[2\alpha \hat {\tilde D}_{rrz} r^{3}{\hat B_z{}^r} 
  -2\alpha \hat D_{zrz} r^{3}{\hat {\tilde B}_r{}^r} 
  +2\alpha r{\hat B_z{}^r} \hat H_{rz} 
\]
\cr}$$
$$\displaylines{\qdd
\tilde \S 
\(D_{zzz} 
\)
=
\[2\alpha \hat D_{rzz} r^{2}{\hat B_z{}^r} 
  -2\alpha D_{zzz} r^{2}{\hat {\tilde B}_r{}^r} 
  -D_{zzz} \hat {\beta^r} 
\]
\cr}$$
$$\displaylines{\qdd
\tilde \S 
\(\tilde s_r 
\)
=
\[2\alpha r
  \(2{\hat {\tilde B}_r{}^r} \hat s 
    +{\hat B_r{}^z} \hat s_z 
    -{B_z{}^z} \hat {\tilde s}_r 
  \)
  +2r\hat {\beta^r} \hat {\tilde s}_r 
\]
\cr}$$
$$\displaylines{\qdd
\tilde \S 
\(s_z 
\)
=
\[2\alpha r
  \(-r^{2}{\hat {\tilde B}_r{}^r} \hat s_z 
    +r^{2}{\hat B_z{}^r} \hat {\tilde s}_r 
    +2{\hat B_z{}^r} \hat s 
  \)
\]
\cr}$$
$$\displaylines{\qdd
\tilde \S 
\(A_r 
\)
=
\[2\alpha r
  \(-\hat A_r {B_z{}^z} 
    +A_z {\hat B_r{}^z} 
  \)
\]
\cr}$$
$$\displaylines{\qdd
\tilde \S 
\(A_z 
\)
=
\[2\alpha r^{2}
  \(\hat A_r {\hat B_z{}^r} 
    -A_z {\hat {\tilde B}_r{}^r} 
  \)
  -A_z \hat {\beta^r} 
\]
\cr}$$
$$\displaylines{\qdd
\tilde \S 
\({\tilde B_r{}^r} 
\)
=
\[2\alpha \chi_{rr} r{\hat {\tilde B}_r{}^r} fH_{rr} ^{
  -1}\nl 
  \off{4125582}
  +H^{-1}\alpha \chi_{rr} r^{3}{\hat {\tilde B}_r{}^r} f
  H_{rr} ^{-1}\hat H_{rz} ^{2}
  +H^{-1}\alpha \chi_{zz} r{\hat {\tilde B}_r{}^r} f
  H_{rr} \nl 
  \off{4125582}
  +H^{-1}\alpha \hat D_{rrr} rH_{rr} ^{
  -1}H_{zz} 
  \(\hat A_r d
    -\hat A_r \mu 
    -2d\hat s 
    +4\mu \hat s 
  \)
  \nl 
  \off{4125582}
  +H^{-2}\alpha \hat D_{rrr} rH_{rr} ^{
  -1}\hat H_{rz} ^{2}H_{zz} 
  \(\frac{1}{
          2}r^{2}\hat A_r d
    -2r^{2}d\hat s 
    +4r^{2}\mu \hat s 
    +2d
  \)
  \nl 
  \off{4125582}
  +2H^{-3}\alpha \hat D_{rrr} r^{3}dH_{rr} ^{
  -1}\hat H_{rz} ^{4}H_{zz} \nl 
  \off{4125582}
  +H^{-1}\alpha \hat {\tilde D}_{rrz} rH_{rr} ^{
  -1}\hat H_{rz} 
  \(-r^{2}\hat A_r d
    +4r^{2}d\hat s 
    -8r^{2}\mu \hat s 
    -2d
  \)
  \nl 
  \off{4125582}
  +H^{-2}\alpha \hat {\tilde D}_{rrz} r^{3}H_{rr} ^{
  -1}\hat H_{rz} ^{3}
  \(-r^{2}\hat A_r d
    +4r^{2}d\hat s 
    -8r^{2}\mu \hat s 
    -6d
  \)
  \nl 
  \off{4125582}
  -4H^{-3}\alpha \hat {\tilde D}_{rrz} r^{5}dH_{rr} ^{
  -1}\hat H_{rz} ^{5}
  +H^{-1}\alpha \hat D_{rzz} r\hat A_r 
  \(\frac{1}{
          2}d
    -\mu 
  \)
  \nl 
  \off{4125582}
  +H^{-2}\alpha \hat D_{rzz} r\hat H_{rz} ^{2}
  \(\frac{1}{
          2}r^{2}\hat A_r d
    -2r^{2}d\hat s 
    +4r^{2}\mu \hat s 
    +d
  \)
  \nl 
  \off{4125582}
  +2H^{-3}\alpha \hat D_{rzz} r^{3}d\hat H_{rz} ^{4}
  +H^{-1}\alpha D_{zrr} r\hat A_r H_{rr} ^{
  -1}\hat H_{rz} 
  \(-d
    +\mu 
  \)
  \nl 
  \off{4125582}
  -
  \frac{1}{
        2}H^{-2}\alpha D_{zrr} r^{3}\hat A_r dH_{rr} ^{
  -1}\hat H_{rz} ^{3}
  +H^{-1}\alpha \hat D_{zrz} r\hat A_r \mu \nl 
  \off{4125582}
  +H^{-2}\alpha \hat D_{zrz} r^{3}\hat A_r d\hat H_{rz} ^{
  2}
  -
  \frac{1}{
        2}H^{-2}\alpha D_{zzz} r\hat A_r dH_{rr} 
  \hat H_{rz} \nl 
  \off{4125582}
  +\alpha r
  \(\frac{1}{
          2}r^{2}\hat A_r dH_{rr} ^{
    -1}\hat {\tilde s}_r 
    -r^{2}\hat A_r H_{rr} ^{-1}\mu \hat {\tilde s}_r 
    +r^{2}{\hat {\tilde B}_r{}^r} f\hat Y \nl 
    \off{5662113}
    -
    \frac{1}{
          2}a\hat A_r ^{2}H_{rr} ^{
    -1}
    +2\hat A_r dH_{rr} ^{-1}\hat s 
    -4\hat A_r H_{rr} ^{-1}\mu \hat s \nl 
    \off{5662113}
    +\hat A_r H_{rr} ^{-1}\mu \hat Z_r 
    -2{\hat {\tilde B}_r{}^r} {B_z{}^z} 
    -{\hat {\tilde B}_r{}^r} fm\theta \nl 
    \off{5662113}
    +2{\hat B_r{}^z} {\hat B_z{}^r} 
    +dH_{rr} ^{-1}\hat {\tilde s}_r 
    -2H_{rr} ^{-1}\mu \hat {\tilde s}_r 
  \)
  \nl 
\cr}$$ 
$$\displaylines{\qdd
  \off{4125582}
  +H^{-1}\alpha r\hat H_{rz} 
  \(\frac{1}{
          2}r^{4}\hat A_r dH_{rr} ^{
    -1}\hat H_{rz} \hat {\tilde s}_r 
    -r^{4}\hat A_r H_{rr} ^{-1}\hat H_{rz} \mu 
    \hat {\tilde s}_r \nl 
    \off{8010110}
    -
    \frac{1}{
          2}r^{2}a\hat A_r ^{2}H_{rr} ^{
    -1}\hat H_{rz} 
    +2r^{2}\hat A_r dH_{rr} ^{
    -1}\hat H_{rz} \hat s \nl 
    \off{8010110}
    -
    \frac{1}{
          2}r^{2}\hat A_r d\hat s_z 
    -4r^{2}\hat A_r H_{rr} ^{-1}\hat H_{rz} \mu \hat s 
    \nl 
    \off{8010110}
    +r^{2}\hat A_r H_{rr} ^{-1}\hat H_{rz} \mu \hat Z_r 
    +r^{2}\hat A_r \mu \hat s_z 
    -2r^{2}{\hat {\tilde B}_r{}^r} \hat \chi_{rz} f\nl 
    \off{8010110}
    +r^{2}dH_{rr} ^{-1}\hat H_{rz} \hat {\tilde s}_r 
    -2r^{2}H_{rr} ^{-1}\hat H_{rz} \mu \hat {\tilde s}_r 
    \nl 
    \off{8010110}
    +
    \frac{1}{
          2}a\hat A_r A_z 
    -\hat A_r dH_{rr} ^{-1}\hat H_{rz} 
    -\hat A_r \mu Z_z \nl 
    \off{8010110}
    +2dH_{rr} ^{-1}\hat H_{rz} \hat s 
    -4H_{rr} ^{-1}\hat H_{rz} \mu \hat s 
  \)
  \nl 
  \off{4125582}
  +H^{-2}\alpha rH_{rr} ^{-1}\hat H_{rz} ^{4}
  \(-r^{2}\hat A_r d
    +2r^{2}d\hat s 
    -4r^{2}\mu \hat s 
    -2d
  \)
  \nl 
  \off{4125582}
  -2H^{-3}\alpha r^{3}dH_{rr} ^{
  -1}\hat H_{rz} ^{6}
  +2r\hat {\beta^r} {\hat {\tilde B}_r{}^r} 
\]
\Nl}$$
$$\displaylines{\qdd
\tilde \S 
\({B_r{}^z} 
\)
=
\[2\alpha \chi_{rr} r{\hat B_r{}^z} fH_{rr} ^{
  -1}\nl 
  \off{4137371}
  +H^{-1}\alpha \chi_{rr} r^{3}{\hat B_r{}^z} fH_{rr} ^{
  -1}\hat H_{rz} ^{2}
  +H^{-1}\alpha \chi_{zz} r{\hat B_r{}^z} fH_{rr} \nl 
  \off{4137371}
  +H^{-1}\alpha \hat D_{rrr} rH_{rr} ^{
  -1}\hat H_{rz} 
  \(-r^{2}\hat A_r d
    +r^{2}\hat A_r \mu \nl 
    \off{11161046}
    +2r^{2}d\hat s 
    -4r^{2}\mu \hat s 
    +d
    -3\mu 
  \)
  \nl 
  \off{4137371}
  +H^{-2}\alpha \hat D_{rrr} r^{3}H_{rr} ^{
  -1}\hat H_{rz} ^{3}
  \(-
    \frac{1}{
          2}r^{2}\hat A_r d
    +2r^{2}d\hat s 
    -4r^{2}\mu \hat s 
    -d
    -3\mu 
  \)
  \nl 
  \off{4137371}
  -2H^{-3}\alpha \hat D_{rrr} r^{5}dH_{rr} ^{
  -1}\hat H_{rz} ^{5}\nl 
  \off{4137371}
  +H^{-1}\alpha \hat {\tilde D}_{rrz} r
  \(r^{2}\hat A_r \mu 
    -2r^{2}d\hat s 
    +4r^{2}\mu \hat s 
    -d
    +3\mu 
  \)
  \nl 
  \off{4137371}
  +H^{-2}\alpha \hat {\tilde D}_{rrz} r^{3}\hat H_{rz} ^{2}
  \(r^{2}\hat A_r d
    -4r^{2}d\hat s 
    +8r^{2}\mu \hat s 
    +d
    +6\mu 
  \)
  \nl 
  \off{4137371}
  +4H^{-3}\alpha \hat {\tilde D}_{rrz} r^{5}d\hat H_{rz} 
  ^{4}\nl 
  \off{4137371}
  +H^{-2}\alpha \hat D_{rzz} rH_{rr} \hat H_{rz} 
  \(-
    \frac{1}{
          2}r^{2}\hat A_r d
    +2r^{2}d\hat s 
    -4r^{2}\mu \hat s 
    +d
    -3\mu 
  \)
  \nl 
  \off{4137371}
  -2H^{-3}\alpha \hat D_{rzz} r^{3}dH_{rr} \hat H_{rz} ^{
  3}
  +H^{-1}\alpha D_{zrr} r\hat A_r 
  \(d
    -2\mu 
  \)
  \nl 
  \off{4137371}
  +
  \frac{1}{
        2}H^{-2}\alpha D_{zrr} r^{3}\hat A_r d
  \hat H_{rz} ^{2}
  -H^{-2}\alpha \hat D_{zrz} r^{3}\hat A_r dH_{rr} 
  \hat H_{rz} \nl 
  \off{4137371}
  +
  \frac{1}{
        2}H^{-2}\alpha D_{zzz} r\hat A_r dH_{rr} ^{2}
  +\alpha r{\hat B_r{}^z} f
  \(r^{2}\hat Y 
    -m\theta 
  \)
  \nl 
  \off{4137371}
  +H^{-1}\alpha r
  \(-
    \frac{1}{
          2}r^{4}\hat A_r d\hat H_{rz} \hat {\tilde s}_r 
    +r^{4}\hat A_r \hat H_{rz} \mu \hat {\tilde s}_r 
    +
    \frac{1}{
          2}r^{2}a\hat A_r ^{2}\hat H_{rz} \nl 
    \off{6917720}
    +
    \frac{1}{
          2}r^{2}\hat A_r dH_{rr} \hat s_z 
    -2r^{2}\hat A_r d\hat H_{rz} \hat s 
    -r^{2}\hat A_r H_{rr} \mu \hat s_z \nl 
    \off{6917720}
    +4r^{2}\hat A_r \hat H_{rz} \mu \hat s 
    -r^{2}\hat A_r \hat H_{rz} \mu \hat Z_r 
    -2r^{2}{\hat B_r{}^z} \hat \chi_{rz} f\hat H_{rz} \nl 
    \off{6917720}
    -r^{2}d\hat H_{rz} \hat {\tilde s}_r 
    +2r^{2}\hat H_{rz} \mu \hat {\tilde s}_r 
    -
    \frac{1}{
          2}a\hat A_r A_z H_{rr} 
    -\hat A_r d\hat H_{rz} \nl 
    \off{6917720}
    +\hat A_r H_{rr} \mu Z_z 
    +3\hat A_r \hat H_{rz} \mu 
    -2d\hat H_{rz} \hat s 
    +4\hat H_{rz} \mu \hat s 
  \)
  \nl 
  \off{4137371}
  +H^{-2}\alpha r\hat H_{rz} ^{3}
  \(r^{2}\hat A_r d
    -2r^{2}d\hat s 
    +4r^{2}\mu \hat s 
    +3\mu 
  \)
  +2H^{-3}\alpha r^{3}d\hat H_{rz} ^{5}
\]
\Nl}$$
$$\displaylines{\qdd
\tilde \S 
\({B_z{}^r} 
\)
=
\[2\alpha \chi_{rr} r{\hat B_z{}^r} fH_{rr} ^{
  -1}
  +H^{-1}\alpha \chi_{rr} r^{3}{\hat B_z{}^r} fH_{rr} ^{
  -1}\hat H_{rz} ^{2}\nl 
  \off{4137371}
  +H^{-1}\alpha \chi_{zz} r{\hat B_z{}^r} fH_{rr} 
  +H^{-1}\alpha \hat D_{rrr} rA_z H_{rr} ^{
  -1}H_{zz} 
  \(d
    -\mu 
  \)
  \nl 
  \off{4137371}
  +
  \frac{1}{
        2}H^{-2}\alpha \hat D_{rrr} r^{3}A_z dH_{rr} ^{
  -1}\hat H_{rz} ^{2}H_{zz} \nl 
  \off{4137371}
  -H^{-1}\alpha \hat {\tilde D}_{rrz} r^{3}A_z dH_{rr} ^{
  -1}\hat H_{rz} 
  -H^{-2}\alpha \hat {\tilde D}_{rrz} r^{5}A_z dH_{rr} ^{
  -1}\hat H_{rz} ^{3}\nl 
  \off{4137371}
  +H^{-1}\alpha \hat D_{rzz} rA_z 
  \(\frac{1}{
          2}d
    -\mu 
  \)
  +
  \frac{1}{
        2}H^{-2}\alpha \hat D_{rzz} r^{3}A_z d
  \hat H_{rz} ^{2}\nl 
  \off{4137371}
  +H^{-1}\alpha D_{zrr} rH_{rr} ^{
  -1}
  \(-A_z d\hat H_{rz} 
    +A_z \hat H_{rz} \mu \nl 
    \off{10068626}
    -2dH_{zz} \hat s 
    +4H_{zz} \mu \hat s 
  \)
  \nl 
  \off{4137371}
  +H^{-2}\alpha D_{zrr} rH_{rr} ^{
  -1}\hat H_{rz} ^{2}
  \(-
    \frac{1}{
          2}r^{2}A_z d\hat H_{rz} 
    -2r^{2}dH_{zz} \hat s \nl 
    \off{11449512}
    +4r^{2}H_{zz} \mu \hat s 
    +2dH_{zz} 
  \)
  \nl 
  \off{4137371}
  +2H^{-3}\alpha D_{zrr} r^{3}dH_{rr} ^{
  -1}\hat H_{rz} ^{4}H_{zz} \nl 
  \off{4137371}
  +H^{-1}\alpha \hat D_{zrz} r
  \(4r^{2}dH_{rr} ^{-1}\hat H_{rz} \hat s 
    -8r^{2}H_{rr} ^{-1}\hat H_{rz} \mu \hat s \nl 
    \off{8288915}
    +A_z \mu 
    -2dH_{rr} ^{-1}\hat H_{rz} 
  \)
  \nl 
  \off{4137371}
  +H^{-2}\alpha \hat D_{zrz} r^{3}\hat H_{rz} ^{2}
  \(4r^{2}dH_{rr} ^{-1}\hat H_{rz} \hat s 
    -8r^{2}H_{rr} ^{-1}\hat H_{rz} \mu \hat s \nl 
    \off{9946508}
    +A_z d
    -6dH_{rr} ^{-1}\hat H_{rz} 
  \)
  \nl 
  \off{4137371}
  -4H^{-3}\alpha \hat D_{zrz} r^{5}dH_{rr} ^{
  -1}\hat H_{rz} ^{5}\nl 
  \off{4137371}
  +H^{-2}\alpha D_{zzz} r\hat H_{rz} 
  \(-2r^{2}d\hat H_{rz} \hat s 
    +4r^{2}\hat H_{rz} \mu \hat s 
    -
    \frac{1}{
          2}A_z dH_{rr} \nl 
    \off{9404853}
    +d\hat H_{rz} 
  \)
  +2H^{-3}\alpha D_{zzz} r^{3}d\hat H_{rz} ^{4}\nl 
  \off{4137371}
  +\alpha r
  \(\frac{1}{
          2}r^{2}A_z dH_{rr} ^{
    -1}\hat {\tilde s}_r 
    -r^{2}A_z H_{rr} ^{-1}\mu \hat {\tilde s}_r 
    +r^{2}{\hat B_z{}^r} f\hat Y \nl 
    \off{5673902}
    -
    \frac{1}{
          2}a\hat A_r A_z H_{rr} ^{
    -1}
    +2A_z dH_{rr} ^{-1}\hat s 
    -4A_z H_{rr} ^{-1}\mu \hat s \nl 
    \off{5673902}
    +A_z H_{rr} ^{-1}\mu \hat Z_r 
    -{\hat B_z{}^r} fm\theta 
    +dH_{rr} ^{-1}\hat s_z 
    -2H_{rr} ^{-1}\mu \hat s_z 
  \)
  \nl 
\cr}$$ 
$$\displaylines{\qdd
  \off{4137371}
  +H^{-1}\alpha r\hat H_{rz} 
  \(\frac{1}{
          2}r^{4}A_z dH_{rr} ^{
    -1}\hat H_{rz} \hat {\tilde s}_r 
    -r^{4}A_z H_{rr} ^{-1}\hat H_{rz} \mu 
    \hat {\tilde s}_r \nl 
    \off{8021899}
    -
    \frac{1}{
          2}r^{2}a\hat A_r A_z H_{rr} ^{
    -1}\hat H_{rz} 
    +2r^{2}A_z dH_{rr} ^{-1}\hat H_{rz} \hat s \nl 
    \off{8021899}
    -
    \frac{1}{
          2}r^{2}A_z d\hat s_z 
    -4r^{2}A_z H_{rr} ^{-1}\hat H_{rz} \mu \hat s \nl 
    \off{8021899}
    +r^{2}A_z H_{rr} ^{-1}\hat H_{rz} \mu \hat Z_r 
    +r^{2}A_z \mu \hat s_z 
    -2r^{2}{\hat B_z{}^r} \hat \chi_{rz} f\nl 
    \off{8021899}
    +r^{2}dH_{rr} ^{-1}\hat H_{rz} \hat s_z 
    -2r^{2}H_{rr} ^{-1}\hat H_{rz} \mu \hat s_z \nl 
    \off{8021899}
    +
    \frac{1}{
          2}aA_z ^{2}
    -A_z dH_{rr} ^{-1}\hat H_{rz} 
    -A_z \mu Z_z 
  \)
  \nl 
  \off{4137371}
  -H^{-2}\alpha r^{3}A_z dH_{rr} ^{
  -1}\hat H_{rz} ^{4}
\]
\Nl}$$
$$\displaylines{\qdd
\tilde \S 
\({B_z{}^z} 
\)
=
\[2\alpha \chi_{rr} {B_z{}^z} fH_{rr} ^{
  -1}
  +H^{-1}\alpha \chi_{rr} r^{2}{B_z{}^z} fH_{rr} ^{
  -1}\hat H_{rz} ^{2}\nl 
  \off{4149100}
  +H^{-1}\alpha \chi_{zz} {B_z{}^z} fH_{rr} 
  +H^{-1}\alpha \hat D_{rrr} r^{2}A_z H_{rr} ^{
  -1}\hat H_{rz} 
  \(-d
    +\mu 
  \)
  \nl 
  \off{4149100}
  -
  \frac{1}{
        2}H^{-2}\alpha \hat D_{rrr} r^{4}A_z dH_{rr} ^{
  -1}\hat H_{rz} ^{3}
  +H^{-1}\alpha \hat {\tilde D}_{rrz} r^{2}A_z \mu \nl 
  \off{4149100}
  +H^{-2}\alpha \hat {\tilde D}_{rrz} r^{4}A_z d
  \hat H_{rz} ^{2}
  -
  \frac{1}{
        2}H^{-2}\alpha \hat D_{rzz} r^{2}A_z dH_{rr} 
  \hat H_{rz} \nl 
  \off{4149100}
  +H^{-1}\alpha D_{zrr} 
  \(2r^{2}dH_{rr} ^{-1}\hat H_{rz} \hat s 
    -4r^{2}H_{rr} ^{-1}\hat H_{rz} \mu \hat s 
    +A_z d\nl 
    \off{8011419}
    -2A_z \mu 
    +dH_{rr} ^{-1}\hat H_{rz} 
    -3H_{rr} ^{-1}\hat H_{rz} \mu 
  \)
  \nl 
  \off{4149100}
  +H^{-2}\alpha D_{zrr} r^{2}\hat H_{rz} ^{2}
  \(2r^{2}dH_{rr} ^{-1}\hat H_{rz} \hat s 
    -4r^{2}H_{rr} ^{-1}\hat H_{rz} \mu \hat s \nl 
    \off{9946478}
    +
    \frac{1}{
          2}A_z d
    -dH_{rr} ^{-1}\hat H_{rz} 
    -3H_{rr} ^{-1}\hat H_{rz} \mu 
  \)
  \nl 
  \off{4149100}
  -2H^{-3}\alpha D_{zrr} r^{4}dH_{rr} ^{
  -1}\hat H_{rz} ^{5}\nl 
  \off{4149100}
  +H^{-1}\alpha \hat D_{zrz} 
  \(-2r^{2}d\hat s 
    +4r^{2}\mu \hat s 
    -d
    +3\mu 
  \)
  \nl 
  \off{4149100}
  +H^{-2}\alpha \hat D_{zrz} r^{2}\hat H_{rz} 
  \(-4r^{2}d\hat H_{rz} \hat s 
    +8r^{2}\hat H_{rz} \mu \hat s 
    -A_z dH_{rr} 
    +d\hat H_{rz} \nl 
    \off{9681530}
    +6\hat H_{rz} \mu 
  \)
  +4H^{-3}\alpha \hat D_{zrz} r^{4}d\hat H_{rz} ^{4}\nl 
  \off{4149100}
  +H^{-2}\alpha D_{zzz} H_{rr} 
  \(2r^{2}d\hat H_{rz} \hat s 
    -4r^{2}\hat H_{rz} \mu \hat s 
    +
    \frac{1}{
          2}A_z dH_{rr} 
    +d\hat H_{rz} \nl 
    \off{9127357}
    -3\hat H_{rz} \mu 
  \)
  -2H^{-3}\alpha D_{zzz} r^{2}dH_{rr} \hat H_{rz} ^{3}
  \nl 
  \off{4149100}
  +\alpha 
  \(-2r^{2}{\hat {\tilde B}_r{}^r} {B_z{}^z} 
    +2r^{2}{\hat B_r{}^z} {\hat B_z{}^r} 
    +r^{2}{B_z{}^z} f\hat Y 
    -{B_z{}^z} fm\theta 
  \)
  \nl 
  \off{4149100}
  +H^{-1}\alpha 
  \(-
    \frac{1}{
          2}r^{4}A_z d\hat H_{rz} \hat {\tilde s}_r 
    +r^{4}A_z \hat H_{rz} \mu \hat {\tilde s}_r 
    +
    \frac{1}{
          2}r^{2}a\hat A_r A_z \hat H_{rz} \nl 
    \off{6651983}
    +
    \frac{1}{
          2}r^{2}A_z dH_{rr} \hat s_z 
    -2r^{2}A_z d\hat H_{rz} \hat s 
    -r^{2}A_z H_{rr} \mu \hat s_z \nl 
    \off{6651983}
    +4r^{2}A_z \hat H_{rz} \mu \hat s 
    -r^{2}A_z \hat H_{rz} \mu \hat Z_r 
    -2r^{2}{B_z{}^z} \hat \chi_{rz} f\hat H_{rz} \nl 
    \off{6651983}
    -r^{2}d\hat H_{rz} \hat s_z 
    +2r^{2}\hat H_{rz} \mu \hat s_z 
    -
    \frac{1}{
          2}aA_z ^{2}H_{rr} \nl 
    \off{6651983}
    -A_z d\hat H_{rz} 
    +A_z H_{rr} \mu Z_z 
    +3A_z \hat H_{rz} \mu 
  \)
  \nl 
  \off{4149100}
  +H^{-2}\alpha r^{2}A_z d\hat H_{rz} ^{3}
  -\hat {\beta^r} {B_z{}^z} 
\]
\Nl}$$
$$\displaylines{\qdd
\tilde \S 
\(\chi_{rr} 
\)
=
\[\kappa \alpha H_{rr} 
  \(\frac{1}{
          2}r^{2}\hat {\tilde \tau} 
    -
    \frac{1}{
          2}\rho_K 
    -
    \frac{1}{
          2}\sigma 
  \)
  \nl 
  \off{4125582}
  +H^{-1}\kappa \alpha 
  \(-r^{2}H_{rr} \hat H_{rz} \hat S_{rz} 
    +
    \frac{1}{
          2}r^{2}\hat H_{rz} ^{2}S_{rr} 
    +
    \frac{1}{
          2}H_{rr} ^{2}S_{zz} 
  \)
  \nl 
  \off{4125582}
  -H^{-1}\alpha \chi_{rr} ^{2}r^{2}H_{rr} ^{
  -1}\hat H_{rz} ^{2}
  +H^{-1}\alpha \chi_{rr} \chi_{zz} H_{rr} \nl 
  \off{4125582}
  +\alpha \chi_{rr} 
  \(2r^{2}{\hat {\tilde B}_r{}^r} 
    +r^{2}\hat Y 
    -2{B_z{}^z} 
    -2\theta 
  \)
  \nl 
  \off{4125582}
  +2H^{-1}\alpha \chi_{rr} r^{2}\hat \chi_{rz} 
  \hat H_{rz} 
  +H^{-1}\alpha \hat D_{rrr} ^{2}r^{4}H_{rr} ^{
  -2}\hat H_{rz} ^{2}\nl 
  \off{4125582}
  -2H^{-1}\alpha \hat D_{rrr} \hat {\tilde D}_{rrz} r^{4}
  H_{rr} ^{-1}\hat H_{rz} 
  +H^{-1}\alpha \hat D_{rrr} \hat D_{rzz} r^{2}\nl 
  \off{4125582}
  -H^{-2}\alpha \hat D_{rrr} \hat D_{rzz} r^{4}\hat H_{rz} 
  ^{2}
  -4H^{-1}\alpha \hat D_{rrr} \hat D_{zrz} r^{2}\nl 
  \off{4125582}
  -2H^{-2}\alpha \hat D_{rrr} \hat D_{zrz} r^{4}
  \hat H_{rz} ^{2}
  +3H^{-2}\alpha \hat D_{rrr} D_{zzz} r^{2}H_{rr} 
  \hat H_{rz} \nl 
  \off{4125582}
  +\alpha \hat D_{rrr} H_{rr} ^{
  -1}
  \(-r^{4}\hat {\tilde s}_r 
    +2r^{2}\hat A_r 
    -2r^{2}\hat s 
    -2r^{2}\hat Z_r 
    -1
  \)
  \nl 
  \off{4125582}
  +H^{-1}\alpha \hat D_{rrr} r^{2}\hat H_{rz} 
  \(r^{4}H_{rr} ^{-1}\hat H_{rz} \hat {\tilde s}_r 
    +r^{2}\hat A_r H_{rr} ^{-1}\hat H_{rz} \nl 
    \off{9634494}
    +2r^{2}H_{rr} ^{-1}\hat H_{rz} \hat s 
    -2r^{2}H_{rr} ^{-1}\hat H_{rz} \hat Z_r \nl 
    \off{9634494}
    -r^{2}\hat s_z 
    -2A_z 
    +2Z_z 
  \)
  \nl 
  \off{4125582}
  +2H^{-2}\alpha \hat {\tilde D}_{rrz} \hat D_{rzz} r^{4}
  H_{rr} \hat H_{rz} \nl 
  \off{4125582}
  +4H^{-1}\alpha \hat {\tilde D}_{rrz} D_{zrr} r^{2}
  +4H^{-2}\alpha \hat {\tilde D}_{rrz} D_{zrr} r^{4}
  \hat H_{rz} ^{2}\nl 
  \off{4125582}
  -4H^{-2}\alpha \hat {\tilde D}_{rrz} \hat D_{zrz} r^{4}
  H_{rr} \hat H_{rz} 
  -2H^{-2}\alpha \hat {\tilde D}_{rrz} D_{zzz} r^{2}
  H_{rr} ^{2}\nl 
  \off{4125582}
  +2H^{-1}\alpha \hat {\tilde D}_{rrz} r^{2}
  \(-r^{4}\hat H_{rz} \hat {\tilde s}_r 
    -r^{2}\hat A_r \hat H_{rz} 
    +r^{2}H_{rr} \hat s_z 
    -2r^{2}\hat H_{rz} \hat s \nl 
    \off{8869754}
    +2r^{2}\hat H_{rz} \hat Z_r 
    +A_z H_{rr} 
    -2H_{rr} Z_z 
    -\hat H_{rz} 
  \)
  \nl 
  \off{4125582}
  -H^{-2}\alpha \hat D_{rzz} ^{2}r^{2}H_{rr} ^{2}
  -3H^{-2}\alpha \hat D_{rzz} D_{zrr} r^{2}H_{rr} 
  \hat H_{rz} \nl 
  \off{4125582}
  +4H^{-2}\alpha \hat D_{rzz} \hat D_{zrz} r^{2}H_{rr} ^{2
  }
  +H^{-1}\alpha \hat D_{rzz} r^{2}\hat A_r H_{rr} \nl 
  \off{4125582}
  +H^{-2}\alpha \hat D_{rzz} r^{2}H_{rr} \hat H_{rz} ^{2}
  -H^{-2}\alpha D_{zrr} ^{2}r^{2}\hat H_{rz} ^{2}\nl 
  \off{4125582}
  +2H^{-2}\alpha D_{zrr} \hat D_{zrz} r^{2}H_{rr} 
  \hat H_{rz} 
  -H^{-2}\alpha D_{zrr} D_{zzz} H_{rr} ^{2}\nl 
\cr}$$ 
$$\displaylines{\qdd
  \off{4125582}
  +H^{-1}\alpha D_{zrr} 
  \(r^{4}\hat H_{rz} \hat {\tilde s}_r 
    +2r^{2}\hat A_r \hat H_{rz} 
    -r^{2}H_{rr} \hat s_z 
    +2r^{2}\hat H_{rz} \hat s \nl 
    \off{7987901}
    -2r^{2}\hat H_{rz} \hat Z_r 
    +2H_{rr} Z_z 
    +3\hat H_{rz} 
  \)
  \nl 
  \off{4125582}
  +2H^{-2}\alpha D_{zrr} r^{2}\hat H_{rz} ^{3}
  -2H^{-1}\alpha \hat D_{zrz} r^{2}\hat A_r H_{rr} \nl 
  \off{4125582}
  -2H^{-2}\alpha \hat D_{zrz} r^{2}H_{rr} \hat H_{rz} ^{2}
  -H^{-2}\alpha D_{zzz} H_{rr} ^{2}\hat H_{rz} \nl 
  \off{4125582}
  +\alpha e^{2 r^2 \hat s} r^{4}H_{rr} 
  \(-
    \frac{1}{
          2}\hat {B^\varphi} ^{2}H_{rr} 
    +
    \frac{1}{
          2}\hat {E^r} ^{2}H_{rr} ^{2}\nl 
    \off{8390968}
    +\hat {E^r} {E^z} H_{rr} \hat H_{rz} 
    +
    \frac{1}{
          2}{E^z} ^{2}\hat H_{rz} ^{2}
  \)
  \nl 
  \off{4125582}
  +\alpha 
  \(-r^{6}\hat {\tilde s}_r ^{2}
    +r^{4}\hat A_r \hat {\tilde s}_r 
    -4r^{4}\hat s \hat {\tilde s}_r 
    -2r^{2}\hat A_r \hat Z_r \nl 
    \off{5384647}
    +4r^{2}{\hat B_r{}^z} \hat \chi_{rz} 
    -4r^{2}\hat s ^{2}
    -4r^{2}\hat {\tilde s}_r 
    -6\hat s 
  \)
  \nl 
  \off{4125582}
  +H^{-1}\alpha 
  \(-r^{4}\hat H_{rz} ^{2}\hat {\tilde s}_r 
    -r^{2}\hat A_r \hat H_{rz} ^{2}
    -2r^{2}\hat \chi_{rz} ^{2}H_{rr} \nl 
    \off{6628465}
    +r^{2}H_{rr} \hat H_{rz} \hat s_z 
    -2r^{2}\hat H_{rz} ^{2}\hat s 
    +2r^{2}\hat H_{rz} ^{2}\hat Z_r \nl 
    \off{6628465}
    +A_z H_{rr} \hat H_{rz} 
    -2H_{rr} \hat H_{rz} Z_z 
    -\hat H_{rz} ^{2}
  \)
  +\chi_{rr} \hat {\beta^r} 
\]
\Nl}$$
$$\displaylines{\qdd
\tilde \S 
\(\chi_{rz} 
\)
=
\[\kappa \alpha r
  \(\frac{1}{
          2}r^{2}\hat H_{rz} \hat {\tilde \tau} 
    +H_{rr} ^{-1}\hat H_{rz} S_{rr} 
    -
    \frac{1}{
          2}\hat H_{rz} \rho_K 
    -
    \frac{1}{
          2}\hat H_{rz} \sigma 
    -\hat S_{rz} 
  \)
  \nl 
  \off{4137371}
  +H^{-1}\kappa \alpha r\hat H_{rz} 
  \(\frac{1}{
          2}r^{2}H_{rr} ^{-1}\hat H_{rz} ^{2}S_{rr} 
    -r^{2}\hat H_{rz} \hat S_{rz} 
    +
    \frac{1}{
          2}H_{rr} S_{zz} 
  \)
  \nl 
  \off{4137371}
  +2H^{-1}\alpha \chi_{rr} \chi_{zz} r\hat H_{rz} 
  +2\alpha \chi_{rr} r{\hat B_z{}^r} \nl 
  \off{4137371}
  -H^{-1}\alpha \chi_{rr} r^{3}\hat \chi_{rz} H_{rr} ^{
  -1}\hat H_{rz} ^{2}
  +2\alpha \chi_{zz} r{\hat B_r{}^z} \nl 
  \off{4137371}
  -H^{-1}\alpha \chi_{zz} r\hat \chi_{rz} H_{rr} 
  -2H^{-1}\alpha \hat D_{rrr} \hat D_{rzz} r^{3}H_{rr} ^{
  -1}\hat H_{rz} \nl 
  \off{4137371}
  -H^{-2}\alpha \hat D_{rrr} \hat D_{rzz} r^{5}H_{rr} ^{
  -1}\hat H_{rz} ^{3}\nl 
  \off{4137371}
  +H^{-1}\alpha \hat D_{rrr} D_{zrr} r^{3}H_{rr} ^{
  -2}\hat H_{rz} ^{2}\nl 
  \off{4137371}
  -2H^{-1}\alpha \hat D_{rrr} \hat D_{zrz} r^{3}H_{rr} ^{
  -1}\hat H_{rz} \nl 
  \off{4137371}
  -2H^{-2}\alpha \hat D_{rrr} \hat D_{zrz} r^{5}H_{rr} ^{
  -1}\hat H_{rz} ^{3}\nl 
  \off{4137371}
  +3H^{-2}\alpha \hat D_{rrr} D_{zzz} r^{3}\hat H_{rz} ^{2
  }
  -\alpha \hat D_{rrr} r^{3}H_{rr} ^{
  -1}\hat s_z \nl 
  \off{4137371}
  -
  \frac{1}{
        2}H^{-1}\alpha \hat D_{rrr} r^{3}A_z H_{rr} ^{
  -1}\hat H_{rz} ^{2}
  +2H^{-1}\alpha \hat {\tilde D}_{rrz} \hat D_{rzz} r^{3}
  \nl 
  \off{4137371}
  +2H^{-2}\alpha \hat {\tilde D}_{rrz} \hat D_{rzz} r^{5}
  \hat H_{rz} ^{2}
  +4H^{-1}\alpha \hat {\tilde D}_{rrz} D_{zrr} r^{3}
  H_{rr} ^{-1}\hat H_{rz} \nl 
  \off{4137371}
  +4H^{-2}\alpha \hat {\tilde D}_{rrz} D_{zrr} r^{5}
  H_{rr} ^{-1}\hat H_{rz} ^{3}
  -4H^{-1}\alpha \hat {\tilde D}_{rrz} \hat D_{zrz} r^{3}
  \nl 
  \off{4137371}
  -4H^{-2}\alpha \hat {\tilde D}_{rrz} \hat D_{zrz} r^{5}
  \hat H_{rz} ^{2}
  -2H^{-2}\alpha \hat {\tilde D}_{rrz} D_{zzz} r^{3}
  H_{rr} \hat H_{rz} \nl 
  \off{4137371}
  -H^{-1}\alpha \hat {\tilde D}_{rrz} r^{3}A_z \hat H_{rz} 
  -H^{-2}\alpha \hat D_{rzz} ^{2}r^{3}H_{rr} \hat H_{rz} 
  \nl 
  \off{4137371}
  +H^{-1}\alpha \hat D_{rzz} D_{zrr} r
  -3H^{-2}\alpha \hat D_{rzz} D_{zrr} r^{3}\hat H_{rz} ^{2
  }\nl 
  \off{4137371}
  +4H^{-2}\alpha \hat D_{rzz} \hat D_{zrz} r^{3}H_{rr} 
  \hat H_{rz} \nl 
  \off{4137371}
  +H^{-1}\alpha \hat D_{rzz} r
  \(-r^{4}\hat H_{rz} \hat {\tilde s}_r 
    +r^{2}H_{rr} \hat s_z 
    -2r^{2}\hat H_{rz} \hat s \nl 
    \off{8288915}
    +2r^{2}\hat H_{rz} \hat Z_r 
    +
    \frac{3}{
          2}A_z H_{rr} 
    -2H_{rr} Z_z 
  \)
  \nl 
  \off{4137371}
  +H^{-2}\alpha \hat D_{rzz} r^{3}\hat H_{rz} ^{3}
  -2H^{-1}\alpha D_{zrr} ^{2}rH_{rr} ^{
  -1}\hat H_{rz} \nl 
  \off{4137371}
  -H^{-2}\alpha D_{zrr} ^{2}r^{3}H_{rr} ^{
  -1}\hat H_{rz} ^{3}
  +2H^{-1}\alpha D_{zrr} \hat D_{zrz} r\nl 
  \off{4137371}
  +2H^{-2}\alpha D_{zrr} \hat D_{zrz} r^{3}\hat H_{rz} ^{2
  }
  -H^{-2}\alpha D_{zrr} D_{zzz} rH_{rr} \hat H_{rz} \nl 
  \off{4137371}
  +2\alpha D_{zrr} rH_{rr} ^{-1}
  \(\hat A_r 
    -\hat Z_r 
  \)
  \nl 
\cr}$$ 
$$\displaylines{\qdd
  \off{4137371}
  +H^{-1}\alpha D_{zrr} r\hat H_{rz} 
  \(r^{4}H_{rr} ^{-1}\hat H_{rz} \hat {\tilde s}_r 
    +
    \frac{3}{
          2}r^{2}\hat A_r H_{rr} ^{
    -1}\hat H_{rz} \nl 
    \off{9381335}
    +2r^{2}H_{rr} ^{-1}\hat H_{rz} \hat s 
    -2r^{2}H_{rr} ^{-1}\hat H_{rz} \hat Z_r \nl 
    \off{9381335}
    -r^{2}\hat s_z 
    +2H_{rr} ^{-1}\hat H_{rz} 
    +2Z_z 
  \)
  \nl 
  \off{4137371}
  +H^{-2}\alpha D_{zrr} r^{3}H_{rr} ^{
  -1}\hat H_{rz} ^{4}
  -H^{-1}\alpha \hat D_{zrz} r^{3}\hat A_r \hat H_{rz} \nl 
  \off{4137371}
  -
  \frac{1}{
        2}H^{-1}\alpha D_{zzz} r\hat A_r H_{rr} 
  -2H^{-2}\alpha D_{zzz} rH_{rr} \hat H_{rz} ^{2}\nl 
  \off{4137371}
  +
  \frac{1}{
        2}\alpha e^{2 r^2 \hat s} Hr^{3}{E^z} 
  \(\hat {E^r} H_{rr} 
    +{E^z} \hat H_{rz} 
  \)
  \nl 
  \off{4137371}
  +\alpha e^{2 r^2 \hat s} r^{5}\hat H_{rz} 
  \(-
    \frac{1}{
          2}\hat {B^\varphi} ^{2}H_{rr} 
    +
    \frac{1}{
          2}\hat {E^r} ^{2}H_{rr} ^{2}\nl 
    \off{8414516}
    +\hat {E^r} {E^z} H_{rr} \hat H_{rz} 
    +
    \frac{1}{
          2}{E^z} ^{2}\hat H_{rz} ^{2}
  \)
  \nl 
  \off{4137371}
  +\alpha r
  \(-r^{4}\hat {\tilde s}_r \hat s_z 
    +
    \frac{1}{
          2}r^{2}\hat A_r \hat s_z 
    +
    \frac{1}{
          2}r^{2}A_z \hat {\tilde s}_r 
    +r^{2}\hat \chi_{rz} \hat Y \nl 
    \off{5673902}
    -2r^{2}\hat s \hat s_z 
    -\hat A_r Z_z 
    -A_z \hat Z_r 
    -2\hat \chi_{rz} \theta 
    -2\hat s_z 
  \)
\]
\Nl}$$
$$\displaylines{\qdd
\tilde \S 
\(\chi_{zz} 
\)
=
\[\kappa \alpha HH_{rr} ^{-2}
  \(\frac{1}{
          2}r^{2}H_{rr} \hat {\tilde \tau} 
    -
    \frac{1}{
          2}H_{rr} \rho_K 
    -
    \frac{1}{
          2}H_{rr} \sigma 
    +S_{rr} 
  \)
  \nl 
  \off{4149100}
  +\kappa \alpha 
  \(\frac{1}{
          2}r^{4}H_{rr} ^{-1}\hat H_{rz} ^{2}
    \hat {\tilde \tau} 
    -
    \frac{1}{
          2}r^{2}H_{rr} ^{-1}\hat H_{rz} ^{2}\rho_K 
    -
    \frac{1}{
          2}r^{2}H_{rr} ^{-1}\hat H_{rz} ^{2}\sigma \nl 
    \off{5785755}
    -r^{2}H_{rr} ^{-1}\hat H_{rz} \hat S_{rz} 
    +
    \frac{3}{
          2}r^{2}H_{rr} ^{-2}\hat H_{rz} ^{2}S_{rr} 
    -
    \frac{1}{
          2}S_{zz} 
  \)
  \nl 
  \off{4149100}
  +H^{-1}\kappa \alpha r^{2}\hat H_{rz} ^{2}
  \(-r^{2}H_{rr} ^{-1}\hat H_{rz} \hat S_{rz} 
    +
    \frac{1}{
          2}r^{2}H_{rr} ^{-2}\hat H_{rz} ^{2}S_{rr} \nl 
    \off{8964632}
    +
    \frac{1}{
          2}S_{zz} 
  \)
  +2\alpha \chi_{rr} \chi_{zz} H_{rr} ^{
  -1}\nl 
  \off{4149100}
  +H^{-1}\alpha \chi_{rr} \chi_{zz} r^{2}H_{rr} ^{
  -1}\hat H_{rz} ^{2}
  -H^{-1}\alpha \chi_{zz} ^{2}H_{rr} \nl 
  \off{4149100}
  +\alpha \chi_{zz} 
  \(-2r^{2}{\hat {\tilde B}_r{}^r} 
    +r^{2}\hat Y 
    +2{B_z{}^z} 
    -2\theta 
  \)
  +2H^{-1}\alpha \chi_{zz} r^{2}\hat \chi_{rz} 
  \hat H_{rz} \nl 
  \off{4149100}
  -2\alpha \hat D_{rrr} \hat D_{rzz} r^{2}H_{rr} ^{
  -2}
  -3H^{-1}\alpha \hat D_{rrr} \hat D_{rzz} r^{4}H_{rr} ^{
  -2}\hat H_{rz} ^{2}\nl 
  \off{4149100}
  -H^{-2}\alpha \hat D_{rrr} \hat D_{rzz} r^{6}H_{rr} ^{
  -2}\hat H_{rz} ^{4}\nl 
  \off{4149100}
  -2H^{-1}\alpha \hat D_{rrr} \hat D_{zrz} r^{4}H_{rr} ^{
  -2}\hat H_{rz} ^{2}\nl 
  \off{4149100}
  -2H^{-2}\alpha \hat D_{rrr} \hat D_{zrz} r^{6}H_{rr} ^{
  -2}\hat H_{rz} ^{4}\nl 
  \off{4149100}
  +2H^{-1}\alpha \hat D_{rrr} D_{zzz} r^{2}H_{rr} ^{
  -1}\hat H_{rz} \nl 
  \off{4149100}
  +3H^{-2}\alpha \hat D_{rrr} D_{zzz} r^{4}H_{rr} ^{
  -1}\hat H_{rz} ^{3}\nl 
  \off{4149100}
  +2H^{-1}\alpha \hat {\tilde D}_{rrz} \hat D_{rzz} r^{4}
  H_{rr} ^{-1}\hat H_{rz} \nl 
  \off{4149100}
  +2H^{-2}\alpha \hat {\tilde D}_{rrz} \hat D_{rzz} r^{6}
  H_{rr} ^{-1}\hat H_{rz} ^{3}\nl 
  \off{4149100}
  +4\alpha \hat {\tilde D}_{rrz} D_{zrr} r^{2}H_{rr} ^{
  -2}
  +8H^{-1}\alpha \hat {\tilde D}_{rrz} D_{zrr} r^{4}
  H_{rr} ^{-2}\hat H_{rz} ^{2}\nl 
  \off{4149100}
  +4H^{-2}\alpha \hat {\tilde D}_{rrz} D_{zrr} r^{6}
  H_{rr} ^{-2}\hat H_{rz} ^{4}\nl 
  \off{4149100}
  -4H^{-1}\alpha \hat {\tilde D}_{rrz} \hat D_{zrz} r^{4}
  H_{rr} ^{-1}\hat H_{rz} \nl 
  \off{4149100}
  -4H^{-2}\alpha \hat {\tilde D}_{rrz} \hat D_{zrz} r^{6}
  H_{rr} ^{-1}\hat H_{rz} ^{3}
  -4H^{-1}\alpha \hat {\tilde D}_{rrz} D_{zzz} r^{2}\nl 
  \off{4149100}
  -2H^{-2}\alpha \hat {\tilde D}_{rrz} D_{zzz} r^{4}
  \hat H_{rz} ^{2}
  -2\alpha \hat {\tilde D}_{rrz} r^{2}A_z H_{rr} ^{
  -1}\nl 
  \off{4149100}
  -2H^{-1}\alpha \hat {\tilde D}_{rrz} r^{4}A_z H_{rr} ^{
  -1}\hat H_{rz} ^{2}
  +H^{-1}\alpha \hat D_{rzz} ^{2}r^{2}\nl 
  \off{4149100}
  -H^{-2}\alpha \hat D_{rzz} ^{2}r^{4}\hat H_{rz} ^{2}
  -2H^{-1}\alpha \hat D_{rzz} D_{zrr} r^{2}H_{rr} ^{
  -1}\hat H_{rz} \nl 
  \off{4149100}
  -3H^{-2}\alpha \hat D_{rzz} D_{zrr} r^{4}H_{rr} ^{
  -1}\hat H_{rz} ^{3}\nl 
\cr}$$ 
$$\displaylines{\qdd
  \off{4149100}
  +2H^{-1}\alpha \hat D_{rzz} \hat D_{zrz} r^{2}
  +4H^{-2}\alpha \hat D_{rzz} \hat D_{zrz} r^{4}
  \hat H_{rz} ^{2}\nl 
  \off{4149100}
  +\alpha \hat D_{rzz} H_{rr} ^{
  -1}
  \(-r^{4}\hat {\tilde s}_r 
    -2r^{2}\hat s 
    +2r^{2}\hat Z_r 
    -1
  \)
  \nl 
  \off{4149100}
  +H^{-1}\alpha \hat D_{rzz} r^{2}\hat H_{rz} 
  \(-r^{4}H_{rr} ^{-1}\hat H_{rz} \hat {\tilde s}_r 
    -2r^{2}H_{rr} ^{-1}\hat H_{rz} \hat s \nl 
    \off{9681530}
    +2r^{2}H_{rr} ^{-1}\hat H_{rz} \hat Z_r 
    +r^{2}\hat s_z 
    +2A_z 
    -2Z_z 
  \)
  \nl 
  \off{4149100}
  +H^{-2}\alpha \hat D_{rzz} r^{4}H_{rr} ^{
  -1}\hat H_{rz} ^{4}
  -2\alpha D_{zrr} ^{2}H_{rr} ^{
  -2}\nl 
  \off{4149100}
  -2H^{-1}\alpha D_{zrr} ^{2}r^{2}H_{rr} ^{
  -2}\hat H_{rz} ^{2}
  -H^{-2}\alpha D_{zrr} ^{2}r^{4}H_{rr} ^{
  -2}\hat H_{rz} ^{4}\nl 
  \off{4149100}
  +2H^{-2}\alpha D_{zrr} \hat D_{zrz} r^{4}H_{rr} ^{
  -1}\hat H_{rz} ^{3}\nl 
  \off{4149100}
  +2H^{-1}\alpha D_{zrr} D_{zzz} 
  -H^{-2}\alpha D_{zrr} D_{zzz} r^{2}\hat H_{rz} ^{2}\nl 
  \off{4149100}
  +2\alpha D_{zrr} H_{rr} ^{-1}
  \(-r^{2}\hat s_z 
    +A_z 
  \)
  +H^{-1}\alpha D_{zrr} r^{2}A_z H_{rr} ^{
  -1}\hat H_{rz} ^{2}\nl 
  \off{4149100}
  +2\alpha \hat D_{zrz} H_{rr} ^{
  -1}
  \(r^{4}\hat {\tilde s}_r 
    +r^{2}\hat A_r 
    +2r^{2}\hat s 
    -2r^{2}\hat Z_r 
    +2
  \)
  \nl 
  \off{4149100}
  +2H^{-1}\alpha \hat D_{zrz} r^{2}\hat H_{rz} 
  \(r^{4}H_{rr} ^{-1}\hat H_{rz} \hat {\tilde s}_r 
    +r^{2}\hat A_r H_{rr} ^{-1}\hat H_{rz} \nl 
    \off{10009210}
    +2r^{2}H_{rr} ^{-1}\hat H_{rz} \hat s 
    -2r^{2}H_{rr} ^{-1}\hat H_{rz} \hat Z_r \nl 
    \off{10009210}
    -r^{2}\hat s_z 
    -A_z 
    +3H_{rr} ^{-1}\hat H_{rz} 
    +2Z_z 
  \)
  \nl 
  \off{4149100}
  +2H^{-2}\alpha \hat D_{zrz} r^{4}H_{rr} ^{
  -1}\hat H_{rz} ^{4}\nl 
  \off{4149100}
  +H^{-1}\alpha D_{zzz} 
  \(-r^{4}\hat H_{rz} \hat {\tilde s}_r 
    -2r^{2}\hat A_r \hat H_{rz} 
    +r^{2}H_{rr} \hat s_z 
    -2r^{2}\hat H_{rz} \hat s \nl 
    \off{8034937}
    +2r^{2}\hat H_{rz} \hat Z_r 
    +A_z H_{rr} 
    -2H_{rr} Z_z 
    -3\hat H_{rz} 
  \)
  \nl 
  \off{4149100}
  -3H^{-2}\alpha D_{zzz} r^{2}\hat H_{rz} ^{3}
  +
  \frac{1}{
        2}\alpha e^{2 r^2 \hat s} H^{2}r^{2}{E^z} ^{2}
  H_{rr} ^{-1}\nl 
  \off{4149100}
  +\alpha e^{2 r^2 \hat s} Hr^{4}
  \(-
    \frac{1}{
          2}\hat {B^\varphi} ^{2}
    +\hat {E^r} {E^z} \hat H_{rz} 
    +{E^z} ^{2}H_{rr} ^{-1}\hat H_{rz} ^{2}
  \)
  \nl 
  \off{4149100}
  +\alpha e^{2 r^2 \hat s} r^{6}\hat H_{rz} ^{2}
  \(-
    \frac{1}{
          2}\hat {B^\varphi} ^{2}
    +
    \frac{1}{
          2}\hat {E^r} ^{2}H_{rr} \nl 
    \off{8702952}
    +\hat {E^r} {E^z} \hat H_{rz} 
    +
    \frac{1}{
          2}{E^z} ^{2}H_{rr} ^{-1}\hat H_{rz} ^{2}
  \)
  \nl 
  \off{4149100}
  +\alpha 
  \(-r^{4}\hat s_z ^{2}
    +r^{2}A_z \hat s_z 
    +4r^{2}{\hat B_z{}^r} \hat \chi_{rz} 
    -2r^{2}\hat \chi_{rz} ^{2}H_{rr} ^{
    -1}\nl 
    \off{5408165}
    -2A_z Z_z 
  \)
  -2H^{-1}\alpha r^{4}\hat \chi_{rz} ^{2}H_{rr} ^{
  -1}\hat H_{rz} ^{2}
  -\chi_{zz} \hat {\beta^r} 
\]
\Nl}$$
$$\displaylines{\qdd
\tilde \S 
\(Y 
\)
=
\[-\kappa \alpha r\hat {\tilde \tau} 
  +2H^{-1}\alpha \chi_{rr} ^{2}rH_{rr} ^{
  -2}\hat H_{rz} ^{2}\nl 
  \off{3693986}
  +2\alpha \chi_{rr} rH_{rr} ^{
  -2}
  \(2{\hat B_r{}^z} \hat H_{rz} 
    +H_{rr} \hat Y 
  \)
  \nl 
  \off{3693986}
  +H^{-1}\alpha \chi_{rr} rH_{rr} ^{
  -1}\hat H_{rz} 
  \(r^{2}\hat H_{rz} \hat Y 
    -4\hat \chi_{rz} 
  \)
  \nl 
  \off{3693986}
  +H^{-1}\alpha \chi_{zz} rH_{rr} \hat Y 
  -4H^{-1}\alpha \hat D_{rrr} ^{2}r^{3}H_{rr} ^{
  -3}\hat H_{rz} ^{2}\nl 
  \off{3693986}
  -H^{-2}\alpha \hat D_{rrr} ^{2}r^{5}H_{rr} ^{
  -3}\hat H_{rz} ^{4}
  +4H^{-1}\alpha \hat D_{rrr} \hat {\tilde D}_{rrz} r^{3}
  H_{rr} ^{-2}\hat H_{rz} \nl 
  \off{3693986}
  +2H^{-2}\alpha \hat D_{rrr} \hat {\tilde D}_{rrz} r^{5}
  H_{rr} ^{-2}\hat H_{rz} ^{3}
  +4H^{-1}\alpha \hat D_{rrr} D_{zrr} rH_{rr} ^{
  -2}\hat H_{rz} \nl 
  \off{3693986}
  +2H^{-2}\alpha \hat D_{rrr} D_{zrr} r^{3}H_{rr} ^{
  -2}\hat H_{rz} ^{3}
  -2H^{-2}\alpha \hat D_{rrr} D_{zzz} r\hat H_{rz} \nl 
  \off{3693986}
  +H^{-1}\alpha \hat D_{rrr} rH_{rr} ^{
  -2}\hat H_{rz} 
  \(-5r^{4}\hat H_{rz} \hat {\tilde s}_r 
    -r^{2}\hat A_r \hat H_{rz} 
    +4r^{2}H_{rr} \hat s_z \nl 
    \off{10717661}
    -2r^{2}\hat H_{rz} \hat s 
    +4r^{2}\hat H_{rz} \hat Z_r \nl 
    \off{10717661}
    +2A_z H_{rr} 
    -4H_{rr} Z_z 
  \)
  \nl 
  \off{3693986}
  +H^{-2}\alpha \hat D_{rrr} r^{3}H_{rr} ^{
  -2}\hat H_{rz} ^{3}
  \(-r^{4}\hat H_{rz} \hat {\tilde s}_r 
    +r^{2}H_{rr} \hat s_z \nl 
    \off{11271075}
    +2r^{2}\hat H_{rz} \hat s 
    +2\hat H_{rz} 
  \)
  \nl 
  \off{3693986}
  -2H^{-2}\alpha \hat {\tilde D}_{rrz} \hat D_{rzz} r^{3}
  \hat H_{rz} 
  -4H^{-1}\alpha \hat {\tilde D}_{rrz} D_{zrr} rH_{rr} ^{
  -1}\nl 
  \off{3693986}
  -6H^{-2}\alpha \hat {\tilde D}_{rrz} D_{zrr} r^{3}
  H_{rr} ^{-1}\hat H_{rz} ^{2}\nl 
  \off{3693986}
  +4H^{-2}\alpha \hat {\tilde D}_{rrz} \hat D_{zrz} r^{3}
  \hat H_{rz} 
  +2H^{-2}\alpha \hat {\tilde D}_{rrz} D_{zzz} rH_{rr} 
  \nl 
  \off{3693986}
  +2H^{-1}\alpha \hat {\tilde D}_{rrz} r
  \(2r^{4}H_{rr} ^{-1}\hat H_{rz} \hat {\tilde s}_r 
    +r^{2}\hat A_r H_{rr} ^{-1}\hat H_{rz} \nl 
    \off{8161451}
    -2r^{2}H_{rr} ^{-1}\hat H_{rz} \hat Z_r 
    -r^{2}\hat s_z 
    -A_z 
    +2Z_z 
  \)
  \nl 
  \off{3693986}
  +2H^{-2}\alpha \hat {\tilde D}_{rrz} r^{3}\hat H_{rz} ^{2
  }
  \(r^{4}H_{rr} ^{-1}\hat H_{rz} \hat {\tilde s}_r 
    -2r^{2}H_{rr} ^{-1}\hat H_{rz} \hat s \nl 
    \off{9819044}
    -r^{2}\hat s_z 
    -H_{rr} ^{-1}\hat H_{rz} 
  \)
  \nl 
  \off{3693986}
  +H^{-2}\alpha \hat D_{rzz} ^{2}rH_{rr} 
  +4H^{-2}\alpha \hat D_{rzz} D_{zrr} r\hat H_{rz} \nl 
  \off{3693986}
  -4H^{-2}\alpha \hat D_{rzz} \hat D_{zrz} rH_{rr} 
  +H^{-1}\alpha \hat D_{rzz} r
  \(-r^{2}\hat {\tilde s}_r 
    -\hat A_r 
    -2\hat s 
  \)
  \nl 
  \off{3693986}
  +H^{-2}\alpha \hat D_{rzz} r^{3}\hat H_{rz} 
  \(-r^{2}\hat H_{rz} \hat {\tilde s}_r 
    +H_{rr} \hat s_z 
    +2\hat H_{rz} \hat s 
  \)
  \nl 
  \off{3693986}
  +2H^{-1}\alpha D_{zrr} r
  \(r^{2}H_{rr} ^{-1}\hat H_{rz} \hat {\tilde s}_r 
    -\hat A_r H_{rr} ^{-1}\hat H_{rz} 
    -\hat s_z 
  \)
  \nl 
\cr}$$ 
$$\displaylines{\qdd
  \off{3693986}
  +H^{-2}\alpha D_{zrr} r\hat H_{rz} ^{2}
  \(r^{4}H_{rr} ^{-1}\hat H_{rz} \hat {\tilde s}_r 
    -2r^{2}H_{rr} ^{-1}\hat H_{rz} \hat s 
    -r^{2}\hat s_z \nl 
    \off{9214657}
    -4H_{rr} ^{-1}\hat H_{rz} 
  \)
  +2H^{-1}\alpha \hat D_{zrz} r
  \(\hat A_r 
    +2\hat s 
  \)
  \nl 
  \off{3693986}
  +2H^{-2}\alpha \hat D_{zrz} r\hat H_{rz} 
  \(-r^{4}\hat H_{rz} \hat {\tilde s}_r 
    +r^{2}H_{rr} \hat s_z 
    +2r^{2}\hat H_{rz} \hat s 
    +2\hat H_{rz} 
  \)
  \nl 
  \off{3693986}
  +H^{-2}\alpha D_{zzz} rH_{rr} 
  \(r^{2}\hat H_{rz} \hat {\tilde s}_r 
    -H_{rr} \hat s_z 
    -2\hat H_{rz} \hat s 
  \)
  -
  \frac{1}{
        2}\alpha e^{2 r^2 \hat s} Hr{E^z} ^{2}\nl 
  \off{3693986}
  +\alpha e^{2 r^2 \hat s} r^{3}
  \(\hat {B^\varphi} ^{2}H_{rr} 
    -\hat {E^r} ^{2}H_{rr} ^{2}
    -2\hat {E^r} {E^z} H_{rr} \hat H_{rz} 
    -{E^z} ^{2}\hat H_{rz} ^{2}
  \)
  \nl 
  \off{3693986}
  +\alpha r
  \(-r^{2}\hat A_r H_{rr} ^{-1}\hat {\tilde s}_r 
    -2r^{2}{\hat {\tilde B}_r{}^r} \hat Y 
    +2r^{2}H_{rr} ^{-1}\hat {\tilde s}_r \hat Z_r \nl 
    \off{5230517}
    +r^{2}\hat Y ^{2}
    -2\hat A_r H_{rr} ^{-1}\hat s 
    +2\hat A_r H_{rr} ^{-1}\hat Z_r \nl 
    \off{5230517}
    -4{\hat B_r{}^z} \hat \chi_{rz} H_{rr} ^{
    -1}
    -2{B_z{}^z} \hat Y 
    +4H_{rr} ^{-1}\hat s \hat Z_r 
    -2\theta \hat Y 
  \)
  \nl 
  \off{3693986}
  +H^{-1}\alpha r
  \(-r^{6}H_{rr} ^{-1}\hat H_{rz} ^{2}\hat {\tilde s}_r ^{2}
    -4r^{4}H_{rr} ^{-1}\hat H_{rz} ^{2}\hat s 
    \hat {\tilde s}_r \nl 
    \off{6474335}
    +2r^{4}H_{rr} ^{-1}\hat H_{rz} ^{2}\hat {\tilde s}_r 
    \hat Z_r 
    +2r^{4}\hat H_{rz} \hat {\tilde s}_r \hat s_z \nl 
    \off{6474335}
    -2r^{2}\hat A_r H_{rr} ^{-1}\hat H_{rz} ^{2}\hat s 
    -2r^{2}\hat \chi_{rz} \hat H_{rz} \hat Y 
    -r^{2}H_{rr} \hat s_z ^{2}\nl 
    \off{6474335}
    -4r^{2}H_{rr} ^{-1}\hat H_{rz} ^{2}\hat s ^{2}
    +4r^{2}H_{rr} ^{-1}\hat H_{rz} ^{2}\hat s \hat Z_r \nl 
    \off{6474335}
    -3r^{2}H_{rr} ^{-1}\hat H_{rz} ^{2}\hat {\tilde s}_r 
    +4r^{2}\hat H_{rz} \hat s \hat s_z 
    -2r^{2}\hat H_{rz} \hat {\tilde s}_r Z_z \nl 
    \off{6474335}
    -2r^{2}\hat H_{rz} \hat s_z \hat Z_r 
    +2A_z \hat H_{rz} \hat s 
    +2\hat \chi_{rz} ^{2}
    +2H_{rr} \hat s_z Z_z \nl 
    \off{6474335}
    -6H_{rr} ^{-1}\hat H_{rz} ^{2}\hat s 
    -4\hat H_{rz} \hat s Z_z 
    +4\hat H_{rz} \hat s_z 
  \)
  \nl 
  \off{3693986}
  +H^{-2}\alpha r\hat H_{rz} ^{3}
  \(r^{4}H_{rr} ^{-1}\hat H_{rz} \hat {\tilde s}_r 
    -2r^{2}H_{rr} ^{-1}\hat H_{rz} \hat s \nl 
    \off{7855221}
    -r^{2}\hat s_z 
    -H_{rr} ^{-1}\hat H_{rz} 
  \)
\]
\Nl}$$
$$\displaylines{\qdd
\tilde \S 
\({E^r} 
\)
=
\[-2\kappa \alpha rH_{rr} ^{-1}\hat S_r 
  +2H^{-1}\kappa \alpha r\hat H_{rz} 
  \(-r^{2}H_{rr} ^{-1}\hat H_{rz} \hat S_r 
    +S_z 
  \)
  \nl 
  \off{3979643}
  +4\alpha \chi_{rr} r\hat {E^r} H_{rr} ^{
  -1}
  +H^{-1}\alpha \chi_{rr} r^{3}\hat {E^r} H_{rr} ^{
  -1}\hat H_{rz} ^{2}\nl 
  \off{3979643}
  +H^{-1}\alpha \chi_{zz} r\hat {E^r} H_{rr} 
  +4H^{-1}\alpha \hat D_{rrr} rH_{rr} ^{
  -1}H_{zz} {Z^\varphi} \nl 
  \off{3979643}
  +4H^{-2}\alpha \hat D_{rrr} r^{3}H_{rr} ^{
  -1}\hat H_{rz} ^{2}H_{zz} {Z^\varphi} \nl 
  \off{3979643}
  -8H^{-1}\alpha \hat {\tilde D}_{rrz} r^{3}H_{rr} ^{
  -1}\hat H_{rz} {Z^\varphi} 
  -8H^{-2}\alpha \hat {\tilde D}_{rrz} r^{5}H_{rr} ^{
  -1}\hat H_{rz} ^{3}{Z^\varphi} \nl 
  \off{3979643}
  +4H^{-2}\alpha \hat D_{rzz} r^{3}\hat H_{rz} ^{2}
  {Z^\varphi} 
  -4H^{-1}\alpha D_{zrr} rH_{rr} ^{
  -1}\hat H_{rz} {Z^\varphi} \nl 
  \off{3979643}
  -4H^{-2}\alpha D_{zrr} r^{3}H_{rr} ^{
  -1}\hat H_{rz} ^{3}{Z^\varphi} 
  +4H^{-1}\alpha \hat D_{zrz} r{Z^\varphi} \nl 
  \off{3979643}
  +8H^{-2}\alpha \hat D_{zrz} r^{3}\hat H_{rz} ^{2}
  {Z^\varphi} 
  -4H^{-2}\alpha D_{zzz} rH_{rr} \hat H_{rz} {Z^\varphi} 
  \nl 
  \off{3979643}
  +\alpha r
  \(-4r^{2}{\hat {\tilde B}_r{}^r} \hat {E^r} 
    +3r^{2}\hat {E^r} \hat Y 
    -2\hat A_r H_{rr} ^{-1}{Z^\varphi} \nl 
    \off{5516174}
    -2{\hat B_z{}^r} {E^z} 
    -2{B_z{}^z} \hat {E^r} 
    -2\hat {E^r} \theta 
  \)
  \nl 
  \off{3979643}
  +2H^{-1}\alpha r\hat H_{rz} 
  \(-r^{2}\hat A_r H_{rr} ^{-1}\hat H_{rz} {Z^\varphi} 
    -r^{2}\hat \chi_{rz} \hat {E^r} \nl 
    \off{8191851}
    +A_z {Z^\varphi} 
    -2H_{rr} ^{-1}\hat H_{rz} {Z^\varphi} 
  \)
  \nl 
  \off{3979643}
  -4H^{-2}\alpha r^{3}H_{rr} ^{
  -1}\hat H_{rz} ^{4}{Z^\varphi} 
  -2r\hat {\beta^r} \hat {E^r} \nl 
  \off{3979643}
  +4
  \(\sqrt{H}
  \)
  ^{-1}\alpha D_{zrr} r\hat {B^\varphi} H_{rr} ^{
  -1}\nl 
  \off{3979643}
  +H^{-1}
  \(\sqrt{H}
  \)
  ^{-1}\alpha D_{zrr} r^{3}\hat {B^\varphi} H_{rr} ^{
  -1}\hat H_{rz} ^{2}\nl 
  \off{3979643}
  -2H^{-1}
  \(\sqrt{H}
  \)
  ^{-1}\alpha \hat D_{zrz} r^{3}\hat {B^\varphi} 
  \hat H_{rz} \nl 
  \off{3979643}
  +H^{-1}
  \(\sqrt{H}
  \)
  ^{-1}\alpha D_{zzz} r\hat {B^\varphi} H_{rr} 
  +
  \(\sqrt{H}
  \)
  ^{-1}\alpha r\hat {B^\varphi} 
  \(3r^{2}\hat s_z 
    -2Z_z 
  \)
\]
\Nl}$$
$$\displaylines{\qdd
\tilde \S 
\({E^z} 
\)
=
\[2H^{-1}\kappa \alpha 
  \(r^{2}\hat H_{rz} \hat S_r 
    -H_{rr} S_z 
  \)
  +4\alpha \chi_{rr} {E^z} H_{rr} ^{
  -1}\nl 
  \off{3991402}
  +H^{-1}\alpha \chi_{rr} r^{2}{E^z} H_{rr} ^{
  -1}\hat H_{rz} ^{2}
  +H^{-1}\alpha \chi_{zz} {E^z} H_{rr} \nl 
  \off{3991402}
  -4H^{-1}\alpha \hat D_{rrr} r^{2}H_{rr} ^{
  -1}\hat H_{rz} {Z^\varphi} 
  -4H^{-2}\alpha \hat D_{rrr} r^{4}H_{rr} ^{
  -1}\hat H_{rz} ^{3}{Z^\varphi} \nl 
  \off{3991402}
  +4H^{-1}\alpha \hat {\tilde D}_{rrz} r^{2}{Z^\varphi} 
  +8H^{-2}\alpha \hat {\tilde D}_{rrz} r^{4}\hat H_{rz} ^{2
  }{Z^\varphi} \nl 
  \off{3991402}
  -4H^{-2}\alpha \hat D_{rzz} r^{2}H_{rr} \hat H_{rz} 
  {Z^\varphi} 
  +4H^{-2}\alpha D_{zrr} r^{2}\hat H_{rz} ^{2}{Z^\varphi} 
  \nl 
  \off{3991402}
  -8H^{-2}\alpha \hat D_{zrz} r^{2}H_{rr} \hat H_{rz} 
  {Z^\varphi} 
  +4H^{-2}\alpha D_{zzz} H_{rr} ^{2}{Z^\varphi} \nl 
  \off{3991402}
  +\alpha 
  \(-2r^{2}{\hat {\tilde B}_r{}^r} {E^z} 
    -2r^{2}{\hat B_r{}^z} \hat {E^r} 
    +3r^{2}{E^z} \hat Y 
    -4{B_z{}^z} {E^z} 
    -2{E^z} \theta 
  \)
  \nl 
  \off{3991402}
  +2H^{-1}\alpha 
  \(r^{2}\hat A_r \hat H_{rz} {Z^\varphi} 
    -r^{2}\hat \chi_{rz} {E^z} \hat H_{rz} 
    -A_z H_{rr} {Z^\varphi} 
    +\hat H_{rz} {Z^\varphi} 
  \)
  \nl 
  \off{3991402}
  +4H^{-2}\alpha r^{2}\hat H_{rz} ^{3}{Z^\varphi} 
  -\hat {\beta^r} {E^z} 
  -4
  \(\sqrt{H}
  \)
  ^{-1}\alpha \hat D_{rrr} r^{2}\hat {B^\varphi} H_{rr} ^{
  -1}\nl 
  \off{3991402}
  -H^{-1}
  \(\sqrt{H}
  \)
  ^{-1}\alpha \hat D_{rrr} r^{4}\hat {B^\varphi} H_{rr} ^{
  -1}\hat H_{rz} ^{2}\nl 
  \off{3991402}
  +2H^{-1}
  \(\sqrt{H}
  \)
  ^{-1}\alpha \hat {\tilde D}_{rrz} r^{4}\hat {B^\varphi} 
  \hat H_{rz} 
  -H^{-1}
  \(\sqrt{H}
  \)
  ^{-1}\alpha \hat D_{rzz} r^{2}\hat {B^\varphi} H_{rr} 
  \nl 
  \off{3991402}
  +
  \(\sqrt{H}
  \)
  ^{-1}\alpha \hat {B^\varphi} 
  \(-3r^{4}\hat {\tilde s}_r 
    -6r^{2}\hat s 
    +2r^{2}\hat Z_r 
    -3
  \)
  \nl 
  \off{3991402}
  +H^{-1}
  \(\sqrt{H}
  \)
  ^{-1}\alpha r^{2}\hat {B^\varphi} \hat H_{rz} ^{2}
\]
\Nl}$$
$$\displaylines{\qdd
\tilde \S 
\({B^\varphi} 
\)
=
\[H^{-1}\alpha \chi_{rr} r\hat {B^\varphi} H_{zz} 
  +H^{-1}\alpha \chi_{zz} r\hat {B^\varphi} H_{rr} \nl 
  \off{4075325}
  -2\alpha r\hat {B^\varphi} 
  \(r^{2}{\hat {\tilde B}_r{}^r} 
    +{B_z{}^z} 
  \)
  -2H^{-1}\alpha r^{3}\hat {B^\varphi} \hat \chi_{rz} 
  \hat H_{rz} \nl 
  \off{4075325}
  -r\hat {\beta^r} \hat {B^\varphi} 
  -H^{-1}
  \(\sqrt{H}
  \)
  ^{-1}\alpha \hat D_{rrr} rH_{zz} 
  \(r^{2}\hat {E^r} \hat H_{rz} 
    +{E^z} H_{zz} 
  \)
  \nl 
  \off{4075325}
  +2H^{-1}
  \(\sqrt{H}
  \)
  ^{-1}\alpha \hat {\tilde D}_{rrz} r^{3}\hat H_{rz} 
  \(r^{2}\hat {E^r} \hat H_{rz} 
    +{E^z} H_{zz} 
  \)
  \nl 
  \off{4075325}
  -
  \(\sqrt{H}
  \)
  ^{-1}\alpha \hat D_{rzz} r{E^z} \nl 
  \off{4075325}
  -H^{-1}
  \(\sqrt{H}
  \)
  ^{-1}\alpha \hat D_{rzz} r^{3}\hat H_{rz} 
  \(\hat {E^r} H_{rr} 
    +{E^z} \hat H_{rz} 
  \)
  \nl 
  \off{4075325}
  +
  \(\sqrt{H}
  \)
  ^{-1}\alpha D_{zrr} r\hat {E^r} \nl 
  \off{4075325}
  +H^{-1}
  \(\sqrt{H}
  \)
  ^{-1}\alpha D_{zrr} r\hat H_{rz} 
  \(r^{2}\hat {E^r} \hat H_{rz} 
    +{E^z} H_{zz} 
  \)
  \nl 
  \off{4075325}
  -2H^{-1}
  \(\sqrt{H}
  \)
  ^{-1}\alpha \hat D_{zrz} r^{3}\hat H_{rz} 
  \(\hat {E^r} H_{rr} 
    +{E^z} \hat H_{rz} 
  \)
  \nl 
  \off{4075325}
  +H^{-1}
  \(\sqrt{H}
  \)
  ^{-1}\alpha D_{zzz} rH_{rr} 
  \(\hat {E^r} H_{rr} 
    +{E^z} \hat H_{rz} 
  \)
  \nl 
  \off{4075325}
  +H^{-1}
  \(\sqrt{H}
  \)
  ^{-1}\alpha r\hat H_{rz} ^{2}
  \(r^{2}\hat {E^r} \hat H_{rz} 
    +{E^z} H_{zz} 
  \)
\]
\Nl}$$
$$\displaylines{\qdd
\tilde \S 
\(\theta 
\)
=
\[-\kappa \alpha 
  \(\rho_K 
    +\sigma 
  \)
  +\alpha \chi_{rr} ^{2}H_{rr} ^{
  -2}
  +H^{-1}\alpha \chi_{rr} ^{2}r^{2}H_{rr} ^{
  -2}\hat H_{rz} ^{2}\nl 
  \off{3493737}
  +2H^{-1}\alpha \chi_{rr} \chi_{zz} 
  +\alpha \chi_{rr} H_{rr} ^{-1}
  \(r^{2}\hat Y 
    -2\theta 
  \)
  \nl 
  \off{3493737}
  +H^{-1}\alpha \chi_{rr} r^{2}H_{rr} ^{
  -1}\hat H_{rz} 
  \(r^{2}\hat H_{rz} \hat Y 
    -2\hat \chi_{rz} 
    -\hat H_{rz} \theta 
  \)
  \nl 
  \off{3493737}
  +H^{-1}\alpha \chi_{zz} H_{rr} 
  \(r^{2}\hat Y 
    -\theta 
  \)
  -2H^{-1}\alpha \hat D_{rrr} ^{2}r^{2}H_{rr} ^{
  -2}H_{zz} \nl 
  \off{3493737}
  -H^{-2}\alpha \hat D_{rrr} ^{2}r^{4}H_{rr} ^{
  -2}\hat H_{rz} ^{2}H_{zz} 
  +2H^{-1}\alpha \hat D_{rrr} \hat {\tilde D}_{rrz} r^{4}
  H_{rr} ^{-2}\hat H_{rz} \nl 
  \off{3493737}
  +2H^{-2}\alpha \hat D_{rrr} \hat {\tilde D}_{rrz} r^{6}
  H_{rr} ^{-2}\hat H_{rz} ^{3}
  -2H^{-1}\alpha \hat D_{rrr} \hat D_{rzz} r^{2}H_{rr} ^{
  -1}\nl 
  \off{3493737}
  -2H^{-2}\alpha \hat D_{rrr} \hat D_{rzz} r^{4}H_{rr} ^{
  -1}\hat H_{rz} ^{2}\nl 
  \off{3493737}
  +4H^{-1}\alpha \hat D_{rrr} D_{zrr} r^{2}H_{rr} ^{
  -2}\hat H_{rz} 
  +2H^{-2}\alpha \hat D_{rrr} D_{zrr} r^{4}H_{rr} ^{
  -2}\hat H_{rz} ^{3}\nl 
  \off{3493737}
  -2H^{-2}\alpha \hat D_{rrr} \hat D_{zrz} r^{4}H_{rr} ^{
  -1}\hat H_{rz} ^{2}
  +2H^{-2}\alpha \hat D_{rrr} D_{zzz} r^{2}\hat H_{rz} 
  \nl 
  \off{3493737}
  +\alpha \hat D_{rrr} H_{rr} ^{
  -2}
  \(-3r^{4}\hat {\tilde s}_r 
    +r^{2}\hat A_r 
    -2r^{2}\hat s 
    +2r^{2}\hat Z_r 
    -1
  \)
  \nl 
  \off{3493737}
  +H^{-1}\alpha \hat D_{rrr} r^{2}H_{rr} ^{
  -2}\hat H_{rz} 
  \(-4r^{4}\hat H_{rz} \hat {\tilde s}_r 
    +r^{2}\hat A_r \hat H_{rz} \nl 
    \off{10794119}
    +3r^{2}H_{rr} \hat s_z 
    +3r^{2}\hat H_{rz} \hat Z_r \nl 
    \off{10794119}
    -A_z H_{rr} 
    -2H_{rr} Z_z 
    +\hat H_{rz} 
  \)
  \nl 
  \off{3493737}
  +H^{-2}\alpha \hat D_{rrr} r^{4}H_{rr} ^{
  -2}\hat H_{rz} ^{3}
  \(-r^{4}\hat H_{rz} \hat {\tilde s}_r 
    +r^{2}H_{rr} \hat s_z 
    +2r^{2}\hat H_{rz} \hat s \nl 
    \off{11070826}
    +r^{2}\hat H_{rz} \hat Z_r 
    -H_{rr} Z_z 
    +2\hat H_{rz} 
  \)
  \nl 
  \off{3493737}
  +2H^{-2}\alpha \hat {\tilde D}_{rrz} \hat D_{rzz} r^{4}
  \hat H_{rz} \nl 
  \off{3493737}
  +2H^{-1}\alpha \hat {\tilde D}_{rrz} D_{zrr} r^{2}
  H_{rr} ^{-1}
  -4H^{-2}\alpha \hat {\tilde D}_{rrz} \hat D_{zrz} r^{4}
  \hat H_{rz} \nl 
  \off{3493737}
  +H^{-1}\alpha \hat {\tilde D}_{rrz} r^{2}
  \(2r^{4}H_{rr} ^{-1}\hat H_{rz} \hat {\tilde s}_r 
    -4r^{2}H_{rr} ^{-1}\hat H_{rz} \hat s \nl 
    \off{7910229}
    -2r^{2}H_{rr} ^{-1}\hat H_{rz} \hat Z_r 
    -A_z 
    -2H_{rr} ^{-1}\hat H_{rz} 
  \)
  \nl 
  \off{3493737}
  +2H^{-2}\alpha \hat {\tilde D}_{rrz} r^{4}\hat H_{rz} ^{2
  }
  \(r^{4}H_{rr} ^{-1}\hat H_{rz} \hat {\tilde s}_r 
    -2r^{2}H_{rr} ^{-1}\hat H_{rz} \hat s \nl 
    \off{9618795}
    -r^{2}H_{rr} ^{-1}\hat H_{rz} \hat Z_r 
    -r^{2}\hat s_z \nl 
    \off{9618795}
    -H_{rr} ^{-1}\hat H_{rz} 
    +Z_z 
  \)
  \nl 
  \off{3493737}
  -H^{-2}\alpha \hat D_{rzz} ^{2}r^{2}H_{rr} 
  +2H^{-2}\alpha \hat D_{rzz} \hat D_{zrz} r^{2}H_{rr} 
  \nl 
  \off{3493737}
  +H^{-1}\alpha \hat D_{rzz} 
  \(-r^{4}\hat {\tilde s}_r 
    +r^{2}\hat A_r 
    -2r^{2}\hat s 
    +r^{2}\hat Z_r 
    -1
  \)
\cr}$$ 
$$\displaylines{\qdd
  \nl 
  \off{3493737}
  +H^{-2}\alpha \hat D_{rzz} r^{2}\hat H_{rz} 
  \(-r^{4}\hat H_{rz} \hat {\tilde s}_r 
    +r^{2}H_{rr} \hat s_z 
    +2r^{2}\hat H_{rz} \hat s \nl 
    \off{9026167}
    +r^{2}\hat H_{rz} \hat Z_r 
    -H_{rr} Z_z 
    +2\hat H_{rz} 
  \)
  \nl 
  \off{3493737}
  -3H^{-1}\alpha D_{zrr} ^{2}H_{rr} ^{
  -1}
  -2H^{-2}\alpha D_{zrr} ^{2}r^{2}H_{rr} ^{
  -1}\hat H_{rz} ^{2}\nl 
  \off{3493737}
  +4H^{-2}\alpha D_{zrr} \hat D_{zrz} r^{2}\hat H_{rz} 
  -2H^{-2}\alpha D_{zrr} D_{zzz} H_{rr} \nl 
  \off{3493737}
  +H^{-1}\alpha D_{zrr} 
  \(3r^{4}H_{rr} ^{-1}\hat H_{rz} \hat {\tilde s}_r 
    -r^{2}\hat A_r H_{rr} ^{-1}\hat H_{rz} \nl 
    \off{7356056}
    +2r^{2}H_{rr} ^{-1}\hat H_{rz} \hat s 
    -2r^{2}H_{rr} ^{-1}\hat H_{rz} \hat Z_r \nl 
    \off{7356056}
    -3r^{2}\hat s_z 
    +2A_z 
    +H_{rr} ^{-1}\hat H_{rz} 
    +2Z_z 
  \)
  \nl 
  \off{3493737}
  +H^{-2}\alpha D_{zrr} r^{2}\hat H_{rz} ^{2}
  \(r^{4}H_{rr} ^{-1}\hat H_{rz} \hat {\tilde s}_r 
    -2r^{2}H_{rr} ^{-1}\hat H_{rz} \hat s \nl 
    \off{9291115}
    -r^{2}H_{rr} ^{-1}\hat H_{rz} \hat Z_r 
    -r^{2}\hat s_z \nl 
    \off{9291115}
    -2H_{rr} ^{-1}\hat H_{rz} 
    +Z_z 
  \)
  \nl 
  \off{3493737}
  +H^{-1}\alpha \hat D_{zrz} 
  \(-r^{2}\hat A_r 
    +4r^{2}\hat s 
    +3
  \)
  \nl 
  \off{3493737}
  +2H^{-2}\alpha \hat D_{zrz} r^{2}\hat H_{rz} 
  \(-r^{4}\hat H_{rz} \hat {\tilde s}_r 
    +r^{2}H_{rr} \hat s_z 
    +2r^{2}\hat H_{rz} \hat s \nl 
    \off{9353847}
    +r^{2}\hat H_{rz} \hat Z_r 
    -H_{rr} Z_z 
    +\hat H_{rz} 
  \)
  \nl 
  \off{3493737}
  +H^{-2}\alpha D_{zzz} H_{rr} 
  \(r^{4}\hat H_{rz} \hat {\tilde s}_r 
    -r^{2}H_{rr} \hat s_z 
    -2r^{2}\hat H_{rz} \hat s 
    -r^{2}\hat H_{rz} \hat Z_r \nl 
    \off{8471994}
    +H_{rr} Z_z 
    -2\hat H_{rz} 
  \)
  -
  \frac{1}{
        4}\alpha e^{2 r^2 \hat s} Hr^{2}{E^z} ^{2}\nl 
  \off{3493737}
  +\alpha e^{2 r^2 \hat s} r^{4}
  \(-
    \frac{1}{
          4}\hat {B^\varphi} ^{2}H_{rr} 
    -
    \frac{1}{
          4}\hat {E^r} ^{2}H_{rr} ^{2}
    -
    \frac{1}{
          2}\hat {E^r} {E^z} H_{rr} \hat H_{rz} \nl 
    \off{6666703}
    -
    \frac{1}{
          4}{E^z} ^{2}\hat H_{rz} ^{2}
  \)
  \nl 
  \off{3493737}
  +\alpha 
  \(-r^{6}H_{rr} ^{-1}\hat {\tilde s}_r ^{2}
    +r^{4}\hat A_r H_{rr} ^{-1}\hat {\tilde s}_r 
    -4r^{4}H_{rr} ^{-1}\hat s \hat {\tilde s}_r \nl 
    \off{4752802}
    +r^{4}H_{rr} ^{-1}\hat {\tilde s}_r \hat Z_r 
    -2r^{2}\hat A_r H_{rr} ^{-1}\hat Z_r 
    -2r^{2}{\hat {\tilde B}_r{}^r} \theta \nl 
    \off{4752802}
    -4r^{2}H_{rr} ^{-1}\hat s ^{2}
    +2r^{2}H_{rr} ^{-1}\hat s \hat Z_r 
    -4r^{2}H_{rr} ^{-1}\hat {\tilde s}_r \nl 
    \off{4752802}
    -r^{2}\theta \hat Y 
    -2{B_z{}^z} \theta 
    -6H_{rr} ^{-1}\hat s 
    +H_{rr} ^{-1}\hat Z_r 
  \)
  \nl 
\cr}$$ 
$$\displaylines{\qdd
  \off{3493737}
  +H^{-1}\alpha 
  \(-r^{8}H_{rr} ^{-1}\hat H_{rz} ^{2}\hat {\tilde s}_r ^{2}
    +r^{6}\hat A_r H_{rr} ^{-1}\hat H_{rz} ^{2}
    \hat {\tilde s}_r \nl 
    \off{5996620}
    -4r^{6}H_{rr} ^{-1}\hat H_{rz} ^{2}\hat s 
    \hat {\tilde s}_r 
    +r^{6}H_{rr} ^{-1}\hat H_{rz} ^{2}\hat {\tilde s}_r 
    \hat Z_r \nl 
    \off{5996620}
    +2r^{6}\hat H_{rz} \hat {\tilde s}_r \hat s_z 
    -2r^{4}\hat A_r H_{rr} ^{-1}\hat H_{rz} ^{2}\hat Z_r 
    -r^{4}\hat A_r \hat H_{rz} \hat s_z \nl 
    \off{5996620}
    -r^{4}A_z \hat H_{rz} \hat {\tilde s}_r 
    -2r^{4}\hat \chi_{rz} \hat H_{rz} \hat Y 
    -r^{4}H_{rr} \hat s_z ^{2}\nl 
    \off{5996620}
    -4r^{4}H_{rr} ^{-1}\hat H_{rz} ^{2}\hat s ^{2}
    +2r^{4}H_{rr} ^{-1}\hat H_{rz} ^{2}\hat s \hat Z_r \nl 
    \off{5996620}
    -3r^{4}H_{rr} ^{-1}\hat H_{rz} ^{2}\hat {\tilde s}_r 
    +4r^{4}\hat H_{rz} \hat s \hat s_z 
    -r^{4}\hat H_{rz} \hat {\tilde s}_r Z_z \nl 
    \off{5996620}
    -r^{4}\hat H_{rz} \hat s_z \hat Z_r 
    +2r^{2}\hat A_r \hat H_{rz} Z_z 
    +r^{2}A_z H_{rr} \hat s_z \nl 
    \off{5996620}
    +2r^{2}A_z \hat H_{rz} \hat Z_r 
    -r^{2}\hat \chi_{rz} ^{2}
    +2r^{2}\hat \chi_{rz} \hat H_{rz} \theta \nl 
    \off{5996620}
    +r^{2}H_{rr} \hat s_z Z_z 
    -8r^{2}H_{rr} ^{-1}\hat H_{rz} ^{2}\hat s 
    -2r^{2}\hat H_{rz} \hat s Z_z \nl 
    \off{5996620}
    +4r^{2}\hat H_{rz} \hat s_z 
    -2A_z H_{rr} Z_z 
    -H_{rr} ^{-1}\hat H_{rz} ^{2}
    -\hat H_{rz} Z_z 
  \)
  \nl 
  \off{3493737}
  +H^{-2}\alpha r^{2}\hat H_{rz} ^{3}
  \(r^{4}H_{rr} ^{-1}\hat H_{rz} \hat {\tilde s}_r 
    -2r^{2}H_{rr} ^{-1}\hat H_{rz} \hat s \nl 
    \off{7931679}
    -r^{2}H_{rr} ^{-1}\hat H_{rz} \hat Z_r 
    -r^{2}\hat s_z \nl 
    \off{7931679}
    -H_{rr} ^{-1}\hat H_{rz} 
    +Z_z 
  \)
  -\hat {\beta^r} \theta 
\]
\Nl}$$
$$\displaylines{\qdd
\tilde \S 
\(Z_r 
\)
=
\[-\kappa \alpha r\hat J_r 
  -
  \frac{1}{
        2}
  \sqrt{H}\alpha e^{2 r^2 \hat s} r^{3}\hat {B^\varphi} 
  {E^z} H_{rr} 
  +H^{-1}\alpha \chi_{rr} \hat D_{rrr} r^{3}H_{rr} ^{
  -2}\hat H_{rz} ^{2}\nl 
  \off{3905462}
  +H^{-1}\alpha \chi_{rr} \hat D_{rzz} r
  -2H^{-1}\alpha \chi_{rr} D_{zrr} rH_{rr} ^{
  -1}\hat H_{rz} \nl 
  \off{3905462}
  -H^{-2}\alpha \chi_{rr} D_{zrr} r^{3}H_{rr} ^{
  -1}\hat H_{rz} ^{3}
  +2H^{-2}\alpha \chi_{rr} \hat D_{zrz} r^{3}\hat H_{rz} 
  ^{2}\nl 
  \off{3905462}
  -H^{-2}\alpha \chi_{rr} D_{zzz} rH_{rr} \hat H_{rz} 
  +\alpha \chi_{rr} rH_{rr} ^{-1}
  \(\hat A_r 
    -2\hat Z_r 
  \)
  \nl 
  \off{3905462}
  +H^{-1}\alpha \chi_{rr} r\hat H_{rz} 
  \(r^{4}H_{rr} ^{-1}\hat H_{rz} \hat {\tilde s}_r 
    +2r^{2}H_{rr} ^{-1}\hat H_{rz} \hat s \nl 
    \off{8747695}
    -2r^{2}H_{rr} ^{-1}\hat H_{rz} \hat Z_r 
    -r^{2}\hat s_z \nl 
    \off{8747695}
    +A_z 
    +H_{rr} ^{-1}\hat H_{rz} 
    +2Z_z 
  \)
  \nl 
  \off{3905462}
  -H^{-2}\alpha \chi_{zz} \hat D_{rrr} r^{3}\hat H_{rz} ^{2
  }
  +2H^{-2}\alpha \chi_{zz} \hat {\tilde D}_{rrz} r^{3}
  H_{rr} \hat H_{rz} \nl 
  \off{3905462}
  -H^{-2}\alpha \chi_{zz} \hat D_{rzz} rH_{rr} ^{2}
  +H^{-1}\alpha \chi_{zz} r\hat A_r H_{rr} \nl 
  \off{3905462}
  +H^{-2}\alpha \chi_{zz} rH_{rr} \hat H_{rz} ^{2}
  -\alpha \hat D_{rrr} r^{3}H_{rr} ^{
  -1}\hat Y \nl 
  \off{3905462}
  +H^{-2}\alpha \hat D_{rrr} r^{5}\hat \chi_{rz} H_{rr} ^{
  -1}\hat H_{rz} ^{3}
  -2H^{-1}\alpha \hat {\tilde D}_{rrz} r^{3}\hat \chi_{rz} 
  \nl 
  \off{3905462}
  -2H^{-2}\alpha \hat {\tilde D}_{rrz} r^{5}\hat \chi_{rz} 
  \hat H_{rz} ^{2}
  +H^{-2}\alpha \hat D_{rzz} r^{3}\hat \chi_{rz} H_{rr} 
  \hat H_{rz} \nl 
  \off{3905462}
  +2H^{-1}\alpha D_{zrr} r\hat \chi_{rz} 
  +H^{-2}\alpha D_{zrr} r^{3}\hat \chi_{rz} \hat H_{rz} ^{2
  }\nl 
  \off{3905462}
  -2H^{-2}\alpha \hat D_{zrz} r^{3}\hat \chi_{rz} H_{rr} 
  \hat H_{rz} 
  +H^{-2}\alpha D_{zzz} r\hat \chi_{rz} H_{rr} ^{2}\nl 
  \off{3905462}
  +\alpha r
  \(-r^{4}\hat {\tilde s}_r \hat Y 
    +r^{2}\hat A_r \hat Y 
    -2r^{2}\hat s \hat Y 
    -2\hat A_r \theta \nl 
    \off{5441993}
    +2{\hat B_r{}^z} Z_z 
    -2{B_z{}^z} \hat Z_r 
    -\hat Y 
  \)
  \nl 
  \off{3905462}
  +H^{-1}\alpha r\hat \chi_{rz} 
  \(-r^{4}\hat H_{rz} \hat {\tilde s}_r 
    -r^{2}\hat A_r \hat H_{rz} 
    +r^{2}H_{rr} \hat s_z 
    -2r^{2}\hat H_{rz} \hat s \nl 
    \off{7655305}
    +2r^{2}\hat H_{rz} \hat Z_r 
    -A_z H_{rr} 
    -2H_{rr} Z_z \nl 
    \off{7655305}
    -2\hat H_{rz} 
  \)
  -H^{-2}\alpha r^{3}\hat \chi_{rz} \hat H_{rz} ^{3}
\]
\Nl}$$
$$\displaylines{\qdd
\tilde \S 
\(Z_z 
\)
=
\[-\kappa \alpha J_z 
  +
  \frac{1}{
        2}
  \sqrt{H}\alpha e^{2 r^2 \hat s} r^{4}\hat {B^\varphi} 
  \hat {E^r} H_{rr} 
  -2\alpha \chi_{rr} D_{zrr} H_{rr} ^{
  -2}\nl 
  \off{3917221}
  -2H^{-1}\alpha \chi_{rr} D_{zrr} r^{2}H_{rr} ^{
  -2}\hat H_{rz} ^{2}
  -H^{-2}\alpha \chi_{rr} D_{zrr} r^{4}H_{rr} ^{
  -2}\hat H_{rz} ^{4}\nl 
  \off{3917221}
  +2H^{-1}\alpha \chi_{rr} \hat D_{zrz} r^{2}H_{rr} ^{
  -1}\hat H_{rz} 
  +2H^{-2}\alpha \chi_{rr} \hat D_{zrz} r^{4}H_{rr} ^{
  -1}\hat H_{rz} ^{3}\nl 
  \off{3917221}
  -H^{-2}\alpha \chi_{rr} D_{zzz} r^{2}\hat H_{rz} ^{2}
  +\alpha \chi_{rr} H_{rr} ^{-1}
  \(-r^{2}\hat s_z 
    +2A_z 
  \)
  \nl 
  \off{3917221}
  +H^{-1}\alpha \chi_{rr} r^{2}A_z H_{rr} ^{
  -1}\hat H_{rz} ^{2}
  -2H^{-1}\alpha \chi_{zz} \hat D_{rrr} r^{2}H_{rr} ^{
  -1}\hat H_{rz} \nl 
  \off{3917221}
  -H^{-2}\alpha \chi_{zz} \hat D_{rrr} r^{4}H_{rr} ^{
  -1}\hat H_{rz} ^{3}
  +2H^{-2}\alpha \chi_{zz} \hat {\tilde D}_{rrz} r^{4}
  \hat H_{rz} ^{2}\nl 
  \off{3917221}
  -H^{-2}\alpha \chi_{zz} \hat D_{rzz} r^{2}H_{rr} 
  \hat H_{rz} 
  +2H^{-1}\alpha \chi_{zz} D_{zrr} \nl 
  \off{3917221}
  +H^{-1}\alpha \chi_{zz} 
  \(-r^{4}\hat H_{rz} \hat {\tilde s}_r 
    +r^{2}\hat A_r \hat H_{rz} 
    +r^{2}H_{rr} \hat s_z 
    -2r^{2}\hat H_{rz} \hat s \nl 
    \off{7401327}
    +2r^{2}\hat H_{rz} \hat Z_r 
    -2H_{rr} Z_z 
    -\hat H_{rz} 
  \)
  \nl 
  \off{3917221}
  +H^{-2}\alpha \chi_{zz} r^{2}\hat H_{rz} ^{3}
  +2\alpha \hat D_{rrr} r^{2}\hat \chi_{rz} H_{rr} ^{
  -2}\nl 
  \off{3917221}
  +3H^{-1}\alpha \hat D_{rrr} r^{4}\hat \chi_{rz} H_{rr} 
  ^{
  -2}\hat H_{rz} ^{2}\nl 
  \off{3917221}
  +H^{-2}\alpha \hat D_{rrr} r^{6}\hat \chi_{rz} H_{rr} ^{
  -2}\hat H_{rz} ^{4}
  -2H^{-1}\alpha \hat {\tilde D}_{rrz} r^{4}\hat \chi_{rz} 
  H_{rr} ^{-1}\hat H_{rz} \nl 
  \off{3917221}
  -2H^{-2}\alpha \hat {\tilde D}_{rrz} r^{6}\hat \chi_{rz} 
  H_{rr} ^{-1}\hat H_{rz} ^{3}
  +H^{-1}\alpha \hat D_{rzz} r^{2}\hat \chi_{rz} \nl 
  \off{3917221}
  +H^{-2}\alpha \hat D_{rzz} r^{4}\hat \chi_{rz} 
  \hat H_{rz} ^{2}
  -\alpha D_{zrr} r^{2}H_{rr} ^{
  -1}\hat Y \nl 
  \off{3917221}
  +H^{-2}\alpha D_{zrr} r^{4}\hat \chi_{rz} H_{rr} ^{
  -1}\hat H_{rz} ^{3}
  -2H^{-1}\alpha \hat D_{zrz} r^{2}\hat \chi_{rz} \nl 
  \off{3917221}
  -2H^{-2}\alpha \hat D_{zrz} r^{4}\hat \chi_{rz} 
  \hat H_{rz} ^{2}
  +H^{-2}\alpha D_{zzz} r^{2}\hat \chi_{rz} H_{rr} 
  \hat H_{rz} \nl 
  \off{3917221}
  +\alpha 
  \(r^{4}\hat \chi_{rz} H_{rr} ^{
    -1}\hat {\tilde s}_r 
    -r^{4}\hat s_z \hat Y 
    -r^{2}\hat A_r \hat \chi_{rz} H_{rr} ^{
    -1}\nl 
    \off{5176286}
    +r^{2}A_z \hat Y 
    -2r^{2}{\hat {\tilde B}_r{}^r} Z_z 
    +2r^{2}{\hat B_z{}^r} \hat Z_r 
    +2r^{2}\hat \chi_{rz} H_{rr} ^{
    -1}\hat s \nl 
    \off{5176286}
    -2r^{2}\hat \chi_{rz} H_{rr} ^{
    -1}\hat Z_r 
    -2A_z \theta 
    +\hat \chi_{rz} H_{rr} ^{-1}
  \)
  \nl 
  \off{3917221}
  +H^{-1}\alpha r^{2}\hat \chi_{rz} \hat H_{rz} 
  \(r^{4}H_{rr} ^{-1}\hat H_{rz} \hat {\tilde s}_r 
    -r^{2}\hat A_r H_{rr} ^{-1}\hat H_{rz} \nl 
    \off{9047950}
    +2r^{2}H_{rr} ^{-1}\hat H_{rz} \hat s 
    -2r^{2}H_{rr} ^{-1}\hat H_{rz} \hat Z_r \nl 
    \off{9047950}
    -r^{2}\hat s_z 
    -A_z 
    +2Z_z 
  \)
  \nl 
  \off{3917221}
  -H^{-2}\alpha r^{4}\hat \chi_{rz} H_{rr} ^{
  -1}\hat H_{rz} ^{4}
  -\hat {\beta^r} Z_z 
\]
\Nl}$$
$$\displaylines{\qdd
\tilde \S 
\({Z^\varphi} 
\)
=
\[-\kappa \alpha {J^\varphi} 
  +2\alpha \hat D_{rrr} r^{2}\hat {E^r} H_{rr} ^{
  -1}
  +
  \frac{1}{
        2}H^{-1}\alpha \hat D_{rrr} r^{4}\hat {E^r} 
  H_{rr} ^{-1}\hat H_{rz} ^{2}\nl 
  \off{4039599}
  -H^{-1}\alpha \hat {\tilde D}_{rrz} r^{4}\hat {E^r} 
  \hat H_{rz} 
  +
  \frac{1}{
        2}H^{-1}\alpha \hat D_{rzz} r^{2}\hat {E^r} 
  H_{rr} \nl 
  \off{4039599}
  +2\alpha D_{zrr} {E^z} H_{rr} ^{
  -1}
  +
  \frac{1}{
        2}H^{-1}\alpha D_{zrr} r^{2}{E^z} H_{rr} ^{
  -1}\hat H_{rz} ^{2}\nl 
  \off{4039599}
  -H^{-1}\alpha \hat D_{zrz} r^{2}{E^z} \hat H_{rz} 
  +
  \frac{1}{
        2}H^{-1}\alpha D_{zzz} {E^z} H_{rr} \nl 
  \off{4039599}
  +\alpha 
  \(\frac{3}{
          2}r^{4}\hat {E^r} \hat {\tilde s}_r 
    -
    \frac{1}{
          2}r^{2}\hat A_r \hat {E^r} 
    -2r^{2}{\hat {\tilde B}_r{}^r} {Z^\varphi} 
    +3r^{2}\hat {E^r} \hat s \nl 
    \off{5298664}
    -r^{2}\hat {E^r} \hat Z_r 
    +
    \frac{3}{
          2}r^{2}{E^z} \hat s_z 
    -
    \frac{1}{
          2}A_z {E^z} 
    -2{B_z{}^z} {Z^\varphi} \nl 
    \off{5298664}
    +
    \frac{3}{
          2}\hat {E^r} 
    -{E^z} Z_z 
  \)
  -
  \frac{1}{
        2}H^{-1}\alpha r^{2}\hat {E^r} \hat H_{rz} ^{2}
  -\hat {\beta^r} {Z^\varphi} 
\]
\Nl}$$



\bibliography{refs}

\begin{thebibliography}{100}

\bibitem{Abrahams93}
A.M. Abrahams and C.R. Evans.
\newblock Critical {B}ehaviour and {S}caling in {V}acuum {A}xisymmetric
  {G}ravitational {C}ollapse.
\newblock {\em Phys. Rev. Lett.}, 70(20):2980--2983, 1993.

\bibitem{Abrahams94}
A.M. Abrahams and C.R. Evans.
\newblock Universality in axisymmetric vacuum collapse.
\newblock {\em Phys. Rev. D}, 49(8):3998--4003, 1994.

\bibitem{Abrahams92}
A.M. Abrahams, K.R. Heiderich, S.L. Shapiro, and S.A. Teukolsky.
\newblock Vacuum initial data, singularities, and cosmic censorship.
\newblock {\em Phys. Rev. D}, 46(6):2452--2463, 1992.

\bibitem{Alcubierre97}
M.~Alcubierre.
\newblock Appearance of coordinate shocks in hyperbolic formalisms of general
  relativity.
\newblock {\em Phys. Rev. D}, 55(10):5981--5991, 1997.

\bibitem{Alcubierre03}
M.~Alcubierre.
\newblock Hyperbolic slicings of spacetime: singularity avoidance and gauge
  shocks.
\newblock {\em Class. Quantum Grav.}, 20:607--623, 2003.

\bibitem{Alcubierre00}
M.~Alcubierre, G.~Allen, B.~Br\"ugmann, G.~Lanfermann, W.~Seidel, W.-M. Suen,
  and M.~Tobias.
\newblock Gravitational collapse of gravitational waves in {3D} numerical
  relativity.
\newblock {\em Phys. Rev. D}, 61(041501), 2000.

\bibitem{Alcubierre01}
M.~Alcubierre, S.~Brandt, B.~Br\"ugmann, D.~Holz, E.~Seidel, R.~Takahashi, and
  J.~Thornburg.
\newblock Symmetry without symmetry: {N}umerical simulation of axisymmetric
  systems using {C}artesian grids.
\newblock {\em Int. J. Mod. Phys.}, 10(3):273--290, 2001.

\bibitem{AlcubierreMasso98}
M.~Alcubierre and J.~Mass\'o.
\newblock Pathologies of hyperbolic gauges in general relativity amd other
  field theories.
\newblock {\em Phys. Rev. D}, 57(8):R4511--15, 1998.

\bibitem{Anninos95}
P.~Anninos, K.~Camarda, J.~Mass\'o, E.~Seidel, W.-M. Suen, and J.~Towns.
\newblock Three-dimensional numerical relativity: {T}he evolution of black
  holes.
\newblock {\em Phys. Rev. D}, 52(4):2059--2082, 1995.

\bibitem{TRI}
W.~Antweiler, A.~Strotmann, and V.~Winkelmann.
\newblock {\em Typesetting {REDUCE} output with {T}e{X} -- {A
  REDUCE-T}e{X}-{I}nterface}.
\newblock University of Cologne, Computer Centre, 1995.

\bibitem{Arnold97}
D.N. Arnold, A.~Mukherjee, and L.~Pouly.
\newblock Adaptive finite elements and colliding black holes.
\newblock {\em Preprint, arXiv::gr-qc}, (9709038), 1997.

\bibitem{ADM62}
R.~Arnowitt, S.~Deser, and C.~W. Misner.
\newblock The dynamics of general relativity.
\newblock In L.~Witten, editor, {\em Gravitation: an introduction to current
  research}, chapter~7, pages 227--265. Wiley, 1962.

\bibitem{Banyuls97}
F.~Banyuls, J.A. Font, J.M. Ib\'a\~nez, J.M. Mart\'i, and J.A. Miralles.
\newblock Numerical {3+1} general relativistic hydrodynamics: A local
  characteristic approach.
\newblock {\em Astrophys. J.}, 476:221--231, 1997.

\bibitem{BarnesPhD}
A.~Barnes.
\newblock {\em Numerical {R}elativistic {H}ydrodynamics in {P}lanar and
  {A}xisymmetric {S}pacetimes}.
\newblock PhD thesis, University of {C}ambridge, August 2004.

\bibitem{Barnes04}
A.P. Barnes, P.G. Lefloch, B.G. Schmidt, and J.M. Stewart.
\newblock The {G}limm scheme for perfect fluids on plane-symmetric {G}owdy
  spacetimes.
\newblock {\em Class. Quantum Grav.}, 21:5043--5074, 2004.

\bibitem{Baumgarte98}
T.~W. Baumgarte and S.~Shapiro.
\newblock Numerical integration of {E}instein's field equations.
\newblock {\em Phys. Rev. D}, 59(024007), 1998.

\bibitem{BergerRigoutsos91}
M.~Berger and I.~Rigoutsos.
\newblock {An Algorithm for Point Clustering and Grid Generation}.
\newblock {\em IEEE Transactions on Systems, Man, and Cybernetics},
  21(5):1278--1286, 1991.

\bibitem{BergerColella88}
M.J. Berger and P.~Colella.
\newblock {Local Adaptive Mesh Refinement for Shock Hydrodynamics}.
\newblock {\em J. Comp. Phys}, 82:64--84, 1988.

\bibitem{BergerOliger84}
M.J. Berger and J.~Oliger.
\newblock Adaptive mesh refinement for hyperbolic partial differential
  equations.
\newblock {\em J. Comput. Phys.}, 53:484--512, 1984.

\bibitem{Birkhoff}
G.D. Birkhoff.
\newblock {\em Relativity and {M}odern {P}hysics}.
\newblock Harvard University Press, 1923.

\bibitem{BicakSchmidt83}
J.~Bi\v{c}\'ak and B.G. Schmidt.
\newblock Isometries compatible with gravitational radiation.
\newblock {\em J. Math. Phys.}, 25(3):606--606, 1984.

\bibitem{BicakSchmidt89}
J.~Bi\v{c}\'ak and B.G. Schmidt.
\newblock Asymptotically flat radiative space-times with boost-rotation
  symmetry.
\newblock {\em Phys. Rev. D}, 40:1827--1853, 1989.

\bibitem{Bona03a}
C.~Bona, T.~Ledvinka, C.~Palenzuela, and M.~\v{Z}\'a\v{c}ek.
\newblock General-covariant evolution formalism for numerical relativity.
\newblock {\em Phys. Rev. D}, 67(104005), 2003.

\bibitem{Bona03b}
C.~Bona, T.~Ledvinka, C.~Palenzuela, and M.~\v{Z}\'a\v{c}ek.
\newblock A symmetry-breaking mechanism for the {Z}4 general-covariant
  evolution system.
\newblock {\em Phys. Rev. D}, 69(064036), 2004.

\bibitem{Bona05}
C.~Bona, T.~Ledvinka, C.~Palenzuela-Luque, and M.~\v{Z}\'a\v{c}ek.
\newblock Constraint-preserving boundary conditions in the {Z}4 numerical
  relativity formalism.
\newblock {\em Class. Quantum Grav.}, 22(13):2615--2633, 2005.

\bibitem{BonaMasso88}
C.~Bona and J.~Mass\'o.
\newblock Harmonic synchronizations of spacetime.
\newblock {\em Phys. Rev. D}, 38(8):2419--2422, 1988.

\bibitem{Bona04a}
C.~Bona and C.~Palenzuela.
\newblock Dynamical shift conditions for the {Z}4 and {BSSN} formalisms.
\newblock {\em Phys. Rev. D}, 69(104003), 2004.

\bibitem{Bondi62}
H.~Bondi, M.G.J. van~der Burg, and A.W.K Metzner.
\newblock {Gravitational Waves in General Relativity: VII. Waves from
  Axi-Symmetric Isolated Systems}.
\newblock {\em Proc. R. Soc. London A}, 269:21--52, 1962.

\bibitem{Bramble88}
J.H. Bramble, J.E. Pasciak, and J.~Xu.
\newblock The {A}nalysis of {M}ultigrid {A}lgorithms for {N}onsymmetric and
  {I}ndefinite {P}roblems.
\newblock {\em Math. Comp.}, 51(184):289--414, 1988.

\bibitem{Brandt77}
A.~Brandt.
\newblock Multilevel adaptive solutions to boundary value problems.
\newblock {\em Math. Comput.}, 31:333--390, 1977.

\bibitem{Briggs}
W.~Briggs, V.E. Henson, and S.F. McCormick.
\newblock {\em A Multigrid Tutorial}.
\newblock SIAM, 1999.

\bibitem{Brill59}
D.~Brill.
\newblock {On the Positive Definite Mass of the Bondi-Weber-Wheeler
  Time-Symmetric Gravitational Waves}.
\newblock {\em Ann. Phys.}, 7:466--483, 1959.

\bibitem{Brodbeck99}
O.~Brodbeck, S.~Frittelli, P.~H\"ubner, and O.~A. Reula.
\newblock Einstein's equations with asymptotically stable constraint
  propagation.
\newblock {\em J. Math. Phys.}, 40(2):909--923, 1999.

\bibitem{Bruhat52}
Y.~Bruhat.
\newblock Th\'eor\`eme d'existence pour certaines syst\`emes d'\'equations aux
  d\'eriv\'ees partielles nonlin\'eaires.
\newblock {\em Acta Math.}, 88:141--225, 1952.

\bibitem{Burke71}
W.~L. Burke.
\newblock Gravitational {R}adiation {D}amping of {S}lowly {M}oving {S}ystems
  {C}alculated {U}sing {M}atched {A}symptotic {E}xpansions.
\newblock {\em J. Math. Phys.}, 12(3):401--418, 1971.

\bibitem{ButcherODEs}
J.C. Butcher.
\newblock {\em Numerical {M}ethods for {O}rdinary {D}ifferential {E}quations}.
\newblock Wiley, 2003.

\bibitem{Calabrese02}
G.~Calabrese, L.~Lehner, and M.~Tiglio.
\newblock Constraint-preserving boundary conditions in numerical relativity.
\newblock {\em Phys. Rev. D}, 65(104031), 2002.

\bibitem{Calabrese03}
G.~Calabrese, J.~Pullin, O.~Sarbach, M.~Tiglio, and O.~Reula.
\newblock Well posed constraint-preserving boundary conditions for the
  linearized {E}instein equations.
\newblock {\em Comm. Math. Phys.}, 240:377--395, 2003.

\bibitem{Campbell77}
S.J. Campbell and J.~Wainwright.
\newblock Algebraic computing and the {N}ewman-{P}enrose formalism in general
  relativity.
\newblock {\em {GRG}}, 8:987--1001, 1977.

\bibitem{Choptuik93}
M.W. Choptuik.
\newblock Universality and scaling in gravitational collapse of a massless
  scalar field.
\newblock {\em Phys. Rev. Lett.}, 70:9--12, 1993.

\bibitem{Choptuik03a}
M.W. Choptuik, E.W. Hirschmann, S.L. Liebling, and F.~Pretorius.
\newblock An axisymmetric gravitational collapse code.
\newblock {\em Class. Quantum Grav.}, 20:1857--1878, 2003.

\bibitem{Choptuik03b}
M.W. Choptuik, E.W. Hirschmann, S.L. Liebling, and F.~Pretorius.
\newblock Critical collapse of the massless scalar field in axisymmetry.
\newblock {\em Phys. Rev. D}, 68(044007), 2003.

\bibitem{Davidson97}
W.~Davidson.
\newblock Barotropic perfect fluid in steady cylindrically symmetric rotation.
\newblock {\em Class. Quantum Grav.}, 14:119--127, 1997.

\bibitem{DeDonder21}
T.~De~Donder.
\newblock {\em La Gravifique Einsteinienne}.
\newblock Gauthier-Villars, Paris, 1921.

\bibitem{Dreyer03}
O.~Dreyer, B.~Krishnan, D.~Shoemaker, and E.~Schnetter.
\newblock Introduction to isolated horizons in numerical relativity.
\newblock {\em Phys. Rev. D}, 67(024018), 2003.

\bibitem{Eardley79}
D.M. Eardley and L.~Smarr.
\newblock Time functions in numerical relativity: Marginally bound dust
  collapse.
\newblock {\em Phys. Rev. D}, 19:2239--2259, 1979.

\bibitem{Elman01}
H.C. Elman, O.G. Ernst, and D.P. O'Leary.
\newblock A {M}ultigrid {M}ethod {E}nhanced by {K}rylov {S}ubspace {I}teration
  for the {D}iscrete {H}elmholtz {E}quations.
\newblock {\em SIAM J. Sci. Comput.}, 23(4):1291--1315, 2001.

\bibitem{Eppley77}
K.~Eppley.
\newblock Evolution of time-symmetric gravitational waves: Initial data and
  apparent horizons.
\newblock {\em Phys. Rev. D}, 16(6):1909--1614, 1977.

\bibitem{Eppley79}
K.~Eppley.
\newblock Pure gravitational waves.
\newblock In L.~Smarr, editor, {\em Sources of Gravitational Radiation}, pages
  275--291. Cambridge University Press, 1979.

\bibitem{Estabrook72}
F.~Estabrook, H.~Wahlquist, S.~Christensen, B.~DeWitt, L.~Smarr, and E.~Tsiang.
\newblock Maximally slicing a black hole.
\newblock {\em Phys. Rev. D}, 7(10):2814--2817, 1972.

\bibitem{Evans94}
C.R. Evans and J.S. Coleman.
\newblock Critical {P}henomena and {S}elf-{S}imilarity in the {G}ravitational
  {C}ollapse of {R}adiation {F}luid.
\newblock {\em Phys. Rev. Lett.}, 72(12):1782--1785, 1994.

\bibitem{EvansPDEs}
L.C. Evans.
\newblock {\em Partial {D}ifferential {E}quations}.
\newblock American {M}athematical {S}ociety, 1998.

\bibitem{Fock59}
V.A. Fock.
\newblock {\em The Theory of Space Time and Gravitation}.
\newblock Pergamon, London, 1959.

\bibitem{Font00}
J.A. Font.
\newblock Numerical hydrodynamics in general relativity.
\newblock {\em Living Reviews in Relativity}, 3, May 2000.

\bibitem{Frauendiener02}
J.~Frauendiener.
\newblock Discretizations of axisymmetric systems.
\newblock {\em Phys. Rev. D}, 66(104027), 2002.

\bibitem{Friedrich81}
H.~Friedrich.
\newblock On the regular and the asymptotic characteristic initial value
  problem for {E}instein's vacuum field equations.
\newblock {\em Proc. R. Soc. London A}, 375:169, 1981.

\bibitem{Friedrich96}
H.~Friedrich.
\newblock Hyperbolic reductions for {E}instein's equations.
\newblock {\em Class. Quantum Grav.}, 13:1451--1469, 1996.

\bibitem{Friedrich98}
H.~Friedrich and G.~Nagy.
\newblock {The Initial Boundary Value Problem for Einstein's Vacuum Field
  Equations}.
\newblock {\em Comm. Math. Phys.}, 201:619--655, 1999.

\bibitem{Frittelli96}
S.~Frittelli and O.A. Reula.
\newblock First-{O}rder {S}ymmetric {H}yperbolic {E}instein {E}quations with
  {A}rbitrary {F}ixed {G}auge.
\newblock {\em Phys. Rev. Lett.}, 76(25):4667--4670, 1996.

\bibitem{Fryxell00}
B.~Fryxell, K.~Olson, P.~Ricker, F.X. Timmes, M.~Zingale, D.Q. Lamb,
  P.~MacNeice, R.~Rosner, Truran J.W., and H.~Tufo.
\newblock {FLASH: An Adaptive Mesh Hydrodynamics Code for Modeling
  Astrophysical Thermonuclear Flashes}.
\newblock {\em Astr. J. Suppl.}, 131:273--334, 2000.

\bibitem{Garfinkle02}
D.~Garfinkle.
\newblock Harmonic coordinate method for simulating generic singularities.
\newblock {\em Phys. Rev. D}, 65(044029), 2002.

\bibitem{Garfinkle00}
D.~Garfinkle and G.C. Duncan.
\newblock Numerical {E}volution of {B}rill {W}aves.
\newblock {\em Phys. Rev. D}, 63(044011), 2000.

\bibitem{Garfinkle97}
D.~Garfinkle, E.N. Glass, and J.P. Krisch.
\newblock Solution generating with perfect fluids.
\newblock {\em Gen. Rel. Grav.}, 29(4):467--480, 1997.

\bibitem{GENTRAN}
B.L. Gates and M.C. Dewar.
\newblock {\em {GENTRAN U}ser's {M}anual -- {REDUCE V}ersion}.
\newblock RAND, Santa Monica and University of Bath, 1991.

\bibitem{Geroch71}
R.~Geroch.
\newblock A {M}ethod for {G}enerating {S}olutions of {E}instein's {E}quations.
\newblock {\em J. Math. Phys.}, 12(6):918--924, 1971.

\bibitem{GilbargTrudinger}
D.~Gilbarg and N.S. Trudinger.
\newblock {\em Elliptic {P}artial {D}fferential {E}quations of {S}econd
  {O}rder}.
\newblock Springer, 1977.

\bibitem{Gourgoulhon02}
E.~Gourgoulhon, P.~Grandclement, and S.~Bonazzola.
\newblock {Binary black holes in circular orbits. I. A global spacetime
  approach}.
\newblock {\em Phys. Rev. D}, 65(044020), 2002.

\bibitem{Gundlach98}
C.~Gundlach.
\newblock Angular momentum at the black hole threshold.
\newblock {\em Phys. Rev. D}, 57:7080--7083, 1998.

\bibitem{Gundlach99}
C.~Gundlach.
\newblock Critical phenomena in gravitational collapse.
\newblock {\em Living Reviews in Relativity}, 2, 1999.

\bibitem{Gundlach05}
C.~Gundlach, G.~Calabrese, I.~Hinder, and J.~M. Mart\'in-Garc\'ia.
\newblock Constraint damping in the {Z}4 formulation and harmonic gauge.
\newblock {\em Class. Quantum Grav.}, 22:3767--3773, 2005.

\bibitem{Gundlach04a}
C.~Gundlach and J.~M. Mart\'in-Garc\'ia.
\newblock Symmetric hyperbolic form of systems of second-order evolution
  equations subject to constraints.
\newblock {\em Phys. Rev. D}, 70(044031), 2004.

\bibitem{Gundlach04b}
C.~Gundlach and J.~M. Mart\'in-Garc\'ia.
\newblock Symmetric hyperbolicity and consistent boundary conditions for
  second-order {E}instein equations.
\newblock {\em Phys. Rev. D}, 70(044032), 2004.

\bibitem{GKO}
B.~Gustafssohn, H.-O. Kreiss, and J.~Oliger.
\newblock {\em Time dependent problems and difference methods}.
\newblock Wiley, 1995.

\bibitem{Hackbusch}
W.~Hackbusch.
\newblock {\em Multi-Grid Methods and Applications}.
\newblock Computational Mathematics. Springer, 1985.

\bibitem{HahnLindquist64}
S.~Hahn and R.~Lindquist.
\newblock The two-body problem in geometrodynamics.
\newblock {\em Ann. Phys.}, 29:304, 1964.

\bibitem{Hamade96}
R.~Hamade and J.M. Stewart.
\newblock The spherically symmetric collapse of a massless scalar field.
\newblock {\em Class. Quantum Grav.}, 13:497--512, 1996.

\bibitem{HawkePhD}
I.~Hawke.
\newblock {\em Computational {U}ltrarelativistic {H}ydrodynamics}.
\newblock Ph{D} thesis, University of Cambridge, 2001.

\bibitem{Hawke02}
I.~Hawke and J.M. Stewart.
\newblock The dynamics of primordial black hole formation.
\newblock {\em Class. Quantum Grav.}, 19:3687--3707, 2002.

\bibitem{HawkingEllis}
S.W. Hawking and G.F.R. Ellis.
\newblock {\em The Large Scale Structure of Spacetime}.
\newblock Cambridge University Press, 1973.

\bibitem{REDUCE}
A.C. Hearn.
\newblock {\em {REDUCE U}ser's {M}anual 3.5}.
\newblock Konrad-{Z}use-{Z}entrum {B}erlin, 1993.

\bibitem{HernPhD}
S.D. Hern.
\newblock {\em {Numerical relativity and Inhomogeneous Cosmologies}}.
\newblock PhD thesis, University of Cambridge, 1999.
\newblock Available on arXiv::gr-qc/0004036.

\bibitem{Holst05}
M.~Holst and D.~Bernstein.
\newblock Adaptive finite element solution of the constraint equations in
  relativity.
\newblock {\em Preprint}, 2005.

\bibitem{Holst04}
M.~Holst, L.~Lindblom, R.~Owen, H.~Pfeiffer, M.A. Scheel, and L.E. Kidder.
\newblock Optimal constraint projection for hyperbolic evolution systems.
\newblock {\em Phys. Rev. D}, 70(085017), 2004.

\bibitem{JEK60}
P.~Jordan, J.~Ehlers, and W.~Kundt.
\newblock Exact solutions of the field equations of general relativity.
\newblock {\em Akad. Wiss. Lit. (Mainz) Abhandl. Math.-Nat. Kl. 2}, 21, 1960.
\newblock In German.

\bibitem{Kaluza21}
T.~Kaluza.
\newblock On the problem of unity in physics.
\newblock {\em Sitzungsber. Preuss. Akad. Wiss., Berlin, Math. Phys.}, K1:966,
  1921.
\newblock In German.

\bibitem{Kasner21}
E.~Kasner.
\newblock Geometrical theorems on {E}instein's cosmological equations.
\newblock {\em Am. J. Math.}, 43:217--221, 1921.

\bibitem{Kidder01}
L.~E. Kidder, M.~A. Scheel, and S.~A. Teukolsky.
\newblock Extending the lifetime of 3d black hole computations with a new
  hyperbolic system of evolution equations.
\newblock {\em Phys. Rev. D}, 64(064017), 2001.

\bibitem{Kidder05}
L.E. Kidder, L.~Lindblom, M.A. Scheel, L.T. Buchman, and H.P. Pfeiffer.
\newblock Boundary conditions for the {E}instein evolution system.
\newblock {\em Phys. Rev. D}, 71(064020), 2005.

\bibitem{Klein26}
O.~Klein.
\newblock Quantum theory and five-dimensional relativity theory.
\newblock {\em Z. Phys.}, 37:895, 1926.
\newblock In German.

\bibitem{Komar59}
A.~Komar.
\newblock {Covariant Conservation Laws in General Relativity}.
\newblock {\em Phys. Rev.}, 113(1):934--936, 1959.

\bibitem{Kompaneets58}
A.S. Kompaneets.
\newblock Strong gravitational waves in vacuum.
\newblock {\em Zh. Eksper. Teor. Fiz.}, 34:953, 1958.
\newblock In Russian.

\bibitem{Kramer88}
D.~Kramer.
\newblock Cylindrically symmetric static perfect fluids.
\newblock {\em Class. Quantum Grav.}, 5(2):393--398, 1988.

\bibitem{Kreiss70}
H.-O. Kreiss.
\newblock Initial boundary value problems for hyperbolic systems.
\newblock {\em Commun. Pure Appl. Math.}, 23:277, 1970.

\bibitem{KO}
H.-O. Kreiss and J.~Oliger.
\newblock Methods for the approximate solution of time dependent problems.
\newblock Technical Report~10, Global atmospheric research programme
  publication series, 1973.

\bibitem{Laney}
C.B. Laney.
\newblock {\em Computational Gasdynamics}.
\newblock Cambridge University Press, 1998.

\bibitem{Lehner01}
L.~Lehner.
\newblock Numerical relativity: A review.
\newblock {\em Class. Quantum Grav.}, 18:R25--R86, 2001.
\newblock Topical Review.

\bibitem{LeVeque}
R.~J. LeVeque.
\newblock {\em Numerical Methods for Conservation Laws}.
\newblock Birkh\"auser Verlag, 1992.

\bibitem{Lindblom03}
L.~Lindblom and M.~A. Scheel.
\newblock Dynamical gauge conditions for the {E}instein evolution equations.
\newblock {\em Phys. Rev. D}, 67(124005), 2003.

\bibitem{Loehner87}
R.~L\"ohner.
\newblock An adaptive finite element scheme for transient problems in {CFD}.
\newblock {\em Comp. Meth. App. Mech. Eng.}, 61:323--338, 1987.

\bibitem{Maeda80}
K.~Maeda, M.~Sasaki, T.~Nakamura, and S.~Miyama.
\newblock A {N}ew {F}ormulation of the {E}instein {E}quations for
  {R}elativistic {R}otating {S}ystems.
\newblock {\em Prog. Theor. Phys.}, 63:719--721, 1980.

\bibitem{MajdaOsher75}
A.~Majda and S.~Osher.
\newblock Initial-{B}oundary {V}alue {P}roblems for {H}yperbolic {E}quations
  with {U}niformly {C}haracteristic {B}oundary.
\newblock {\em Comm. Pure \& Appl. Math.}, 28:607--675, 1975.

\bibitem{MTW}
C.W. Misner, K.S. Thorne, and J.A. Wheeler.
\newblock {\em Gravitation}.
\newblock Freeman, 1970.

\bibitem{NOR04}
G.~Nagy, O.~Ortiz, and O.~Reula.
\newblock Strongly hyperbolic second order {E}instein's evolution equations.
\newblock {\em Phys. Rev. D}, 70(044012), 2004.

\bibitem{Nakamura87}
T.~Nakamura, K.~Oohara, and Y.~Kojima.
\newblock General relativistic collapse of axially symmetric stars and 3d time
  evolution of pure gravitational waves.
\newblock {\em Prog. Theor. Phys. Suppl.}, 90:13--109, 1987.

\bibitem{Neilsen00b}
D.W. Neilsen and M.W. Choptuik.
\newblock Critical phenomena in perfect fluids.
\newblock {\em Class. Quantum Grav.}, 17:761--782, 2000.

\bibitem{Neilsen00a}
D.W. Neilsen and M.W. Choptuik.
\newblock Ultrarelativistic fluid dynamics.
\newblock {\em Class. Quantum Grav.}, 17:733--759, 2000.

\bibitem{Oliger}
J.~Oliger.
\newblock {\em Numerical {M}ethods for {P}artial {D}ifferential {E}quations}.
\newblock New York: Academic, 1978.

\bibitem{PenroseRindler}
R.~Penrose and W.~Rindler.
\newblock {\em Spinors and space-time 2: {S}pinor and twistor methods in
  space-time geometry}.
\newblock Cambridge {U}niversity {P}ress, 1986.

\bibitem{NR}
W.~H. Press et~al.
\newblock {\em Numerical Recipes in C++: the art of scientific computing}.
\newblock Cambridge University Press, 2nd edition, 2002.

\bibitem{PretoriusPhD}
F.~Pretorius.
\newblock {\em Numerical Simulations of Gravitational Collapse}.
\newblock Ph{D} thesis, University of British Columbia, 2002.

\bibitem{Pretorius05a}
F.~Pretorius.
\newblock {Evolution of Binary Black Hole Spacetimes}.
\newblock {\em Preprint, arXiv::gr-qc}, (0507014), July 2005.

\bibitem{Pretorius05}
F.~Pretorius.
\newblock Numerical relativity using a generalized harmonic decomposition.
\newblock {\em Class. Quantum Grav.}, 22(2):425--451, 2005.

\bibitem{Pretorius05b}
F.~Pretorius and M.W. Choptuik.
\newblock {Adaptive Mesh Refinement for Coupled Elliptic-Hyperbolic Systems}.
\newblock {\em Preprint, arXiv::gr-qc}, (0508110), August 2005.

\bibitem{ProtterWeinberger}
M.H. Protter and H.F. Weinberger.
\newblock {\em Maximum principles in differential equations}.
\newblock Prentice-Hall, 1967.

\bibitem{Rauch85}
J.~Rauch.
\newblock Symmetric positive systems with boundary characteristic of constant
  multiplicity.
\newblock {\em Trans. Am. Math. Soc.}, 291:167--187, 1996.

\bibitem{ReulaLivRev}
O.~Reula.
\newblock Hyperbolic {M}ethods for {E}instein's {E}quations.
\newblock {\em Living Rev. Relativity}, 1(3), 1998.

\bibitem{ReulaSarbach04}
O.~Reula and O.~Sarbach.
\newblock A model problem for the initial-boundary value formulation of
  {E}instein's field equations.
\newblock {\em Preprint, arXiv::gr-qc}, (0409027), 2004.

\bibitem{RinneEssay}
O.~Rinne.
\newblock Numerical {S}tudies of {A}xisymmetric {S}pacetimes.
\newblock Smith-{K}night {P}rize {E}ssay, {U}niversity of {C}ambridge, January
  2004.

\bibitem{RinneStewart05}
O.~Rinne and J.~M. Stewart.
\newblock A strongly hyperbolic and regular reduction of {E}instein's equations
  for axisymmetric spacetimes.
\newblock {\em Class. Quantum Grav.}, 22(6):1143--1166, 2005.

\bibitem{Sachs61a}
R.~Sachs.
\newblock {Gravitational Waves in General Relatvity: VIII. Waves in
  asymptotically flat space-time}.
\newblock {\em Proc. R. Soc. London A}, page 103, 1961.

\bibitem{SarbachTiglio05}
O.~Sarbach and M.~Tiglio.
\newblock Boundary conditions for {E}instein's field equations: {A}nalytical
  and numerical analysis.
\newblock {\em Preprint, arXiv::gr-qc}, (0412115), 2004.
\newblock To appear in JHDE.

\bibitem{Scheel02}
M.A. Scheel, L.E. Kidder, L.~Lindblom, H.P. Pfeiffer, and S.A. Teukolsky.
\newblock Toward stable 3{D} evolutions of black-hole spacetimes.
\newblock {\em Phys. Rev. D}, 66(124005), 2002.

\bibitem{SchiesserMOL}
W.E. Schiesser.
\newblock {\em The {N}umerical {M}ethod of {L}ines: {I}ntegration of {P}artial
  {D}ifferential {E}quations}.
\newblock Academic Press, New York, 1991.

\bibitem{Secchi96}
P.~Secchi.
\newblock Well-{P}osedness of {C}haracteristic {S}ymmetric {H}yperbolic
  {S}ystems.
\newblock {\em Arch. Ration. Mech. Anal.}, 134:155--197, 1996.

\bibitem{Shewchuk94}
J.R. Shewchuk.
\newblock An {I}ntroduction to the {C}onjugate {G}radient {M}ethod {W}ithout
  the {A}gonizing {P}ain.
\newblock unpublished draft, School of {C}omputer {S}cience, {C}arnegie
  {M}ellon {U}niversity, 1994.

\bibitem{Shu97}
C.-W. Shu.
\newblock Essentially non-oscillatory and weighted essentially non-oscillatory
  schemes for hyperbolic conservation laws.
\newblock Technical Report 97-65, ICASE, 1997.

\bibitem{Shu88}
C.-W. Shu and S.~Osher.
\newblock Efficient implementation of essentially non-oscillatory
  shock-capturing schemes.
\newblock {\em J. Comp. Phys.}, 77:439--471, 1988.

\bibitem{StewartAdvGR}
J.M. Stewart.
\newblock {\em Advanced general relativity}.
\newblock Cambridge {U}niversity {P}ress, 1991.

\bibitem{Stewart98}
J.M. Stewart.
\newblock The {C}auchy problem and the initial boundary value problem in
  numerical relativity.
\newblock {\em Class. Quantum Grav.}, 15:2865--2889, 1998.

\bibitem{StoerBulirsch}
J.~Stoer and R.~Bulirsch.
\newblock {\em Introduction to {N}umerical {A}nalysis}.
\newblock Springer, second edition, 2002.

\bibitem{TabenskyTaub73}
R.~Tabensky and A.H. Taub.
\newblock Plane-symmetric self-gravitating fluids with pressure equal to energy
  density.
\newblock {\em Comm. Math. Phys.}, 29:61--77, 1973.

\bibitem{Teukolsky82}
S.A. Teukolsky.
\newblock Linearized quadrupole waves in general relativity and the motion of
  test particles.
\newblock {\em Phys. Rev. D}, 26(4):745--750, 1982.

\bibitem{Teukolsky00}
S.A. Teukolsky.
\newblock {Stability of the iterated Crank-Nicholson method in numerical
  relativity}.
\newblock {\em Phys. Rev. D}, 61(087501), 2000.

\bibitem{Toro}
E.F. Toro.
\newblock {\em Riemann {S}olvers and {N}umerical {M}ethods for {F}luid
  {D}ynamics}.
\newblock Springer, 1997.

\bibitem{ToroMUSTA2}
E.F. Toro and V.A. Titarev.
\newblock {MUSTA} schemes for multi-dimensional hyperbolic systems: analysis
  and improvements.
\newblock {\em Isaac Newton Institute preprint}, (NI04032-NPA), 2004.

\bibitem{ToroMUSTA1}
E.F. Toro and V.A. Titarev.
\newblock {MUSTA} {S}chemes for {S}ystems of {C}onservation {L}aws.
\newblock {\em Isaac Newton Institute preprint}, (NI04033-NPA), 2004.

\bibitem{Vorst92}
H.A. van~der Vorst.
\newblock {BI-CGSTAB}: a fast and smoothly converging variant of {BI-CG} for
  the solution of nonsymmetric linear systems.
\newblock {\em {SIAM} Journal on Scientific and Statistical Computing},
  13:631--644, 1992.

\bibitem{SCOPE}
J.A. van Hulzen.
\newblock {\em {SCOPE} 1.5 -- {A} {S}ource-{C}ode {O}ptimization Packag{E} for
  {REDUCE} 3.5}.
\newblock University of Twente, Department of Computer Science.

\bibitem{Wald91}
R.M. Wald and V.~Iyer.
\newblock {Trapped surfaces in the Schwarzschild geometry and cosmic
  censorship}.
\newblock {\em Phys. Rev. D}, 44:R3719, 1991.

\bibitem{WeberWheeler57}
J.~Weber and J.A. Wheeler.
\newblock Reality of the cylindrical gravitational waves of {E}instein and
  {R}osen.
\newblock {\em Rev. Mod. Phys.}, 29(3):509--515, 1957.

\bibitem{Wesseling}
P.~Wesseling.
\newblock {\em An Introduction to Multigrid Methods}.
\newblock Pure and Applied Mathematics. Wiley, 1992.

\bibitem{Wilson79}
J.R. Wilson.
\newblock A numerical method for relativistic hydrodynamics.
\newblock In L.~Smarr, editor, {\em Sources of {G}ravitational {R}adiation},
  pages 423--445. Cambridge {U}niversity {P}ress, 1979.

\bibitem{YonedaShinkai01}
G.~Yoneda and H.~Shinkai.
\newblock Constraint propagation in the family of {ADM} systems.
\newblock {\em Phys. Rev. D}, 63(124019), 2001.

\bibitem{York79}
J.M. York.
\newblock Kinematics and dynamics of general relativity.
\newblock In L.~Smarr, editor, {\em Sources of Gravitational Radiation},
  page~83. Cambridge University Press, 1979.

\bibitem{York89}
J.M. York.
\newblock Initial data for collisions of black holes and other gravitational
  miscellany.
\newblock In C.~Evans, L.~Finn, and D.~Hobill, editors, {\em Frontiers in
  Numerical Relativity}, pages 89--109. Cambridge University Press, 1989.

\end{thebibliography}

\end{document}